\documentclass[acmtog]{./ref/acmart}
\AtBeginDocument{%
  }

\usepackage{tikz}
\usepackage[export]{adjustbox}

\usepackage{colortbl}
\usepackage{siunitx}
\DeclareMathOperator*{\argmin}{argmin}
\usepackage{xcolor}
\usepackage{array}
\usepackage{stfloats}
\usepackage{multirow}
\usepackage{url}
\usepackage{wrapfig}
\usepackage{algorithm}
\usepackage{natbib}
\setcitestyle{authoryear,open={[},close={]}} 
\usepackage[noend]{algpseudocode}
\usepackage{sidecap}
\usepackage{tabularx}
\usepackage{gensymb}
\usepackage{mathtools}
\usepackage{ccicons}

\algdef{SE}[DOWHILE]{Do}{doWhile}{\algorithmicdo}[1]{\algorithmicwhile\ #1}
\algnewcommand\algorithmicforeach{\textbf{for each}}
\algdef{S}[FOR]{ForEach}[1]{\algorithmicforeach\ #1\ \algorithmicdo}

\usepackage{prettyref}
\newrefformat{fig}{Fig.~\ref{#1}}
\newrefformat{par}{Section~\ref{#1}}
\newrefformat{appen}{Appendix~\ref{#1}}
\newrefformat{sec}{Section~\ref{#1}}
\newrefformat{sub}{Section~\ref{#1}}
\newrefformat{table}{Table~\ref{#1}}
\newrefformat{tab}{Table~\ref{#1}}
\newrefformat{ass}{Assumption~\ref{#1}}
\newrefformat{alg}{Algorithm~\ref{#1}}
\newrefformat{def}{Definition~\ref{#1}}
\newrefformat{thm}{Theorem~\ref{#1}}
\newrefformat{cor}{Corollary~\ref{#1}}
\newrefformat{lem}{Lemma~\ref{#1}}
\newrefformat{step}{Step~\ref{#1}}
\newrefformat{ln}{Line~\ref{#1}}
\newrefformat{rem}{Remark~\ref{#1}}
\newrefformat{eq}{Equation~\ref{#1}}
\newrefformat{pb}{Problem~\ref{#1}}
\newrefformat{it}{Item~\ref{#1}}
\newrefformat{te}{Term~\ref{#1}}
\def\Eqref Eq:#1:{\eqref{eq:#1}}
\newrefformat{Eq}{Equation~\Eqref#1:}
\newrefformat{App}{Appendix~\ref#1:}

\newcommand\restr[2]{{\left.\kern-\nulldelimiterspace{}#1\right|_{#2}}}

\usepackage{prettyref}
\newrefformat{fig}{Fig.~\ref{#1}}
\newrefformat{par}{Section~\ref{#1}}
\newrefformat{appen}{Appendix~\ref{#1}}
\newrefformat{sec}{Section~\ref{#1}}
\newrefformat{sub}{Section~\ref{#1}}
\newrefformat{table}{Table~\ref{#1}}
\newrefformat{tab}{Table~\ref{#1}}
\newrefformat{ass}{Assumption~\ref{#1}}
\newrefformat{alg}{Algorithm~\ref{#1}}
\newrefformat{def}{Definition~\ref{#1}}
\newrefformat{thm}{Theorem~\ref{#1}}
\newrefformat{cor}{Corollary~\ref{#1}}
\newrefformat{lem}{Lemma~\ref{#1}}
\newrefformat{step}{Step~\ref{#1}}
\newrefformat{ln}{Line~\ref{#1}}
\newrefformat{rem}{Remark~\ref{#1}}
\newrefformat{eq}{Equation~\ref{#1}}
\newrefformat{pb}{Problem~\ref{#1}}
\newrefformat{it}{Item~\ref{#1}}
\newrefformat{te}{Term~\ref{#1}}
\def\Eqref Eq:#1:{\eqref{eq:#1}}
\newrefformat{Eq}{Equation~\Eqref#1:}
\newrefformat{App}{Appendix~\ref#1:}

\begin{document}
\title{Single Edge Collapse Quad-Dominant Mesh Reduction}

\author{Julian Knodt}
\email{julianknodt@gmail.com}
\orcid{1234-5678-9012}
\affiliation{%
  \institution{LightSpeed Studios}
  \city{Bellevue}
  \state{Washington}
  \country{USA}
  \postcode{98004}
}

\begin{abstract}
Mesh reduction using quadric error metrics is the industry standard for producing level-of-detail (LOD) geometry for meshes. Although industry tools produce visually excellent LODs, mesh topology is often ruined during decimation. This is because tools focus on triangle simplification and preserving rendered appearance, whereas artists often produce quad dominant meshes with clean edge topology. Artist created manual LODs preserve both appearance and quad topology. Furthermore, most existing tools for quad decimation only accept pure quad meshes and cannot handle any triangles. The gap between quad and triangular mesh decimation is because they are built on fundamentally different operations, triangle simplification uses single edge collapses, whereas quad decimation requires that entire sets of edges be collapsed atomically. In this work, we demonstrate that single edge collapse can be used to preserve most input quads without degrading geometric quality. Single edge collapse quad preservation is made possible by introducing dihedral-angle weighted quadrics for every edges, allowing optimization to evenly space edges while preserving features. It is further enabled by explicitly ordering edge collapses with nearly equivalent quadric error that preserves quad topology. In addition to quad preservation, we demonstrate that by introducing weights for quadrics on certain edges, our framework can be used to preserve symmetry and joint influences. To demonstrate our approach is suitable for skinned mesh decimation (a key use case of quad meshes), we show that QEM with attributes can preserve joint influences better than prior work. We implement and test our approach on 67 static and 19 animated meshes from Sketchfab. On both static and animated meshes, our approach consistently outperforms prior work with lower Chamfer and Hausdorff distance, while preserving more quad topology.
\end{abstract}

\setcopyright{acmlicensed}
\acmJournal{TOG}
\acmYear{2024} \acmVolume{1} \acmNumber{1} \acmArticle{1} \acmMonth{1} \acmPrice{15.00}\acmDOI{}
\acmSubmissionID{104}

\ccsdesc[500]{Computing methodologies~Computer Graphics}

\keywords{Mesh Reduction, Quad Mesh Reduction, Skinned Mesh Reduction}

\maketitle

\begin{figure*}[!ht]
    \centering
    \begin{tabular}{c c c c}
        Input & QEM & MeshLab~\citep{meshlab} & Ours \\
        
        \put(-0.1, 0.08\linewidth){\frame{\includegraphics[width=0.11\linewidth]{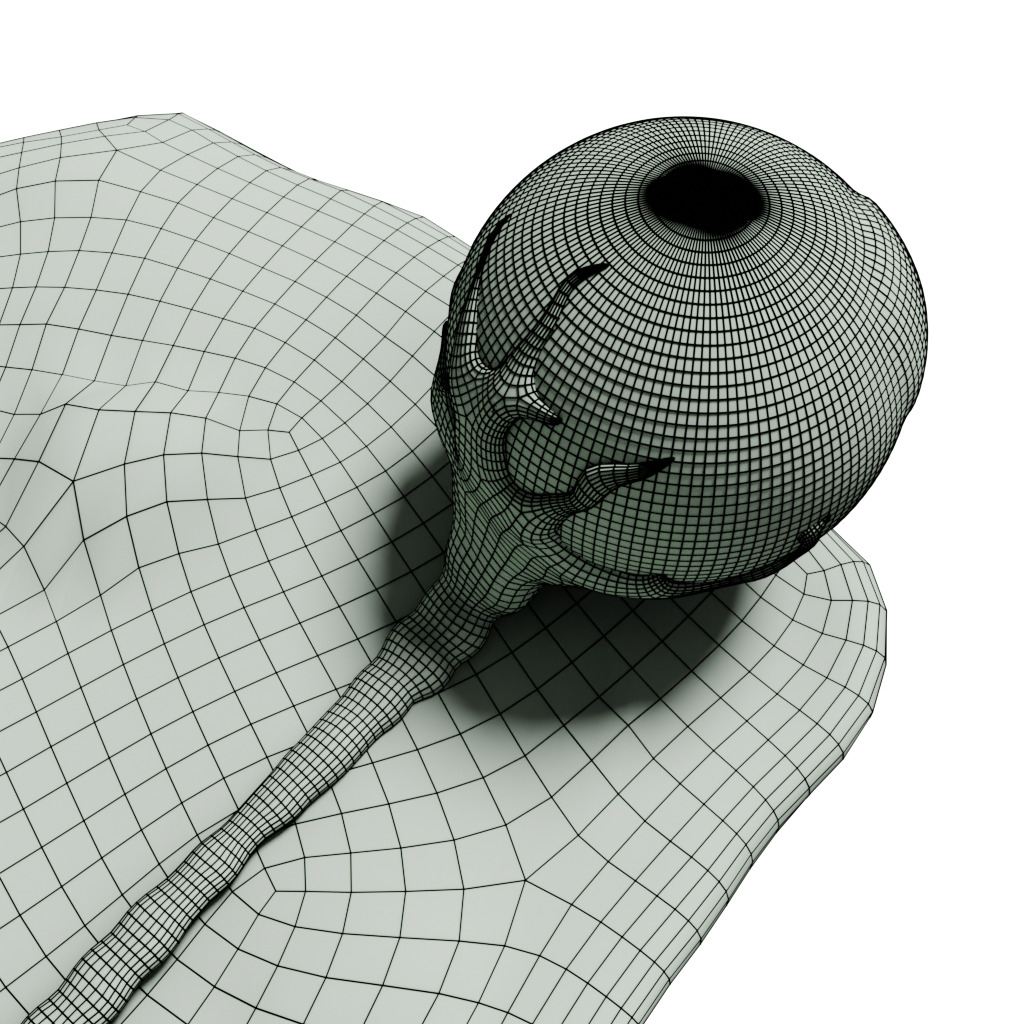}}}
        \includegraphics[width=0.24\linewidth]{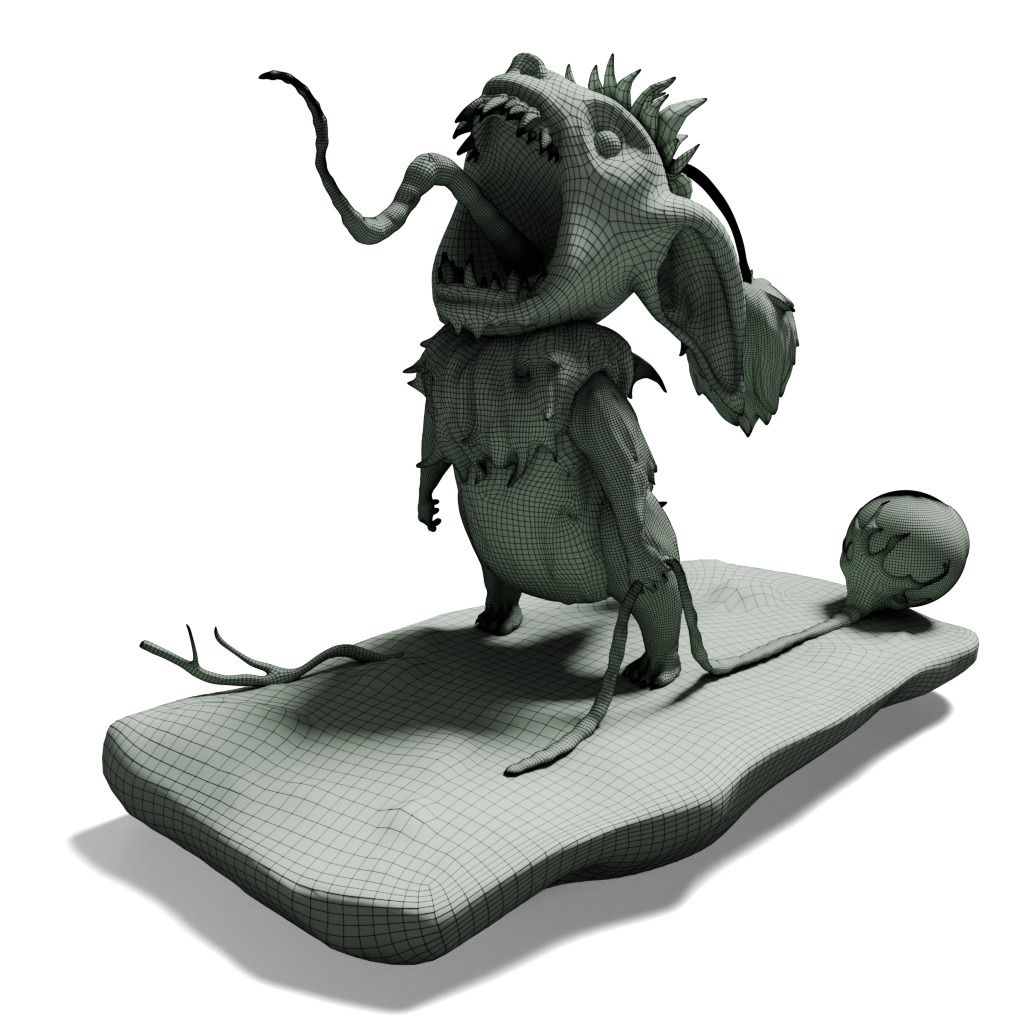}
        &
        \put(-0.1, 0.08\linewidth){\frame{\includegraphics[width=0.11\linewidth]{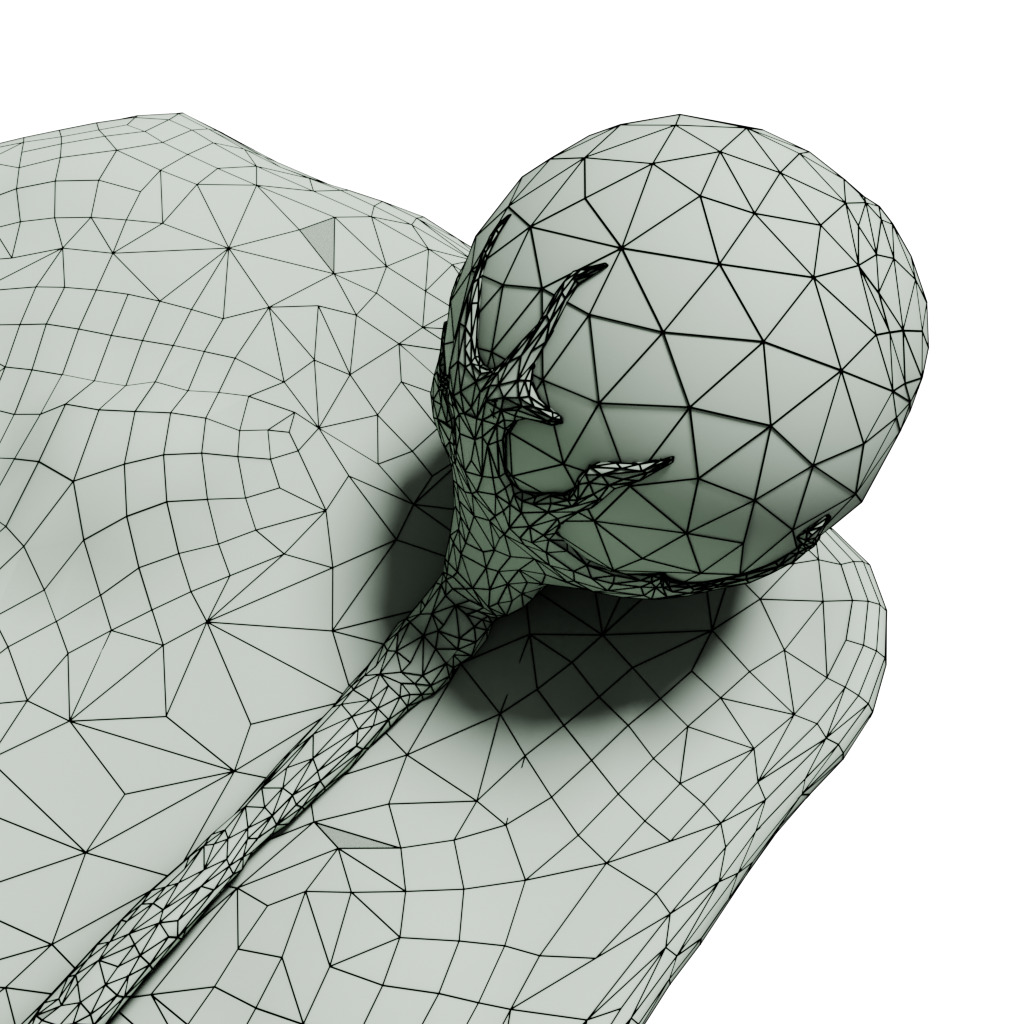}}}
        \includegraphics[width=0.24\linewidth]{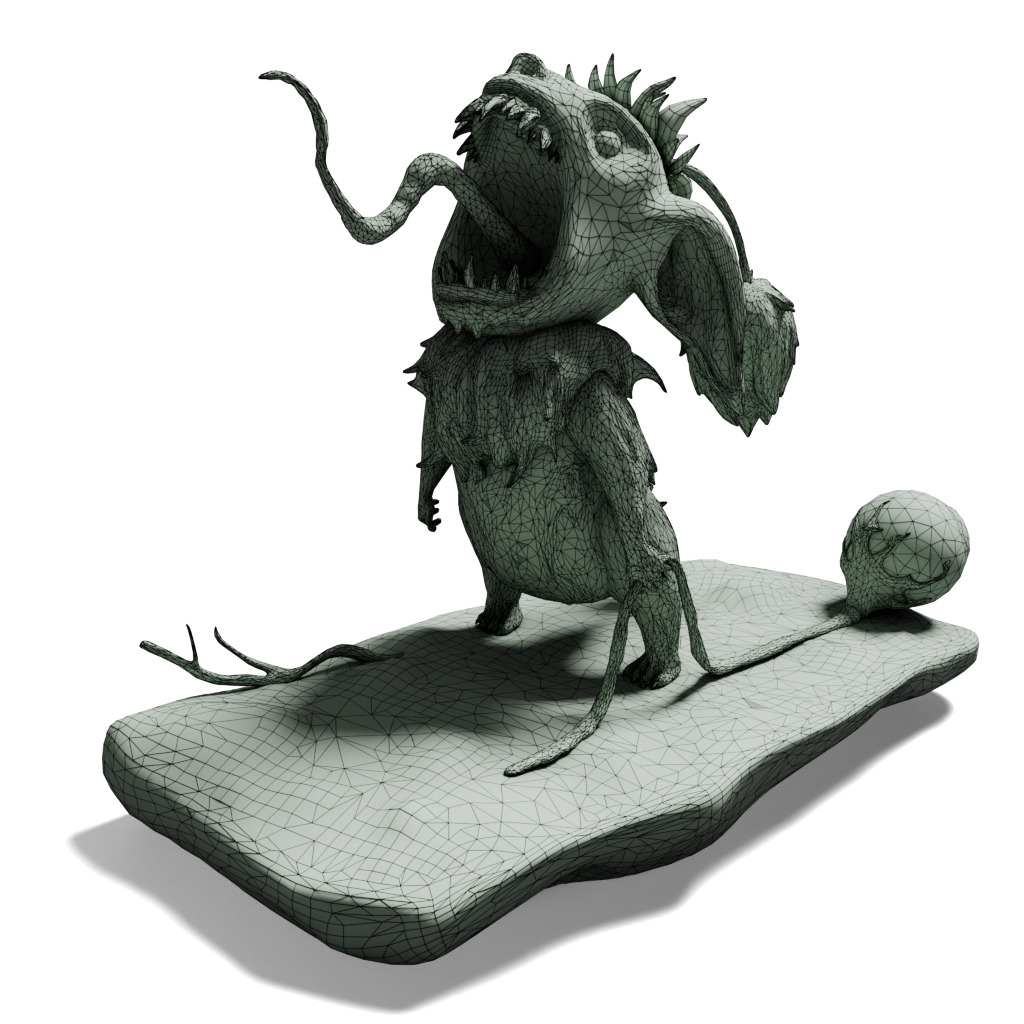}
        &
        \put(-0.1, 0.08\linewidth){\frame{\includegraphics[width=0.11\linewidth]{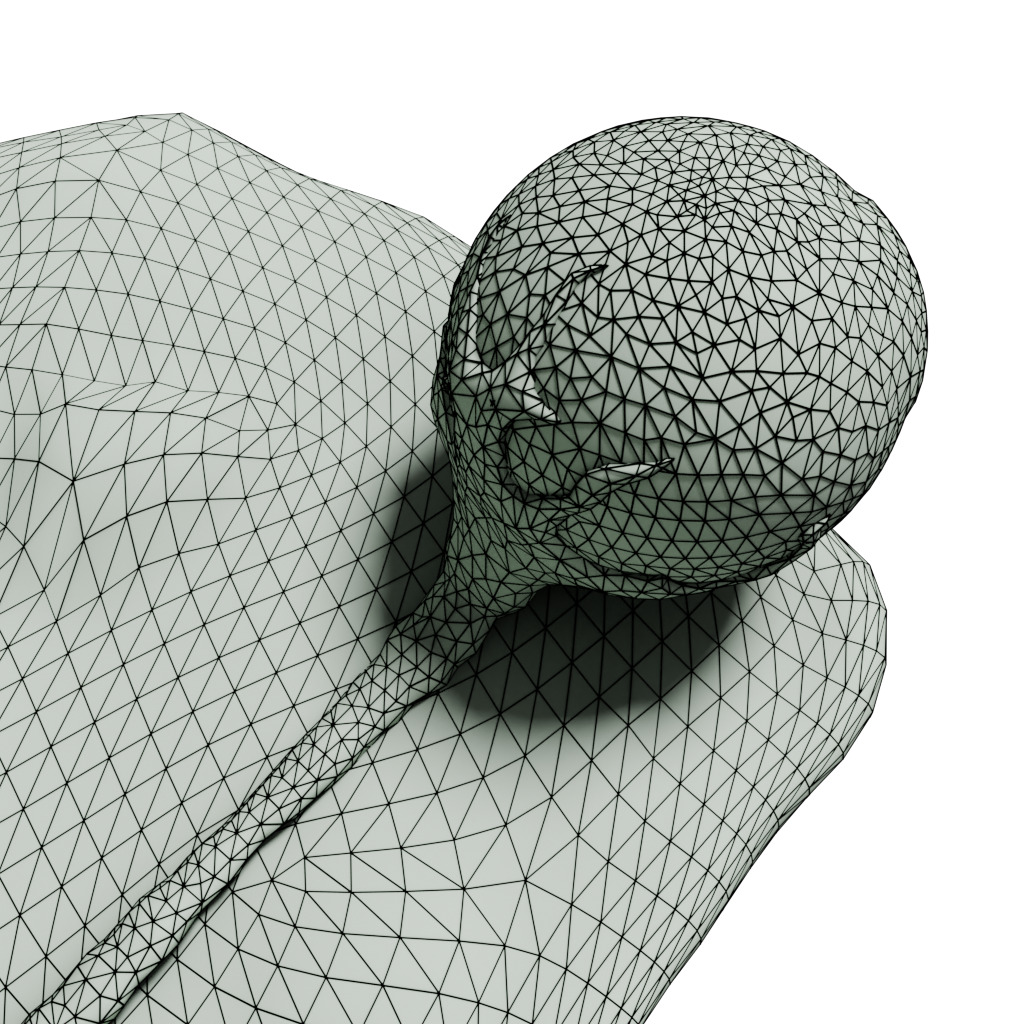}}}
        \includegraphics[width=0.24\linewidth]{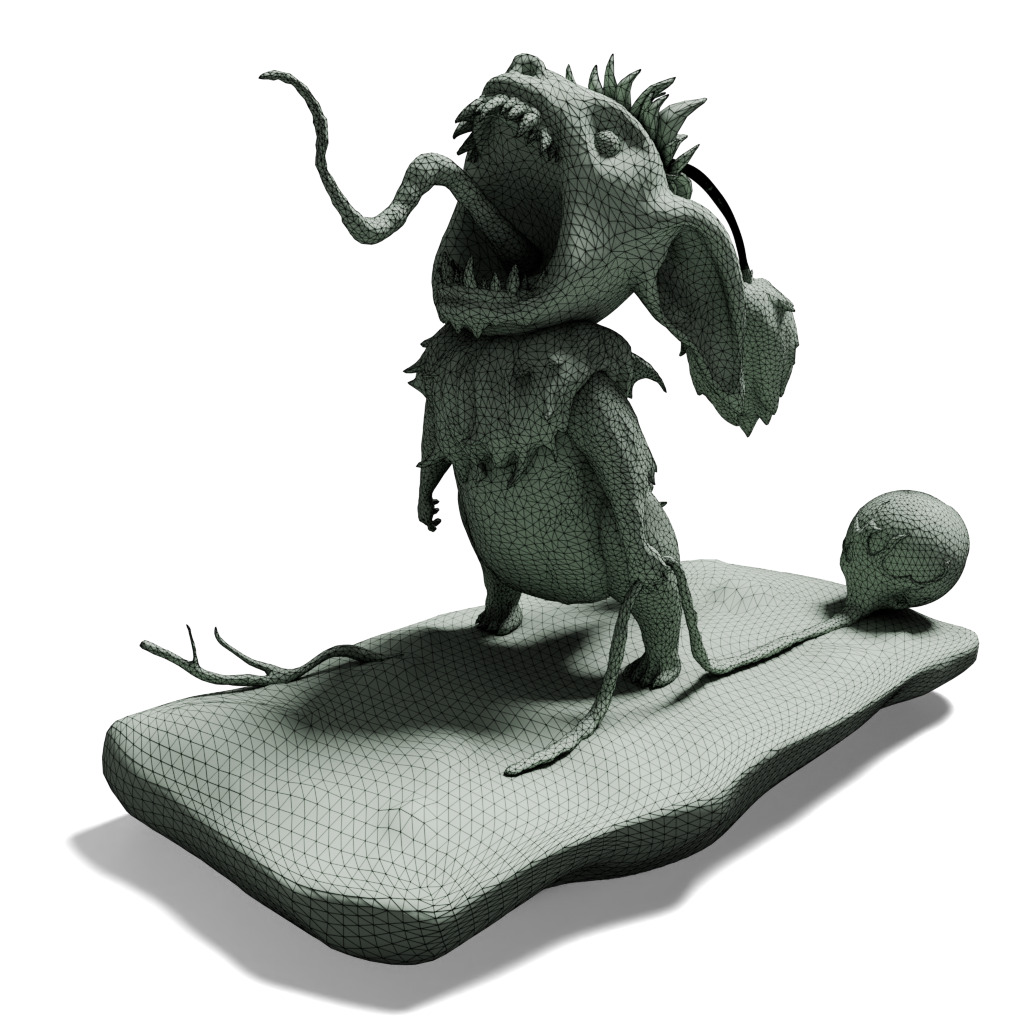}
        &
        \put(-0.1, 0.08\linewidth){\frame{\includegraphics[width=0.11\linewidth]{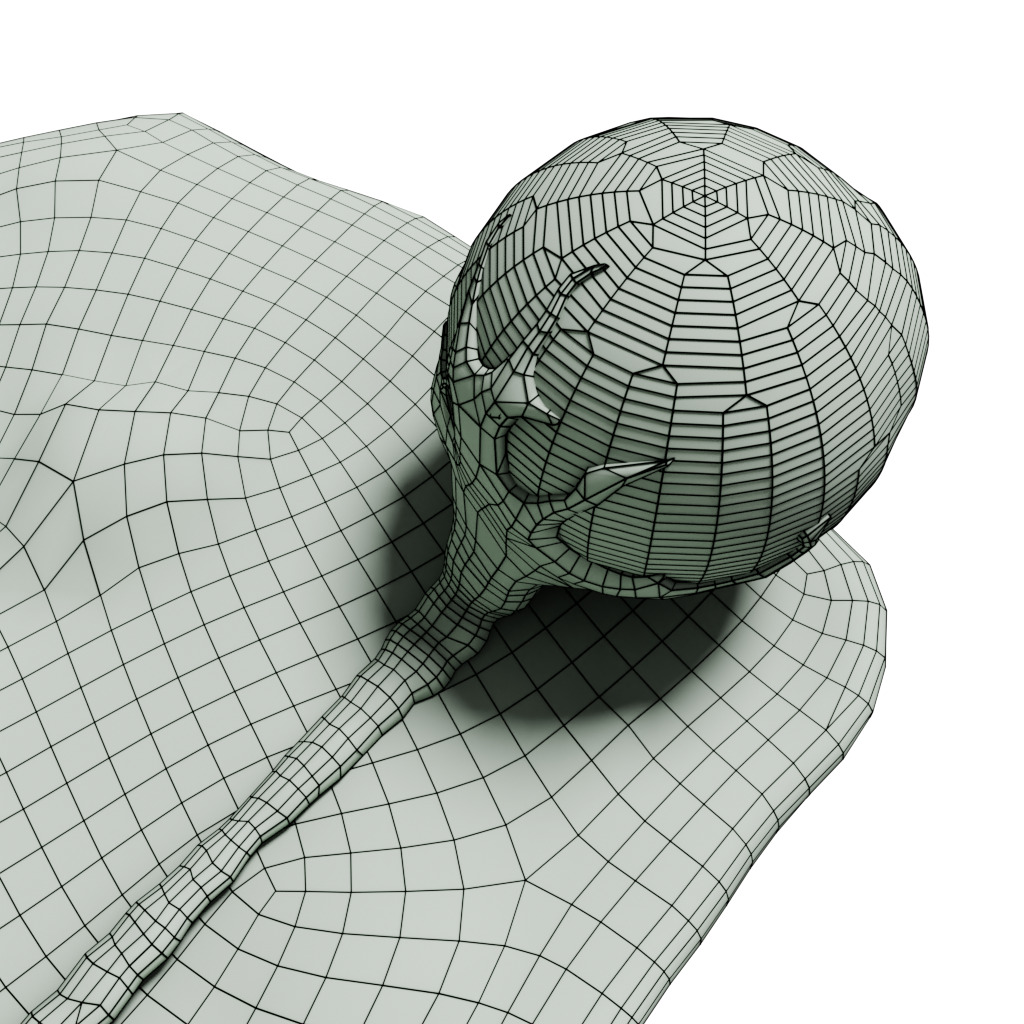}}}
        \includegraphics[width=0.24\linewidth]{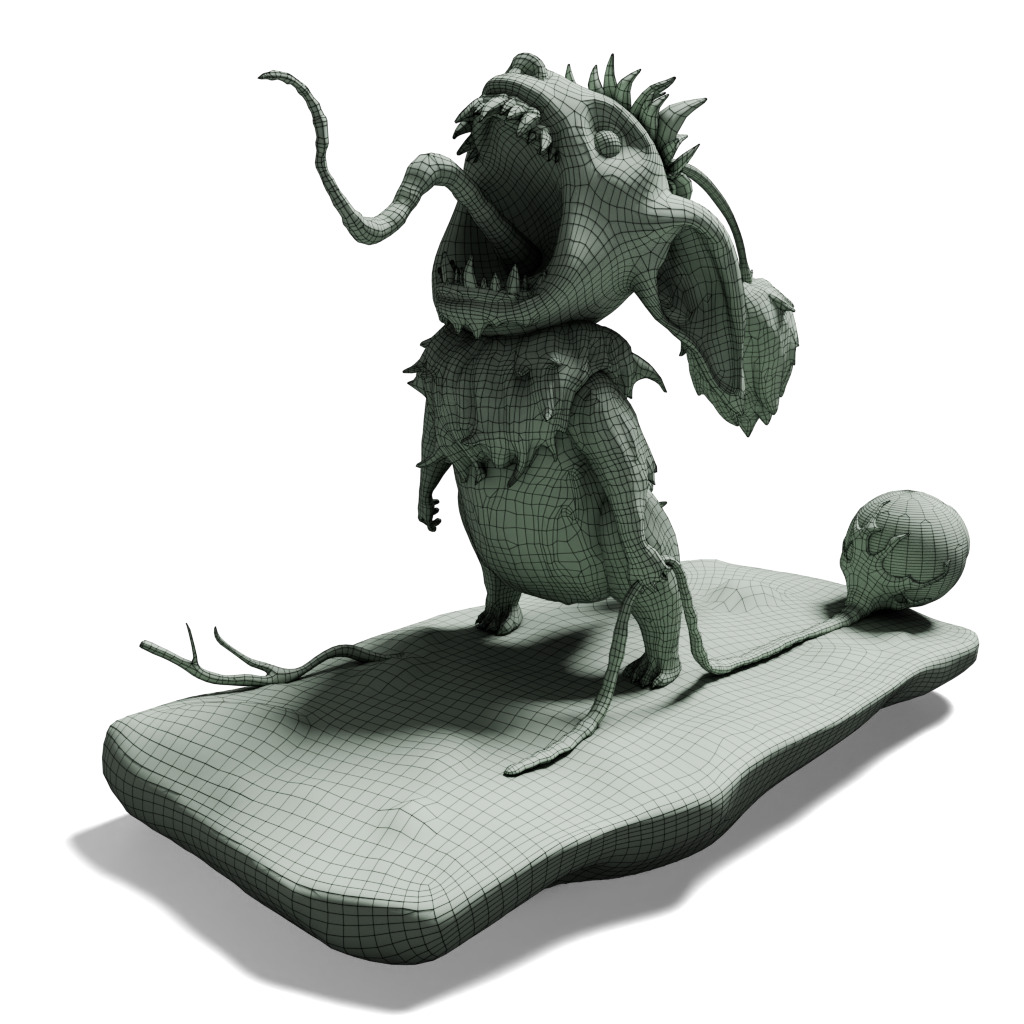} \\
        $\text{Quad}(\square)^\uparrow/\text{Tri}(\triangle)^\downarrow$: 157140/409 & 17892/44753 (80537 total) & 0/78684 (78684 total) & \textbf{38467/1751} (78685 total) \\
        
        Chamfer$^\downarrow$/Hausdorff$^\downarrow$ & \num{1.834e-4}/\num{2.757e-3} & \num{1.538e-4}/\num{2.521e-3} & \textbf{\num{1.430e-4}/\num{2.141e-3}} \\
    \end{tabular}
    \caption{Our approach is able to better preserve input topology and quads than QEM~\citep{qem, qem_hoppe} for hybrid triangle/quad meshes, with comparable geometric similarity to the input mesh. In the zoomed inset, the input sphere becomes triangulated in QEM and the floor has many high valence vertices, whereas our approach preserves quads on the sphere and floor. We also compare to MeshLab~\citep{meshlab}, which requires that the input mesh be triangulated before decimation, making it unsuitable for quad preservation. \ccby Humesh Dilchund.}
    \label{fig:teaser}
    \Description{Teaser figure of our approach. It shows the original mesh, a creature with its tongue sticking out, out on the left. In the middle is QEM, which contains a significant number of triangles in its output. On the right is our approach, which consists mostly of quads. Each figure has a zoomed inset showing the wireframe of a ball the creature is dragging around.}
\end{figure*}

\section{Introduction}

Games rely on mesh simplification to generate multiple levels of detail (LODs) for meshes, so as meshes get further away from the viewer they are substituted with cheaper to render approximations of the original mesh. Mesh simplification reduces the number of triangles in a mesh, and industry implementations are built on Quadric Mesh Reduction~\citep{qem, qem_hoppe, qem_uv_texture}, since it can preserve geometry, UV mappings, vertex normals, skinning weights, and other attributes. Quadric mesh reduction makes it easy to drop automatically generated LODs into rendering pipelines with little manual modification. There are a plethora of industry tools for producing simplified meshes, including Simplygon~\citep{simplygon}, InstaLOD~\citep{Instalod}, and tooling within DCC tools such as Blender~\citep{Blender} and Unreal Engine~\citep{UE5}. Unreal Engine in specific contains ``Nanite'', a system built on top of QEM, which generates LODs for different patches of a mesh and switches between them at runtime. Due to Unreal Engine's ubiquity, Nanite is used by many games as a drop-in LOD system. These tools are built-to-purpose and are able to produce \textit{visually} similar meshes but often operate purely on triangle meshes, whereas many artist created meshes are quad-dominant meshes. In addition to visual similarity, artists often want decimation to preserve edge topology and edge loops, as they affect the animation and deformation of the mesh. With that said, current mesh reduction usually destroys the topology of the output mesh, making it uneven and irregular, independent of the input topology quality. Tools that do preserve quad topology, such as quad reduction in Simplygon often operate entirely on quad meshes, making it brittle to use in practice. In contrast, recent proprietary tooling in Electronic Art's Frostbite Engine~\citep{FrostbiteQuadReduce} supports mixed input quad/triangle meshes, while preserving quad topology and symmetry, simplifying artists' workflow for manual modification. While EA's tool is not public, it was described at a GDC talk to be based on Quadrilateral Mesh Simplification~\citep{quad_mesh_simp}, with additional modifications that allow partial quad-chords to be decimated, where a quad-chord is a set of conjoined quads such that only adjacent quads share a vertex (See Fig.~\ref{fig:quad-diagram}). One key insight from EA's tool is that artists do not need all quads to be preserved, but ask that \textit{most} will be preserved, relaxing the requirement that all quads be kept. While the presentation of Frostbite's mesh reduction shows visually pleasing results, from their description it is a complex system with many heuristics and is not easily replicable. Our motivation in this work is driven by asking if there is a simpler approach to preserve topology and quads in mesh reduction.

This leads us to question why triangle mesh reduction currently cannot preserve quad topology, and we identify three key reasons that quads are not preserved with existing mesh reduction techniques. First, the tangent space for each quadric is ill-conditioned, such that vertices output from edge-collapse can lie anywhere in a face's plane and cannot be distinguished by quadric error. Second, quad meshes and quad-chords contain edges that have nearly equivalent quadric error. Due to floating point error when computing quadrics or negligible but real geometric differences, these edges are treated as geometrically different, which is manifested in the ordering of edge collapses. While the ordering over these equivalent sets of edges is deterministic, it is essentially random, preventing topological preservation for regions where edges have near equivalence. Third, we observe that memoryless simplification discards the constraints of the input mesh, making it difficult to preserve topology.

To remedy these shortcomings, we introduce a number of modifications to quadric mesh reduction. To fix the ill-conditioned tangent space of each face, we introduce a per-edge dihedral-angle weighted quadric orthogonal to the adjacent faces' normals. This quadric leads to an even distribution of vertex positions in the tangent space of each face and preserves sharp geometric features. To resolve ambiguity between approximately equivalent edges and to preserve clean topology, we treat edges as having equivalent error if their quadric errors are within some distance, and within sets of equivalent edges we introduce a stateful ordering based on the recency of collapsed opposing quad edges. Finally, we do not perform memoryless simplification, allowing us to preserve all input constraints, but modify the quadric error to represent the error from the current state, rather than the error from initialization. This change serves the same purpose as memoryless simplification, and is equally performant both in runtime and efficacy. An example of the results from our approach are shown in Fig.~\ref{fig:teaser}.

Since our approach adds per-edge quadrics, it naturally lends itself to introducing importance weights along each edge, also known as weight painting. To demonstrate how importance weights can be used, we test using per-edge symmetry and joint distance to preserve certain edges more. This stands in contrast to the original QEM, which has no explicit tools to preserve joint regions used in animation, and is unable to preserve edge-loops on key axes of symmetry. Our approach can more easily preserve symmetry and joint regions by measuring a per-edge importance, and increasing the weights of the corresponding quadrics.

To validate that our approach is suitable for joint preservation, it is necessary to show our approach can preserve geometric similarity to the original mesh during skinned animation. While reviewing prior work, we found that while all previous approaches had acknowledged the existence of ~\citep{qem_hoppe}, none had tested its ability to preserve joint influences. This is due to the assumption that joint influences are a special attribute which affects geometry unlike other attributes. We test this assumption and directly use ~\citep{qem_hoppe} to preserve joint influences, and find that it outperforms prior work on articulated mesh simplification, without needing animations or poses.

To demonstrate the proposed improvements, we test our approach on 67 static mixed quad/tri meshes and 19 animated triangle meshes from Sketchfab~\citep{sketchfab}. Our approach is able to preserve clean quad topology on the tested meshes, while also maintaining geometric fidelity. To measure the geometric quality of the simplified, we measure the Chamfer and Hausdorff distances from the original mesh. To measure the ability to preserve topology, we compare the difference the ratio of the outputs quads to the total number of triangles to the ratio of input quads to the total number of triangles. Our approach has consistently higher or similar geometric quality, while at the same time is able to preserve a significantly higher fraction of input quads.

To summarize, the contributions of this work are as follows:
\begin{enumerate}
    \item An algorithm that adapts single edge-collapse for quad-dominant mesh reduction while still preserving geometric quality. This includes adding per-edge quadrics, revisiting the equation for quadric error, and specially ordering edges with near equivalent quadric error. This approach is validated by thorough testing on a diverse dataset, where it preserves significantly more quads while maintaining high geometric fidelity.
    \item An extension of per-edge quadrics to demonstrate that our approach supports soft-symmetry preservation along symmetric axes, and can improve joint preservation by increasing weights near key joints.
    \item A test of \citep{qem_hoppe} to show it is suitable for the preservation of joint influences on skeletal meshes.
\end{enumerate}
\section{Related Work}

\paragraph{Static/Animated Triangular Quadric Mesh Reduction}
A large body of work has been devoted to gemoetry processing using quadrics, starting from the original quadric mesh reduction algorithm~\citep{qem}. Quadric mesh reduction was then extended to preserve attributes~\citep{qem_hoppe, qem_uv_texture}, where ``attributes'' are arbitrary variables defined per vertex, which are linearly interpolated using a functional, $\mathbb{R}^3 \to \mathbb{R}$. Quadrics have also been applied to animated mesh simplification~\citep{pose_indep_simp, articulated_mesh_simplification, artic_mesh_smooth_transitions, coarse_view_dependent_lod, temporal_and_spatial_lod}. In \citep{pose_indep_simp}, quadrics are computed at multiple poses of the input mesh, and joint influences are computed as linear interpolations between vertices upon collapse. \citep{articulated_mesh_simplification} takes a different approach, treating joint simplification as an alternating optimization problem over joint weights and vertex position. Outside of articulated mesh simplification, quadrics have been used to generate other simplified representations of meshes, such as sphere meshes~\citep{sqem, anim_sqem}, remeshed surfaces~\citep{variational_quadrics}, simplified Bezier splines~\citep{locally_integ_error_metric}, or with manipulating mass moments~\citep{mass_moment_manipulation}. There has also been a recent preprint which examines using a virtual edge-collapse~\citep{simplifying_wild_tris} and rebaking output textures to a new UV. While this may work well in some cases, such as decimation below 10\%, for downstream applications that require multiple levels of detail it is not possible to generate new UVs as different levels of detail rely on texture reuse.

\paragraph{Quadrilateral Mesh Reduction}
In parallel to triangle mesh reduction, there has been work in quad-only mesh simplification~\citep{quad_mesh_simp, practical_quad_mesh_simp, adaptive_quad_mesh_simp}, and quad mesh generation~\citep{Jakob2015Instant, SimpleQuadDomain, featureLineQuadRemesh}. Pure quad mesh simplification replaces the edge-collapse operator in triangle mesh reduction with coarser operators such as quad-chord removal, quad diagonal collapses, and singlet/doublet collapses. This makes current algorithms for quad mesh simplification incompatible with approaches for triangle mesh simplification, despite their facades appearing similar. A downside of pure quad-chord based decimation is that quad-chords are often non-local and may wind around the mesh surface, which, when decimated, may lead to coarse results. To mitigate this coarseness, there is also mixed quad simplification, using partial quad-chord simplification~\citep{FrostbiteQuadReduce}. This approach decimates parts of quad-chords, at the cost of introducing triangles. Since this work is not public, it is unclear their exact criteria for splitting quad-chords, and how many triangles are introduced by quad-chord splitting.

\paragraph{Near Equality with Floating Point Numbers}
One key component of this work is comparisons of floating point numbers and their approximation errors, so we briefly describe floating point and their presence in geometric algorithms. Floating point numbers are the most common digital representation of the real numbers. Floating point quantizes the set of real numbers into logarithmically spaced values centered around 0, and all floating point arithmetic maps to this finite set. Since the set is finite and arithmetic operations may not fall exactly on quantized values, arithmetic operations introduce approximation error. Exactly how approximation error is introduced is specified by the IEEE-754 standard~\citep{ieee754}, with clearly defined and deterministic rules, making it possible to build algorithms which are aware of and directly handle the error. The primary case where floating point error becomes a problem and must be handled with care is when floating point numbers are \textit{nearly equivalent}, and their relative ordering matters. This is a common problem in areas such as generating manifold meshes~\citep{fast_and_robust_mesh_arrangements, containment_check} and collision detection~\citep{continuous_collision_detection, efficient_geom_exact_ccd}. Due to difficulty in mitigating near equivalent ambiguity, floating point error motivates work that uses ``rationals''~\citep{tetWild, fTetWild, geogram, levyExactPredicates} or integer computations~\citep{ember}, but these works often trade correctness for being significantly slower. Instead of relying on more precise formats, this work identifies and gets around floating point error, joining a larger body of research that is aware of floating point error~\citep{kahanSum, shewchukAdaptivePrecision, numerics_of_gram-schmidt}.
\section{Method}

Our approach is built on quadrics~\citep{qem, qem_hoppe}, similar to previous mesh reduction approaches, where we define a quadric, $(\mathbb{R}^3,\mathbb{S}^3)\to\mathbb{R}^3\to\mathbb{R}$, as: \begin{equation}
    \text{Quadric}(p, n)(x) = x^\top (nn^\top) x - 2(n^\top p) x^\top n + (n^\top p)^2
\end{equation}
where $p,n\in\mathbb{R}^3, n^\top n = 1$, and $x \in\mathbb{R}^3$ is the position that the quadric is evaluated at, which is usually the optimized vertex's position. In \citep{probabilistic_quadrics}, this is referred to as a ``plane quadric'', but we will simply refer to it as a ``quadric''. We also define shorthand for the quadric of a face $f$ as $\text{Quadric}(f) = \text{Quadric}(v, \text{normal}(f))$, where $v$ is any of the vertices of $f$.

\begin{SCfigure}[][!h]
    \centering
    \caption{\label{fig:quad-diagram}
    A visualization of terminology for structure within quad meshes. Part of a quad chord is shown in \textcolor{teal}{teal}, opposing quad edges are shown in \textcolor{red}{red}, and part of an edge loop is shown in \textcolor{green}{green}.
    }
    \includegraphics[width=0.4\linewidth]{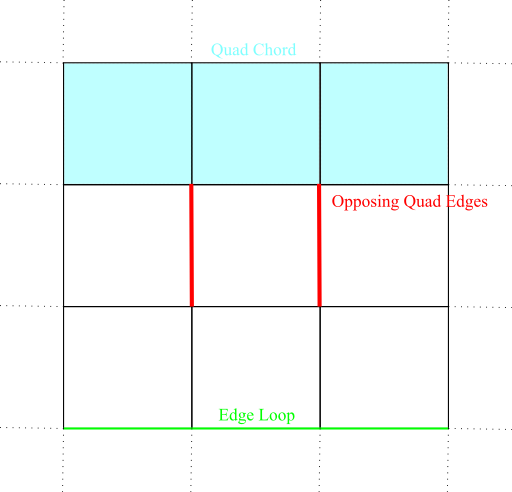}
    \Description{A simple set of quads, color coded to show different quad features. First row is teal, showing it is a quad chord. Two edges which share a quad but share no vertices are marked in red. 3 edges which all share one vertex and are parallel are shown in green.}
\end{SCfigure}

\paragraph{Defining Edge Loops} Within the 3D modeling community, it is common to refer to ``edge loops'', a term not frequently used in academic work. Edge loops are sets of edges where at every vertex with degree 4, the two edges in the edge loop have one edge clockwise and one edge counterclockwise between them. At vertices which do not have degree 4, edge loops are terminated. More loosely, it's a set of connected edges with low amounts of twisting. Under this definition, triangle meshes have almost no edge flow. Edge loops arise naturally from quad meshes with a small number of singularities (vertices which are not degree 4), and are desired because they are more easily animated, and can ``replicate muscle structure of the real object''~\citep{digital_sculpture_techniques}. This provides one key motivation for our work, allowing artists to decimate a high-density quad-dominant mesh and animate the lower density mesh. By preserving quad topology in the decimated mesh, the artist's workflow becomes simpler. A figure showing an edge loop and other terms related to quad topology is shown in Fig.~\ref{fig:quad-diagram}.

\subsection{Per-Edge Weighted Quadric}
One key reason that clean topology and edge loops are not preserved in QEM~\citep{qem} is because there are no constraints in the tangent plane of each face. For faces which are coplanar, there is ambiguity where in the tangent plane the collapsed vertex should be placed. To resolve this ambiguity, we introduce additional quadrics in the tangent space of each face. For each edge $e$ with vertices $e_0, e_1$ and a face $f$ that contains $e$, we introduce an additional quadric orthogonal to each adjacent faces normal: $Q_\text{edge} = \text{Quadric}(e_0,\frac{e_0-e_1}{\lVert e_0 - e_1 \rVert_2} \times \text{normal}(f))$, where $\times$ is the cross product. This quadric is weighed according to the dihedral angle if the edge is manifold, or 1 otherwise: 
\[
w = \left\{\begin{array}{lr}
     \frac{1}{\pi}\arccos(\text{n}(f_0)^\top\text{n}(f_1)) & (e_0,e_1) \in  f_0, f_1  \\
     1 & \text{otherwise (edge non-manifold)} 
\end{array}\right\}
\]
\begin{equation}\label{eq:edge-quadric}
 Q_\text{edge}'= w \lVert e_0 - e_1 \rVert_2 Q_\text{edge}
\end{equation}
where $\text{n}(\cdot)$ is the unit normal of a face, and $\frac{1}{\pi}$ is a normalization constant. In our implementation, we additionally constrain $w$ to $w = \max(w, \num{1e-2})$ to prevent degenerate quadrics.

By introducing weighted quadrics on each edge, our approach can be used to weigh edges by other attributes. We explore increasing weights on edges by a measure of mesh symmetry across each edge plane, and the distance between joint influences on the end of each edge.

\begin{figure}[h!]
    \centering
    \begin{tabular}{c c c}
        Symmetry Matching & Soft Symmetry & No Symmetry \\
        \includegraphics[width=0.3\linewidth]{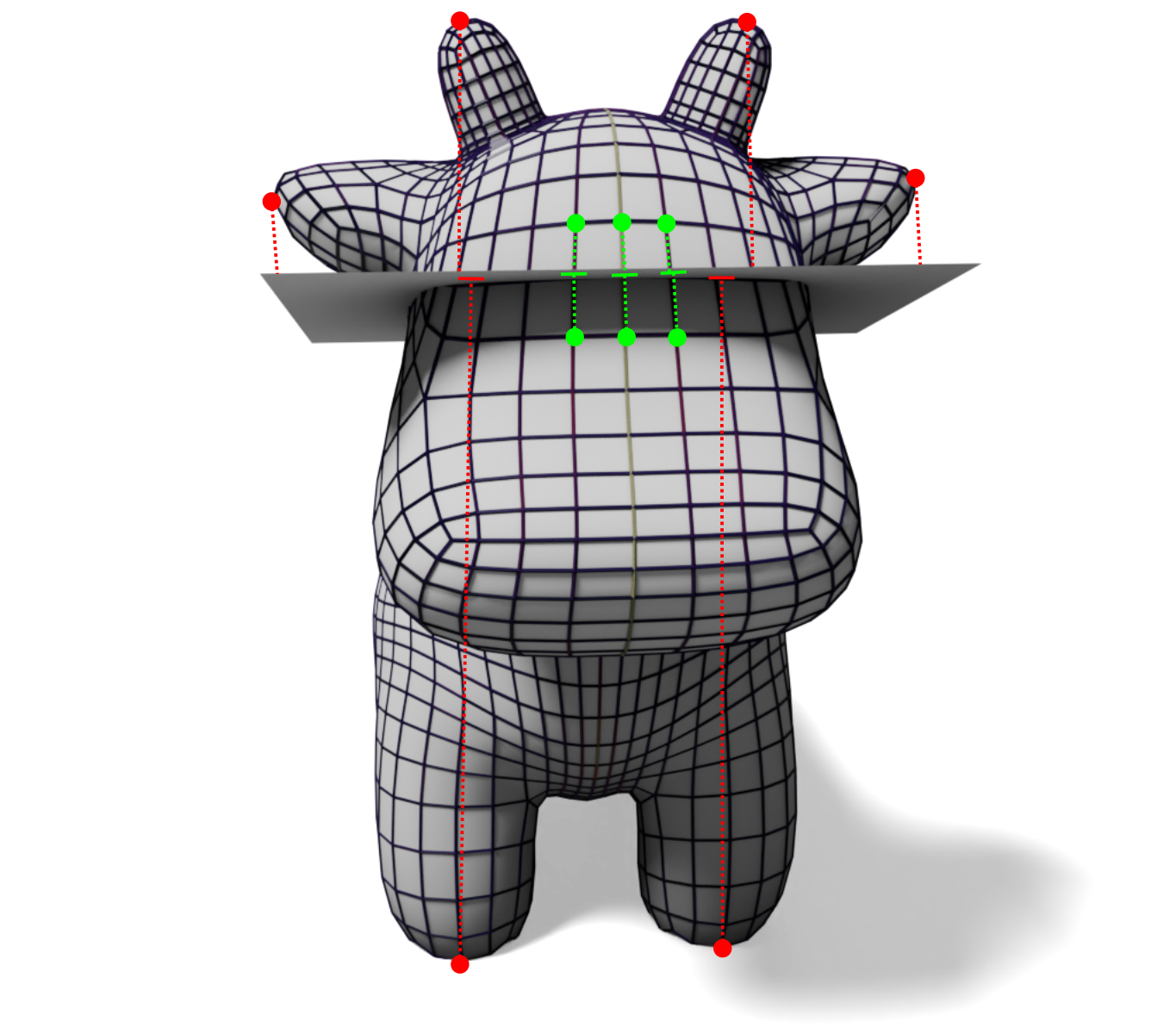} &
        \includegraphics[width=0.3\linewidth]{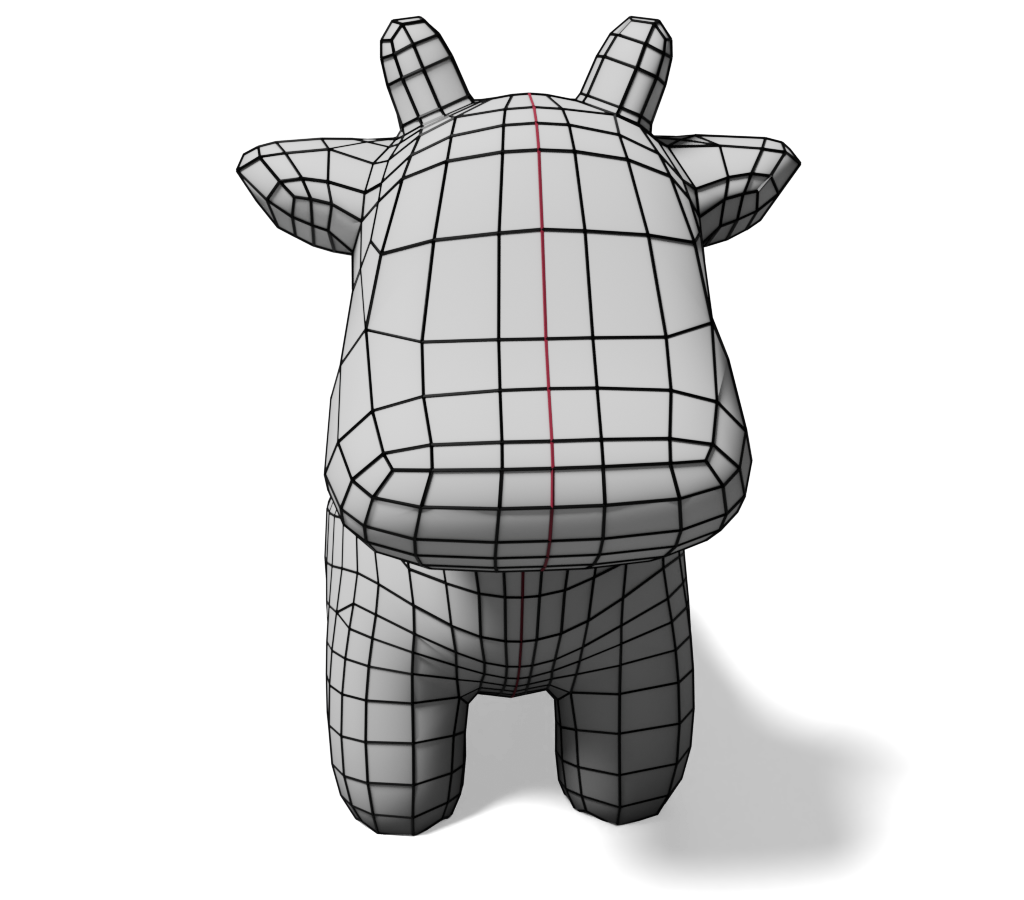} &
        \includegraphics[width=0.3\linewidth]{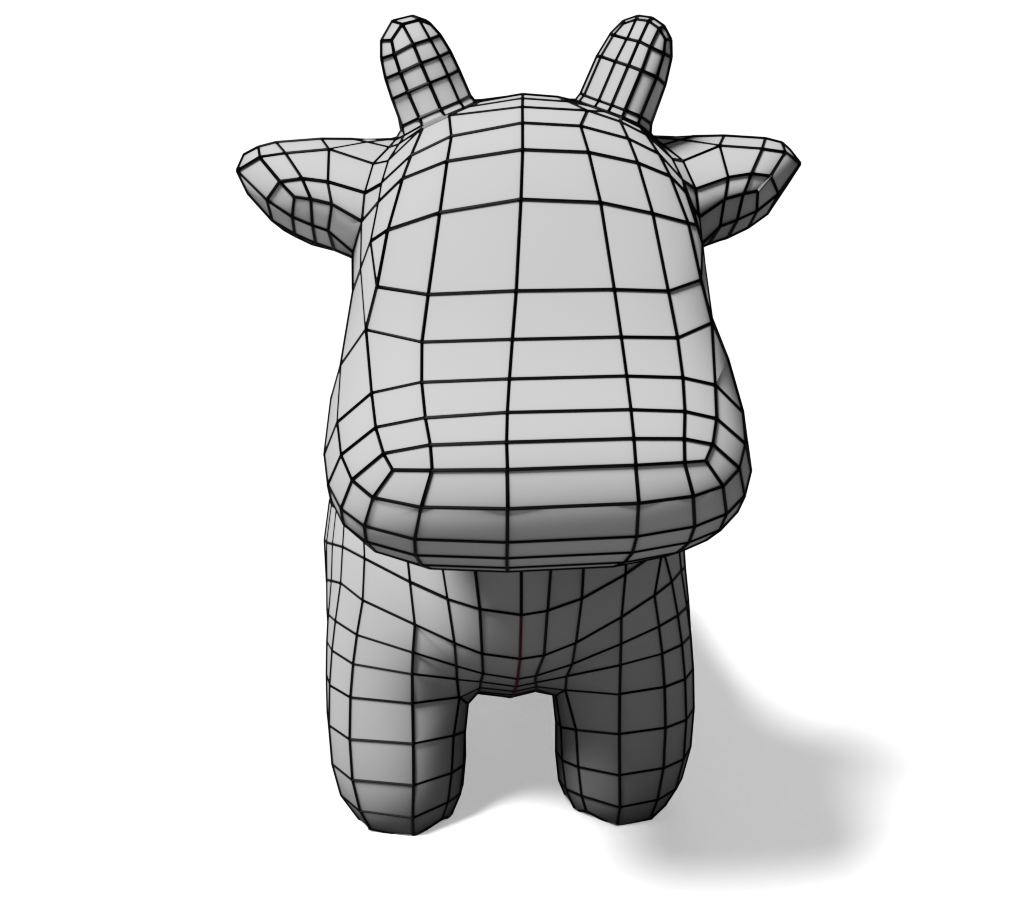} \\
        \includegraphics[width=0.3\linewidth]{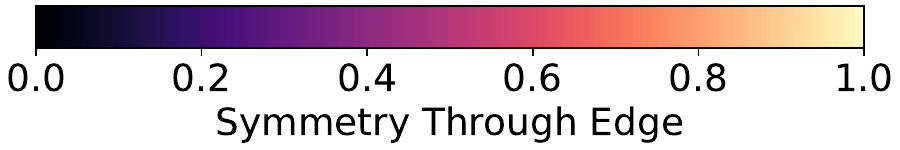} && \\
        $\square/\triangle: 2928/0$ & 1458/14 & 1457/14 \\
    \end{tabular}
    \caption{To measure the symmetry of each edge, we compute a matching of all vertices across the plane that spans each manifold edge and the halfway vector between its two adjacent faces. Left: vertices with a matching are shown in green, and those without are shown in red. Each edge's weight is the fraction of all vertices with a matching. This weight is then used during edge collapse for each edge's quadric. In the middle figure, edges along the axes of symmetry (shown in \textcolor{red}{red}) are preserved with this weighting, but are lost when it is not explicitly included, shown on the right. \cczero Keenan Crane.}
    \label{fig:symmetry-weight}
    \Description{Figure comparing our approach with symmetry and without. In the middle is our approach with symmetry preservation, where the axis of symmetry is exactly preserved. On the right is without symmetry preservation, and vertices have deviated from the center of the mesh.}
\end{figure}

\paragraph{Soft Symmetry} Using the newly introduced per-edge quadrics, we are interested in seeing if they can be used to preserve symmetric topology of the mesh. This is motivated by downstream applications, where some symmetry preservation is often desired as it is easier to manipulate than asymmetric meshes, but not as a hard requirement. To that end, we treat symmetry as an implicit soft constraint, guided by edge collapses. To guide the edge collapses, we increase the weight of each edge's quadrics based on a measure of symmetry across each edge. These per-edge symmetry weights $s \in [0,1]$ are incorporated by setting $w$ in Eq.~\ref{eq:edge-quadric} as $w = \max(w,\lambda_\text{sym} s)$, where $\lambda_\text{sym}$ is a parameter that controls the importance of symmetry.  To compute $s$, we compute a matching of all vertices across the plane that spans the edge and the halfway vector between the edges' two adjacent faces for manifold edges, and the singular face's normal for boundary edges. $s$ is then the ratio of vertices that have a matching vertex across the plane within a distance of $\epsilon_\text{sym}$ to the total number of vertices. An example of the weights for each edge and how they are computed is visualized in Fig.~\ref{fig:symmetry-weight}. This weighting implicitly induces an ordering over edges where edges that are equidistant from the plane of symmetry have equal importance. Pseudocode for our algorithm for symmetry matching is given in the Appendix, Alg.~\ref{alg:symmetric-edge-weights}.

\begin{figure}[h!]
    \setlength{\tabcolsep}{0pt}
    \centering
    \begin{tabular}{c c}
        Joint Distances & Decimated Skinned Animation \\
        \includegraphics[width=0.49\linewidth]{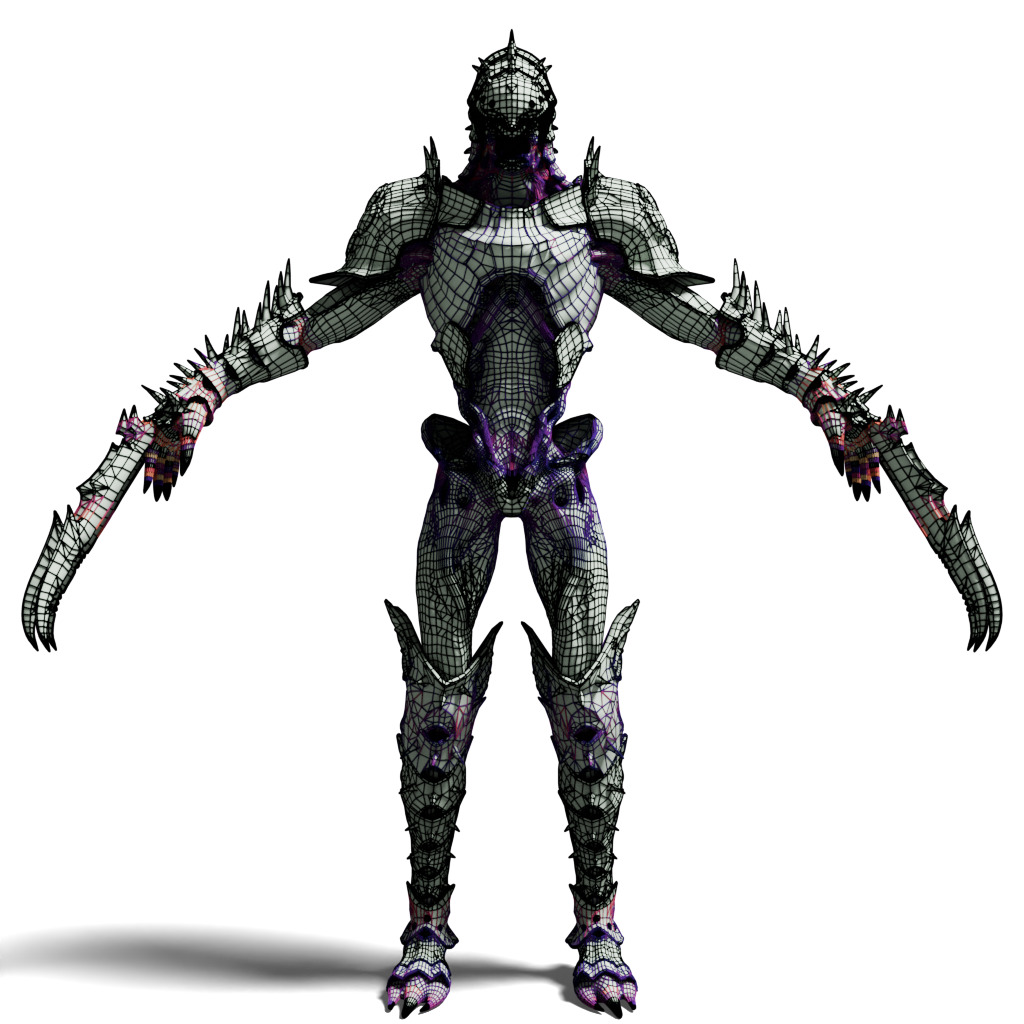} &
        \includegraphics[width=0.49\linewidth]{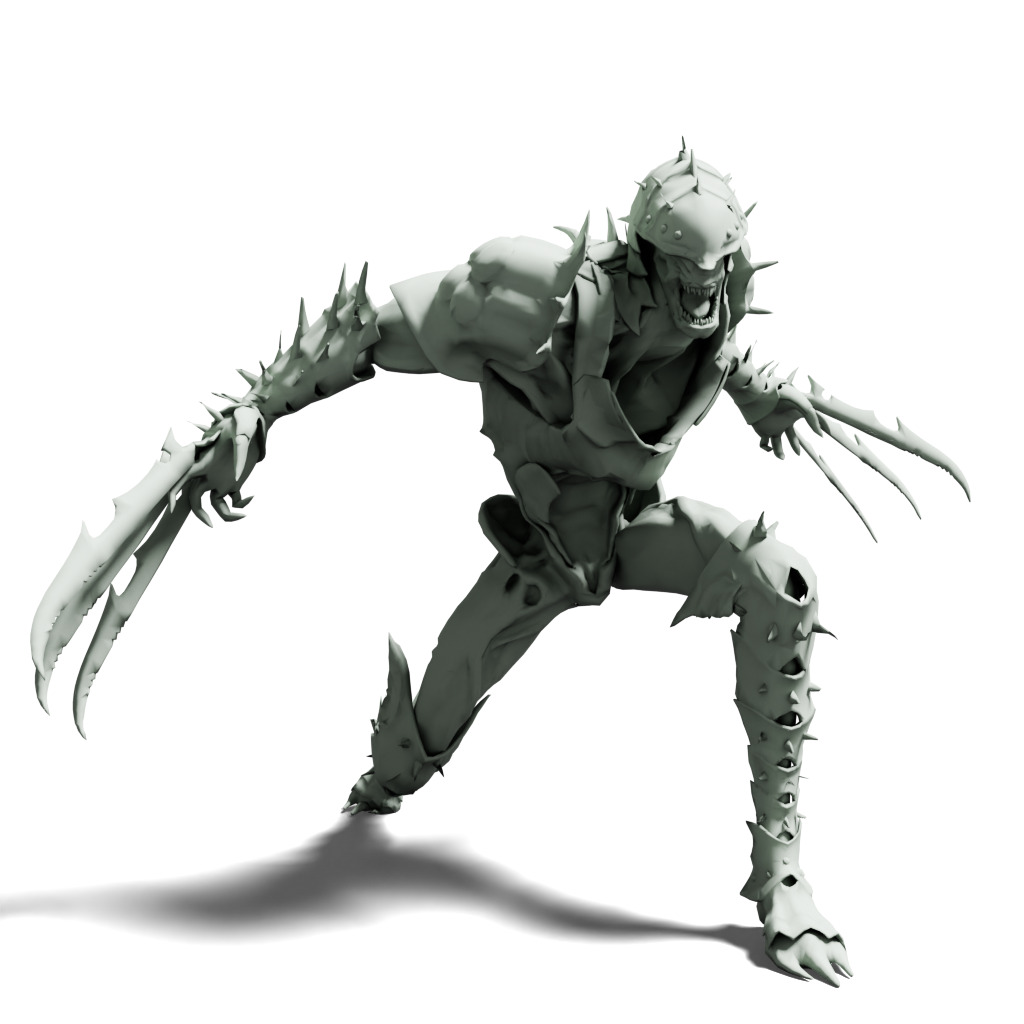} \\
        \includegraphics[width=0.3\linewidth]{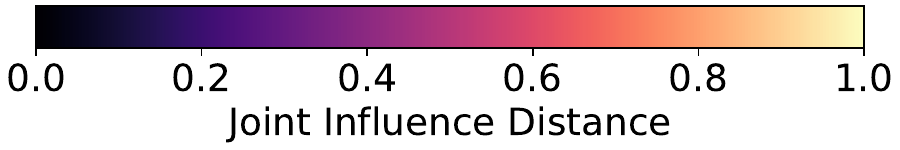} & \\
        $\square/\triangle$: 33499/39122 & 12384/24403 \\
    \end{tabular}
    \caption{Left: mesh in its rest pose, with edges marked with higher joint distances highlighted. Most distances are centered around anatomical joints, such as elbows and fingers. Right: frame of the animation of the mesh, visualizing range of motion after decimation. Animations tend to deform regions where joint influences vary heavily between vertices. \ccbync Catholomew.}
    \label{fig:skinned-mesh}
    \Description{Left: Alien Mesh in T-pose, with areas around joints highlighted due to difference in joint influences. Right: Alien animated, shrieking outwards.}
\end{figure}

\paragraph{Preserving Edges Between Distinct Joints} We also use edge weights to more strongly preserve edges with high gradients between joint influences. More plainly, edges where the endpoints have joint influences that are significantly different should be decimated less. For each edge, we compute the difference $d \in[0,1]$ between joint influence distributions as $\frac{1}{2}\ell_1(j_0, j_1)$, where $j_0, j_1$ are the discrete set of joint influences at the end of each edge, with each influence $j_{i_k}\in[0,1]$, and $\sum\limits_{k} j_{i_k} = 1$. An example of the $\ell_1$ difference between two joint distributions with 2 bone influences each is shown in Eq.~\ref{eq:l1-joint-diff}. We then take $w = \max(w, \lambda_\text{joint} d)$, where $\lambda_\text{joint}$ is a parameter that controls the importance of joint differences. A visualization of regions where joint influences vary is shown in Fig.~\ref{fig:skinned-mesh}. Key joint regions such as elbows and fingers have the largest difference, and our decimation allows for controlling how strongly those regions are preserved.

\begin{align}\label{eq:l1-joint-diff}
    & \text{Example $\ell_1$ diff. of two arbitrary joint influences:} \\\nonumber   
    d_\text{joint}&(j_0,j_1) = \frac{1}{2}\ell_1( \\\nonumber
    & j_0: k = \begin{bmatrix} a & b \end{bmatrix}, j_{0_k} = \begin{bmatrix} \alpha & 1-\alpha\end{bmatrix}, \\\nonumber
    & j_1: k = \begin{bmatrix} b & c \end{bmatrix}, j_{1_k} = \begin{bmatrix} \beta & 1 - \beta\end{bmatrix} \\\nonumber
    )  = & \frac{1}{2}(|\alpha| + |(1-\alpha) - \beta| + |{-(1-\beta)}|)
\end{align}

\subsection{Ordering Approximately Equal Edges}

\begin{figure}[!h]
    \setlength{\tabcolsep}{0pt}
    \centering
    \begin{tabular}{c c c c}
        Input & \multicolumn{3}{c}{Resolution $\rightarrow$} \\
         & \multicolumn{3}{c}{Ordered by QEM \& Recency} \\
        \multirow{2}{*}{\includegraphics[width=0.24\linewidth]{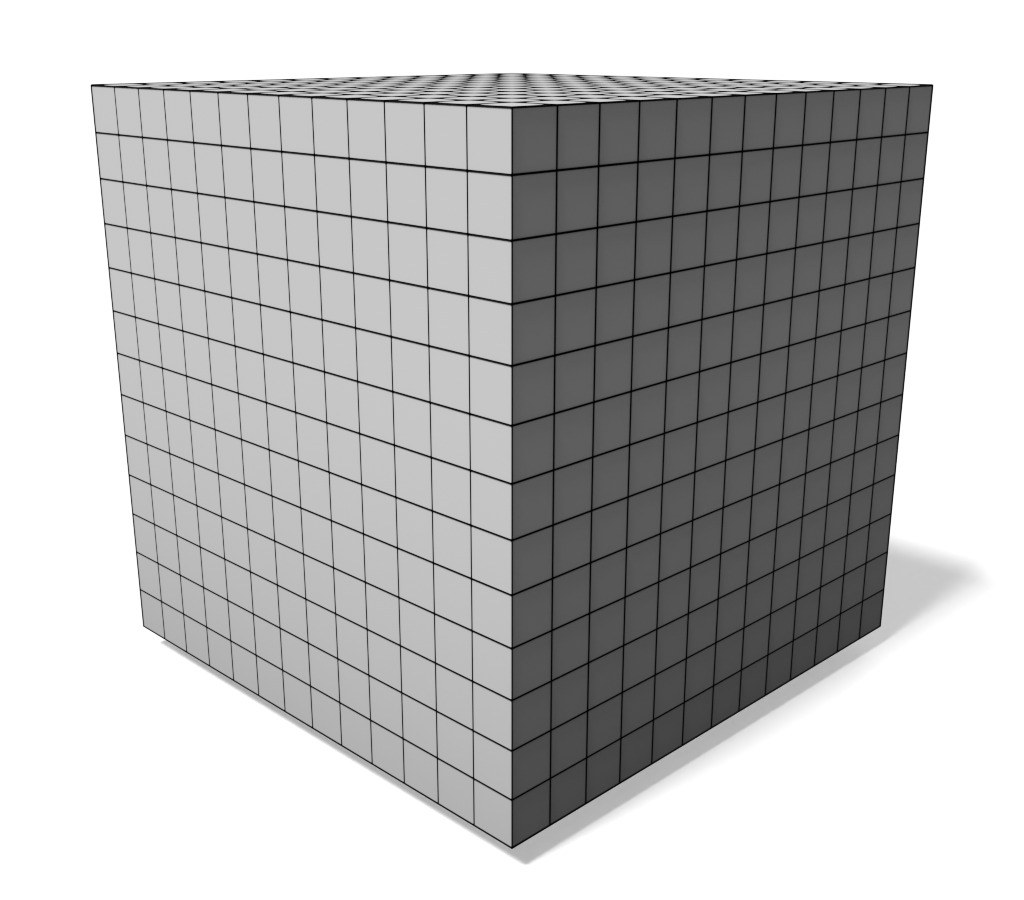}} & 
        \includegraphics[width=0.24\linewidth]{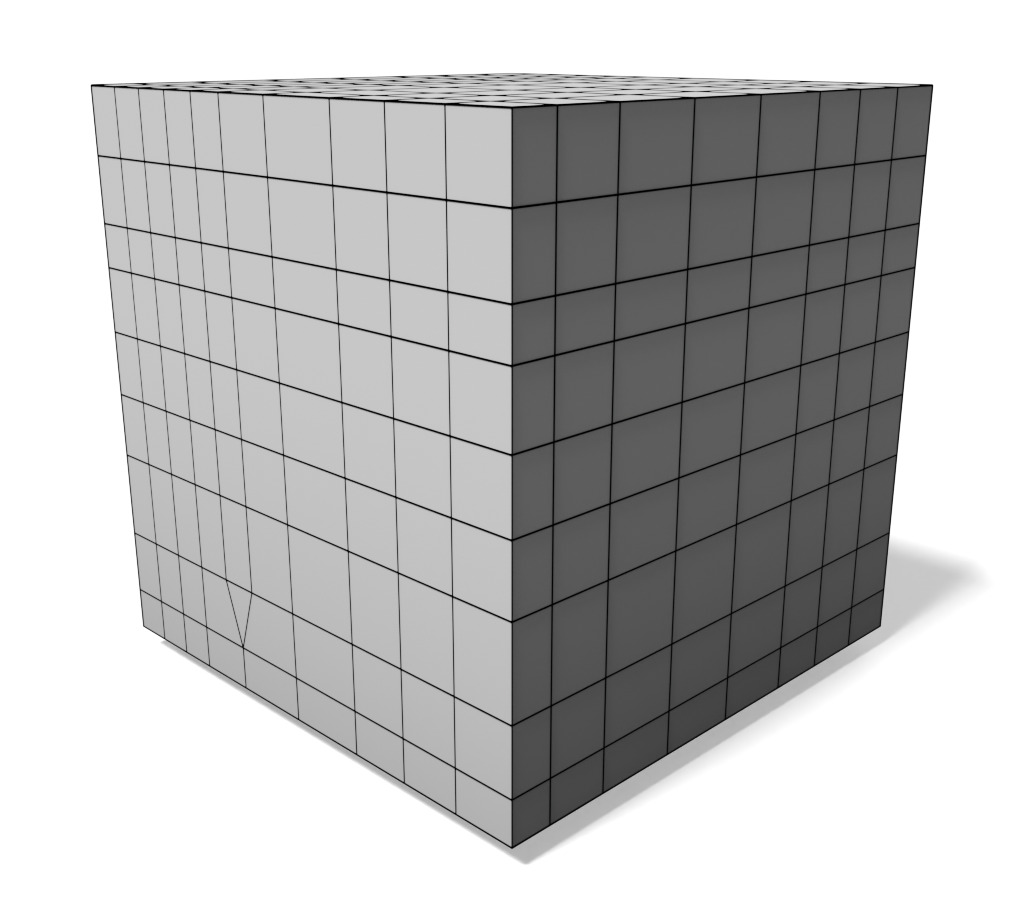} & 
        \includegraphics[width=0.24\linewidth]{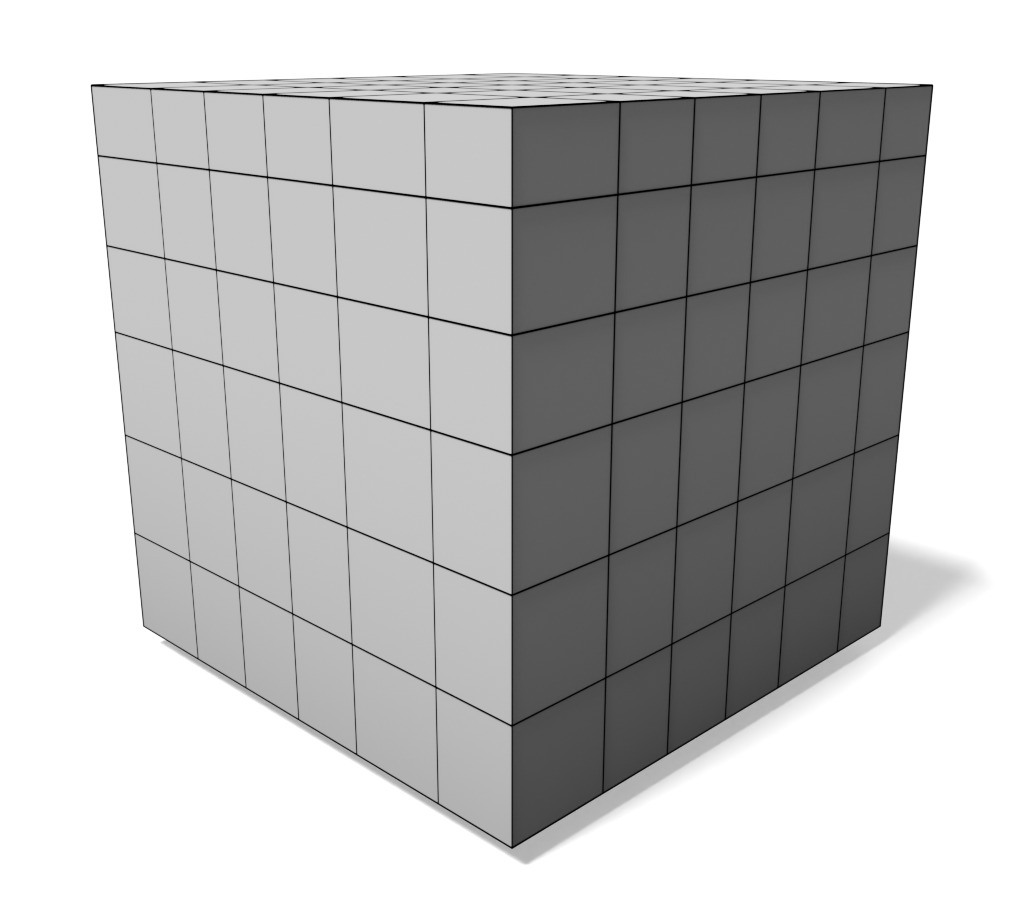} & 
        \includegraphics[width=0.24\linewidth]{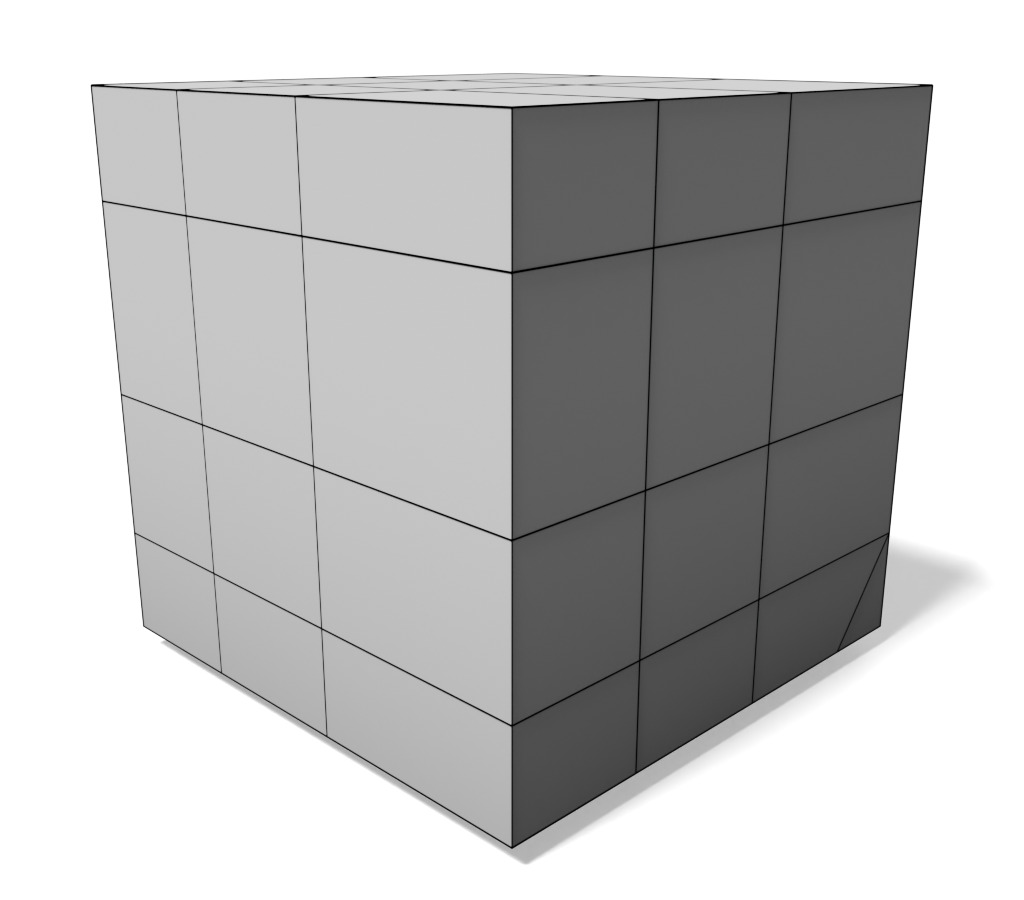} \\
        &
        \includegraphics[width=0.24\linewidth]{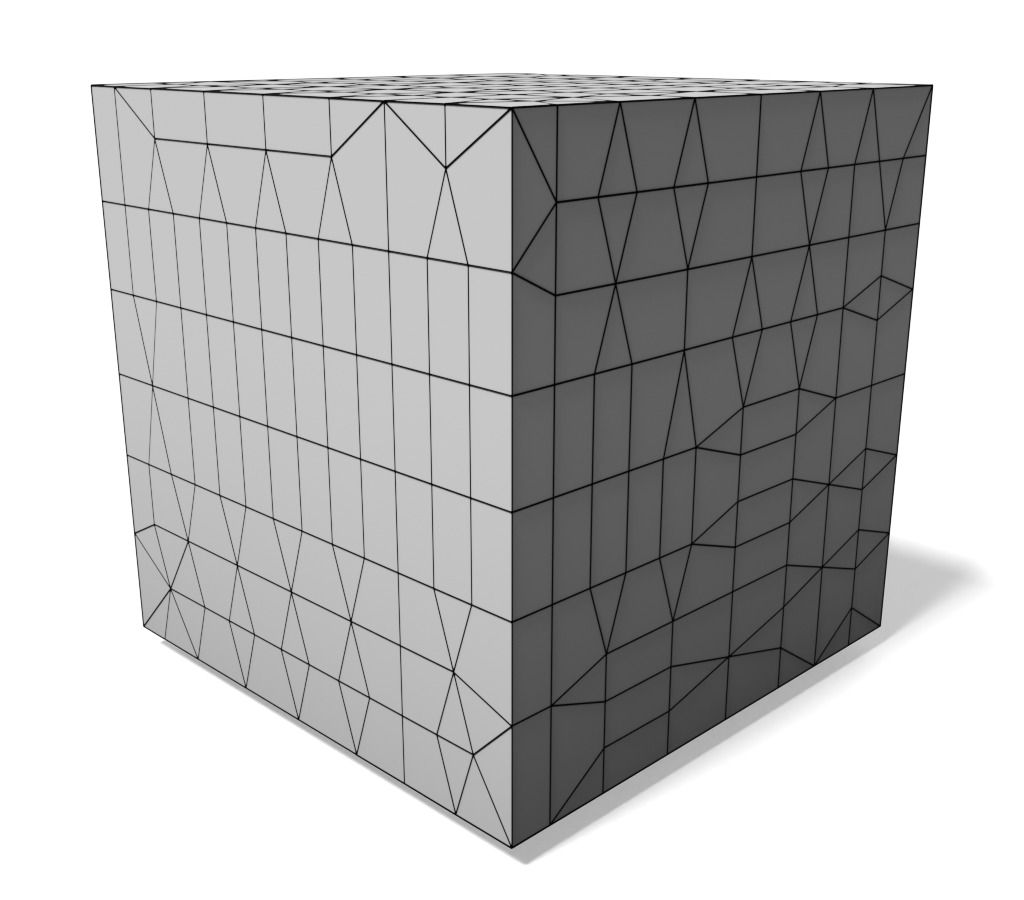} & 
        \includegraphics[width=0.24\linewidth]{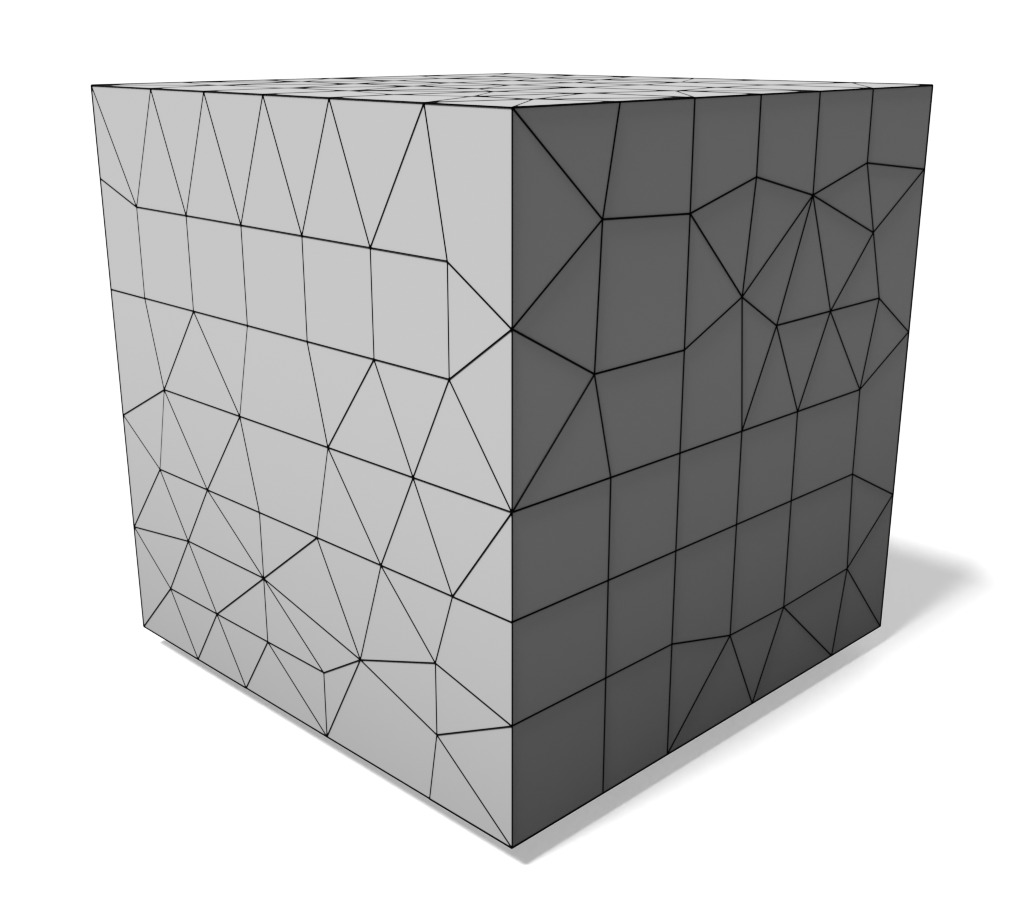} & 
        \includegraphics[width=0.24\linewidth]{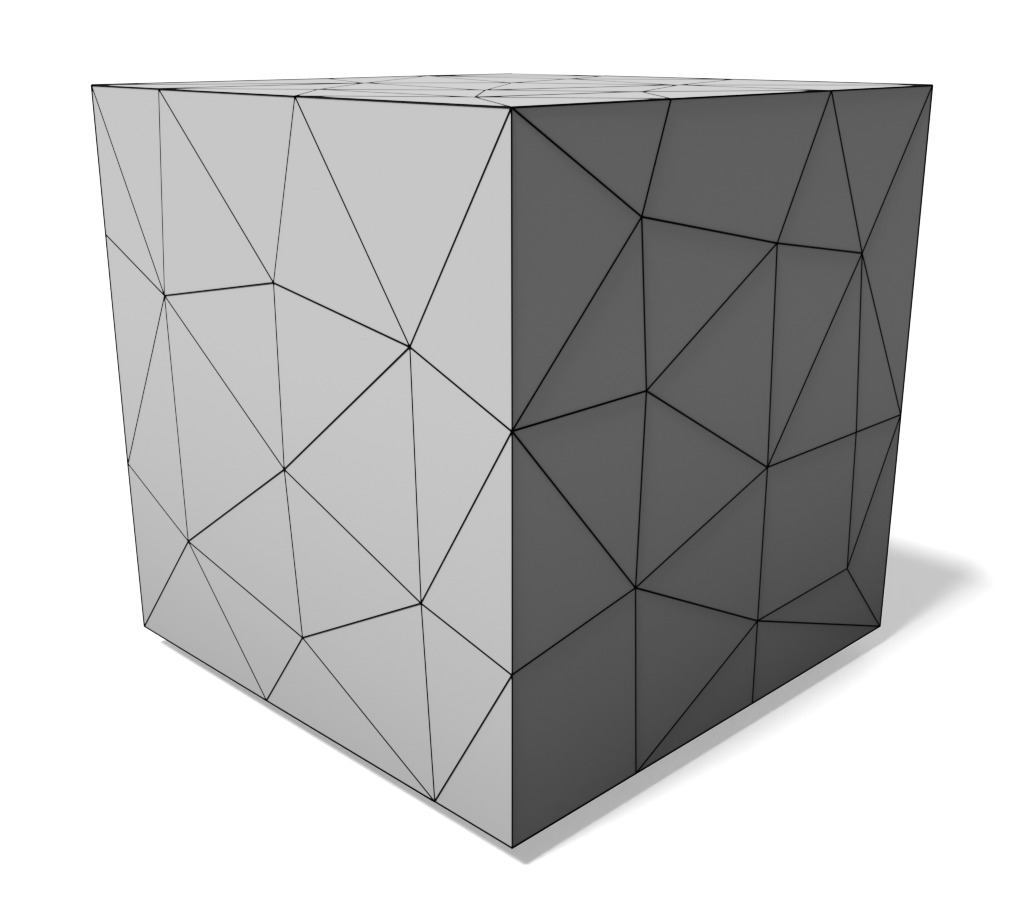} \\
        & \multicolumn{3}{c}{Ordered by QEM only} \\
    \end{tabular}
    \caption{Cube mesh simplified with our approach. All edges in the input cube have equivalent cost, but differ slightly when computed due to floating point error. We group these approximately equivalent edges and provide an explicit ordering based on the recency of collapsed opposing quad edges, which preserves topology. This is in contrast to ordering purely based on QEM, which is unable to preserve topology.}
    \label{fig:dense-cube}
    \Description{Visualization of the effect of recency in our approach. In the leftmost column is a densely subdivided cube. In the next 3 columns on the right are increasingly sparse decimations of the input cube. The top row shows structure in the output, the bottom row shows triangulation which becomes more like a voronoi diagram.}
\end{figure}

\noindent\paragraph{Total Ordering}
The second key aspect of our approach that allows for preserving quads is that the order of edge collapses is not dictated by quadric error alone. The observation that lead us to modify the ordering is that prior quadric mesh reduction imposes a strict total ordering over all edge collapses. A total ordering is where every element of a set is either greater than, equal to, or less than other elements, and there is no ambiguity in pairwise relationships. This is in contrast to a partial ordering, where some elements are incomparable. For quadric mesh reduction~\citep{qem}, there is an implicit assumption that quadric edge costs exactly and correctly dictate the order of edge collapses. While this makes sense assuming that geometric errors are correct and exact, we reject this assumption because it overly constrains the ordering of collapsed edges with edges that have approximately equal error. As an example, consider the input cube mesh in Fig.~\ref{fig:dense-cube}. Each edge on the cube has equal quadric error, with edges periodically spaced along the surface. With floating point approximation though, the ordering of edges is dominated by floating point error. This effectively means the ordering of equivalent edges is deterministically random, and by collapsing these equivalent edges in random order the input topology is destroyed. While floating point error is one motivation of our approach, exact computation will not resolve this problem for two reasons. First, even if edges are found to be exactly equivalent, the ordering over these equivalent edges is unspecified. Second, there may be some real but \textit{negligible} geometric difference between quadric error of edges, which is acceptable to ignore if it helps maintain quad topology. We address this by treating each quadric error as an approximation, and induce a \textit{partial ordering} over quadric errors, treating them as incomparable if they are too close.

\paragraph{Partial Ordering over Edges}
We treat comparisons between quadric errors as a partial ordering, where if two edges are compared to be approximately equal we consider it to be indeterminable if one is greater than the other. We define approximate equality between two quadric errors as: \begin{equation}\label{eq:abs-diff}
    a \approx b \coloneqq |a - b| < \epsilon_\text{abs}
\end{equation} where $\epsilon_\text{abs}$ is a positive threshold for defining equality based on the difference of the two floats~\citep{ComparingFloatingPointNumbers}. The set of collapsible edges is divided into ``equivalence classes'', or sets of edges which are approximately equal to each other.  Within each equivalence class, we introduce an additional ordering on top of the quadric error, and use an ordering that prefers decimating quad chords. By using such an ordering, quads within each equivalence class are preserved. This approach allows itself to tune the trade-off between quad and triangle preservation by modifying $\epsilon_\text{abs}$. Since $\epsilon_\text{abs}$ determines whether two edges are considered equal, a higher $\epsilon_\text{abs}$ preserves more quads at the cost of lower geometric similarity, and vice-versa when lowering $\epsilon_\text{abs}$.

\begin{figure}[h!]
    \centering
    \begin{tabular}{c c}
         Input & Recency After Collapse  \\
        \frame{\includegraphics[width=0.3\linewidth]{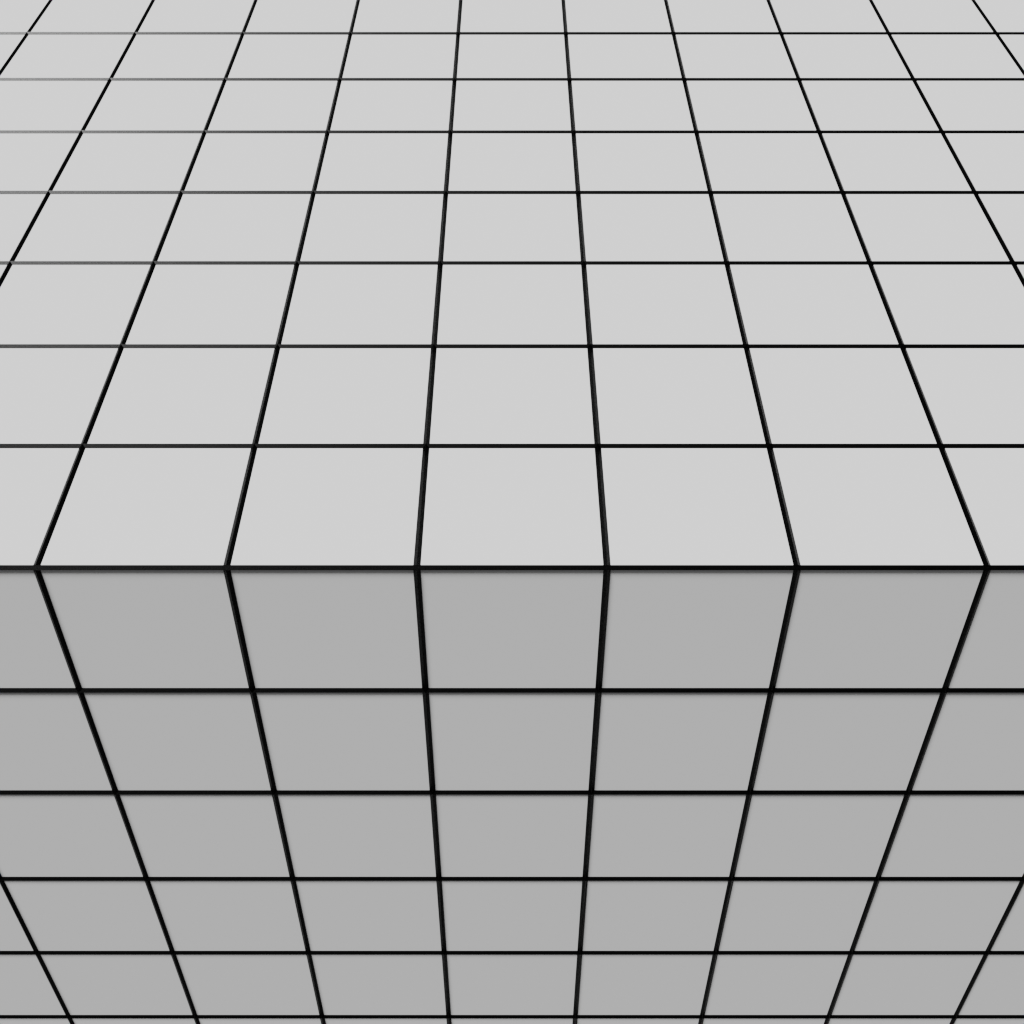}} & 
        \frame{\includegraphics[width=0.3\linewidth]{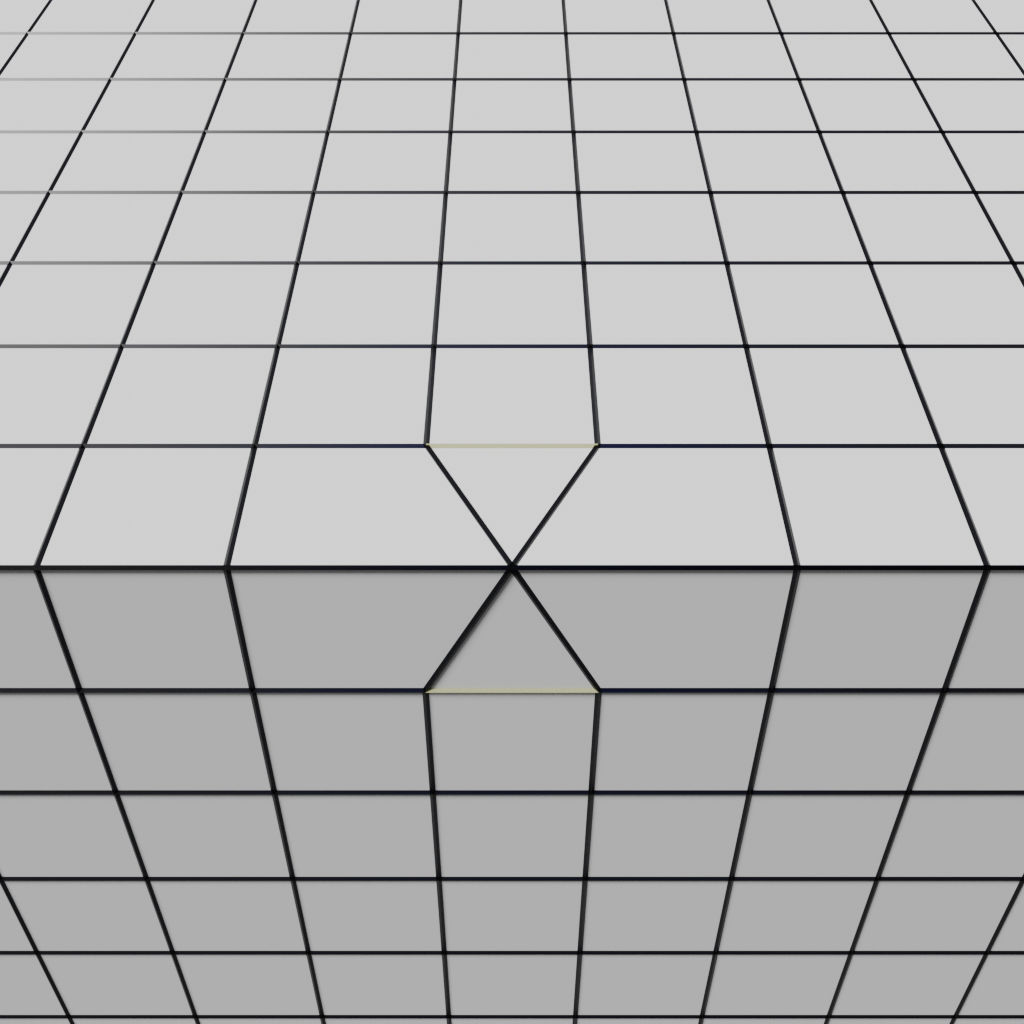}}
        \put(-0.1575\linewidth, 0.065\linewidth){\textcolor{red}{$\uparrow$}}
        \put(-0.1575\linewidth, 0.18\linewidth){\textcolor{red}{$\downarrow$}}
        \\
    \end{tabular}
    \caption{Before edge collapses, all edges have 0 recency (marked in black). After an edge collapse, opposing quad edges have their recency increased (marked in white with red arrows) by the recency of the collapsed edge plus 1.}
    \label{fig:recency-diagram}
    \Description{Left: zoom in of a dense cube mesh. Right: dense cube mesh with single edge collapsed. The quad edges opposite of the collapsed edge are marked in white to indicate they have high priority to be collapsed next.}
\end{figure}

\paragraph{Ordering Approximately Equal Edge Collapses}
For approximately equal edge collapses, we introduce additional state that implicitly collapses quad chords. Specifically, we introduce a metric we call ``recency'' to determine collapse order. For every edge collapsed, we increase the recency of any edges on the opposite side of quads containing the collapsed edge to the recency of the collapsed edge plus one. A visualization of recency is shown in Fig.~\ref{fig:recency-diagram}. We prioritize edges within each equivalence class so that the edge with highest recency is collapsed first. Since opposing quad edges will have higher recency and be collapsed earlier, this leads to an implicit decimation of quad chords. Once an entire set of approximately equivalent edges has been decimated, we reset the recency of all edges to 0, since there is no longer a preference of which edge to decimate.

\paragraph{Implementation} In our implementation, partial ordering and recency are implemented using two priority queues. The first priority queue is ordered entirely by quadric error, identical to prior approaches (line \ref{alg:line:qem}, Alg.~\ref{alg:hier_total_ordering}). The second priority queue is ordered by recency, and if all edges have equivalent recency then it is ordered by quadric error (line \ref{alg:line:recency}, Alg.~\ref{alg:hier_total_ordering}). If the second priority queue is empty, one edge is removed from the first priority queue, and moved into the second priority, with recency 0 (line \ref{alg:line:prerecency}-\ref{alg:line:recency}, Alg.~\ref{alg:hier_total_ordering}). While the second priority queue is not empty, the highest priority edge is removed from the second priority queue and that edge is collapsed (line \ref{alg:line:rec-collapse}-\ref{alg:line:rec-collapse-post}, Alg.~\ref{alg:hier_total_ordering}). For every collapse from the second priority queue, we check if there are edges on the top of the first priority queue which are roughly equivalent, and move them from the first to the second priority queue as long as the top is approximately equivalent (line \ref{alg:line:move}-\ref{alg:line:move-post}, Alg.~\ref{alg:hier_total_ordering}). If an edge has its quadric error updated and it is in the second priority queue, it is moved back to the first priority queue (encapsulated in line \ref{alg:line:adj-updates}, Alg.\ref{alg:hier_total_ordering}). The full algorithm of our approach, including updating recency and selecting edges based on priority is in Alg.~\ref{alg:hier_total_ordering}.

\begin{algorithm}
\caption{Quad mesh preserving QEM \label{alg:hier_total_ordering}}
\begin{algorithmic}[1]
    \Statex \textbf{Input: } vertices $V$, faces: $F$
    \Statex \textbf{Output: } new vertices $V'$, new faces $F' \subseteq F$ 
    \State $Q_i = \text{Quadric}(0)$ \Comment{Initialize zero quadric for all vertices}
    \For{v $\in V$}
        \For{$f \in F$ \textbf{if} $v \in f$}
            \State $Q_{v} \mathrel{+}= \text{area}(f)\text{Quadric}(f)$ \Comment{Add quadric for adj. faces}
        \EndFor
    \EndFor
    \For{$(e_0, e_1) \in \text{uniq}(edges)$}
        \If{$(e_0, e_1)$ adjacent to $f_0, f_1$}
        \State $\angle(e_0, e_1) = \frac{1}{\pi}\arccos(n(f_0)^\top n(f_1))$
        \Comment{Manifold Edges}
        \EndIf
        \State \textbf{else } $\angle(e_0, e_1) = 1$ \Comment{Non-manifold edges}
        \State $w = \max(\lambda_\text{sym}w_\text{sym}, \lambda_\text{j}w_\text{joint}, \angle(e_0, e_1), \num{1e-2})$
        \State $Q_{e_0} \mathrel{+}= w \lVert e_0 - e_1 \rVert_2\text{Quadric}(e_0, e_1)$
        \State $Q_{e_1} \mathrel{+}= w \lVert e_0 - e_1 \rVert_2\text{Quadric}(e_0, e_1)$
    \EndFor 
    \State $\text{pq}_\text{qem}$: Priority Queue, $\text{pq}_\approx$: Priority Queue
    \For{$(e_0, e_1) \in$ E} \Comment{Initialize Priority Queue}
        \State $\text{pq}_\text{qem}$.push(priority = CollapseCost($Q_{e_0}, Q_{e_1}$), e) \label{alg:line:qem}
    \EndFor
    \While{tri count $>$ target}
        \State $\forall e: \text{Recency}(e) = 0$ \Comment{Reset all recencies}
        \State [qem, e] = $\text{pq}_\text{qem}$.pop() \label{alg:line:prerecency}
        \State $\text{pq}_\approx$.push(priority=(0, qem),  e) \label{alg:line:recency}
        \While{not empty($\text{pq}_\approx$)} \label{alg:line:rec-collapse}
            \State [$\text{rec}_e$, qem, e] = $\text{pq}_\approx$.pop()
            \State CollapseEdge(e) \Comment{Identical collapse to QEM} \label{alg:line:rec-collapse-post}
            \For{$(e_0^*, e_1^*)$ opposite to $e$ in any quad}
                \State $\text{Recency}((e_0^*, e_1^*)) \mathrel{+}= \text{rec}_e + 1$
            \EndFor
            \State Update one-ring costs and two-ring recency \label{alg:line:adj-updates}
            \While{len($\text{pq}_\text{qem}$) > 0 and top($\text{pq}_\text{qem}$).prio $\approx$ prio}\label{alg:line:move}
                \State [qem$_n$, e$_n$] = $\text{pq}_\text{qem}$.pop()
                \State $\text{pq}_\approx$.push(priority=(recency(e$_n$), qem$_n$), e$_n$) \label{alg:line:move-post}
            \EndWhile
        \EndWhile
    \EndWhile
\end{algorithmic}
\end{algorithm}

This precise ordering relies on \textit{not using memoryless simplification}, and computing all quadrics during initialization. This is because memoryless simplification affects the 2-ring of each vertex, which causes the costs of opposing edges on a quad to change and removes them from the equivalence class. While it was shown in prior work that memoryless simplification leads to improved results, we are able to remove memory simplification and modify the quadric error metric without impacting the geometric quality of our results.

\paragraph{Minimizing Introduced Error} To replace memoryless simplification, we change the QEM error so it no longer measures the error from the initial mesh, but from the decimated mesh. To measure introduced error, we redefine quadric error using the following definition:
\begin{equation}\label{eq:new-qem}
    \text{QEM}(e, v) = (\text{Q}_{e_0} + \text{Q}_{e_1})(v) - (Q_{e_0}(e_0) + Q_{e_1}(e_1))
\end{equation}

Where $v\in\mathbb{R}^3$ is the position of the new vertex, and $Q_{e_0}, Q_{e_1}$ are the quadrics for $e_0, e_1$ respectively. The first term $(\text{Q}_{e_0} + \text{Q}_{e_1})(v)$ is the original quadric error metric, and the new term $-(Q_{e_0}(e_0) + Q_{e_1}(e_1))$ is the sum of the current quadric errors of the edge's two endpoints. Using this definition, edge collapses which introduced errors previously will not be penalized if future edge collapses will not introduce additional error. This is similar to using memoryless simplification, since both measure error from the \textit{current} state, rather than from initialization. Since this modification was never mentioned in prior work, it is likely a key reason that memoryless simplification was purported to work better.

\paragraph{Skinned Mesh Reduction} Since quad meshes are crucial to animations and skinning, we also implement animated mesh reduction. There are two prior works we focus on for animated mesh reduction: \citep{pose_indep_simp} and \citep{articulated_mesh_simplification}. Both of these works argue that \citep{qem_hoppe} is insufficient for producing high quality joint weights and treat joint influences differently from UVs, normals, and vertex colors. Unfortunately, neither of those works provide evidence that this is true. For our implementation, we adapt \citep{qem_hoppe} for joint influence simplification. We treat joint influences identically to other attributes and store a linear functional for each joint influence: \begin{equation}
    j_{i_b}(p) = g_{j_i}^\top p + d_{j_i}
\end{equation}
Where $j_i$ is the i$^\text{th}$ joint influence and $i$ goes up to the number of bones, $p\in\mathbb{R}^3$ is the new vertex position, and $g_{j_i}, d_{j_i}$ are computed identically to other attributes from \citep{qem_hoppe}. Since joint influences are sparse, we only store non-zero joint influences of each vertex. To ensure that our quadrics are purely stack allocated (which improves performance), we store a maximum of 16 joint influences per vertex, and if there are more than that we drop the influence with the minimum weight: $\max(g_{j_i}^\top p_0 + d_{j_i},g_{j_i}^\top p_1 + d_{j_i}) $, where $p_0,p_1$ are the positions of the two quadrics being combined. Note that this resolving which vertex to keep only occurs at at extreme decimation when the new mesh does not appear close to the original, and does not occur in our dataset. After decimation, we are able to recover joint influences for all bones that influence any of the merged vertices, and we select the top 4 and normalize their influences to sum to 1. This approach, as compared to \citep{articulated_mesh_simplification}, is independent of the order of collapses, so bone influences will not be culled until the mesh is fully decimated.

\paragraph{Attribute Preserving QEM for Polygonal faces} While \citep{qem_hoppe} was designed specifically for triangle meshes, their approach generalizes to faces with an arbitrary number of vertices. This is done by computing the least squares solution for $g,d$ in:
\begin{equation}
    \argmin\limits_{g,d} 
    \begin{bmatrix}
        p^\top_1 & 1 \\
        p^\top_2 & 1 \\
        \cdots & \cdots \\
        p^\top_n & 1 \\
        n^\top & 0
    \end{bmatrix}
    \begin{bmatrix}
        g \\
        d \\
    \end{bmatrix}
    -
    \begin{bmatrix}
        s_1 \\
        s_2 \\
        \cdots \\
        s_n \\
        0
    \end{bmatrix}
\end{equation}
where $p_\bullet$ is the position of each vertex, $s_\bullet$ is the attribute of the corresponding vertex and $g, d$ are terms from the linear functional that determines the attribute at an arbitary position $s(p) = g^\top p  + d$. For our implementation, we use a least squares QR solver to compute $g$ and $d$.


\begin{figure*}
\renewcommand{\arraystretch}{0.85}
\setlength{\tabcolsep}{0pt}
\begin{tabular}{c c c}
    \multicolumn{3}{c}{Visual Comparison of Quad Preservation \& Geometric Similarity} \\
    \put(-0.23\linewidth, 0){\small Input $\square/\triangle$: 67632/64} Ours & QEM & MeshLab~\citep{meshlab} \\
    \put(-0.11\linewidth, 0.15\linewidth){\frame{\includegraphics[width=0.1\linewidth]{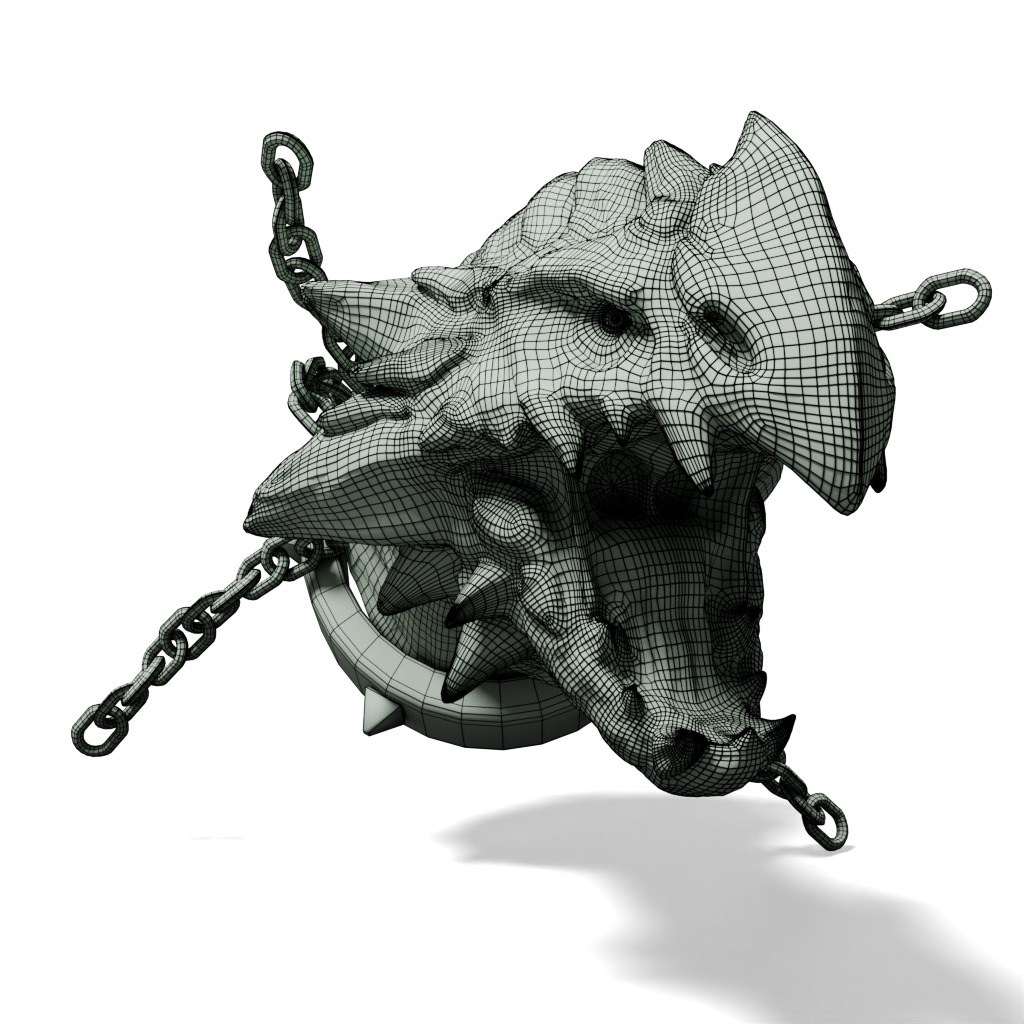}}}
    \put(-0.05\linewidth,0){\rotatebox{90}{50\% Total $\triangle$ Count}}
    \includegraphics[width=0.25\linewidth]{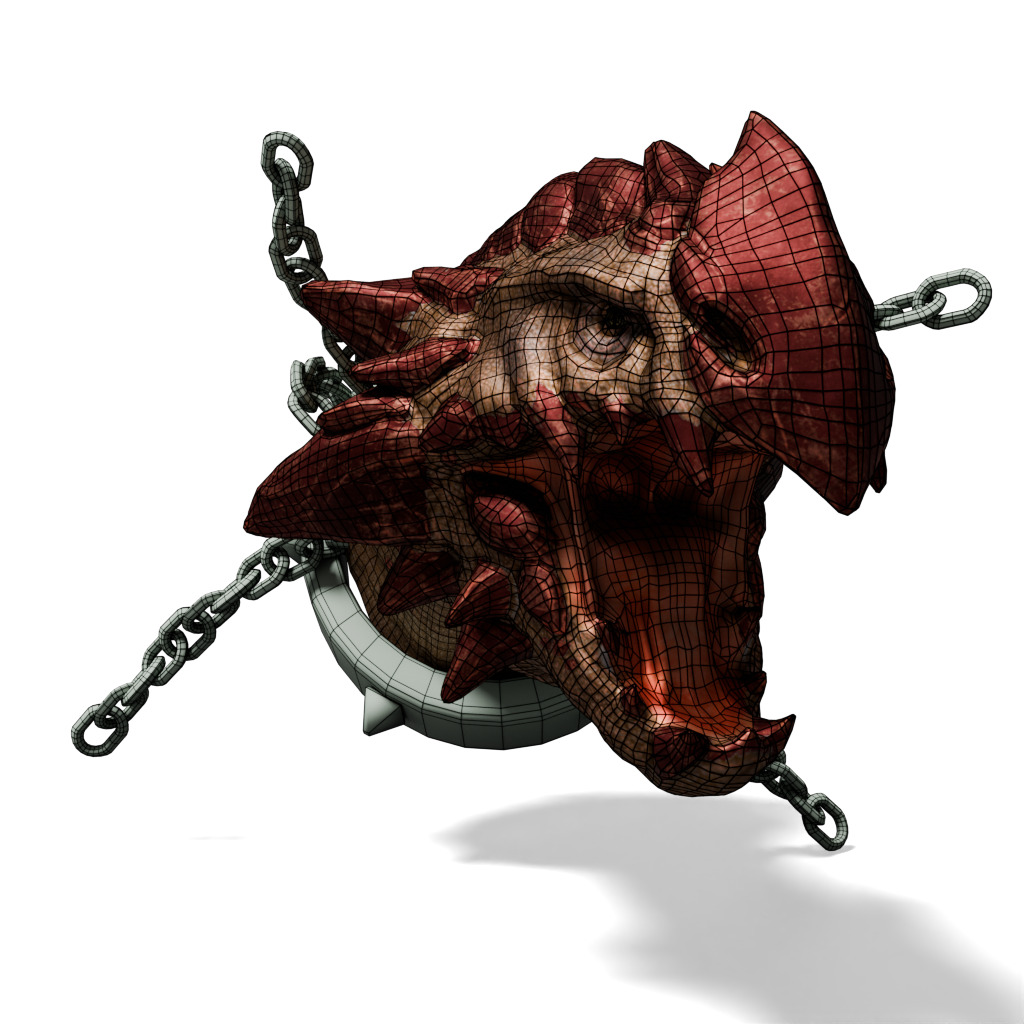}
    &
    \includegraphics[width=0.25\linewidth]{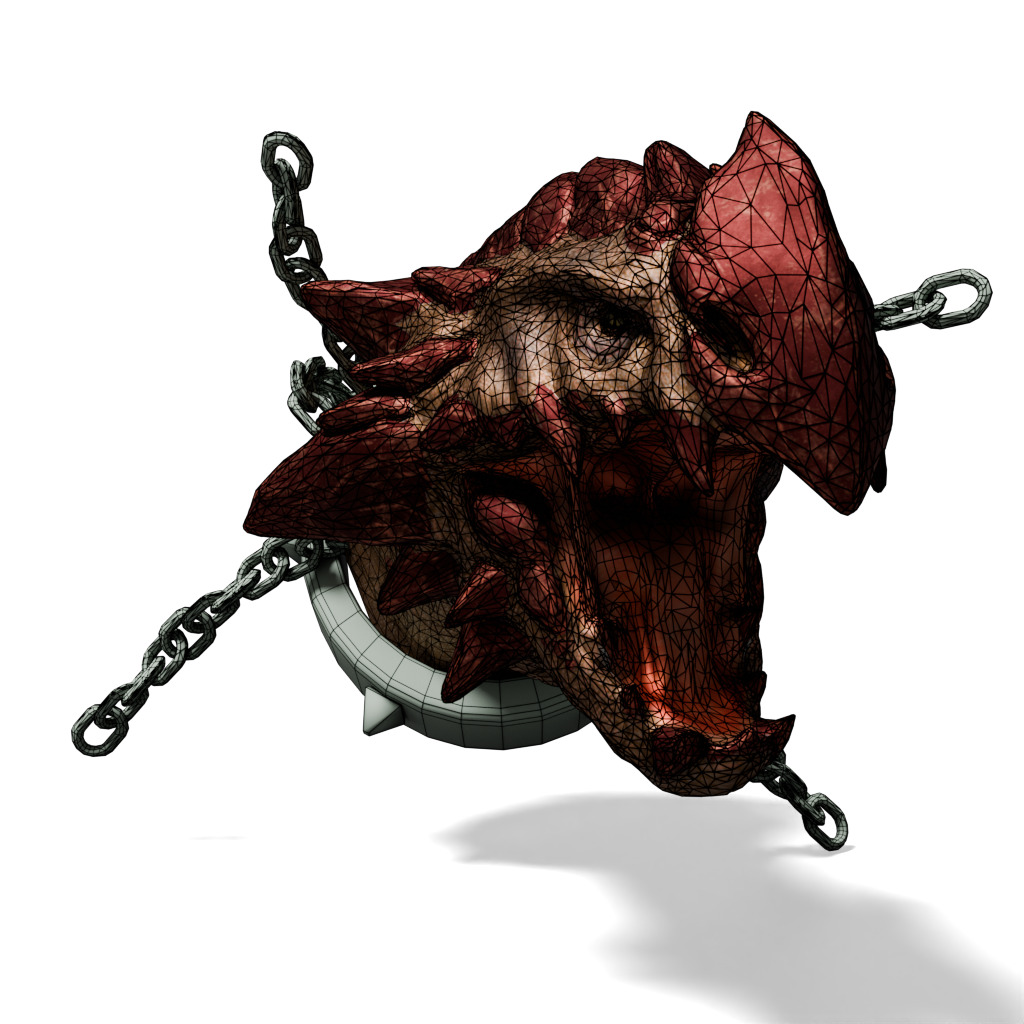}
    &
    \includegraphics[width=0.25\linewidth]{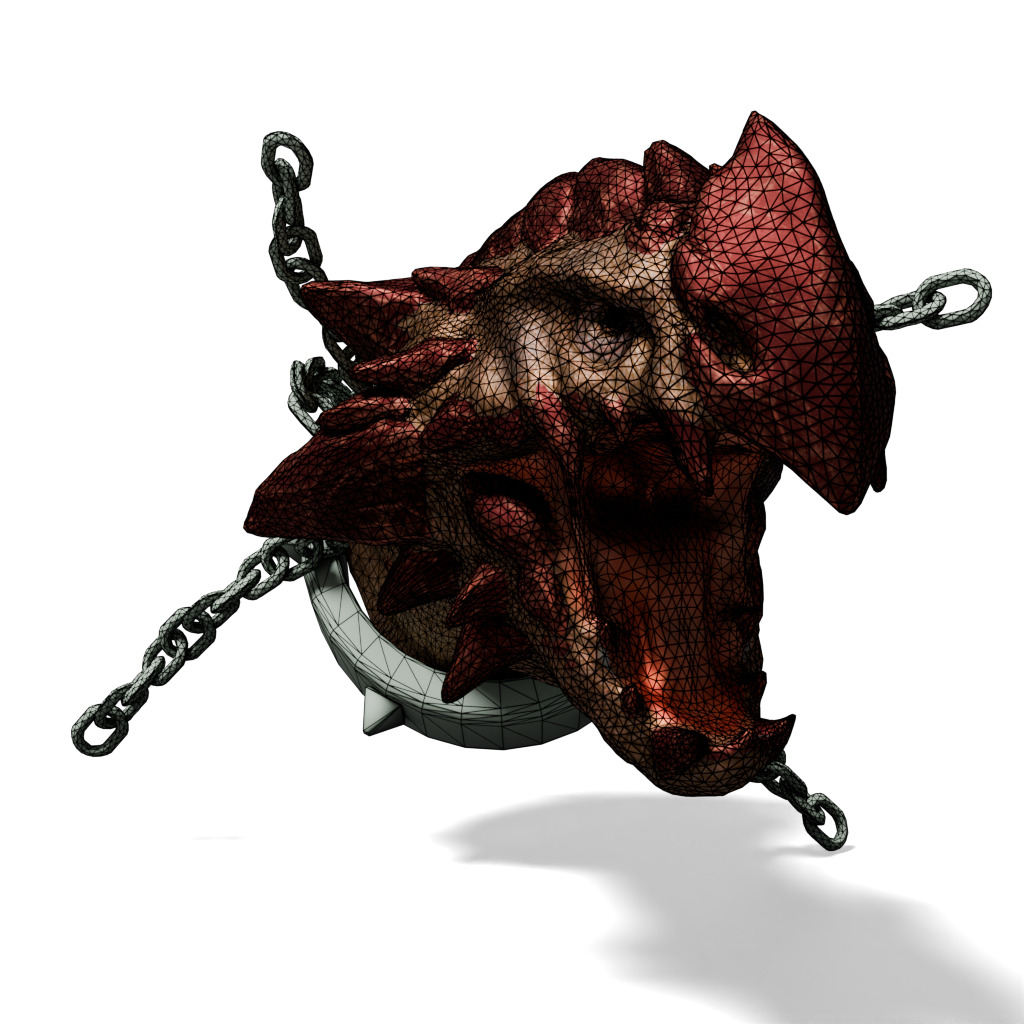}
    \\
    \frame{\includegraphics[width=0.1\linewidth]{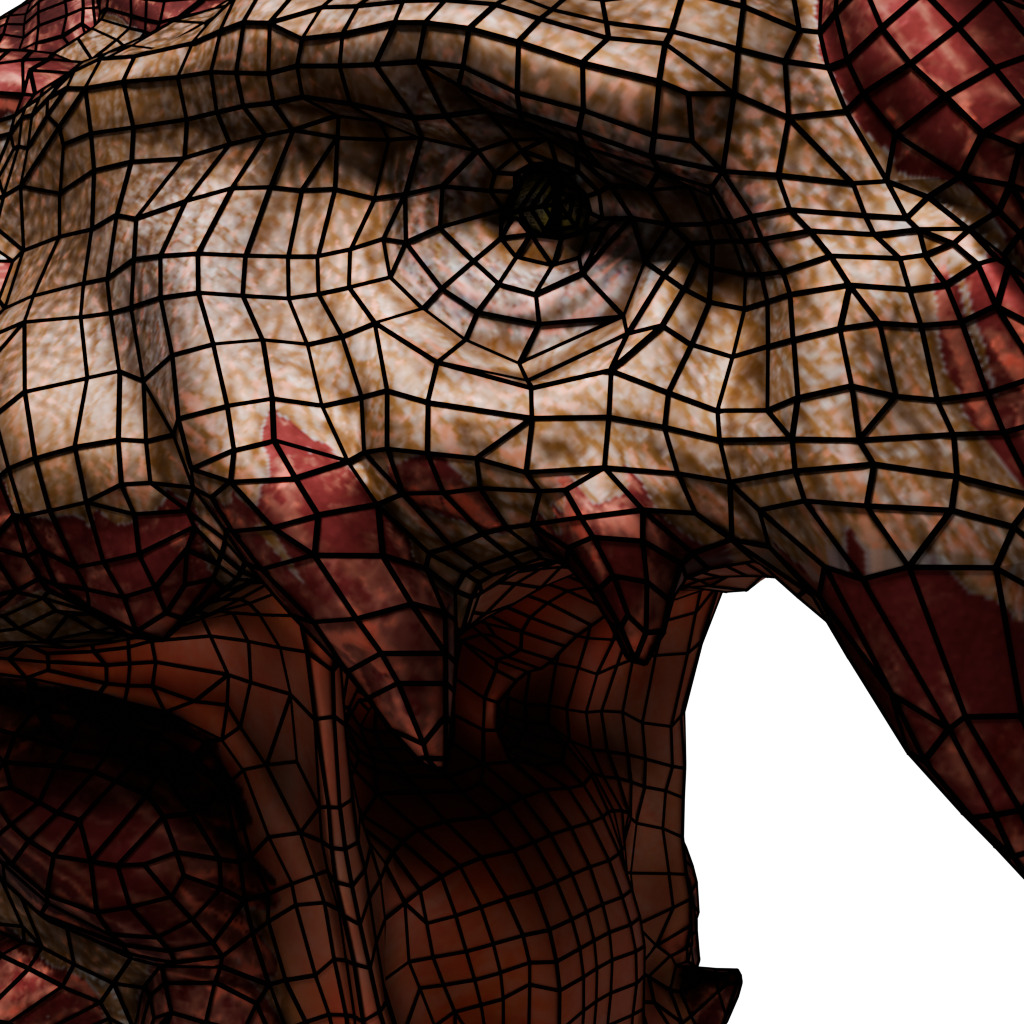}}
    \frame{\includegraphics[width=0.1\linewidth]{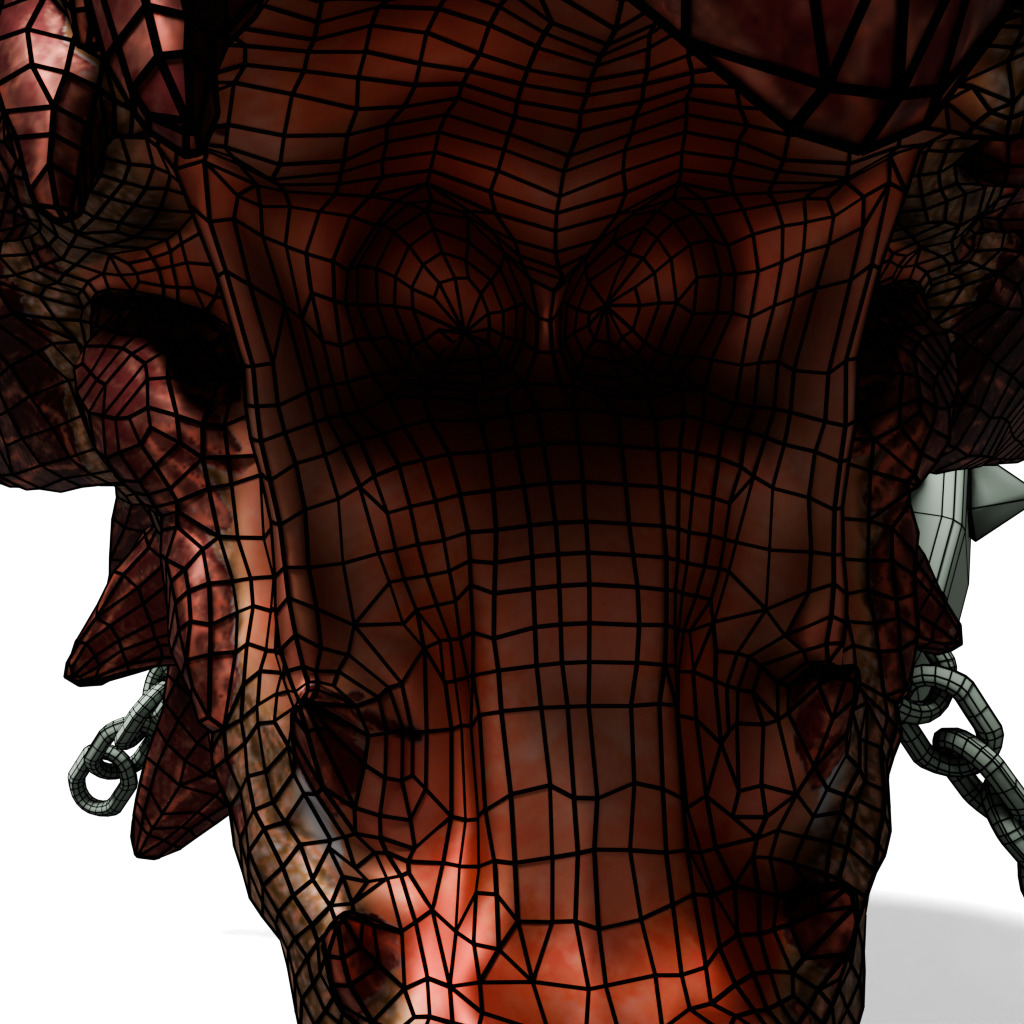}}
    &
    \frame{\includegraphics[width=0.1\linewidth]{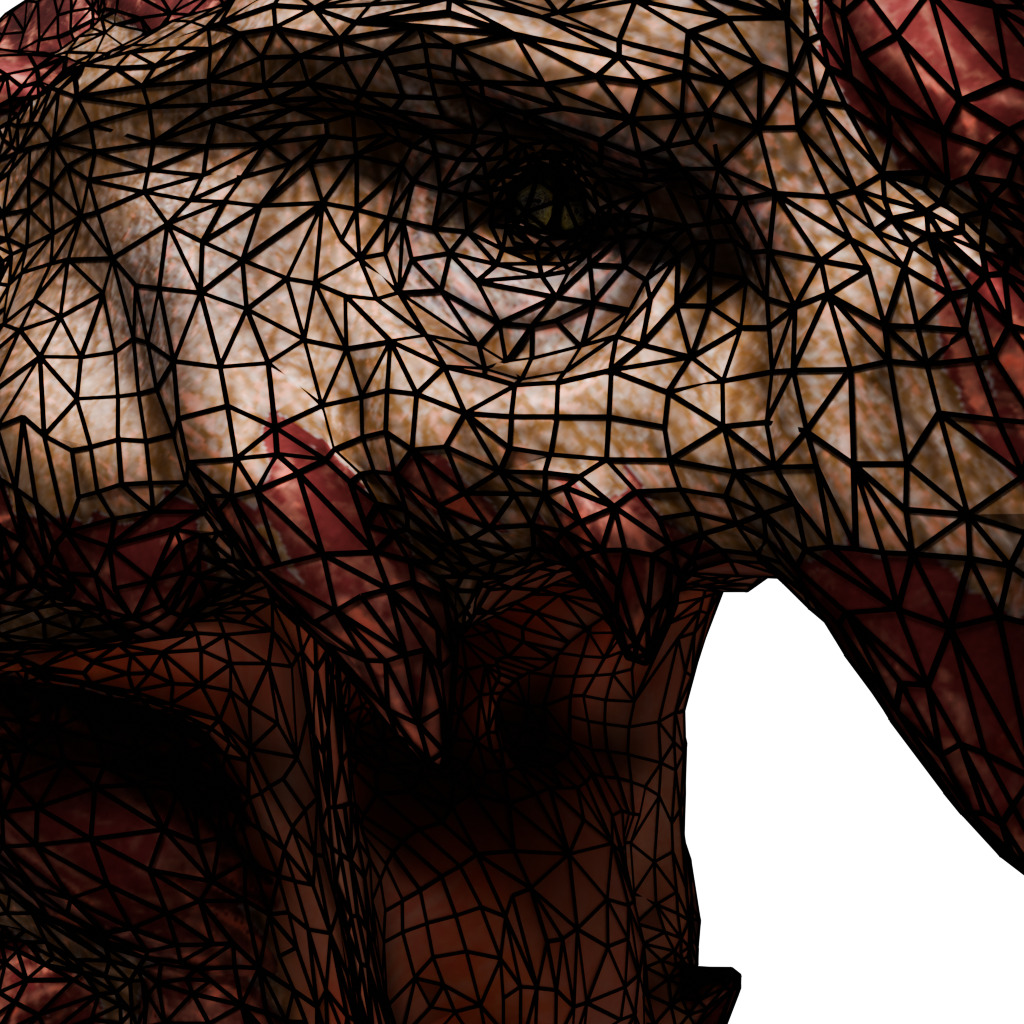}}
    \frame{\includegraphics[width=0.1\linewidth]{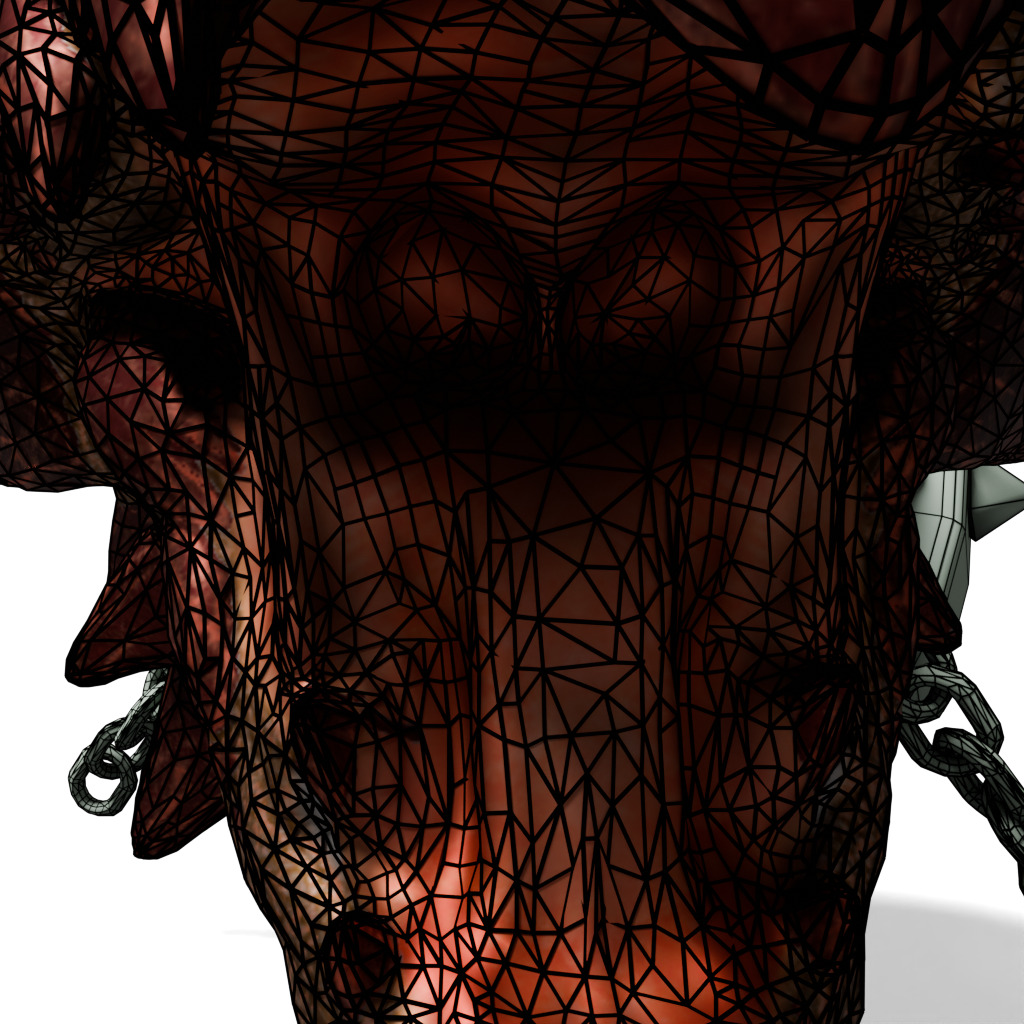}}
    &
    \frame{\includegraphics[width=0.1\linewidth]{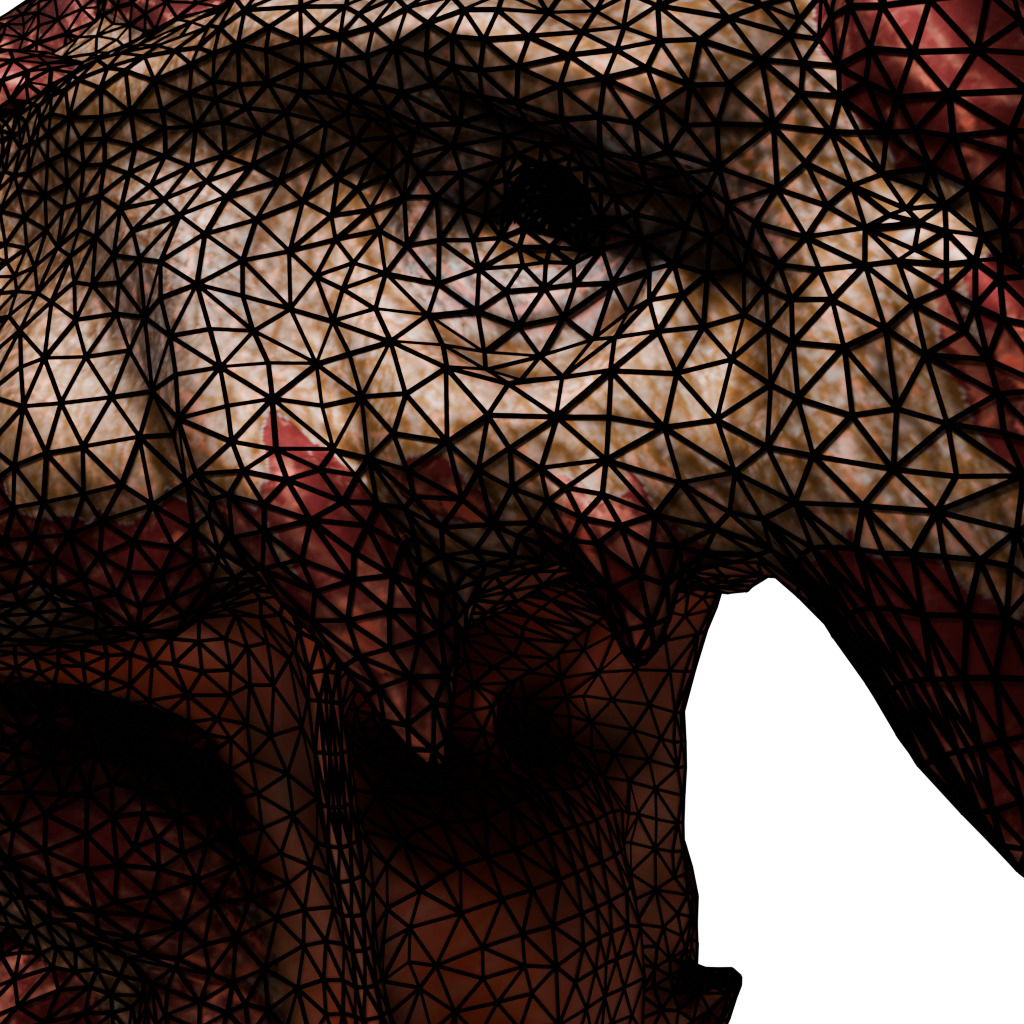}}
    \frame{\includegraphics[width=0.1\linewidth]{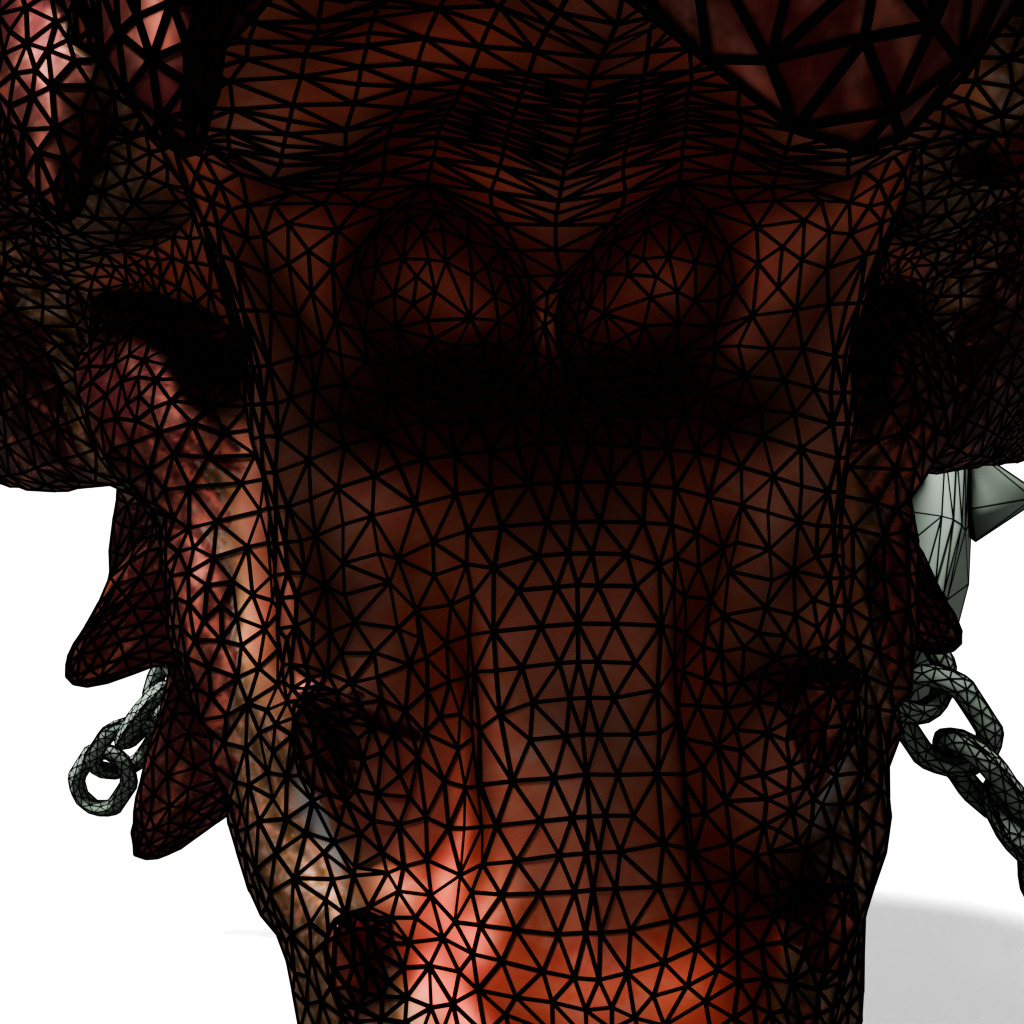}}
    \\
    $\square/\triangle$: 33306/1056 & 21236/27231 & 0/67668 \\
    {\small Chamfer/Hausdorff$^\downarrow$}: $\num{1.251e-4}/\num{1.336e-3}$ & $\num{1.941e-4}/\num{2.260e-3}$ & $\num{1.428e-4}/\num{1.686e-3}$ \\
    
    \put(-0.05\linewidth,0){\rotatebox{90}{25\% Total $\triangle$ Count}}
    \includegraphics[width=0.25\linewidth]{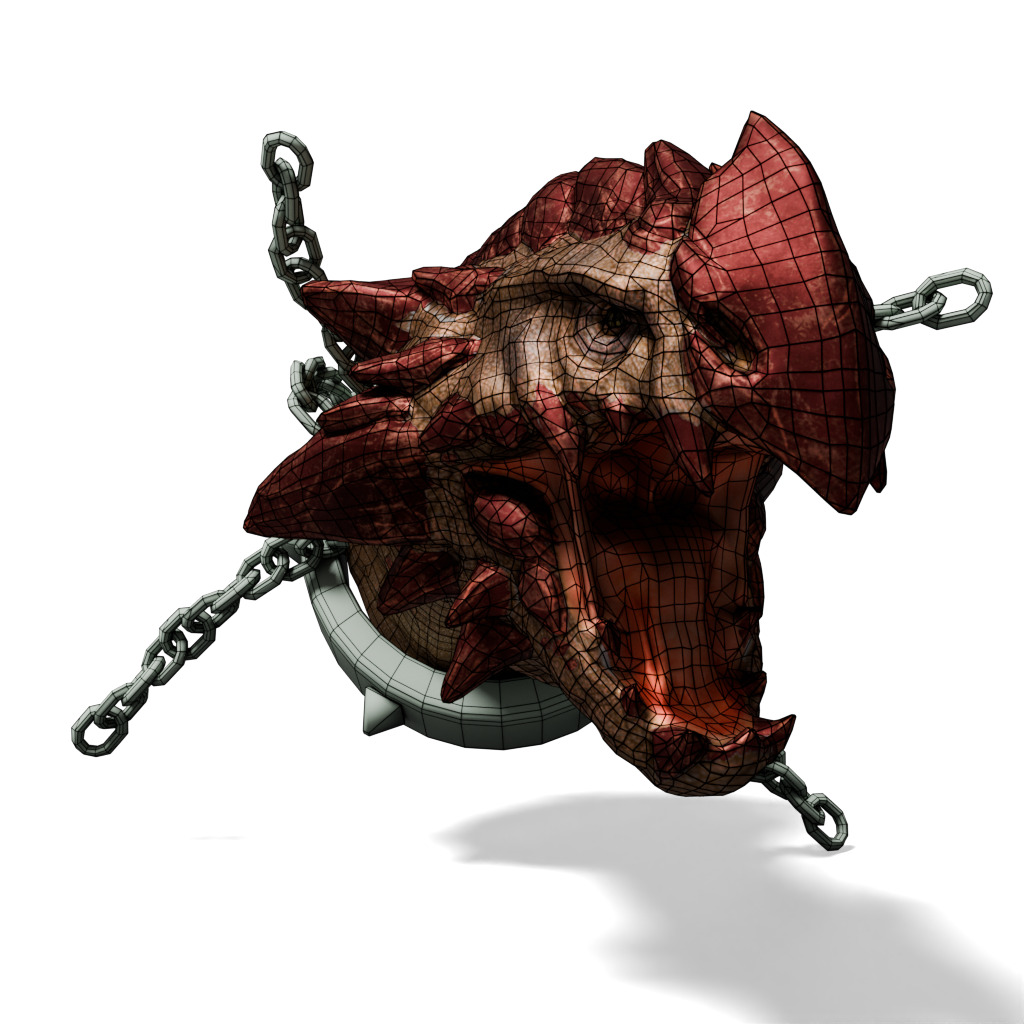}
    &
    \includegraphics[width=0.25\linewidth]{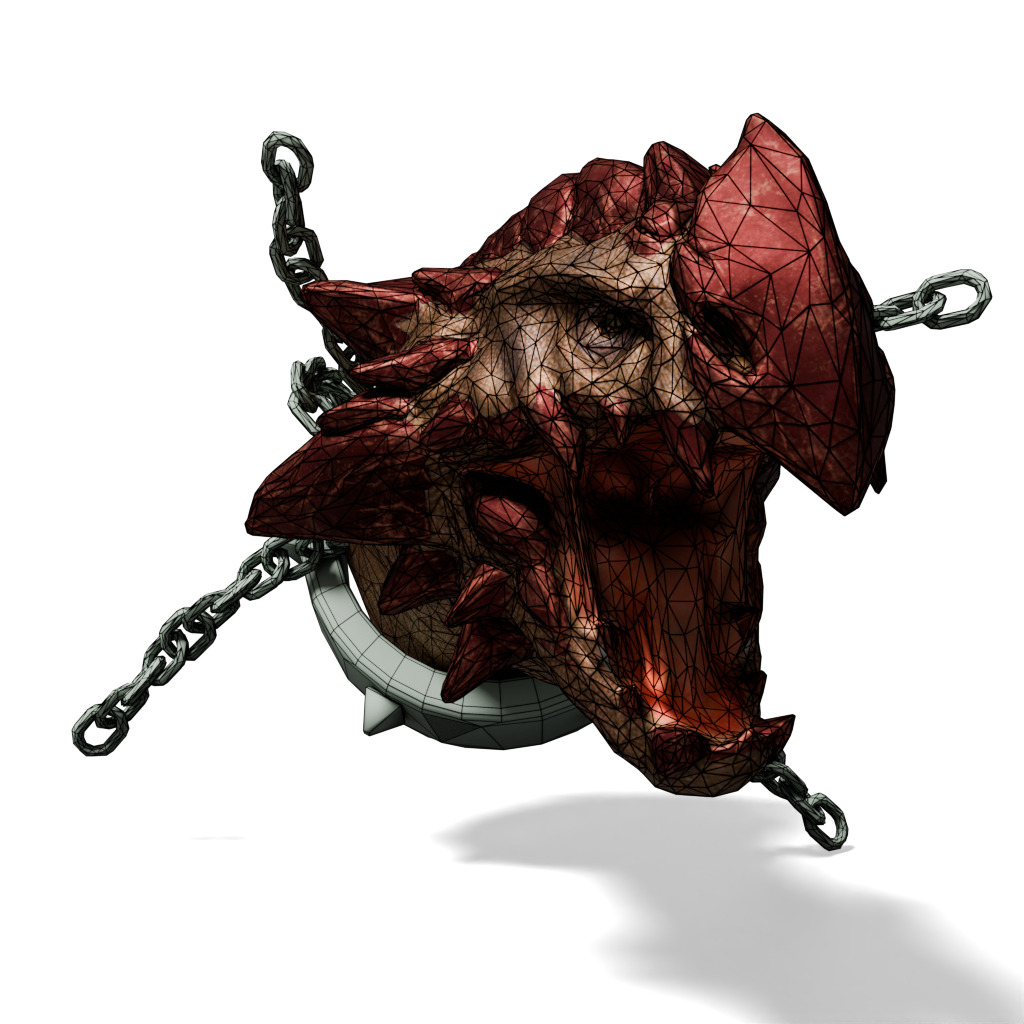}
    &
    \includegraphics[width=0.25\linewidth]{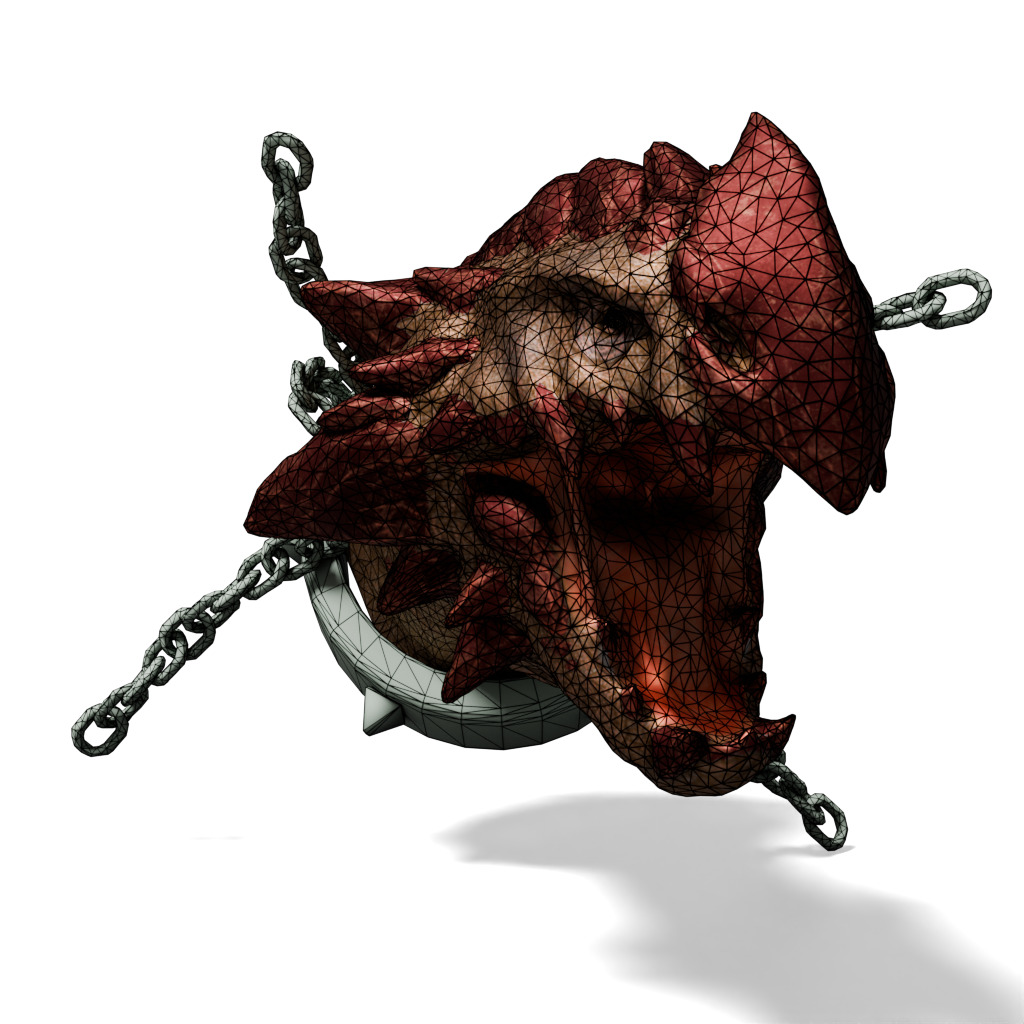}
    \\
    \frame{\includegraphics[width=0.1\linewidth]{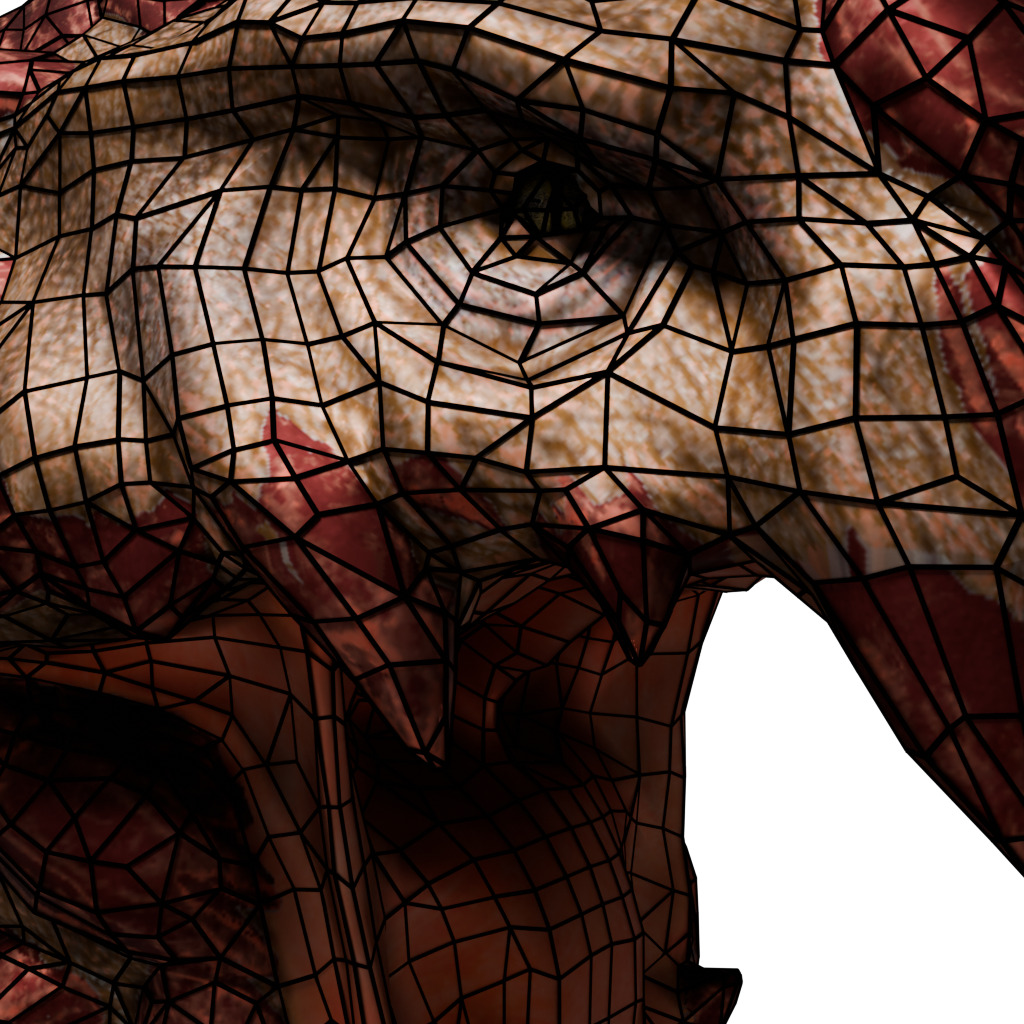}}
    \frame{\includegraphics[width=0.1\linewidth]{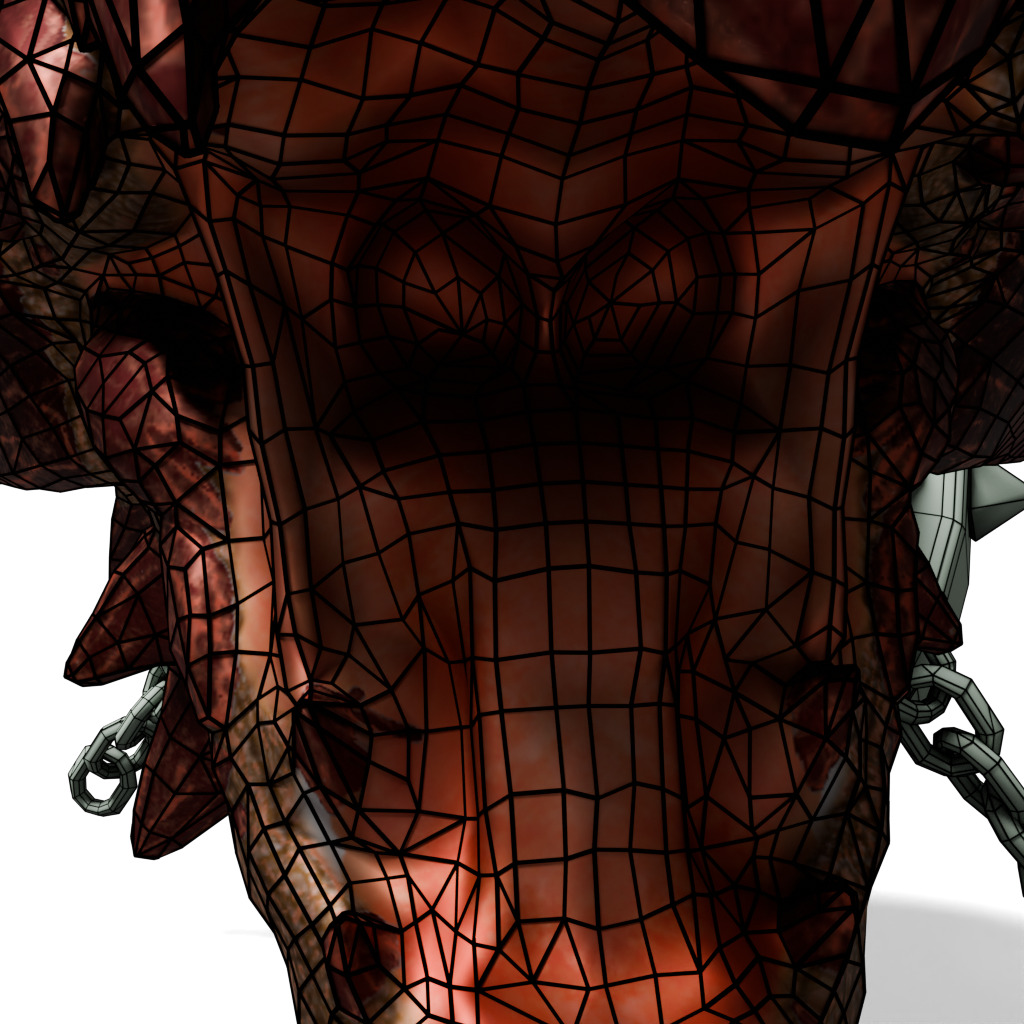}}
    &
    \frame{\includegraphics[width=0.1\linewidth]{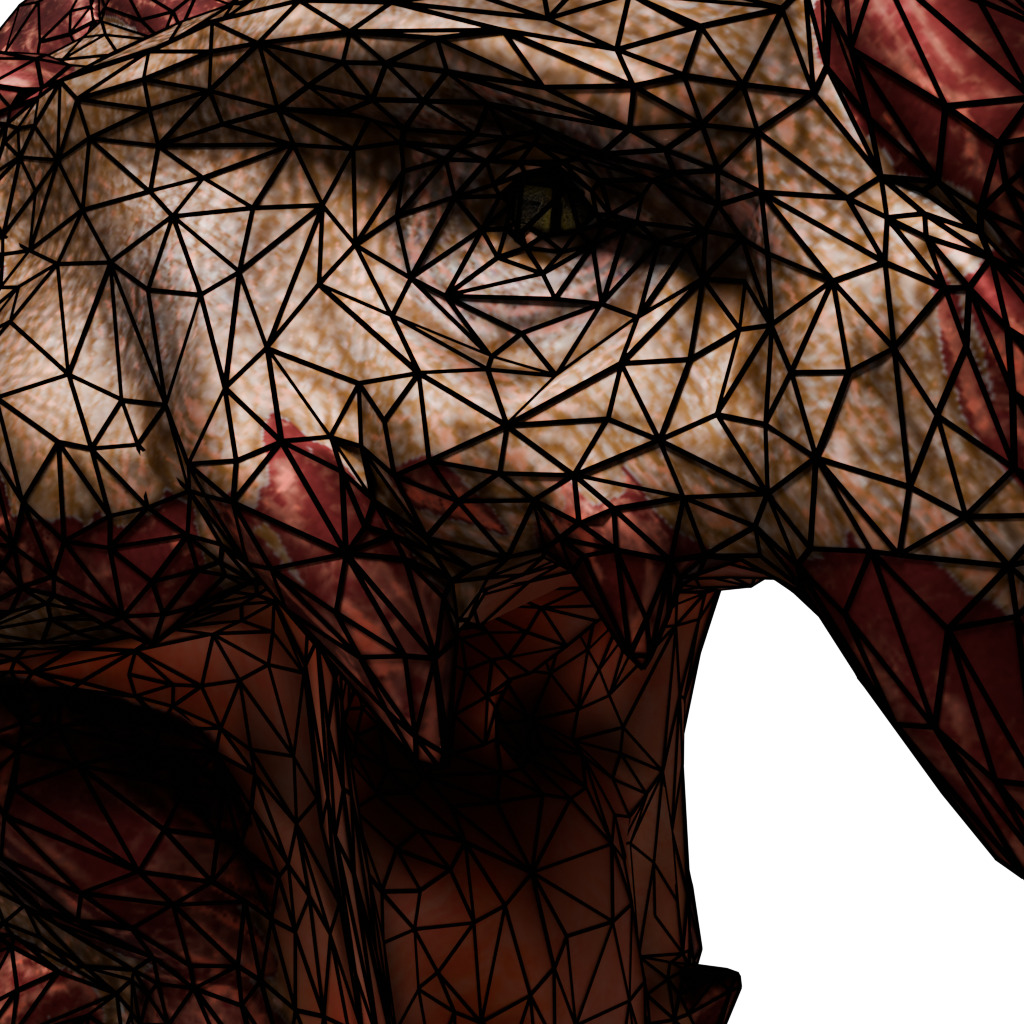}}
    \frame{\includegraphics[width=0.1\linewidth]{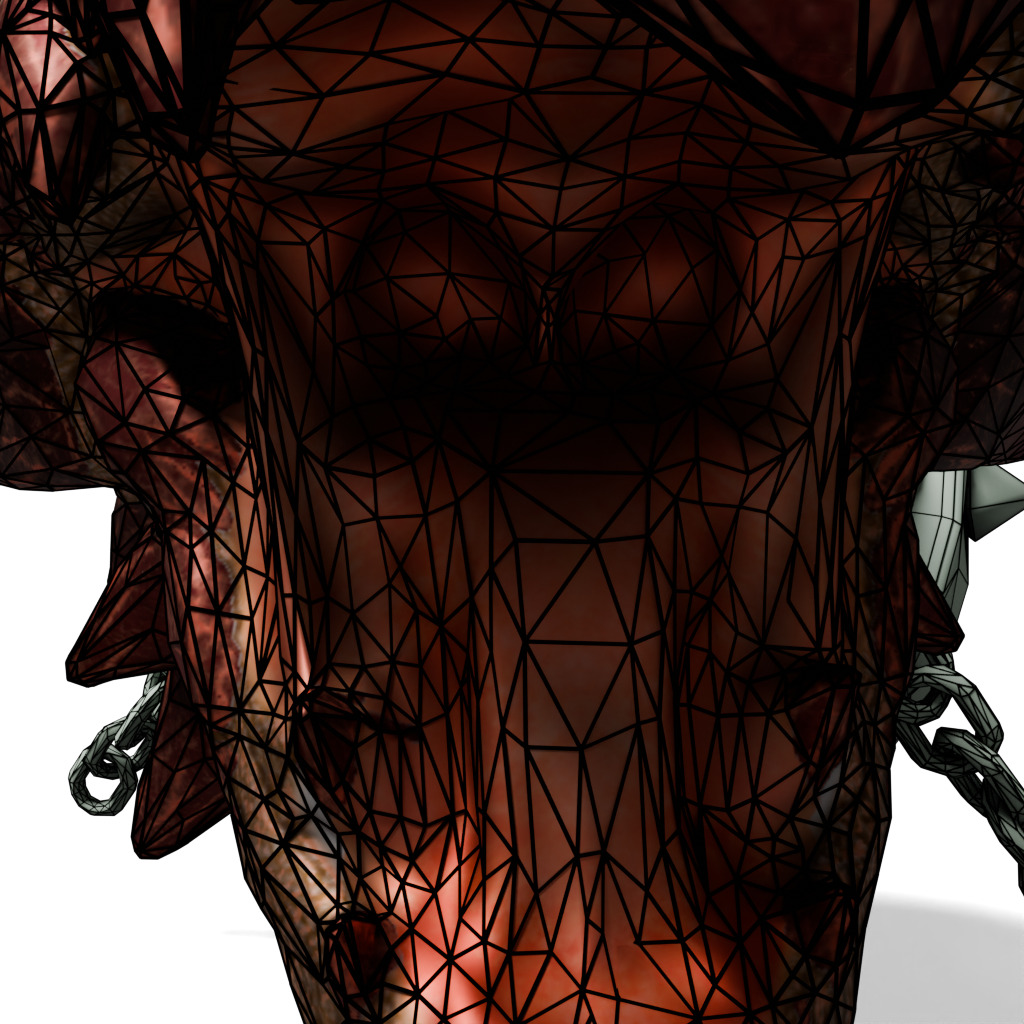}}
    &
    \frame{\includegraphics[width=0.1\linewidth]{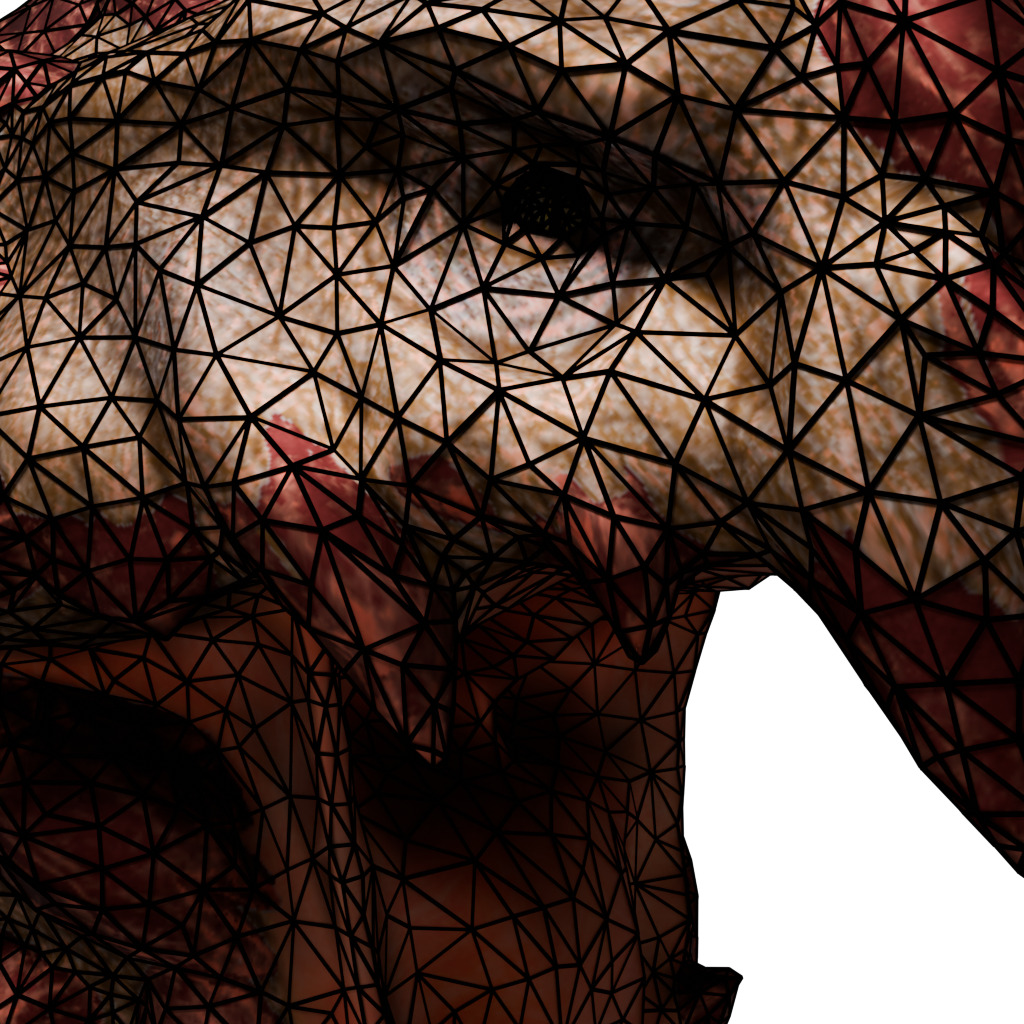}}
    \frame{\includegraphics[width=0.1\linewidth]{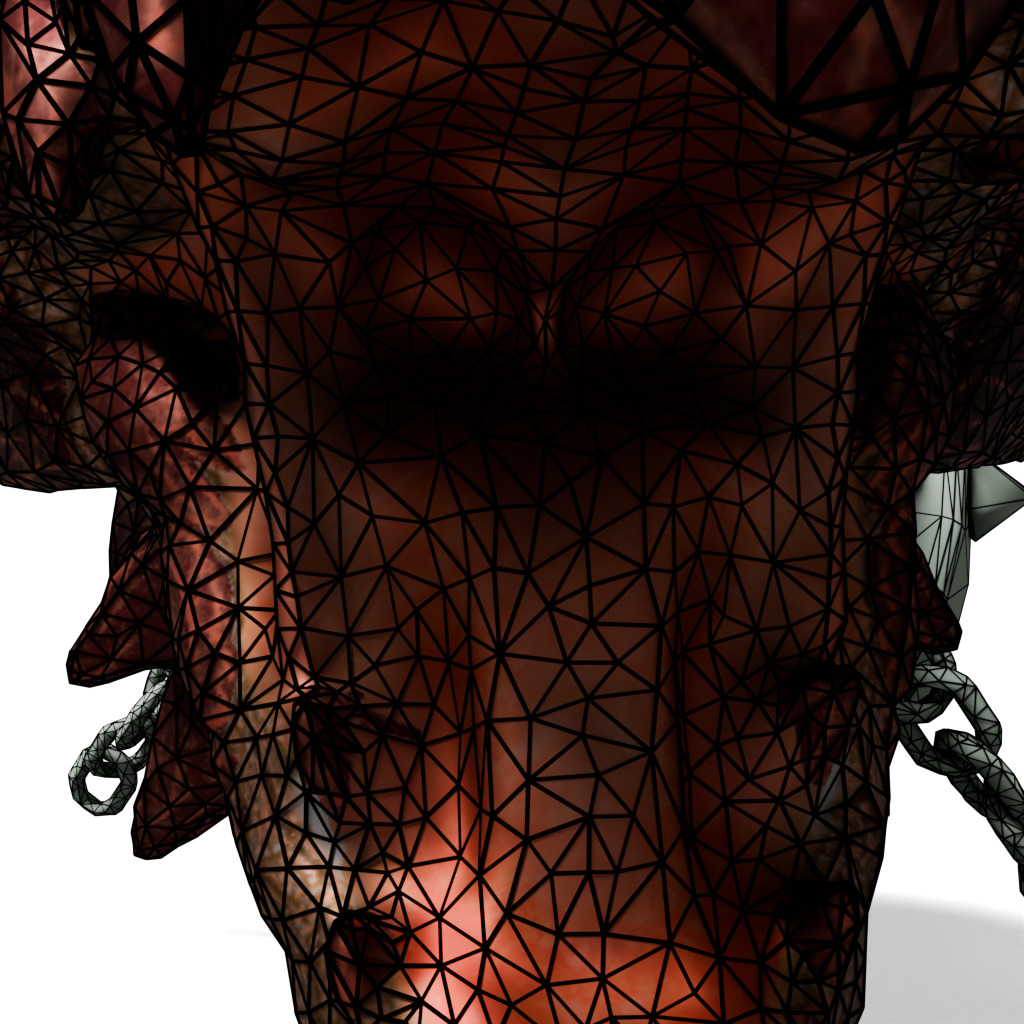}}
    \\
    $\square/\triangle$: 15600/2644 & 6062/22143 & 0/33844 \\
    {\small Chamfer/Hausdorff$^\downarrow$}: $\num{2.929e-4}/\num{1.577e-3}$ & $\num{5.193e-4}/\num{4.169e-3}$ & $\num{4.033e-4}/\num{2.651e-3}$ \\
    
    \put(-0.05\linewidth,0){\rotatebox{90}{10\% Total $\triangle$ Count}}
    \includegraphics[width=0.25\linewidth]{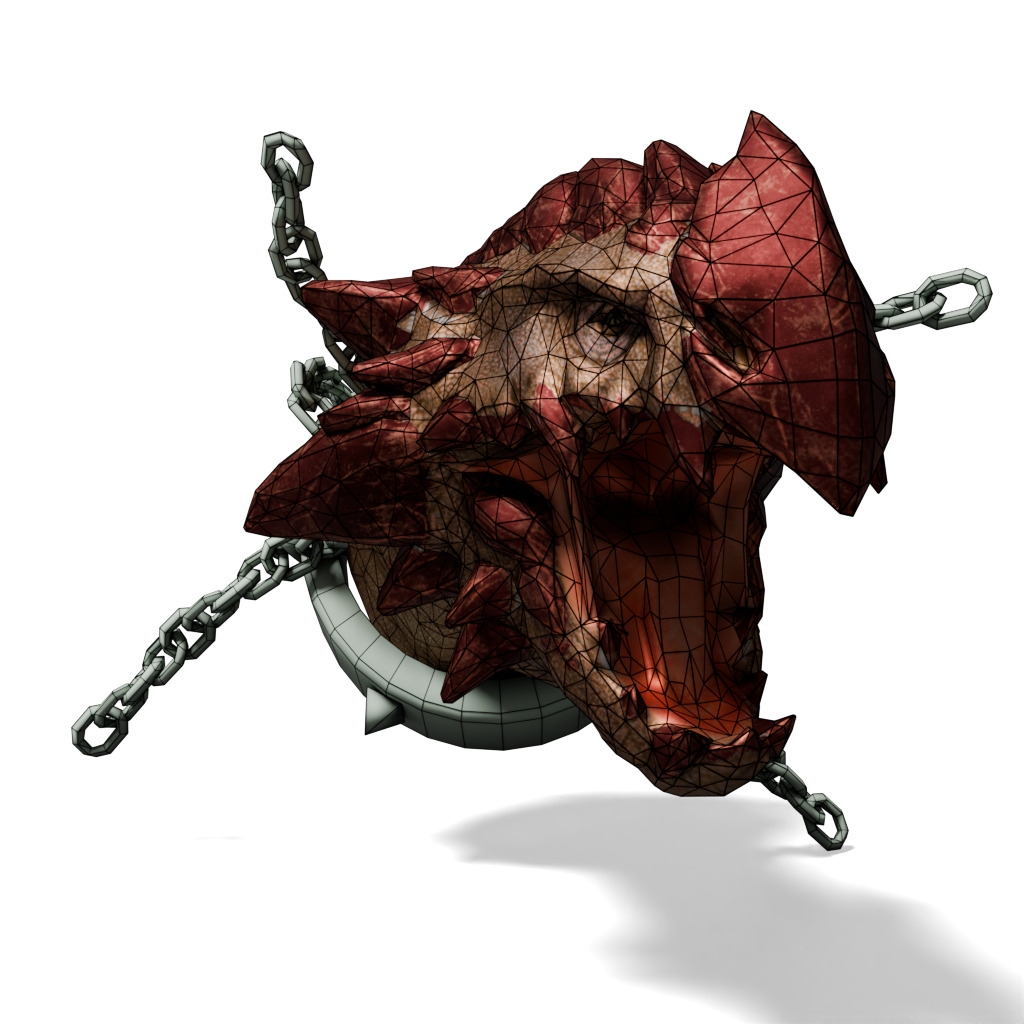}
    &
    \includegraphics[width=0.25\linewidth]{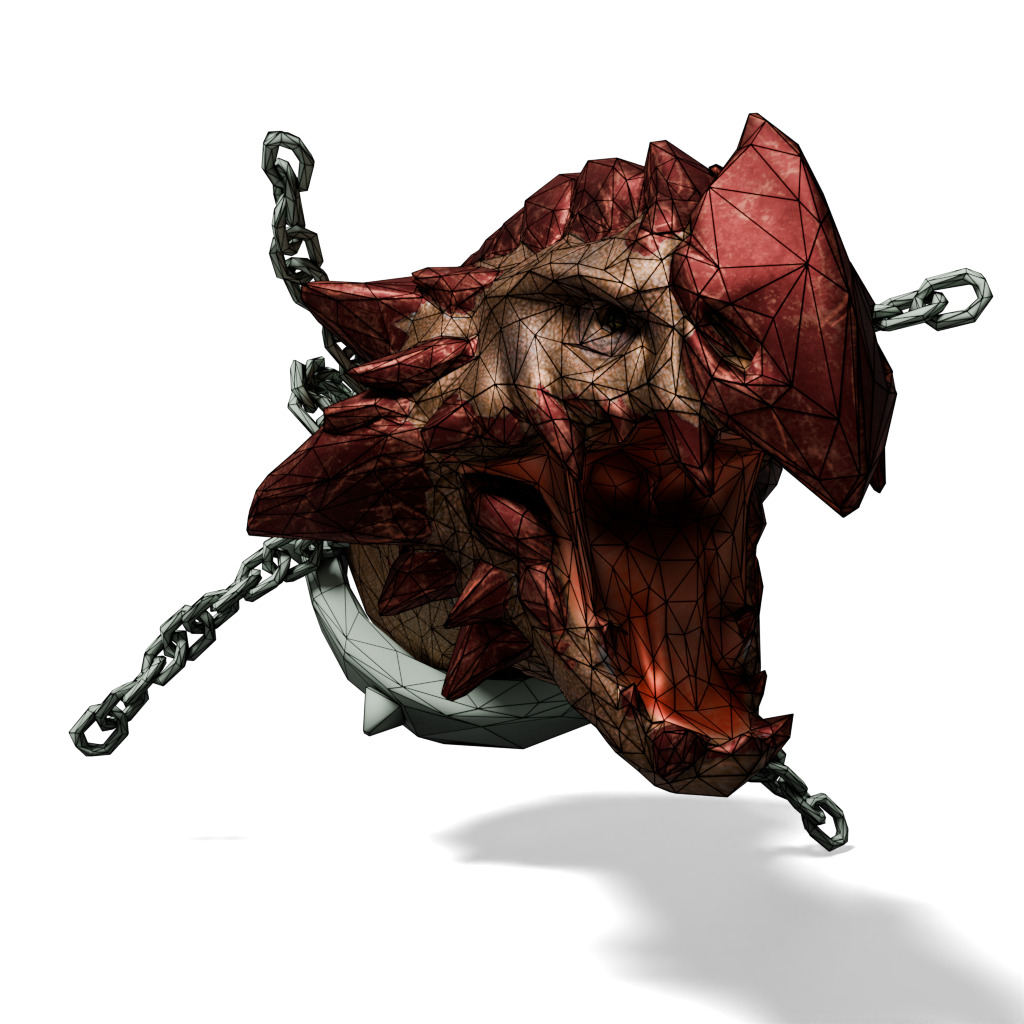}
    &
    \includegraphics[width=0.25\linewidth]{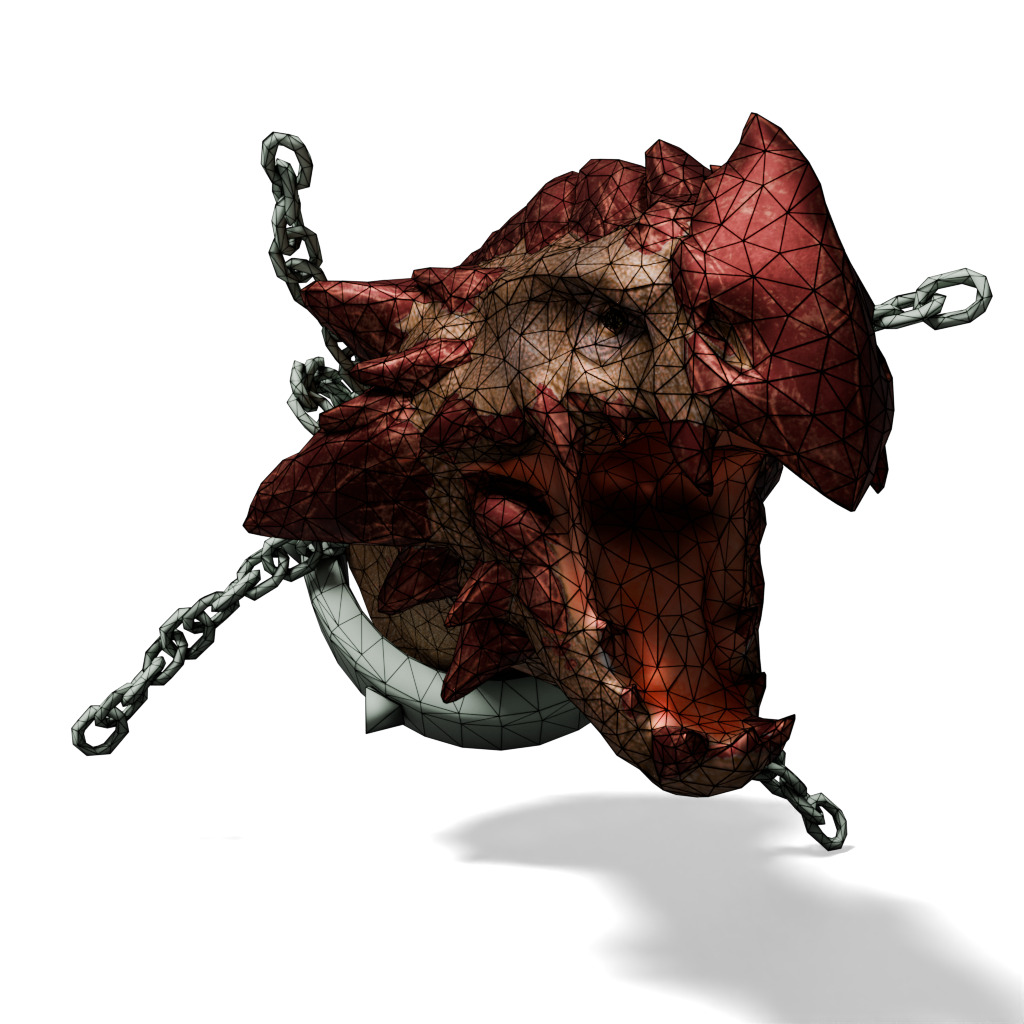}
    \\
    \frame{\includegraphics[width=0.1\linewidth]{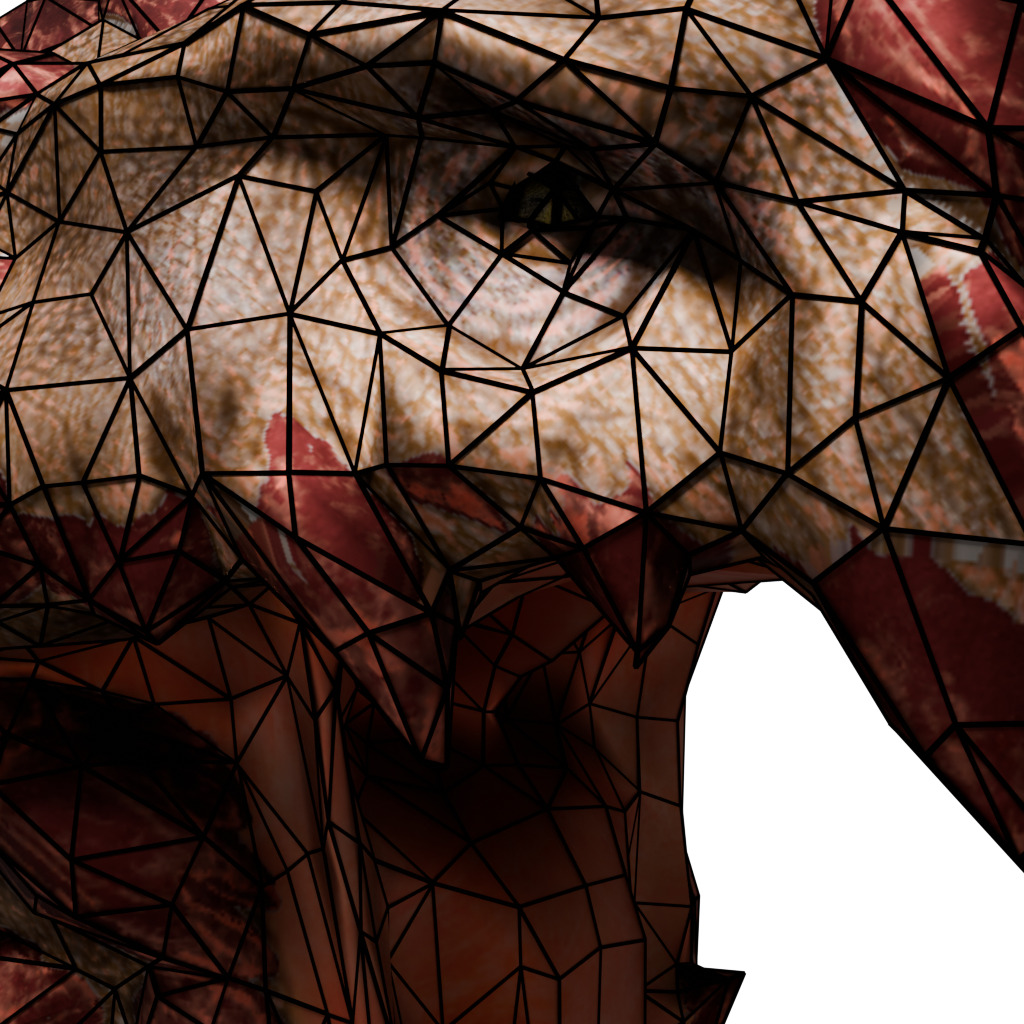}}
    \frame{\includegraphics[width=0.1\linewidth]{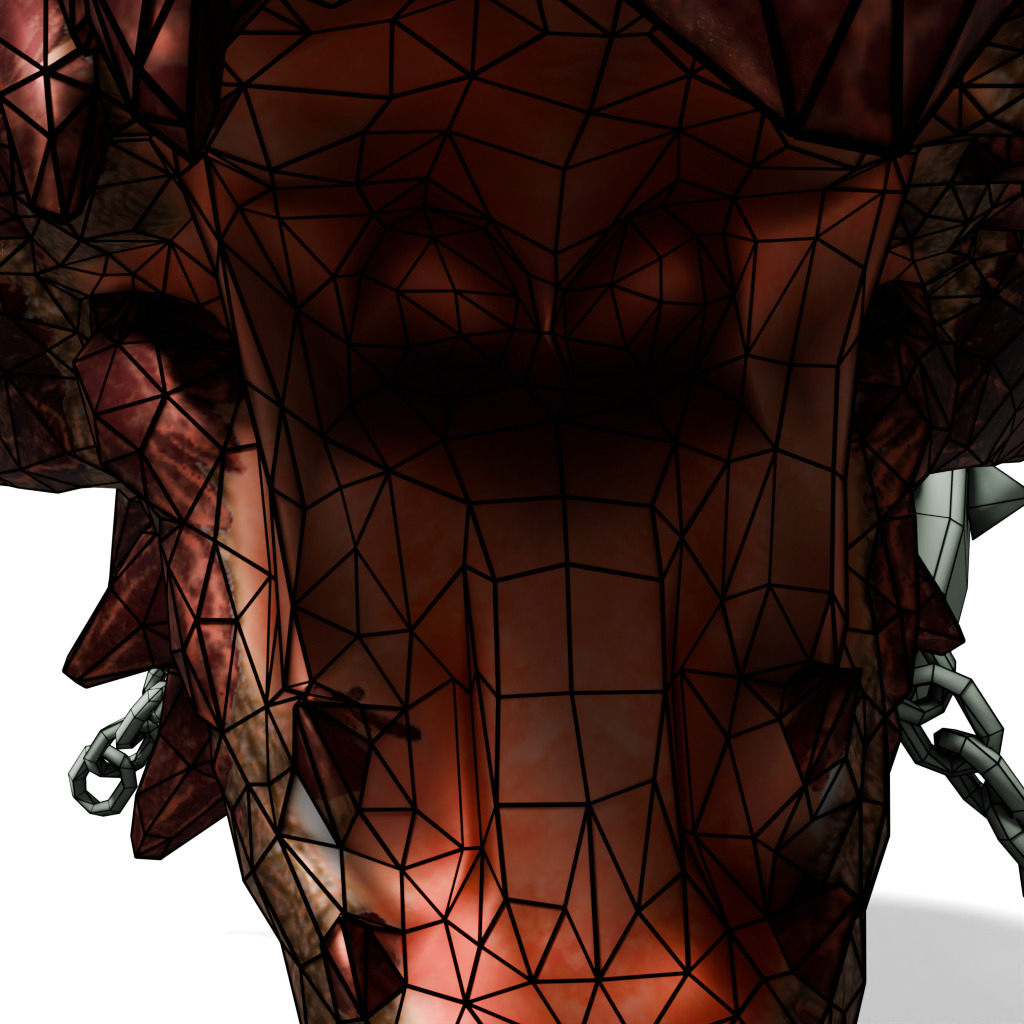}}
    &
    \frame{\includegraphics[width=0.1\linewidth]{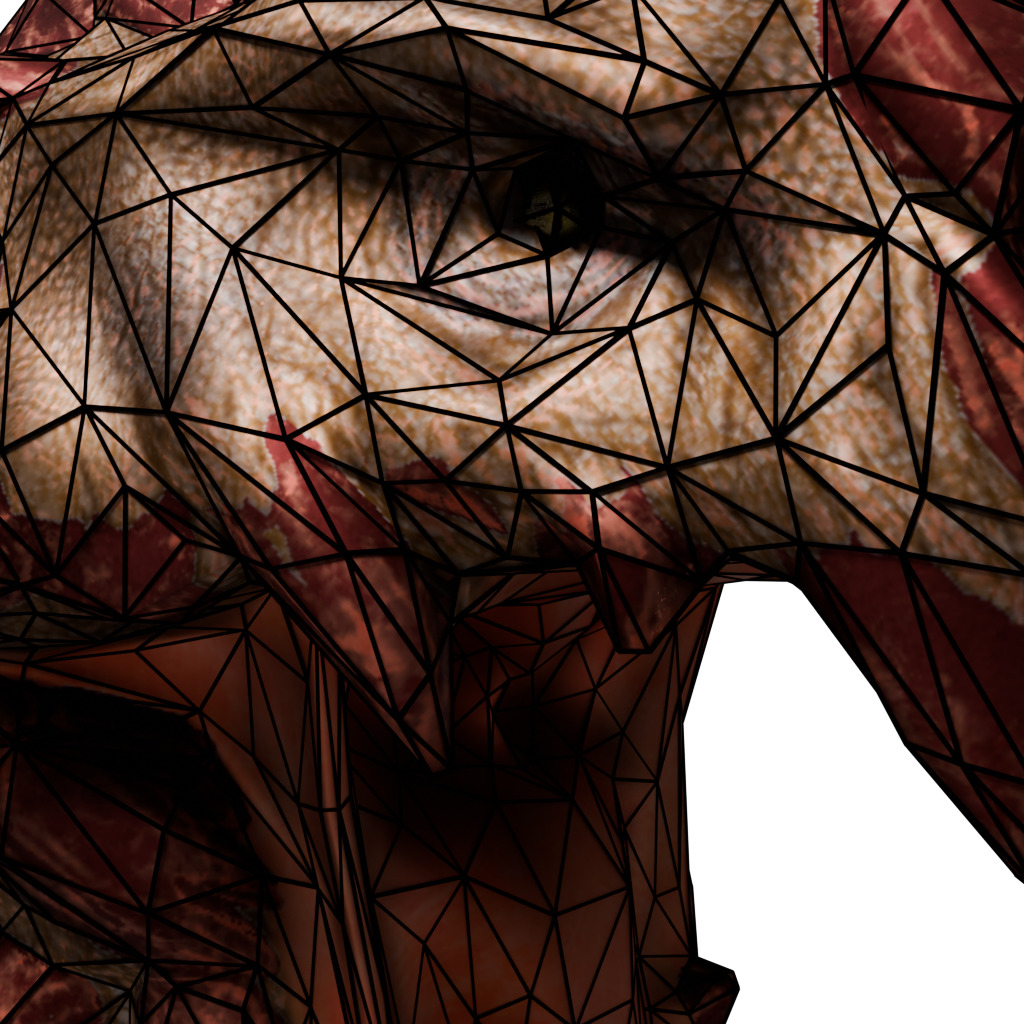}}
    \frame{\includegraphics[width=0.1\linewidth]{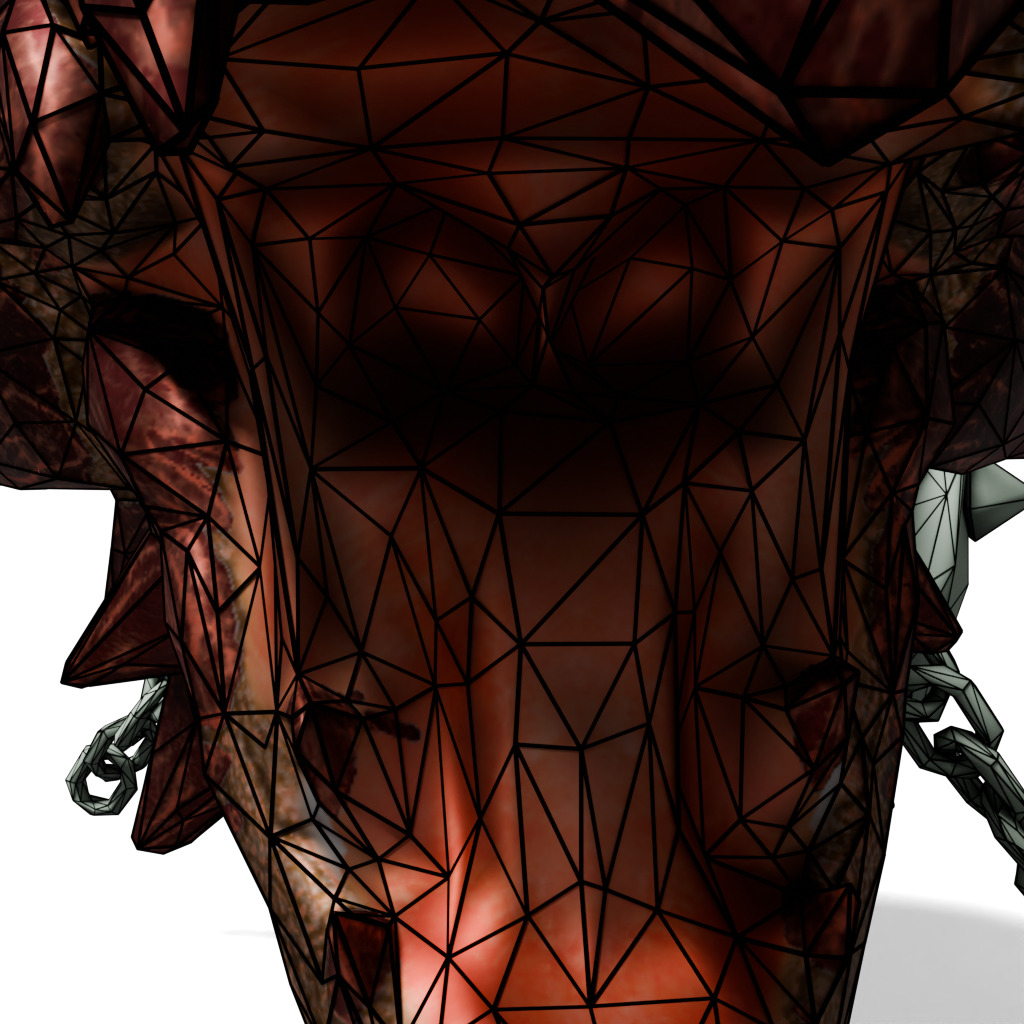}}
    &
    \frame{\includegraphics[width=0.1\linewidth]{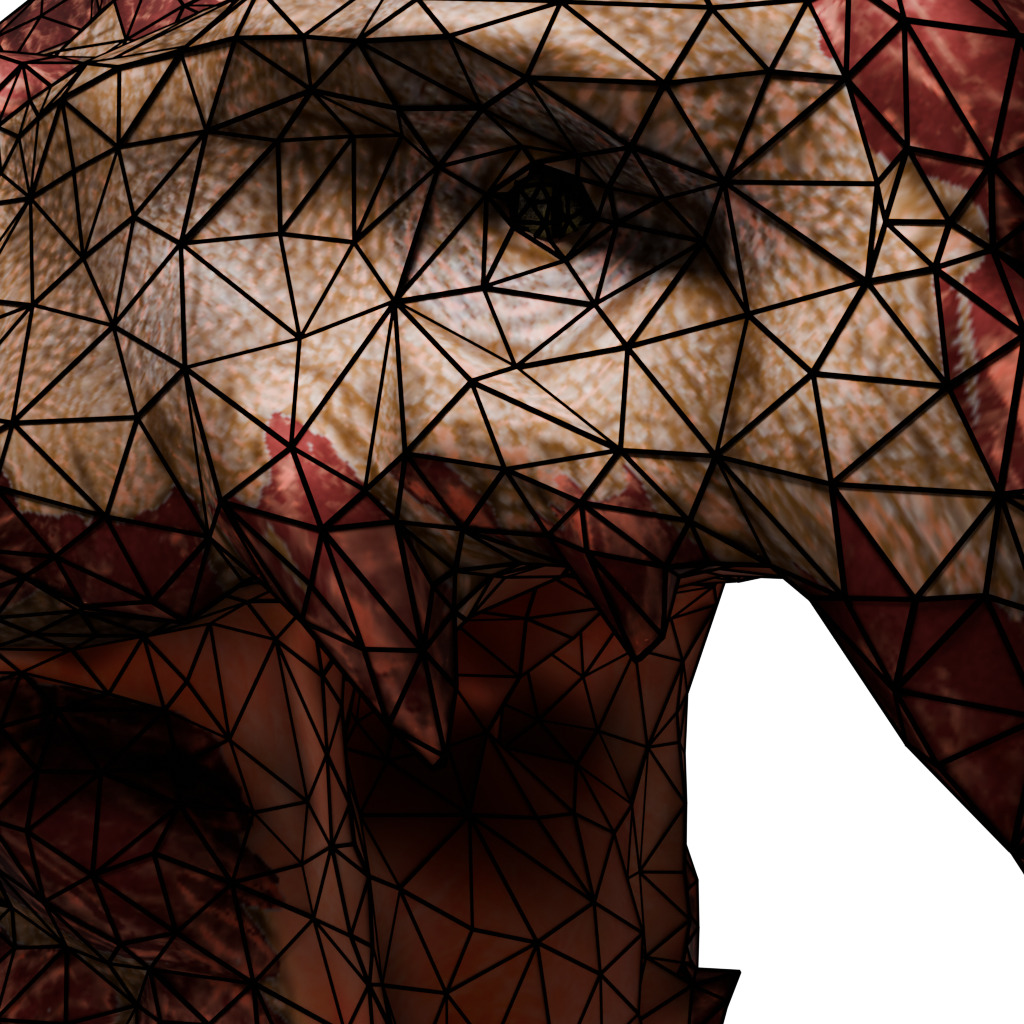}}
    \frame{\includegraphics[width=0.1\linewidth]{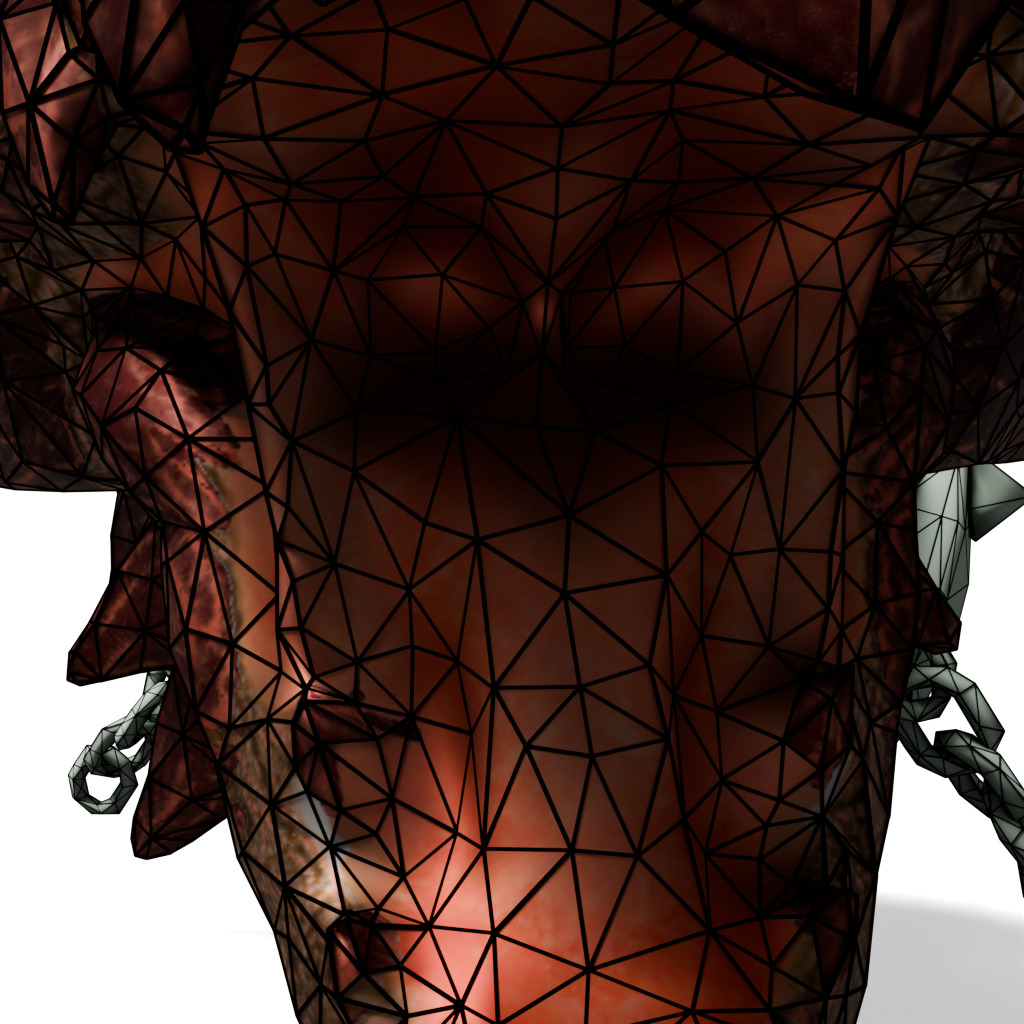}}
    \\
    $\square/\triangle$: 3912/5734 & 1031/11542 & 0/13558 \\
    {\small Chamfer/Hausdorff$^\downarrow$}: $\num{7.414e-4}/\num{2.836e-3}$ & $\num{1.162e-3}/\num{9.739e-3}$ & $\num{8.826e-4}/\num{6.936e-3}$ \\
\end{tabular}
\caption{\label{fig:qual-comp} Our approach compared to an implementation of ~\citep{qem_hoppe} (labeled QEM) and ~\citep{meshlab} on a quad dominant mesh. Our approach is able to preserve more quads, while also having higher geometric similarity than other approaches. \ccby ricocilliers.}
\Description{
Side-by-side comparison of a dragon head with chain at the base of the head. There are two insets, one on the area slightly underneath the eye, and one directly inside the mouth of the dragon. There is a wireframe mesh overlaid on top of each decimation.
}
\end{figure*}

\section{Experiments}

To test our approach, we gather a dataset of 67 static meshes and 19 animated meshes from Sketchfab~\citep{sketchfab} containing UV charts, linearly skinned animations, and per-vertex normals. There is a wide diversity of meshes: some pure quad meshes, some pure triangle meshes, and some hybrid meshes. Full statistics for input meshes are shown in the Appendix, Tab.~\ref{tab:dataset_summary}. We take care to respect the licensing of the artists, and note that our approach is \textit{not generative AI}, as some artists specifically label their art to not be included for training of generative AI.

We compare our approach to internal tooling used for decimation of meshes, labeled ``QEM'' in figures. This tool is a battle-tested implementation of ~\citep{qem_hoppe} and ~\citep{articulated_mesh_simplification}, with additional constraints to prevent degeneracies, and has been used for skinned and static meshes in downstream projects. It can preserve quads by triangulating immediately before decimation and labeling the introduced edges. After decimation, if any of the original labeled edges added during triangulation remain, they are removed and the quads recovered. We also compare to MeshLab's quadric decimation~\citep{meshlab} which implements~\citep{qem_hoppe}. \citep{meshlab} only works for pure triangle meshes and is unable to preserve joint influences, so we only test it on static meshes. In our experiments we compare each approach at triangle ratios of 50\%, 25\%, and 10\%, where triangle ratios are the number of the triangles in a triangulated version of the mesh. We do not compare against any pure quad decimation approach, as such an approach \textbf{would not work} on a large subset of our data, and there are no free, canonical, "compiles \& works out-of-the-box" implementations.

All animated meshes are in the GLB file format, which only supports triangles and no quads or polygons. Thus, we only compare geometric quality for those meshes, and do not compare quad preservation. MeshLab currently does not support animated mesh decimation, so we only compare QEM, which implements ~\citep{articulated_mesh_simplification}. We are shoehorned into using the GLB file format since it is an open standard and has widespread, proper support. Other animated mesh formats such as FBX often introduce subtle bugs due to their closed source, closed specification, and are designed to work within a closed ecosystem. For experiments with animated meshes, we also target triangle ratios of 50\%, 25\% and 10\%. In our experiments we fix $\lambda_\text{joint} = 1, \lambda_\text{sym} = 0, \epsilon_\text{abs} = \num{5e-6}$. We ablate $\lambda_\text{joint}$ and $\epsilon_\text{abs}$ later in Sec.\ref{sec:ablate-ljoint}. We set $\lambda_\text{sym} = 0$ and do not compute symmetric matchings since we assume that the majority of input meshes do not have symmetry.

Through our experiments we demonstrate a few properties:
\begin{enumerate}
    \item Our approach can preserve at least 30\% more quads than the baseline during decimation, measured by the ratio of output quads and triangles
    \item Our approach has higher geometric similarity than prior work as measured by the average Chamfer and Hausdorff distance across the entire dataset.
    \item Our approach can softly preserve topological extrinsic symmetry in meshes, measured qualitatively on a subset of meshes with symmetry.
    \item Using \citep{qem_hoppe} we are able to preserve animation quality better across animations as compared to ~\citep{articulated_mesh_simplification}.
\end{enumerate}

All experiments are run on an AMD Ryzen Threadripper 3970X 32-Core CPU without any GPU usage. QEM, MeshLab~\citep{meshlab} and our approach are single-threaded.

\section{Results}

\begin{table}[!h]
    \centering
    \setlength{\tabcolsep}{0.1pt}
    \sisetup{tight-spacing=true}
    \centering
    \begin{tabular}{|c| c|c|c|}
        \multicolumn{4}{c}{Static Mesh Geometric Summary Stats} \\
        \hline
        Metric & Ours & QEM & MeshLab \\\hline
        \multicolumn{4}{|c|}{50\% $\triangle$} \\\hline
        Avg. Chamfer$^\downarrow$ & \cellcolor{blue!10} $\num{1.775e-04}$ & $\num{2.881e-04}$ & $\num{2.661e-04}$ \\\hline
        Avg. Hausdorff$^\downarrow$ & $\num{7.692e-03}$ & $\num{6.081e-03}$ & \cellcolor{blue!10} $\num{4.977e-03}$ \\\hline
        Med. $\frac{\text{Out Quads}}{\text{Out Tris}} / \frac{\text{In Quads}}{\text{In Tris}}$$^\uparrow$ & \cellcolor{blue!10} 0.949 & 0.674 & \textcolor{red}{$\times$} \\\hline
        \multicolumn{4}{|c|}{25\% $\triangle$} \\\hline
        Avg. Chamfer$^\downarrow$ & \cellcolor{blue!10} $\num{5.830e-04}$ & $\num{7.950e-04}$ & $\num{9.485e-04}$ \\\hline
        Avg. Hausdorff$^\downarrow$ & \cellcolor{blue!10} $\num{1.067e-02}$ & $\num{1.069e-02}$ & $\num{1.221e-02}$ \\\hline
        Med. $\frac{\text{Out Quads}}{\text{Out Tris}} / \frac{\text{In Quads}}{\text{In Tris}}$$^\uparrow$ & \cellcolor{blue!10} 0.761 & 0.437 & \textcolor{red}{$\times$} \\\hline
        \multicolumn{4}{|c|}{10\% $\triangle$} \\\hline
        Avg. Chamfer$^\downarrow$ & \cellcolor{blue!10} $\num{1.811e-03}$ & $\num{2.066e-03}$ & $\num{3.160e-03}$ \\\hline
        Avg. Hausdorff$^\downarrow$ & \cellcolor{blue!10} $\num{1.970e-02}$ & $\num{2.135e-02}$ & $\num{2.366e-02}$ \\\hline
        Med. $\frac{\text{Out Quads}}{\text{Out Tris}} / \frac{\text{In Quads}}{\text{In Tris}}$$^\uparrow$ & \cellcolor{blue!10} 0.622 & 0.256 & \textcolor{red}{$\times$} \\\hline
    \end{tabular}
    \caption{Comparison of our approach to an implementation of QEM and MeshLab~\citep{meshlab}. Our approach has better geometric similarity to the original mesh, while also retaining more quads on median. MeshLab requires that the input mesh be fully triangulated and thus cannot preserve any quads.}
    \label{tab:summary-stats}
\end{table}
\paragraph{Quantitative} We compare our results to QEM and \citep{meshlab} on all static meshes. Summary statistics for our approach are shown in Tab.~\ref{tab:summary-stats}, which shows our approach has comparable or better Chamfer and Hausdorff distance to the baseline approaches while consistently retaining about 30\% more quads. \citep{meshlab} requires that the input mesh be triangulated and because of that is unable to preserve any quads. To visualize the distribution of geometric differences versus topology preservation, we plot the geometric quality against quad preservation of QEM and our approach in Fig.~\ref{fig:geom-vs-quad}, with outliers for both approaches removed from the visualization. Being further right on this chart indicates that the input has more quads, while being further down indicates higher geometric similarity. Our approach has a smaller variance of geometric similarity while having a noticeable increase in the number of quads preserved. This demonstrates that our implementation is able to more consistently preserve quads from the input mesh, with no trade-off in geometric quality. \citep{meshlab} is omitted from this graph since it cannot preserve any quads.

\begin{figure*}[ht!]
    \centering
    \begin{tabular}{c c c c}
         Input (Static Mesh) & \multicolumn{3}{c}{Resolution $\rightarrow$} \\
         \includegraphics[width=0.2\linewidth]{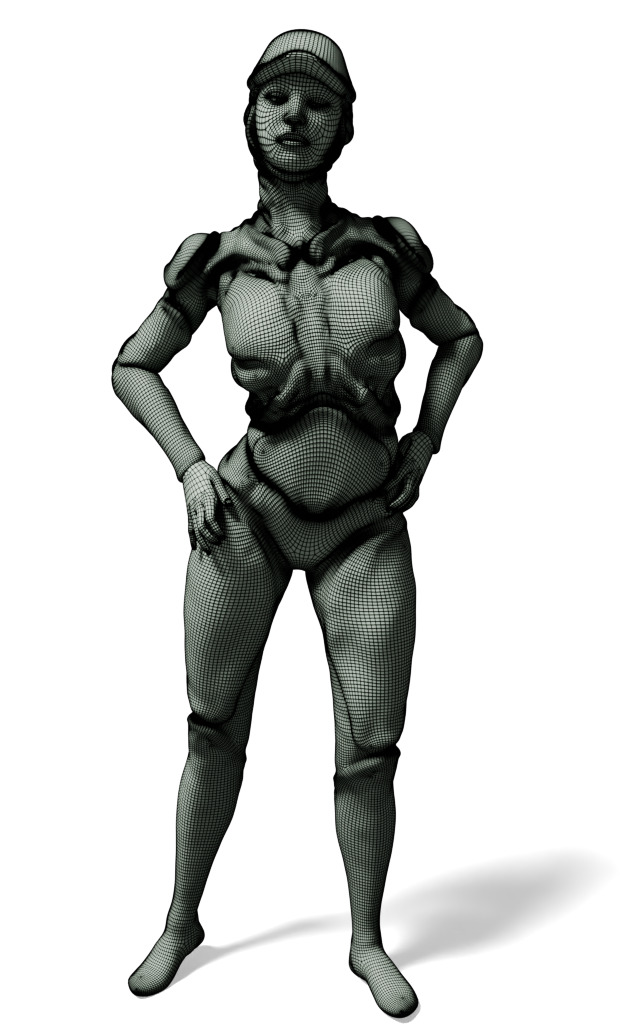} &
         \includegraphics[width=0.2\linewidth]{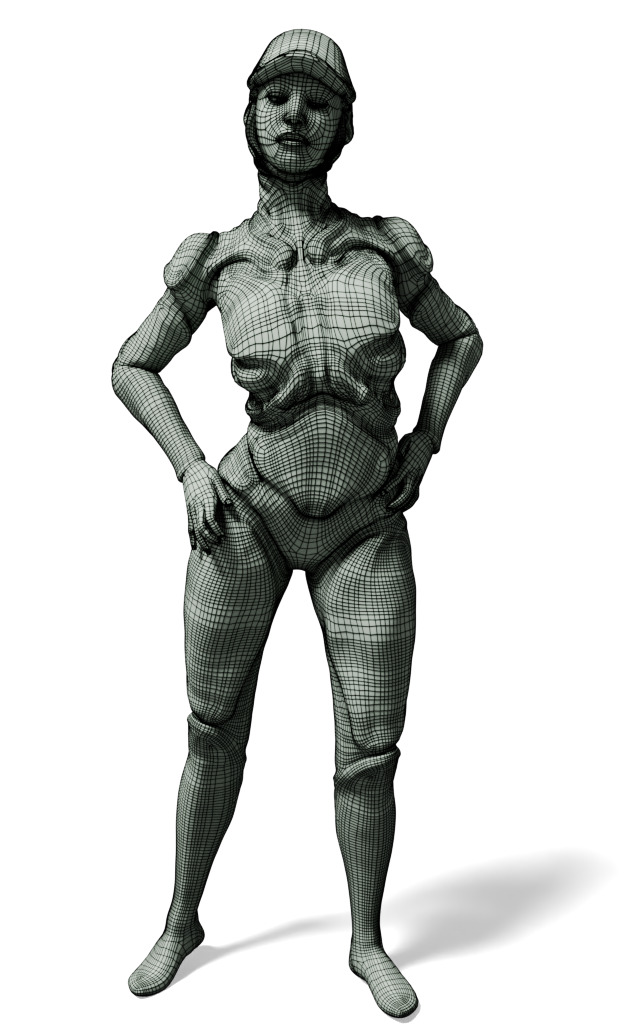} &
         \includegraphics[width=0.2\linewidth]{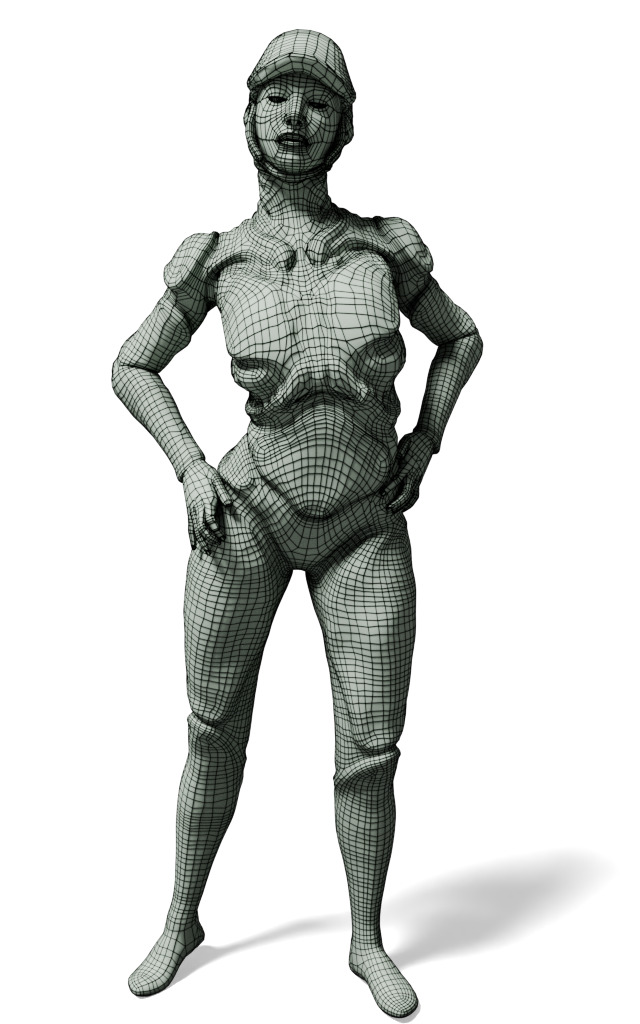} &
         \includegraphics[width=0.2\linewidth]{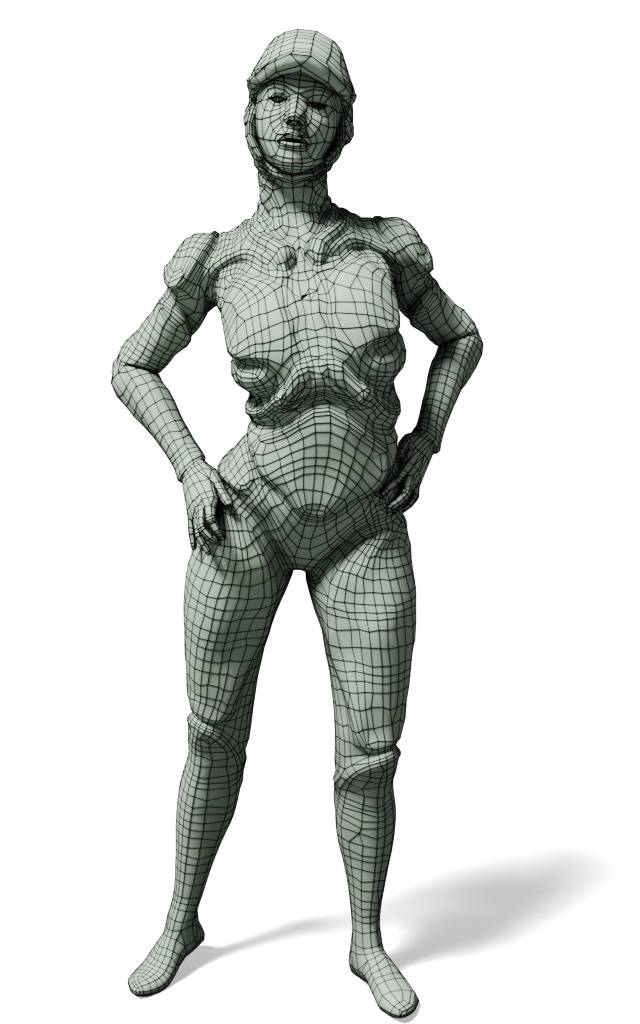} \\
        $\square$/$\triangle$: 97712/482 & 49264/576 & 24160/662 & 9377/845 \\
        Chamfer/Hausdorff$^\downarrow$: & \num{8.451e-5}/\num{1.058e-3} & \num{1.756e-4}/\num{1.303e-3} & \num{3.853e-4}/\num{2.600e-3} \\
    \end{tabular}
    \caption{Decimation of a hybrid quad mesh with many components at multiple resolutions. Our approach preserves most quads without sacrificing geometric quality using the edge collapse operator. Triangles are introduced between regions of varying quadric error, especially at lower resolutions. \ccbync Ploobert.}
    \label{fig:example-output}
    \Description{Example output mesh from our approach at multiple resolutions. As the resolution of the mesh decreases, the original feature lines still appear on the output mesh.}
\end{figure*}

\begin{figure}[!h]
    \centering
    \setlength{\tabcolsep}{0.1pt}
    \begin{tabular}{c c c}
         50\%$\triangle$ & 25\%$\triangle$ & 10\%$\triangle$ \\
         \includegraphics[width=0.33\linewidth]{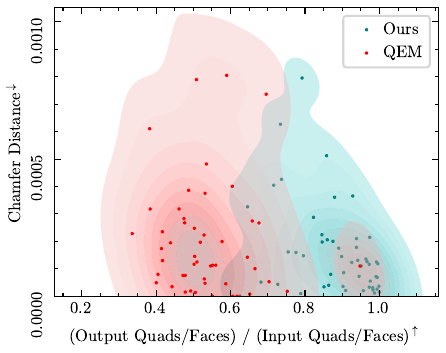} &
         \includegraphics[width=0.33\linewidth]{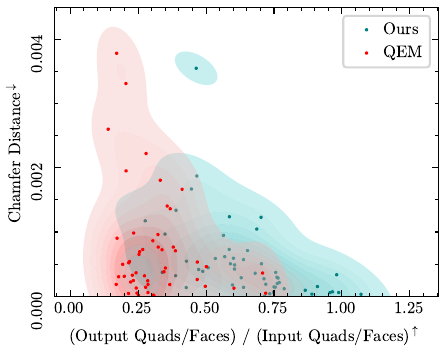} &
         \includegraphics[width=0.33\linewidth]{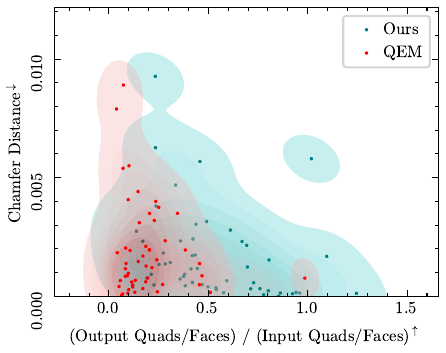} \\
         
         \includegraphics[width=0.33\linewidth]{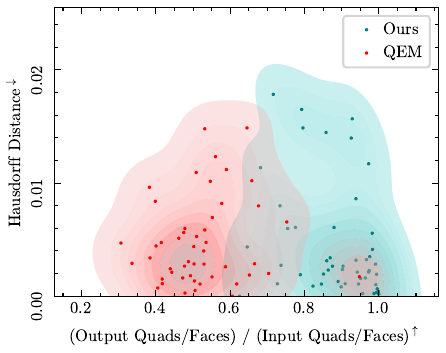} &
         \includegraphics[width=0.33\linewidth]{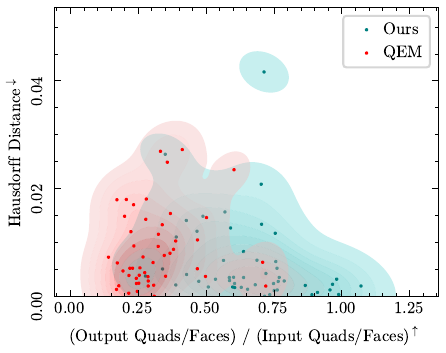} &
         \includegraphics[width=0.33\linewidth]{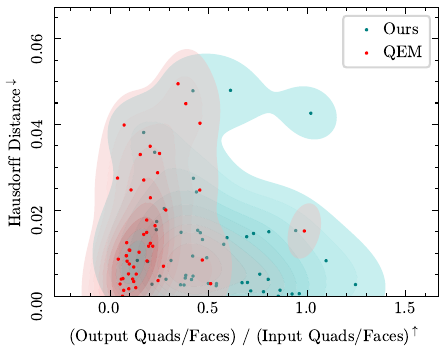} \\
    \end{tabular}
    \caption{Visualization of quad preservation against geometric quality for our approach and ``QEM'' for static meshes with one or more quads. Our approach preserves more quads and has higher geometric fidelity at all decimation levels. MeshLab~\citep{meshlab} is not included since it cannot preserve any quads.}
    \label{fig:geom-vs-quad}
    \Description{3 scatter plots overlapped on kernel density estimation charts, comparing our approach to QEM on geometric preservation and quad quality. Our approach generally falls to the right of QEM, and has higher density regions with lower chamfer regions.}
\end{figure}

\paragraph{Qualitative Comparison} We visually compare our approach to QEM and \citep{meshlab} in Fig.~\ref{fig:qual-comp} for a single static mesh with UVs and texture at multiple resolutions. The input is a quad-dominant character mesh, which contains 67632 quads and 64 triangles. At 50\% of the input total triangle count (2$\times$ \#quads + \#triangles), our approach visually looks to be a quad-dominant mesh, and has 96\% of its faces as quads. One example region where our approach preserves more quads is the nose where our approach retains the original quad structure from the input at 50\% and 25\% triangle ratios. In contrast, QEM and ~\citep{meshlab} destroy the quad topology of the nose. Meshlab preserves some of the input structure, but since it requires that the input be triangulated it is harder to tell. At 25\% of the input triangle count, our approach is able to preserve most quads (86\%), and introduces some triangles between regions of varying quadric error. QEM and \citep{meshlab}, on the other hand at this point have removed most of the input topology, except around the collar. At 10\% of the input total triangle count, our approach introduces a larger number of triangles, with 41\% of all faces being quads. Our approaches introduces these triangle as pure quad preservation would sacrifice geometric quality. While there are still some quad elements in our approach, in QEM or \citep{meshlab} it is difficult to recognize any of the input topology.

We also show an example of only our approach at different resolutions on another hybrid quad mesh in Fig.~\ref{fig:example-output}, where we have increased $\epsilon_\text{abs}$ to $\num{1e-4}$, leading to higher quad preservation. Our approach preserves the original quad topology at all levels of reduction, with visible quad structure at every step.

\paragraph{Animated Meshes} To compare our approach on animated meshes, we decimate each animated mesh to $50\%, 25\%, \text{ and } 10\%$ of the input mesh's triangle count. We then compute the Chamfer and Hausdorff distance at the first 50 frames of the first animation for each mesh. Plots of frame number against geometric similarity for each model are shown in the Appendix, Fig.~\ref{fig:anim-mesh-comparison}. We find that across the dataset, our approach outperforms QEM on both Hausdorff and Chamfer distances for most meshes. Specifically, we find that adapting \citep{qem_hoppe} to preserve joint influences outperforms ~\citep{articulated_mesh_simplification} on the majority of models for all frames, as shown in Tab.~\ref{tab:anim-mesh-summary-stats}. In Tab.~\ref{tab:anim-mesh-summary-stats}, we count the number of models where the average is lower across the first 50 frames. We also show a visual comparison of animated mesh reduction in the Appendix, Fig.~\ref{fig:anim-mesh-visual}.

\begin{table}[!h]
    \centering
    \begin{tabular}{|c|c|c|}
        \multicolumn{3}{c}{Skinned Mesh Reduction Geometric Stats} \\
        \hline
        Metric & QEM & Ours \\\hline
        \multicolumn{3}{|c|}{50\% $\triangle$} \\\hline
        \#Lower Avg. Chamfer & 1 & \cellcolor{blue!10} 18 \\\hline
        \#Lower Avg. Hausdorff & 5 & \cellcolor{blue!10} 14 \\\hline
        \multicolumn{3}{|c|}{25\% $\triangle$} \\\hline
        \#Lower Avg. Chamfer & 3 & \cellcolor{blue!10} 16 \\\hline
        \#Lower Avg. Hausdorff & 5 & \cellcolor{blue!10} 14 \\\hline
        \multicolumn{3}{|c|}{25\% $\triangle$} \\\hline
        \#Lower Avg. Chamfer & 1 & \cellcolor{blue!10} 18 \\\hline
        \#Lower Avg. Hausdorff & 5 & \cellcolor{blue!10} 14 \\\hline 
    \end{tabular}
    \caption{Number of animated models where QEM (which implements \citep{articulated_mesh_simplification}) versus our approach has lower average geometric distance on the first 50 frames. Our approach, which uses ~\citep{qem_hoppe} to preserve joint influence consistently outperforms the baseline. Full plots showing the Chamfer and Hausdorff distance per frame are shown in the Appendix, Fig.~\ref{fig:anim-mesh-comparison}.}
    \label{tab:anim-mesh-summary-stats}
\end{table}

\subsection{Symmetry Preservation}
\begin{figure}[h!]
    \setlength{\tabcolsep}{0.5pt}
    \centering
    \begin{tabular}{c c c}
         Input & QEM & Ours \\
         \includegraphics[width=0.33\linewidth]{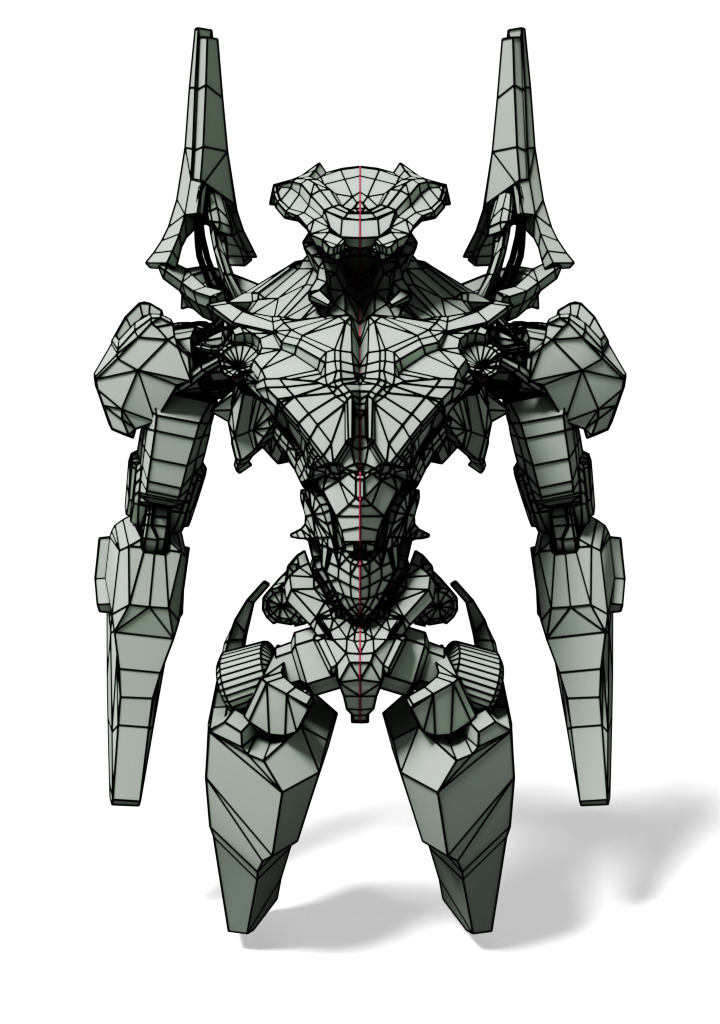} &
         \includegraphics[width=0.33\linewidth]{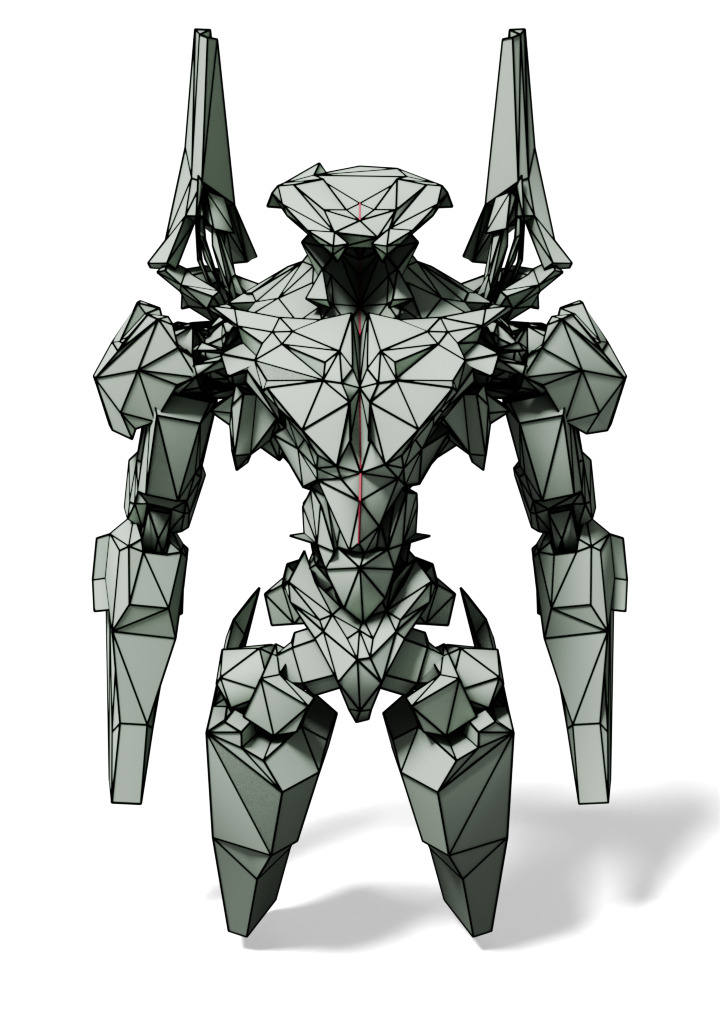} &
         \includegraphics[width=0.33\linewidth]{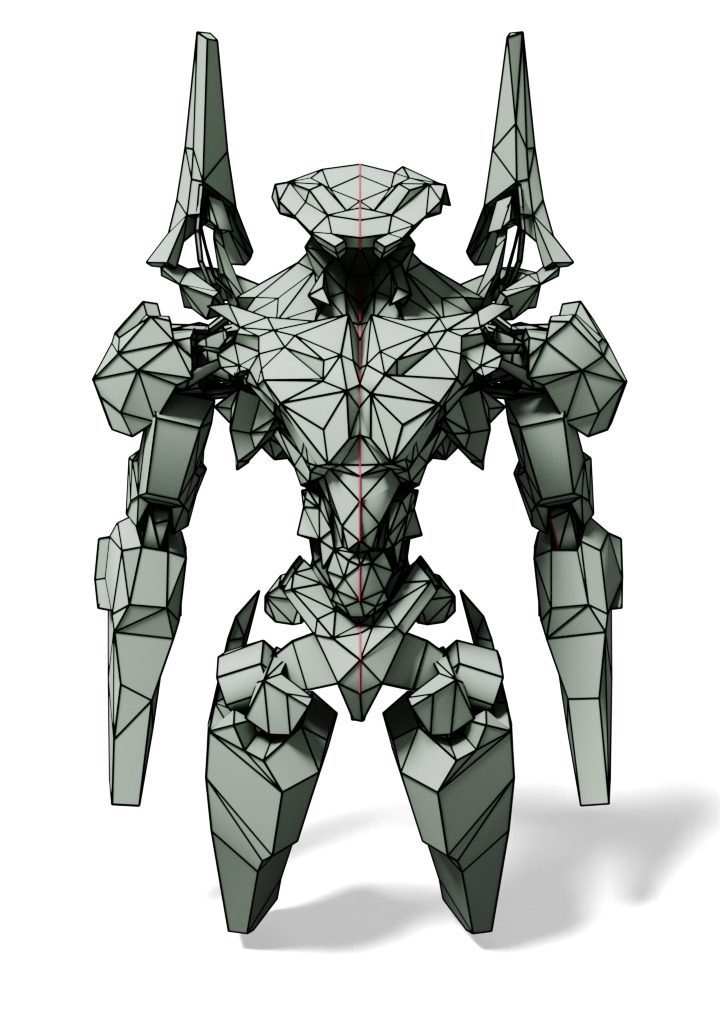}
         \\
        $\square/\triangle$: 5850/2872 & 511/1971 & 670/1590 \\
        Chamfer$^\downarrow$: & $\num{1.822e-3}$ & $\num{1.944e-3}$ \\
        Hausdorff$^\downarrow$: & $\num{1.448e-2}$ & $\num{1.315e-2}$ \\
    \end{tabular}
    \caption{As compared to normal QEM, our approach is able to preserve more topological symmetry, such as on the head of this mesh. The key axis of symmetry is shown in \textcolor{red}{red} for each mesh. \ccby KhoaMinh. }
    \label{fig:mech-symmetry-fig}
    \Description{Left: Input mesh, which is a robot mech, symmetric across its center axis. Middle: Robot mech decimated with QEM, but the symmetric axis is not preserved. Right: mesh decimated with our approach, with axis of symmetry preserved.}
\end{figure}

\begin{figure}[h!]
    \setlength{\tabcolsep}{0.5pt}
    \centering
    \begin{tabular}{c c c}
        Input & No Symmetry & Symmetry \\
        \includegraphics[width=0.33\linewidth]{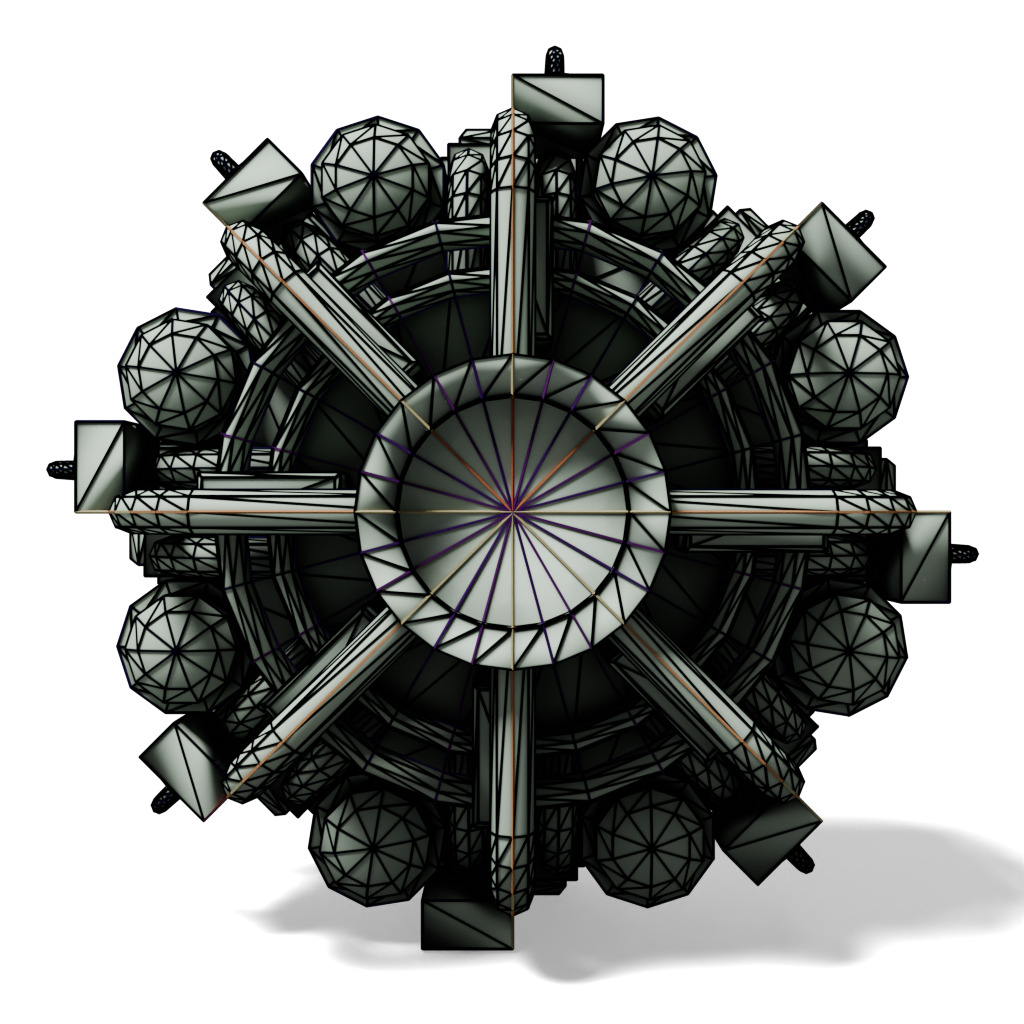} &
        \includegraphics[width=0.33\linewidth]{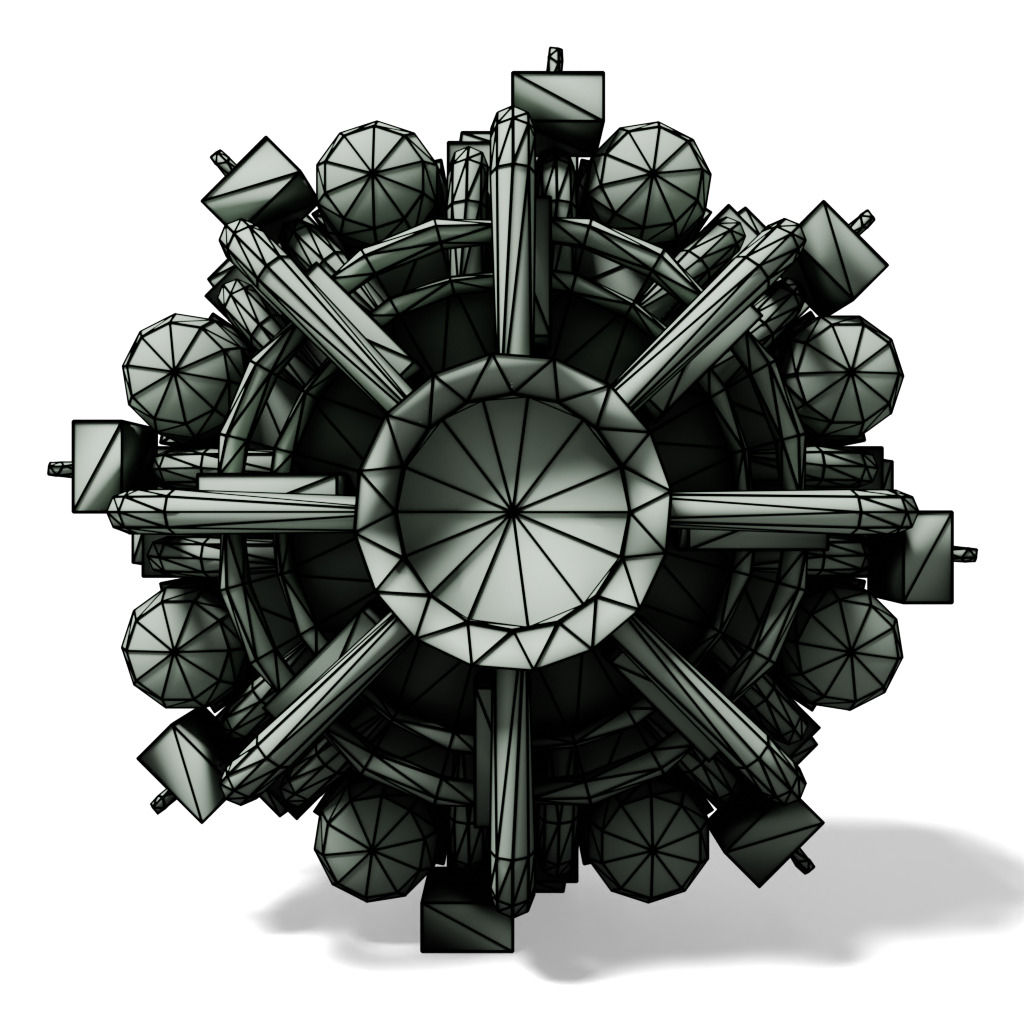} &
        \includegraphics[width=0.33\linewidth]{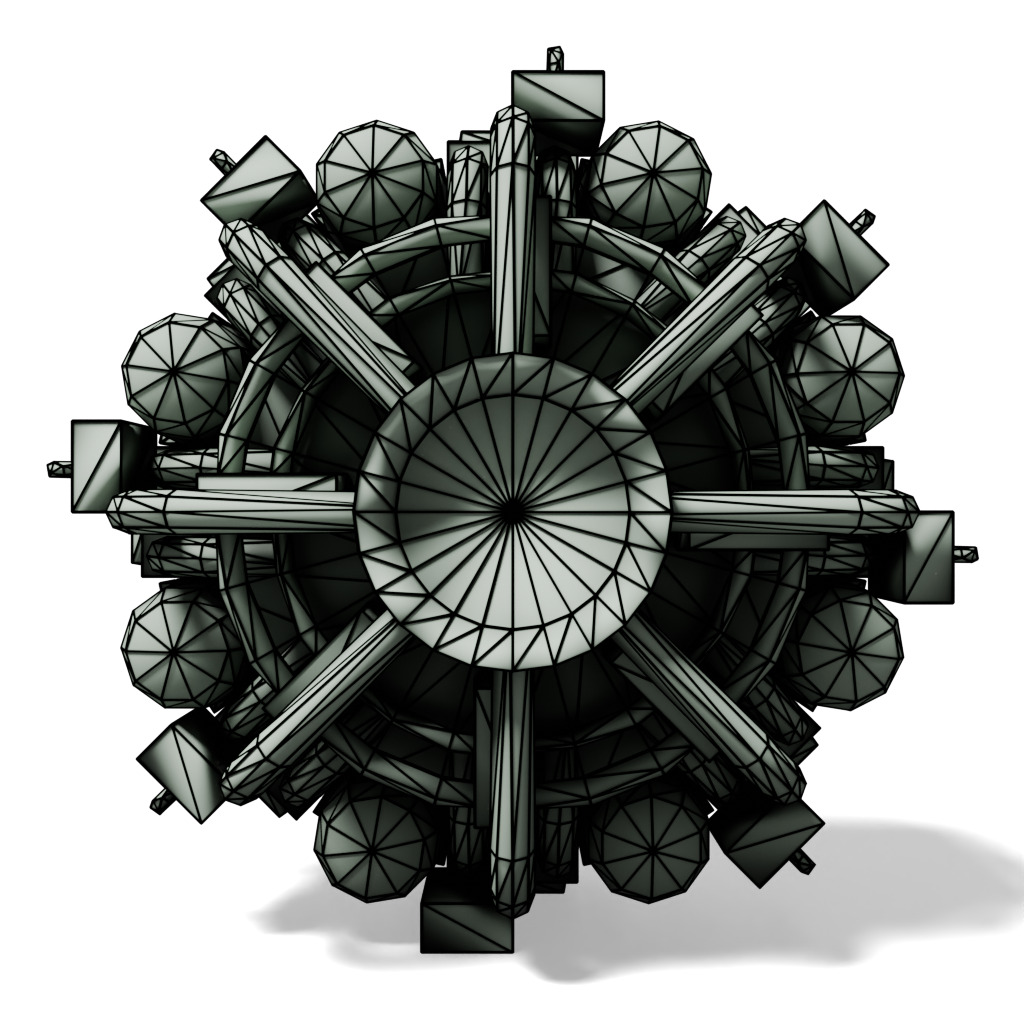}
        \\
        \includegraphics[width=0.33\linewidth]{diagrams/sym_colorbar.pdf} && \\
        $\square/\triangle$: 0/10112 & 0/4174 & 0/4186 \\
        Chamfer$^\downarrow$: & $\num{2.691e-3}$ & $\num{2.641e-3}$ \\
        Hausdorff$^\downarrow$: & $\num{8.388e-3}$ & $\num{8.329e-3}$ \\
    \end{tabular}
    \caption{Our approach identifies and preserves radial symmetry. Left: input mesh. Center: mesh with no symmetry preservation. Right: mesh with symmetry preservation. \ccby Renafox.}
    \label{fig:radial-symmetry-fig}
    \Description{Left: Input mesh, a pipe thing with multiple radial symmetries. Without any symmetry weight, the radial symmetries are not preserved. When including symmetry weights, the radial axes of symmetry are better preserved.}
\end{figure}

We show that our approach is able to identify and preserve key symmetries, by testing our approach on some example meshes with symmetries. This includes reflectional symmetry, as well as radial symmetry. Example outputs for reflectional symmetry are shown in Fig.~\ref{fig:mech-symmetry-fig}, and for radial symmetry is shown in Fig.~\ref{fig:radial-symmetry-fig}. Our approach is able to preserve a key axis of symmetry of the mech mesh, whereas QEM is unable to, especially along the head and lower part of the body. For radial symmetry preservation, the output has cleaner topology, for example the pipes extending from the central ring more closely resemble each other when including symmetry. We show 3 additional examples of symmetry in the Appendix, Fig.~\ref{fig:sym-additional-examples}.

\subsection{Runtime} 
\begin{SCfigure}
    \centering
    \includegraphics[width=0.5\linewidth]{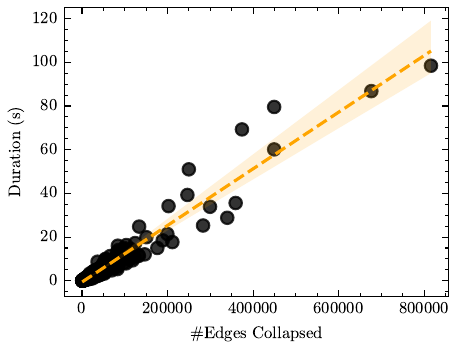}
    \caption{Runtime in seconds as compared to the number of edges collapsed. Our approach appears to be linear in the number of edges collapsed, because the overhead cost for collapsing an edge is higher than the $\text{O}(n\log n)$ of reordering the priority queue.}
    \label{fig:runtime}
    \Description{Scatter plot with regression line through the points. Starts at 0,0, and goes up to 800,000 edges collapsed at about 100 seconds.}
\end{SCfigure}

Quadric mesh reduction~\citep{qem} has $\text{O}(n\log n)$ complexity with $n$ being the number of edge collapses, but in practice has linear runtime, $\text{O}(n)$, as the cost of edge collapse dominates over reordering the priority queue. Our approach has $\text{O}(n\log n + k\log k)$, where $k$ is the size of the second priority queue used to order equivalent edges. In practice $k << n$ and our approach has similar runtime to QEM. We visualize the runtime of our approach in Fig.~\ref{fig:runtime}, which is linear for the same reason that quadric mesh reduction is linear.

\section{Ablations}
We ablate a number of parameters and design choices in our approach. Specifically, we compare our approach with and without priority based on recency of opposing quad edges, our new QEM formulation versus the original, selecting $\lambda_\text{joint}$, the importance of choosing $\epsilon_\text{abs}$ and $\epsilon_\text{rel}$ for comparing quadric errors, and why we use the dihedral angle to weigh the importance of each edge.

\subsection{Recency}
\begin{figure}[!h]
    \setlength{\tabcolsep}{-14pt}
    \centering
    \begin{tabular}{c c c}
        Input & With Recency & No Recency  \\
        \includegraphics[width=0.4\linewidth]{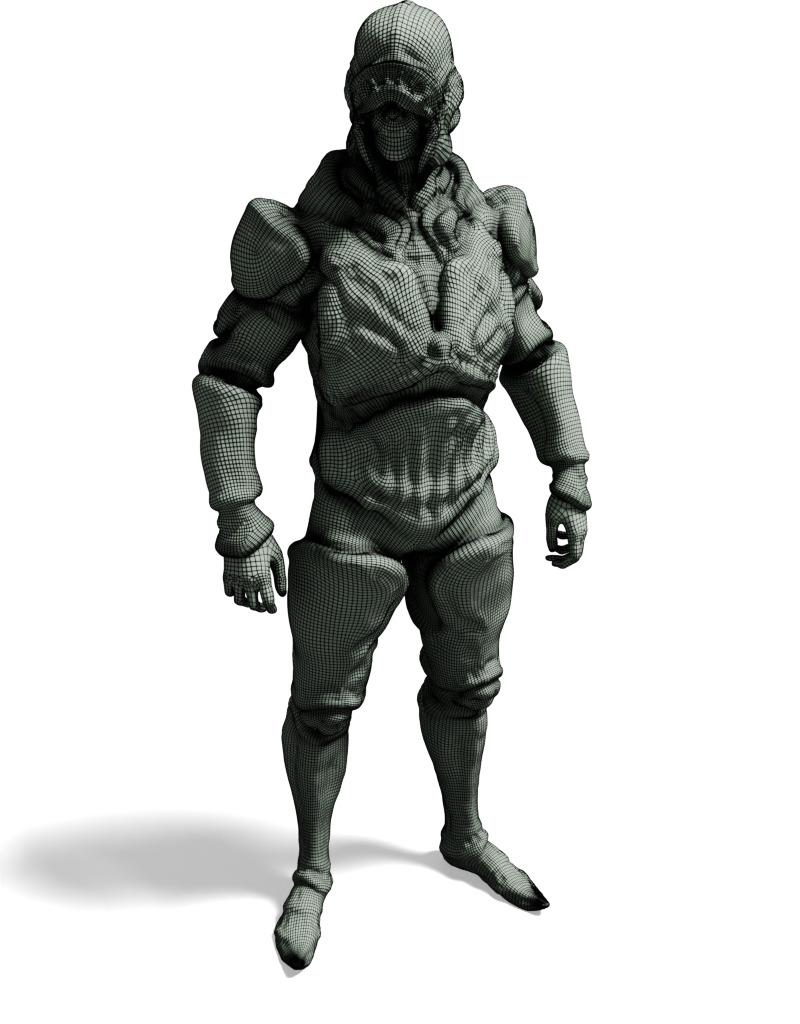} &
        \includegraphics[width=0.4\linewidth]{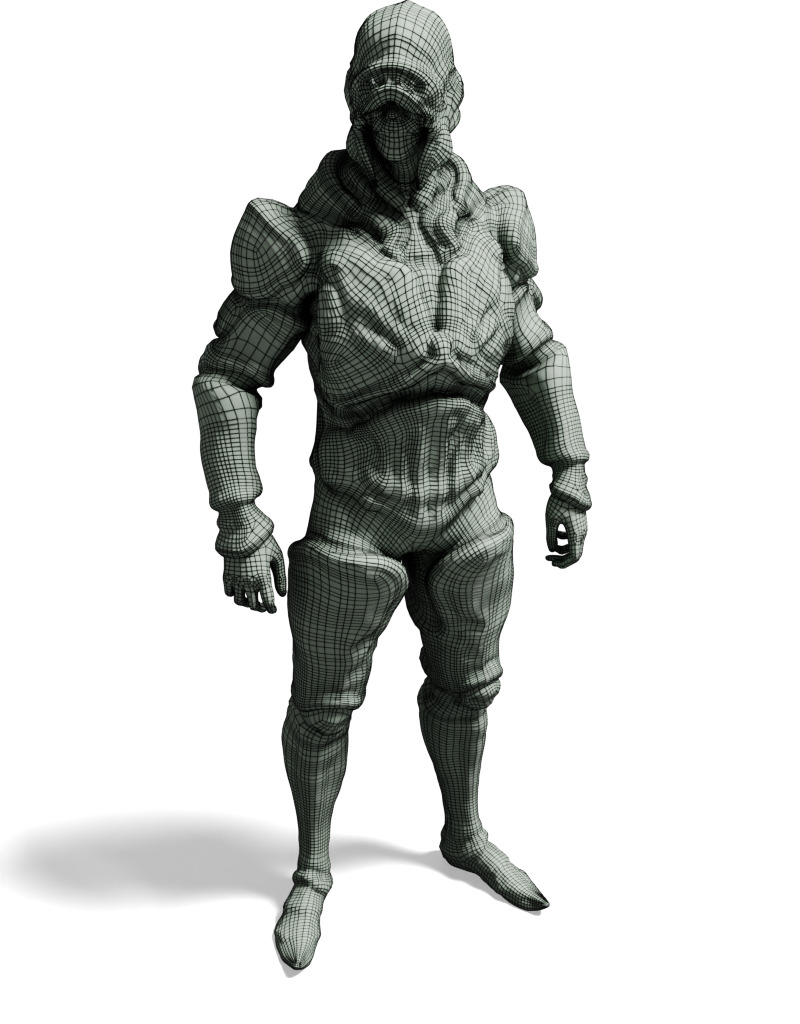} &
        \includegraphics[width=0.4\linewidth]{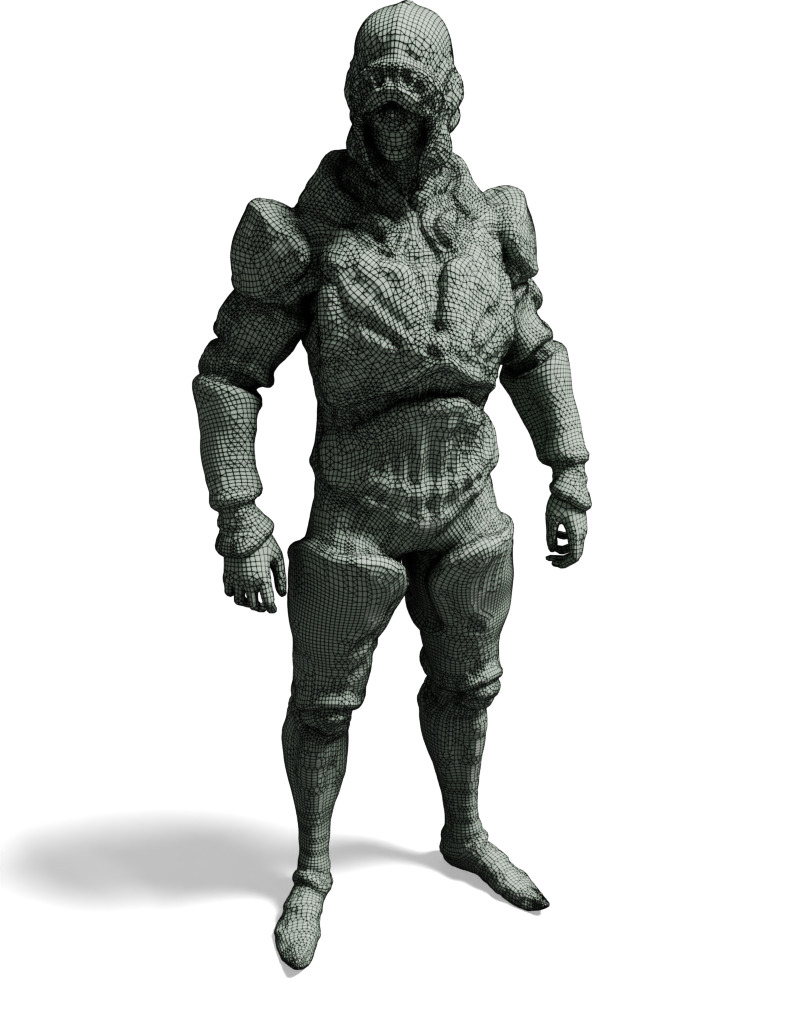} \\
        $\square^\uparrow/\triangle^\downarrow$: 87212/46 & 43199/102 & 26326/33989 \\
        Chamfer$^\downarrow$: & \num{1.830e-4} & \num{1.249e-4} \\
        Hausdorff$^\downarrow$: & \num{3.457e-3} & \num{1.051e-3} \\ 
    \end{tabular}
    \caption{We ablate our approach with and without recency on a complex mesh. When recency is disabled, the output mesh has is slightly more geometrically similar to the input mesh, but with significantly worse topology. Best Viewed Zoomed In. \ccbysa Ploobert.}
    \label{fig:recency-ablation}
    \Description{Ablation of our approach on a character mesh. On the left is the dense input quad-dominantor mesh which is a character with boney armor. Center is our approach which looks to be close to a quad mesh. Right is a mesh with no recency, which appears to be mostly a triangle mesh.}
\end{figure}

To demonstrate the importance of recency, we ablate our approach on a quad-dominant mesh decimated to 50\% of the original triangle count, shown in Fig.~\ref{fig:recency-ablation}. With recency, most quads are preserved, and the original edge flow is strongly preserved. Without recency, the edge flow is destroyed, and a large number of triangles are introduced. Most quads that remain are untouched from the original mesh. On the other hand, the geometric quality suffers slightly when including recency.

\subsection{New QEM Equation vs. Original}
\begin{figure}[!h]
    \setlength{\tabcolsep}{-4pt}
    \centering
    \begin{tabular}{c c c}
        \multicolumn{3}{c}{Comparison without Memoryless Simplification of QEM} \\
        Input & Original QEM & New QEM (Eq. ~\ref{eq:new-qem})  \\
        \includegraphics[width=0.33\linewidth]{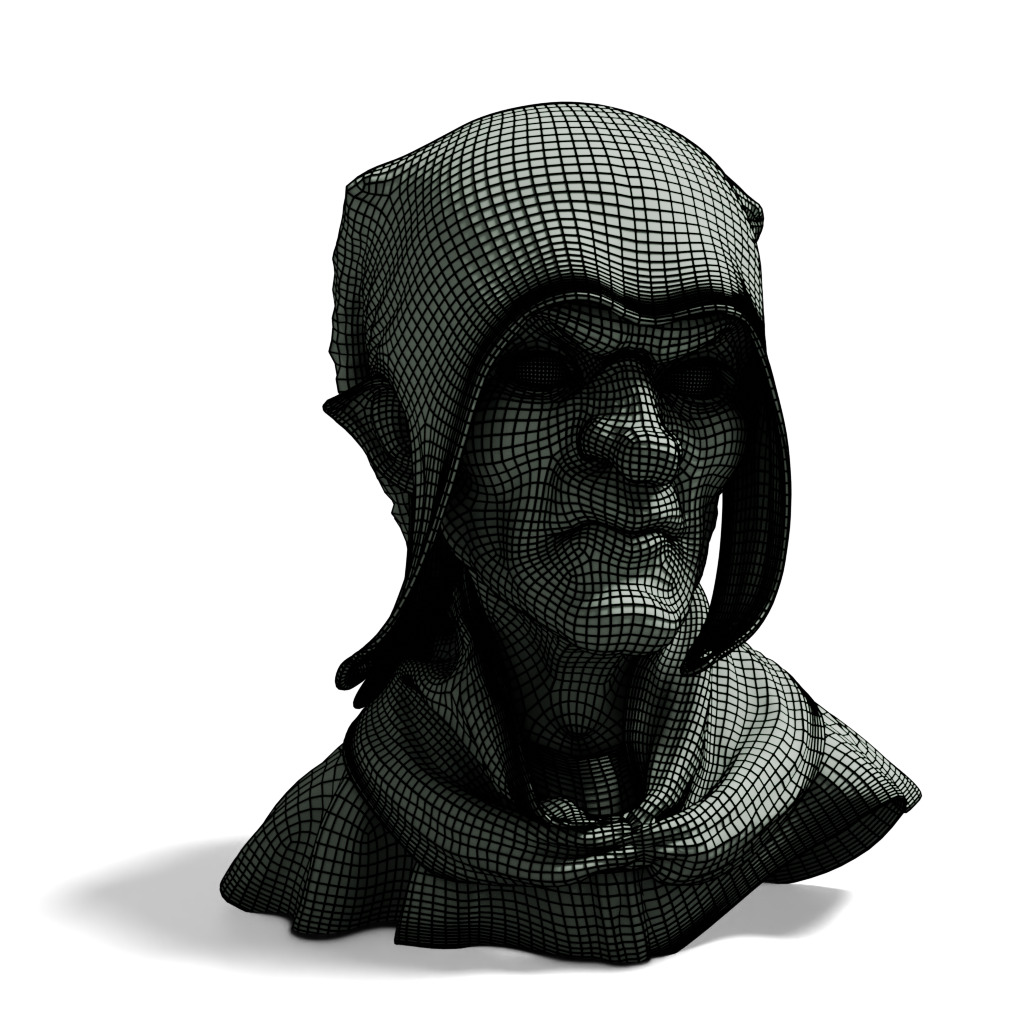} &
        \includegraphics[width=0.33\linewidth]{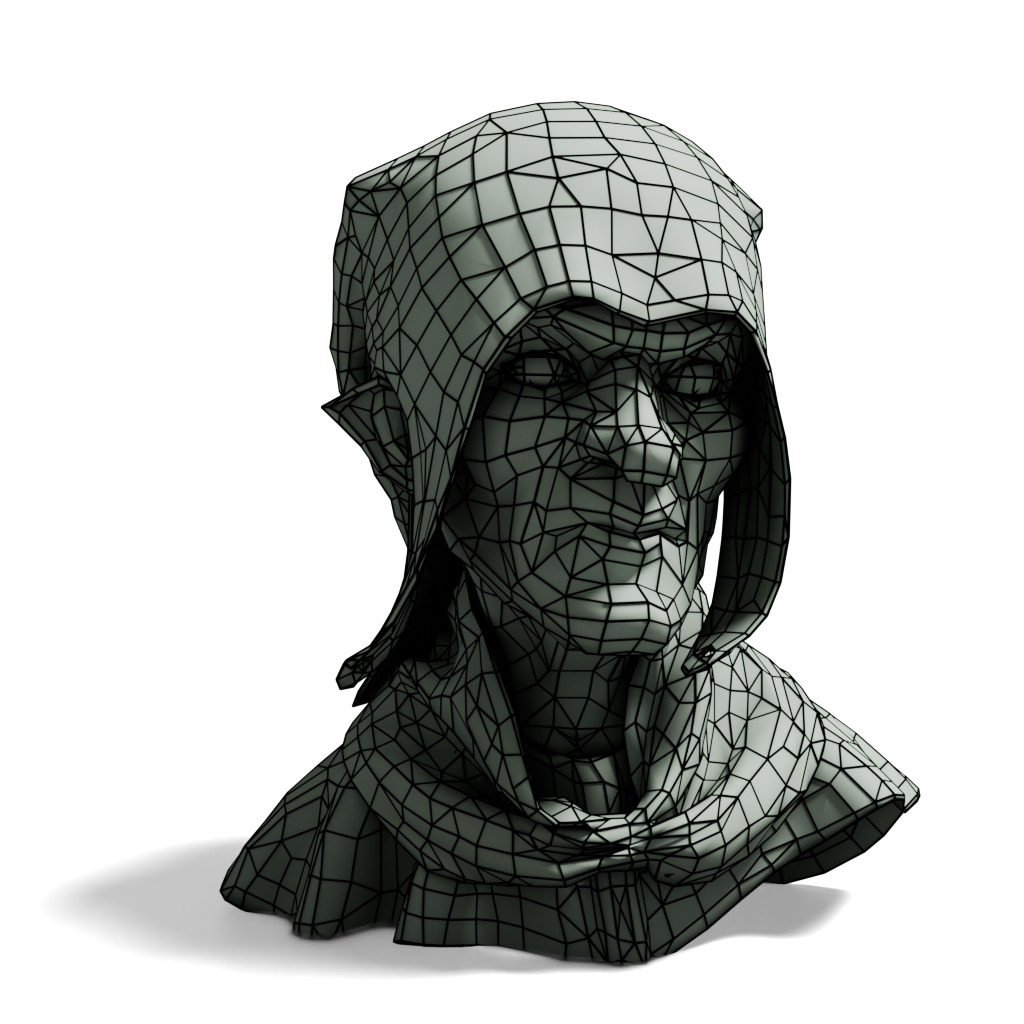} &
        \includegraphics[width=0.33\linewidth]{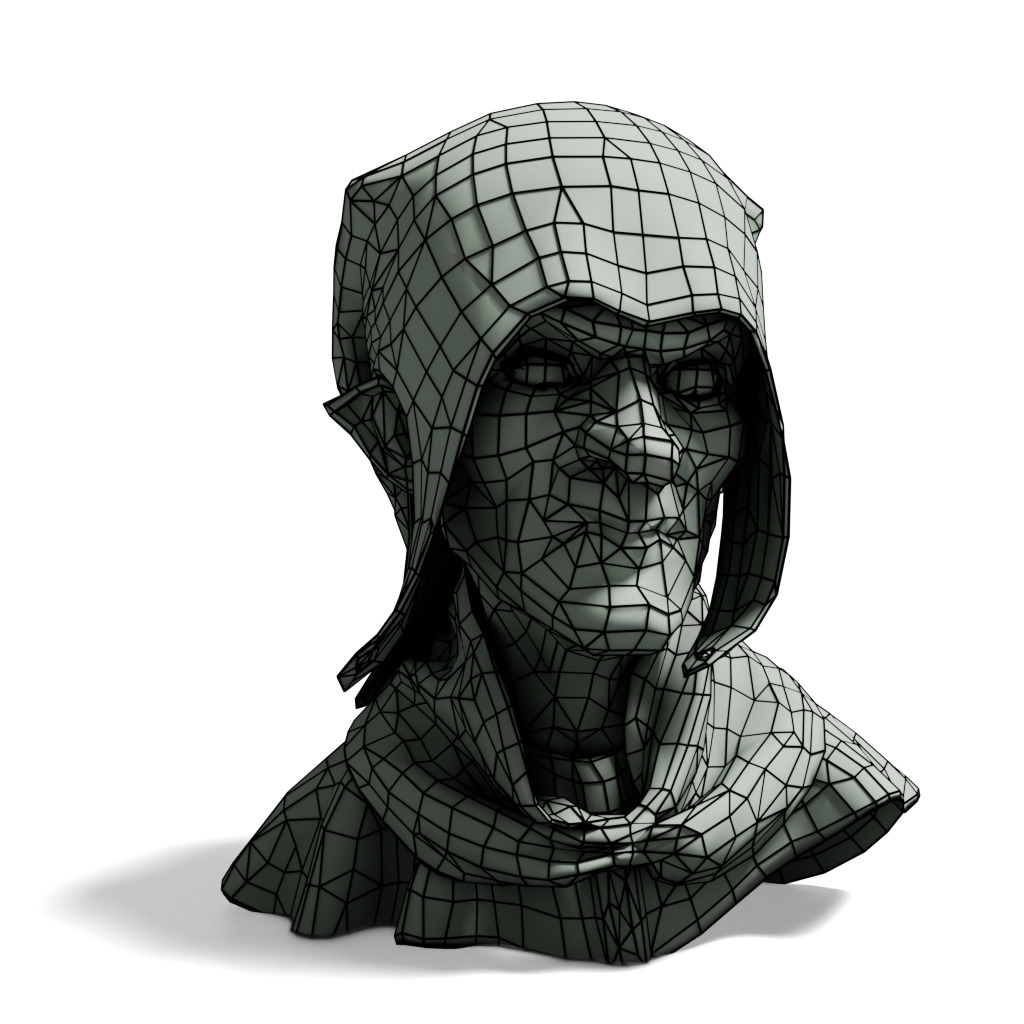} \\
        $\square/\triangle$: 54388/131 & 4153/2585 & 4482/1939 \\
        Chamfer$^\downarrow$: & \num{1.031e-3} & \num{1.003e-3} \\
        Hausdorff$^\downarrow$: & \num{1.574e-2} & \num{1.059e-2} \\
    \end{tabular}
    \caption{\label{fig:delta-cost-ablation} Ablation of the new QEM formulation from Eq. ~\ref{eq:new-qem} against the original QEM formulation from \citep{qem} without memoryless simplification. The new QEM formulation improves over the original when comparing geometric similarity. With our simplification approach, the new QEM leads to topology that more closely matches the input.}
    \Description{Left: Input mesh, dense quad mesh of a goblin. Middle/Right: decimated version of the goblin mesh with low-density topology.}
\end{figure}

We ablate our new QEM formulation (Eq.~\ref{eq:new-qem}) against the original QEM formulation on a single mesh in Fig.~\ref{fig:delta-cost-ablation}, both without memoryless simplification. There is geometric improvement using Eq.~\ref{eq:new-qem} as compared to the original QEM~\citep{qem}, and the topology of the result is closer to the original.

\subsection{Edge Weighting (Eq.~\ref{eq:edge-quadric})}
\begin{figure}[!h]
    \setlength{\tabcolsep}{-0pt}
    \centering
    \begin{tabular}{c c c c}
        \multicolumn{4}{c}{Tangent Space Quadric Weighting Ablation} \\
        Distribution of Angles & No Weights & Uniform & Dihedral  \\
        \includegraphics[width=0.25\linewidth]{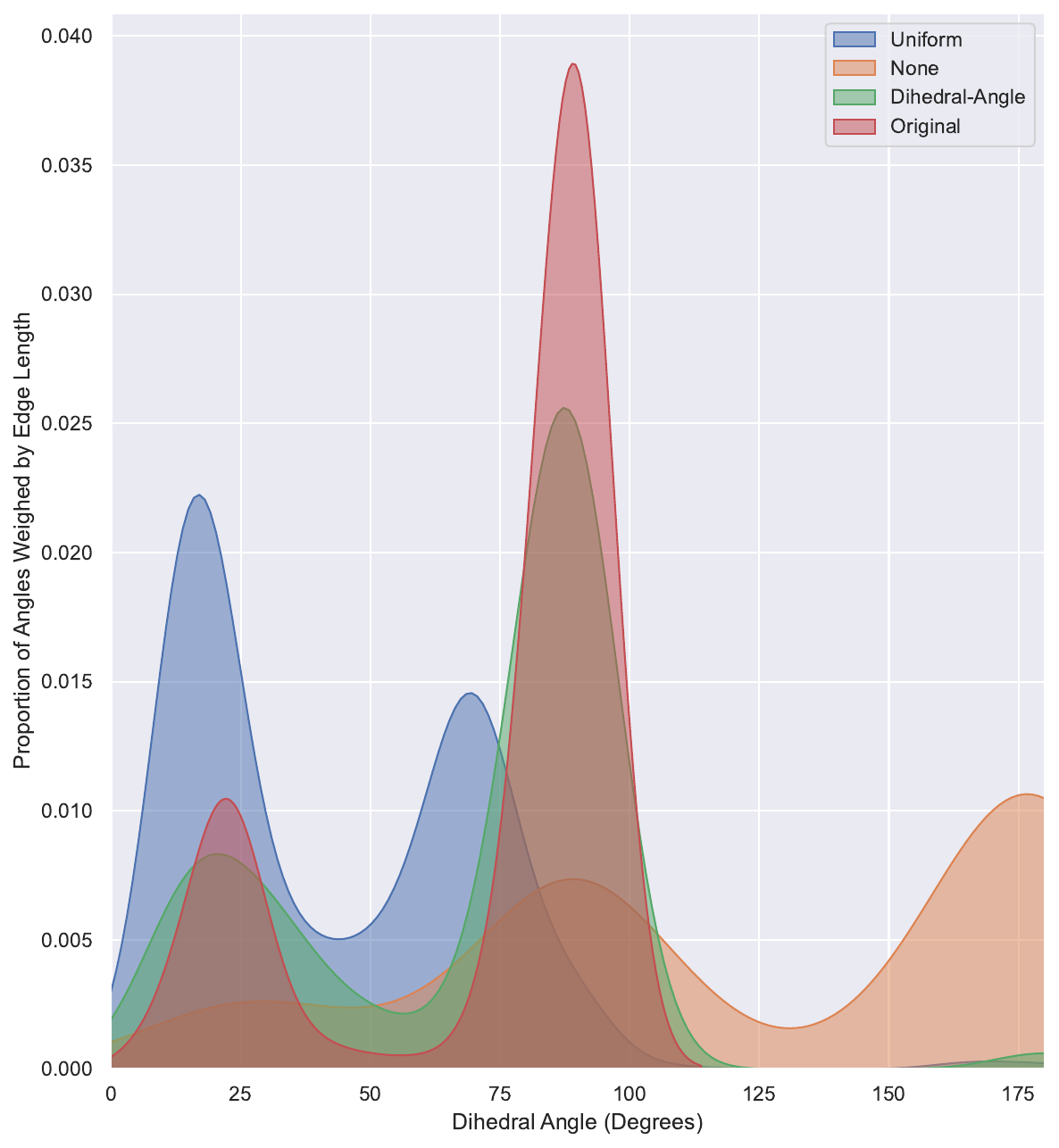} &
        \includegraphics[width=0.24\linewidth]{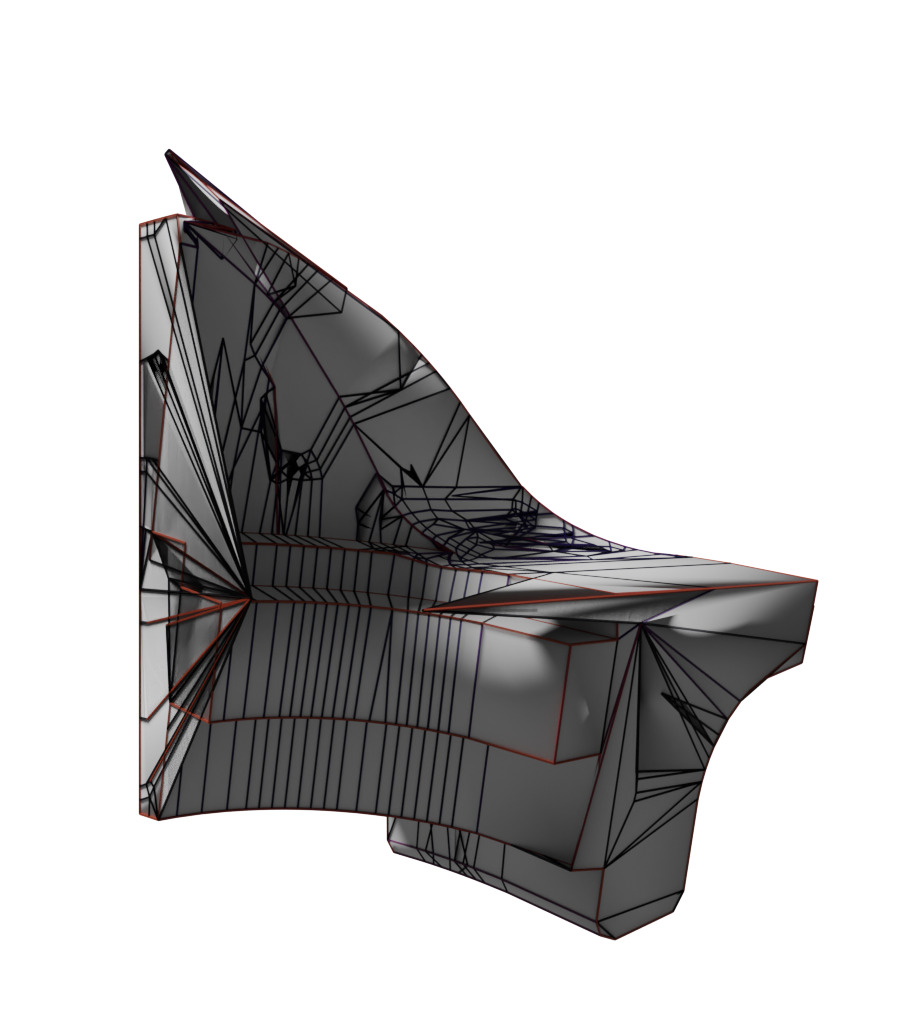} &
        \includegraphics[width=0.24\linewidth]{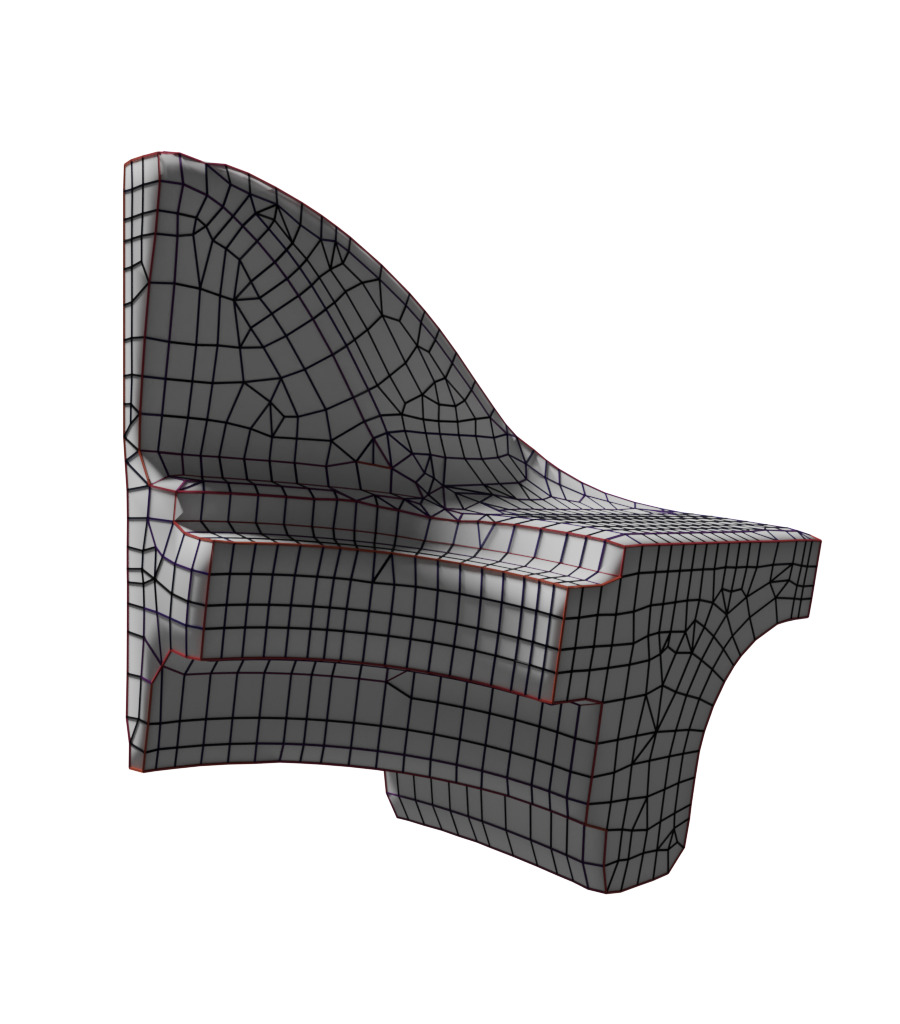} &
        \includegraphics[width=0.24\linewidth]{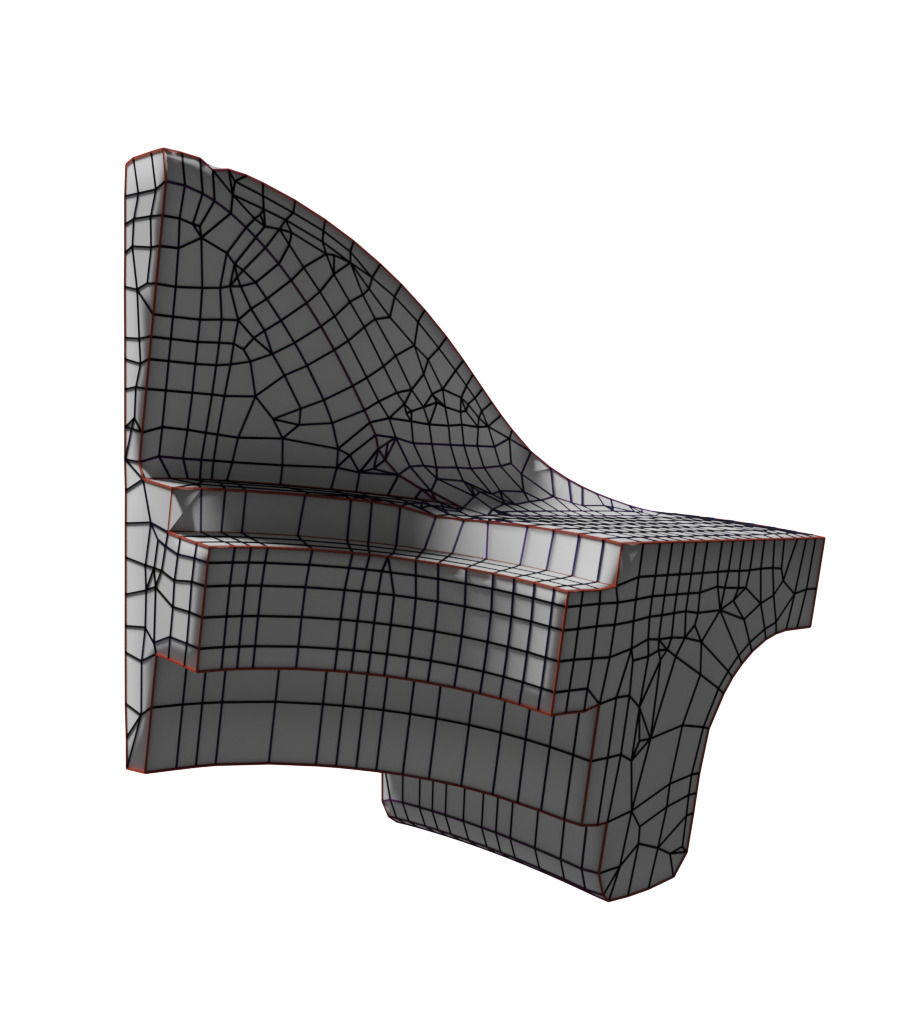} \\
        & \multicolumn{3}{c}{\includegraphics[width=0.4\linewidth]{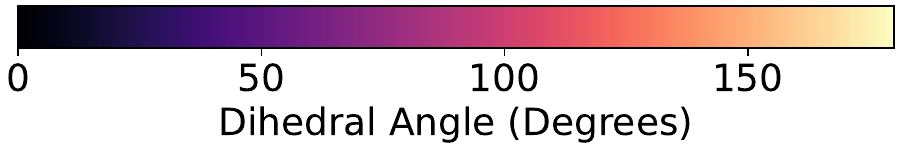}} \\
        Chamfer$^\downarrow$: & $\num{5.587e-2}$ & $\num{7.257e-4}$ & \num{3.433e-4} \\
        Hausdorff$^\downarrow$: & \num{8.129e-1} & $\num{1.409e-2}$ & $\num{2.797e-3}$ \\
    \end{tabular}
    \caption{We ablate our approach with different weights on tangent space quadric. The color of wireframe edges indicates dihedral angle, with redder being larger angle. The distribution shown does not include angles below $0.5\degree$ as they do not define geometric features and obscure the distribution. The dihedral-angle weighted distribution most closely matches the input mesh's dihedral angle distribution.}
    \label{fig:tangent-space-weighting-ablation}
    \Description{Left column, fandisk with degenerate faces all over it. Second from left, fandisk with slightly rounded corners and quad topology. Second from right, fandisk with sharper corners and quad topology. Right, a graph showing the distribution of different angle values of each mesh. The dihedral-angle weighted mesh most closely matches the input.}
\end{figure}

To demonstrate the importance of using the dihedral angle to weigh the importance of the per edge quadric, we test our approach by fixing $w$ from Eq.~\ref{eq:edge-quadric} to $w=0$ (no weights), $w=1$ (uniform weights), and compare those to using dihedral angles to select a weight in Fig.~\ref{fig:tangent-space-weighting-ablation}. When including no weights, the position of quadrics in the tangent space of coplanar faces are degenerate, leading to flipped, degenerate faces. When $w = 1$, sharp edges often become over-smoothed. With dihedral angles, sharp edges are kept sharp, as measured by the geometric similarity. On the left of Fig.~\ref{fig:tangent-space-weighting-ablation}, we visualize the distribution of dihedral angles of manifold edges in each mesh, not including angles below $0.5\degree$. When using dihedral angles as weights, the output mesh's distribution of angles more accurately matches the input mesh's distribution.

\subsection{Joint Distance $\lambda_\text{joint}$}\label{sec:ablate-ljoint}
To demonstrate the importance of weighing edges by joint distance, we compare our approach with $\lambda_\text{joint} = 0, \lambda_\text{joint} = 1, \lambda_\text{joint} = 10$. We focus primarily on the monster's leg joint, which are bulbous regions with slightly higher density of edges than the rest of the leg. As $\lambda_\text{joint}$ is increased, we see an increase in joint density in that region, at the cost of higher chamfer distance in the rest pose. When $\lambda_\text{joint} = 1$, there is a lower chamfer and Hausdorff distance across all frames as compared to $\lambda_\text{joint} = 0 \text{ and } \lambda_\text{joint} = 10$. This makes it evident that some importance of quadrics along edges between vertices with highly variable joint influences help preserves animation quality.

\begin{figure}[!h]
    \setlength{\tabcolsep}{-12pt}
    \centering
    \begin{tabular}{c c c c}
        \multicolumn{4}{c}{Ablation of $\lambda_\text{joint}$} \\
        Input & $\lambda_\text{joint} = 0$ & $\lambda_\text{joint} = 1$ & $\lambda_\text{joint} = 10$ \\
        
        \includegraphics[width=0.35\linewidth]{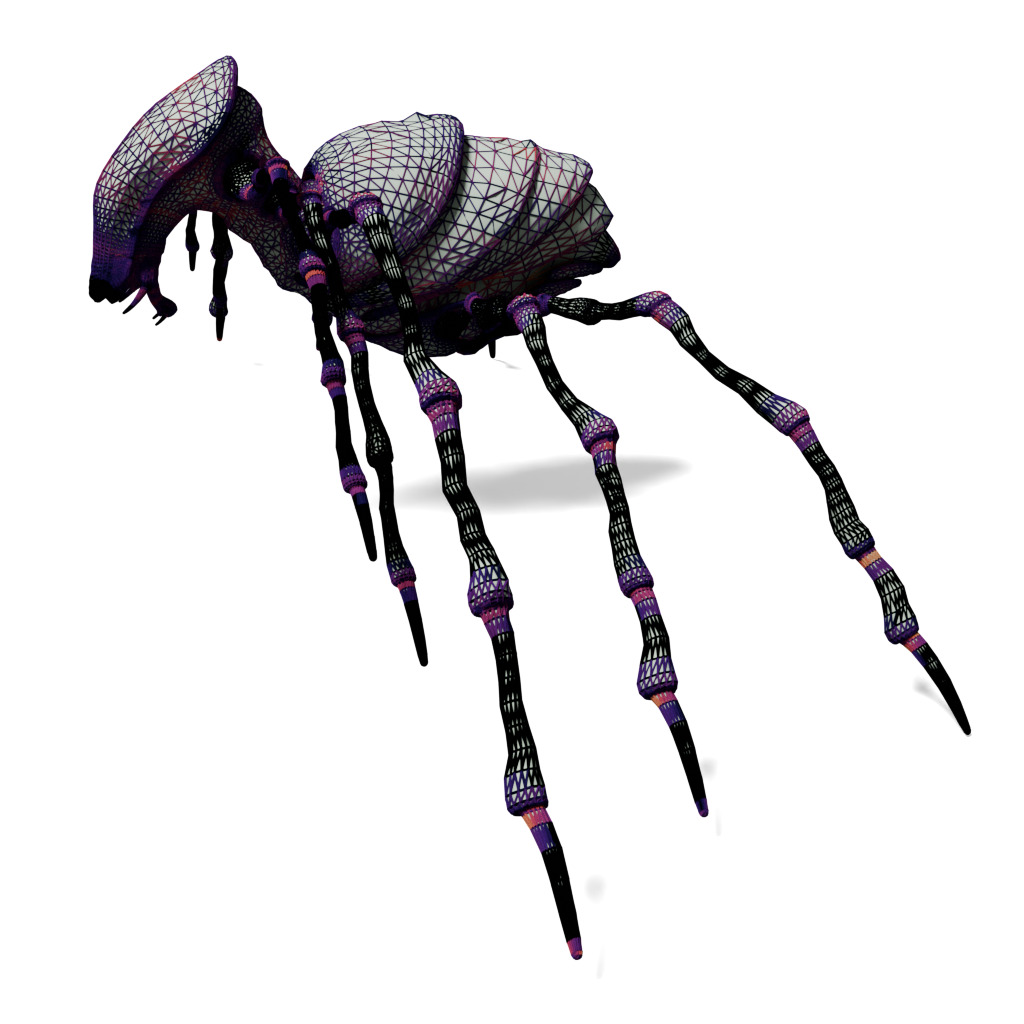} &
        \includegraphics[width=0.35\linewidth]{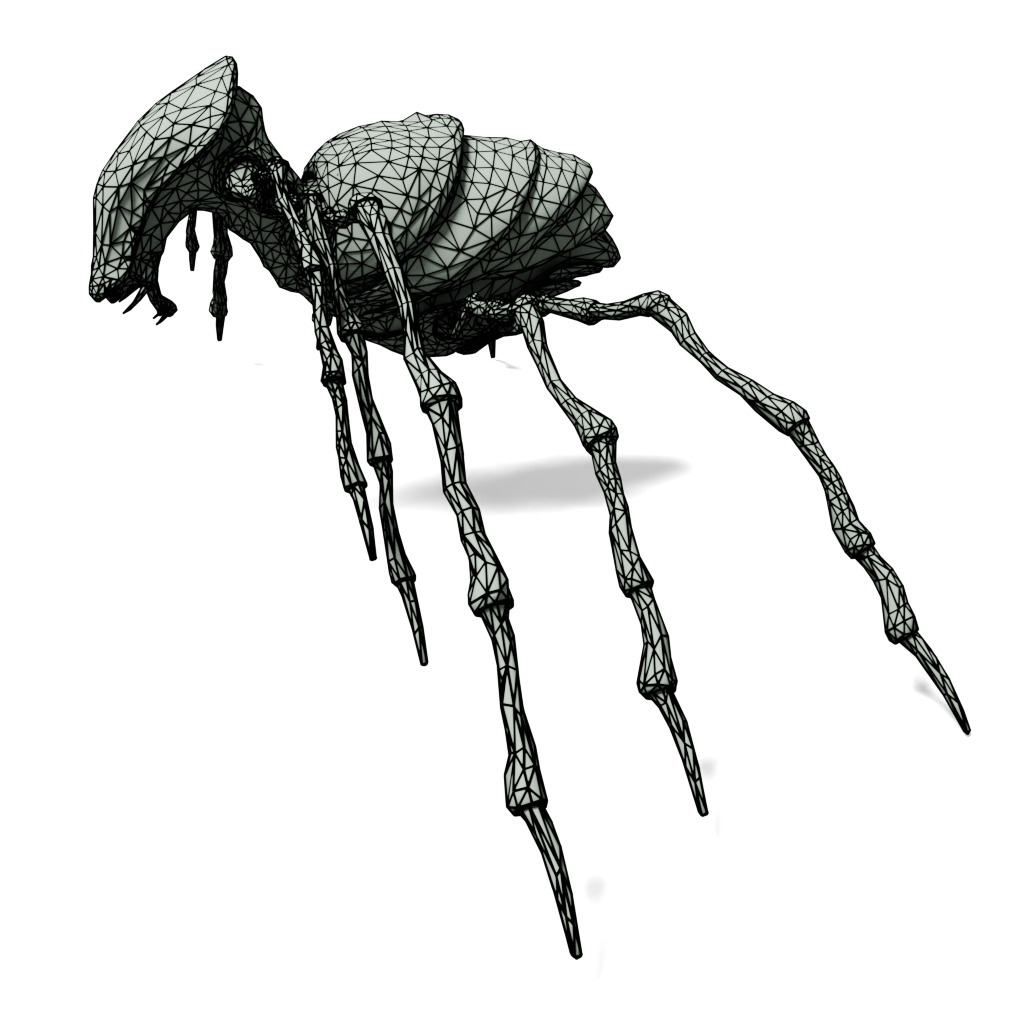} &
        \includegraphics[width=0.35\linewidth]{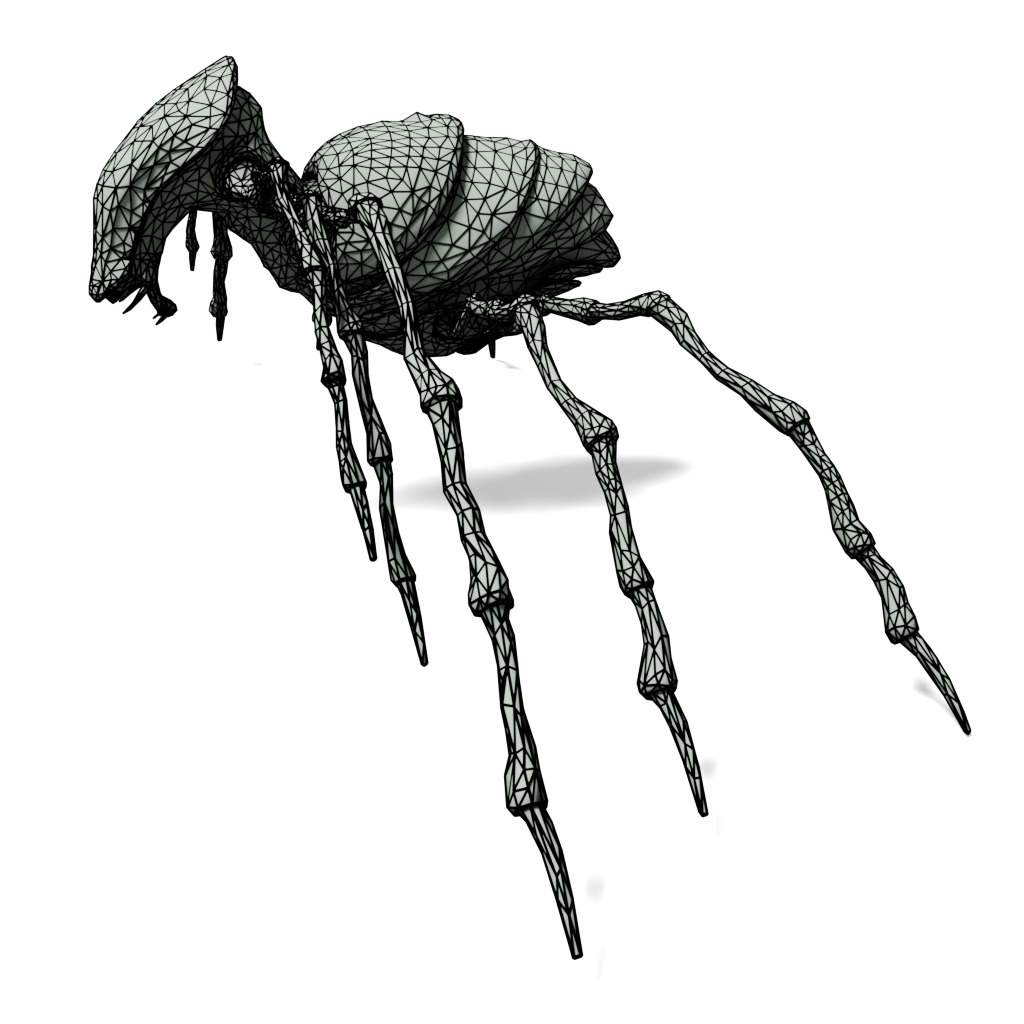} &
        \includegraphics[width=0.35\linewidth]{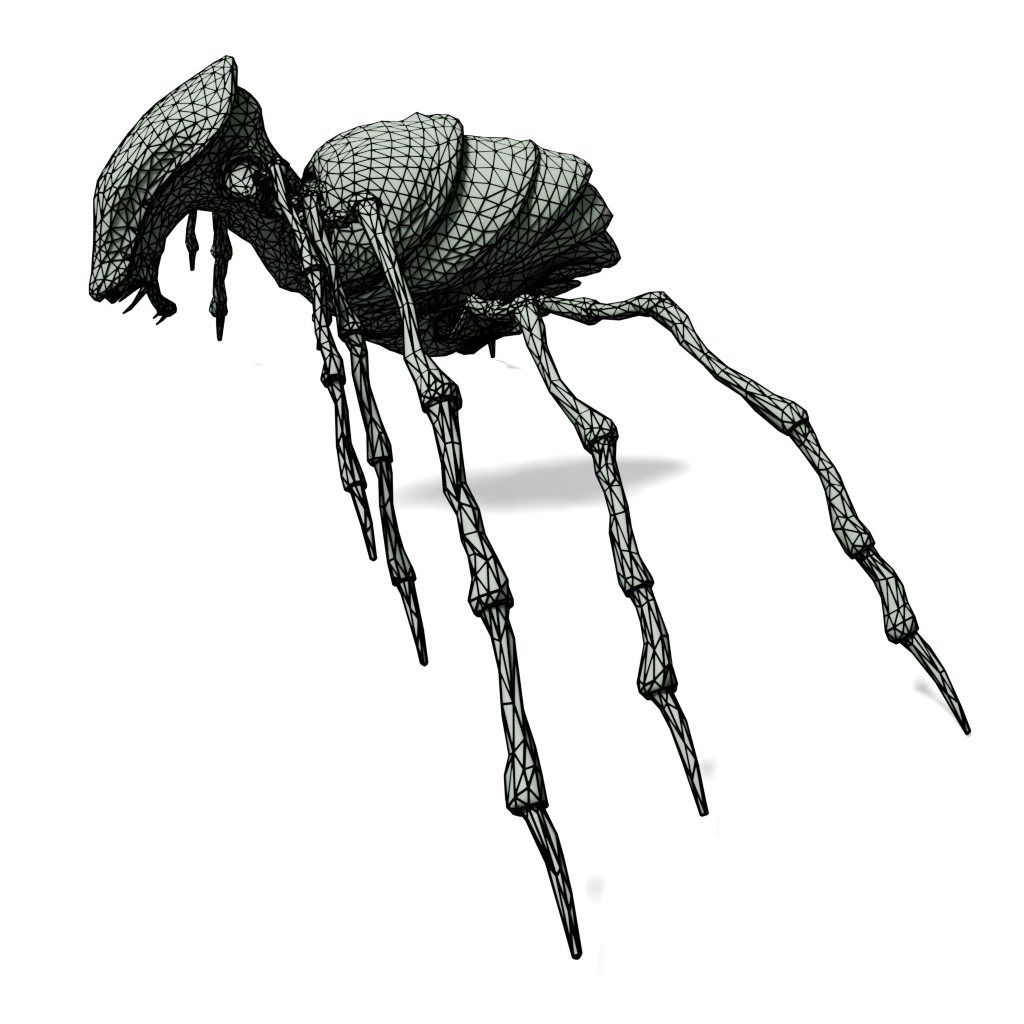} \\

        \frame{\includegraphics[width=0.2\linewidth]{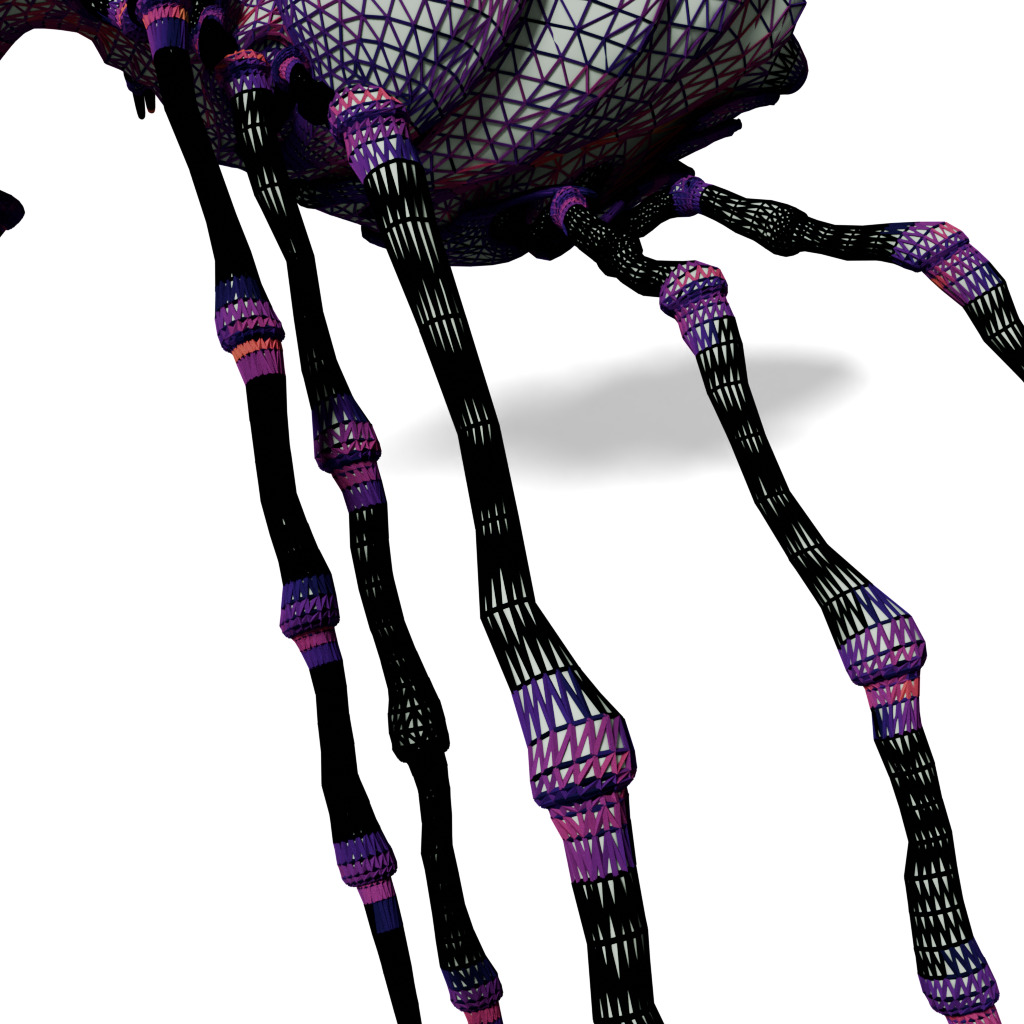}} &
        \frame{\includegraphics[width=0.2\linewidth]{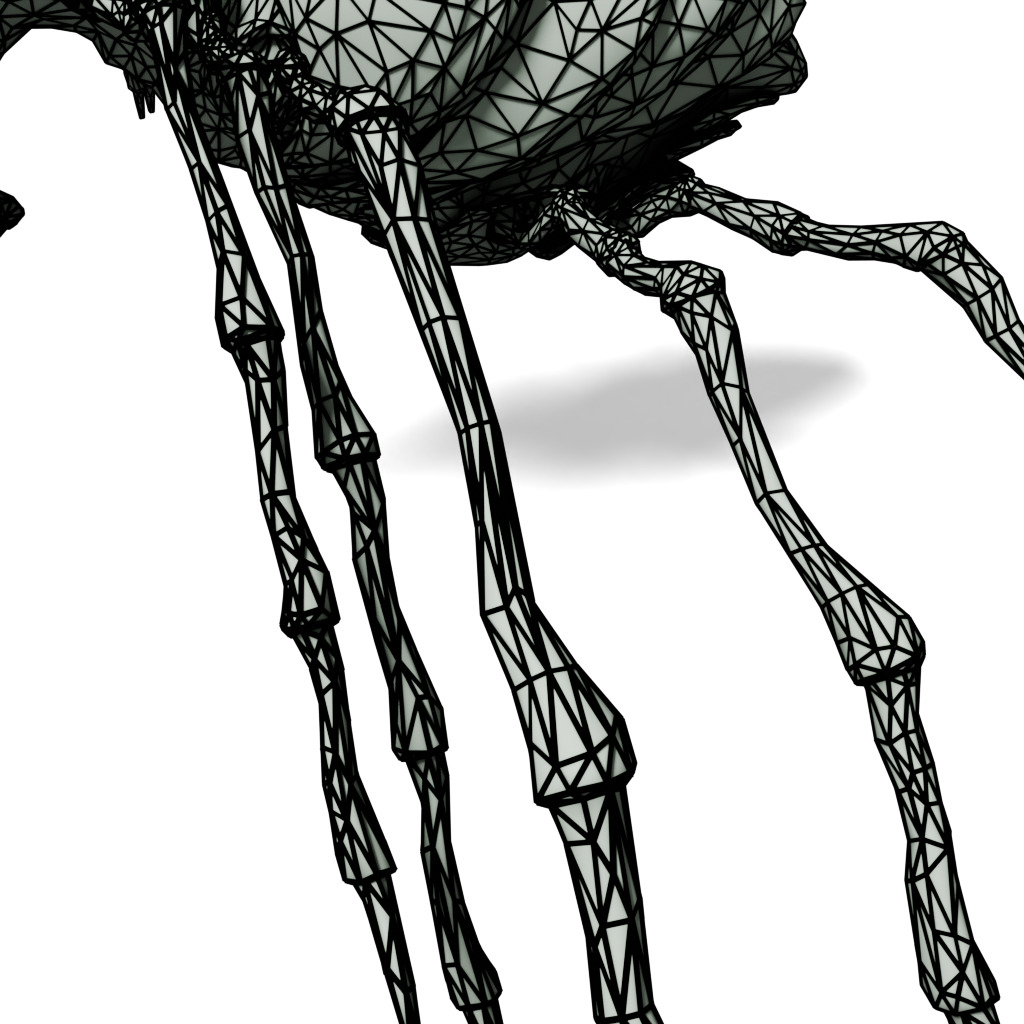}} &
        \frame{\includegraphics[width=0.2\linewidth]{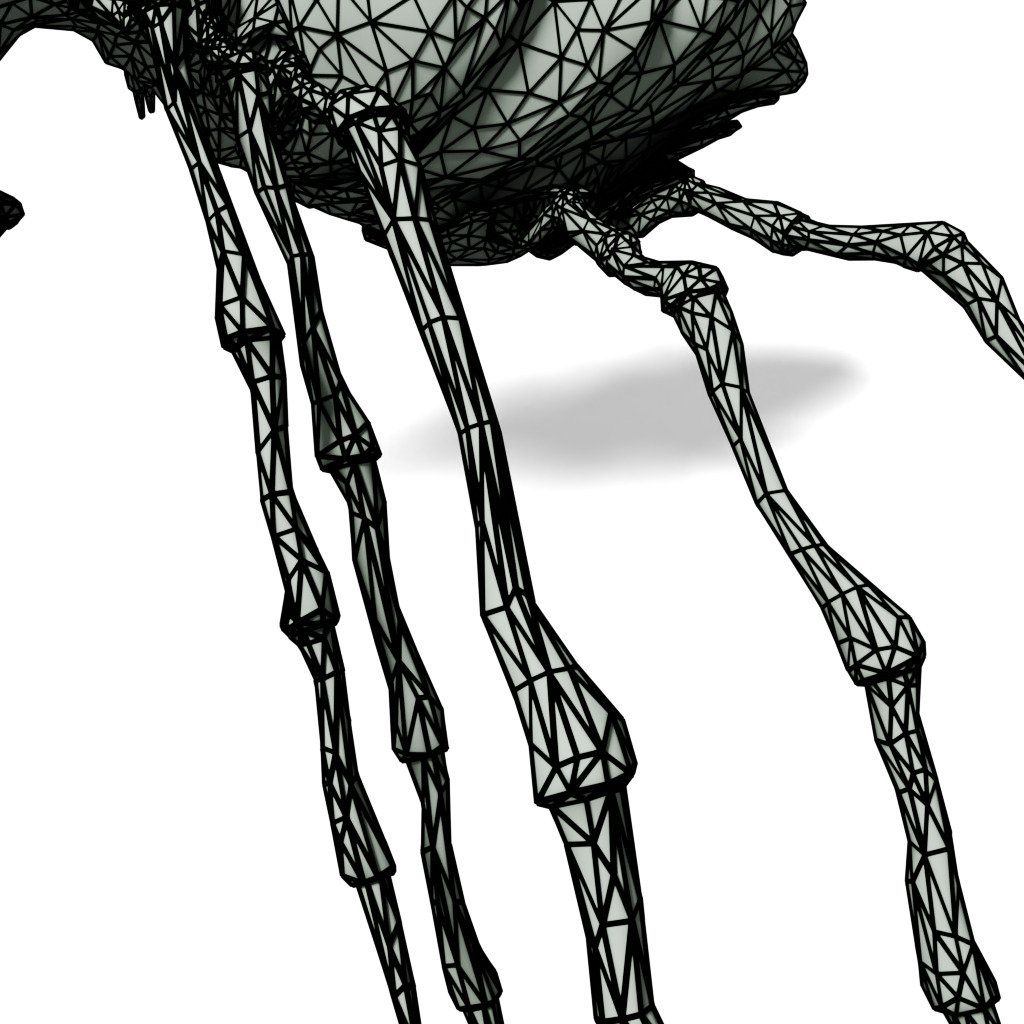}} &
        \frame{\includegraphics[width=0.2\linewidth]{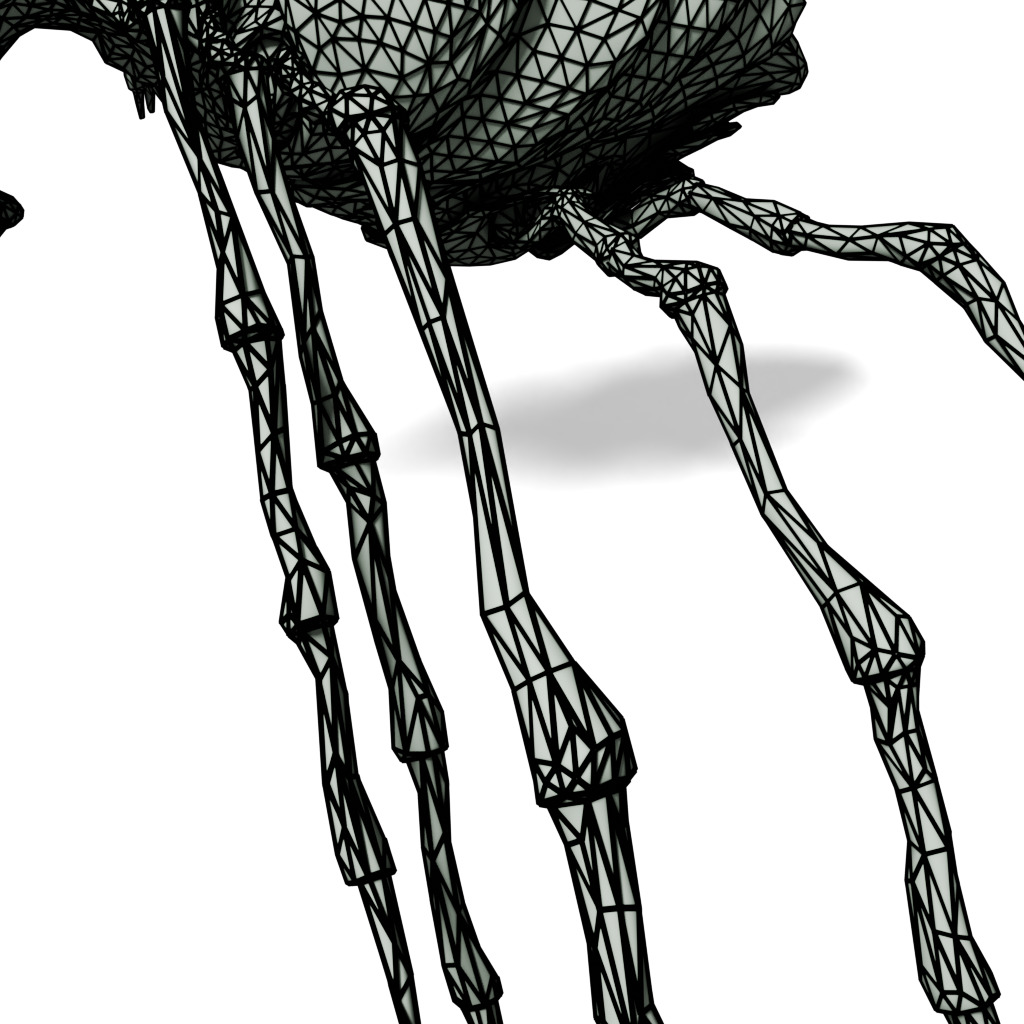}} \\

        \includegraphics[width=0.2\linewidth]{diagrams/joint_colorbar.pdf} && \\
        $\triangle$: 67972 & 13842 & 13884 & 13821 \\
        Rest Chamfer$^\downarrow$: & $\num{5.241e-4}$ & $\num{5.140e-4}$ & $\num{6.143e-4}$ \\
        Rest Hausdorff$^\downarrow$: & $\num{3.311e-3}$ & $\num{3.332e-3}$ & $\num{2.848e-3}$ \\
        \multicolumn{2}{c}{\includegraphics[width=0.49\linewidth]{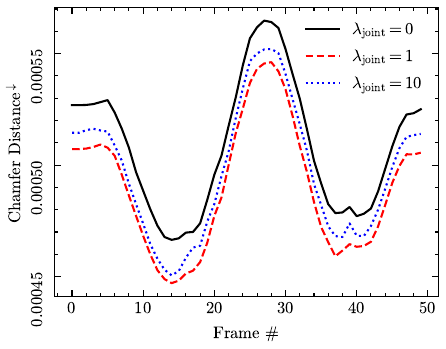}} &
        \multicolumn{2}{c}{\includegraphics[width=0.49\linewidth]{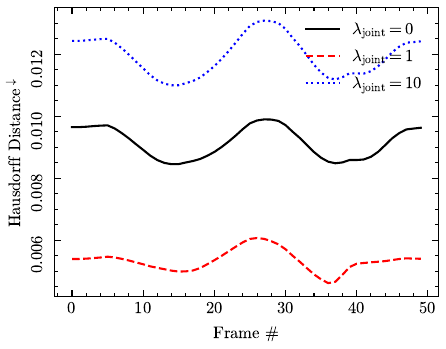}} \\
    \end{tabular}
    \caption{Ablation of our approach with varying $\lambda_\text{joint}$. As $\lambda_\text{joint}$ increases, the area around joints is more strongly preserved, at the cost of higher decimation in other areas. Despite the rest pose chamfer distance increasing, we see that at $\lambda_\text{joint} = 1$ there is an improvement to both the hausdorff and chamfer distance during animation, whereas when $\lambda_\text{joint} = 10$ there is too much preservation at joints, leading to lower geometric similarity during animation. \ccby Rasmus. \label{fig:joint-distance-ablation}}
    \Description{Left column: input animated spider mesh, with boney edges near joints. Next 3 columns, decimated mesh with different joint distance weights. There are more edges preserved around the joints as the weight increases.}
\end{figure}

\subsection{Approximate Equality $\epsilon_\text{abs}$}
We also ablate our approach while varying $\epsilon_\text{abs}$ to demonstrate its impact on quad preservation and geometric quality.

\begin{figure}[!h]
    \setlength{\tabcolsep}{0.5pt}
    \centering
    \begin{tabular}{c c c c}
        & \multicolumn{3}{c}{Varying $\epsilon_\text{abs}$} \\
        Input & $\num{1e-4}$ & $\num{1e-5}$ & $\num{1e-6}$ \\ 
        \includegraphics[width=0.245\linewidth]{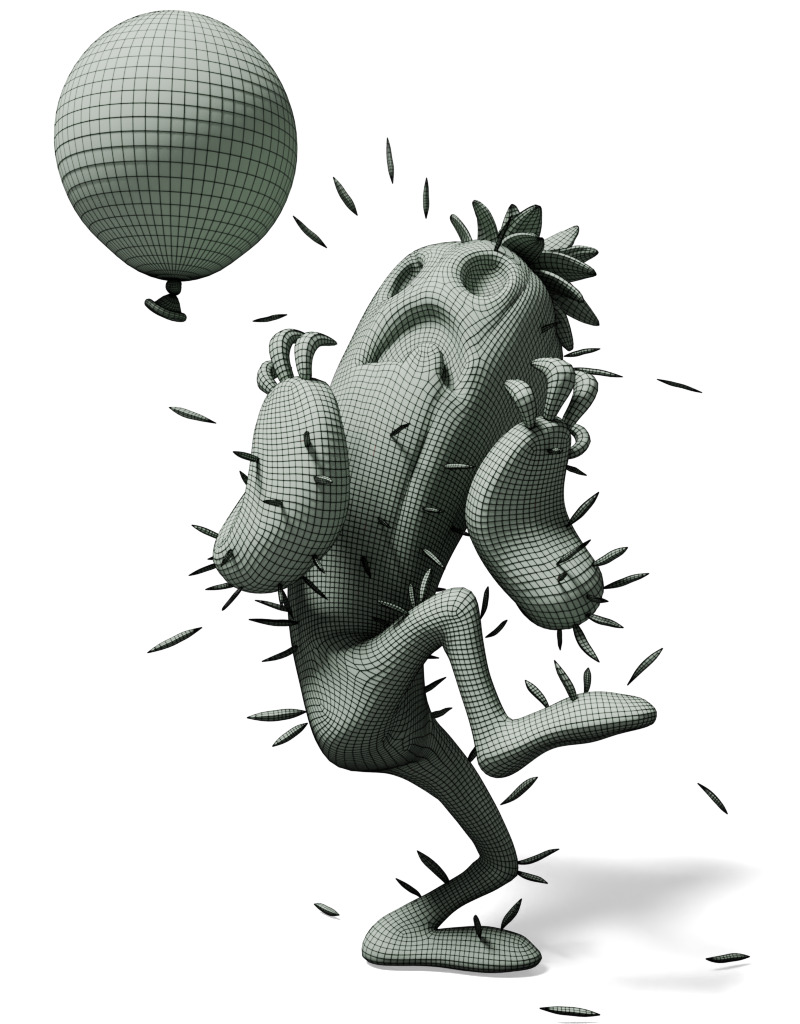}
        & \includegraphics[width=0.245\linewidth]{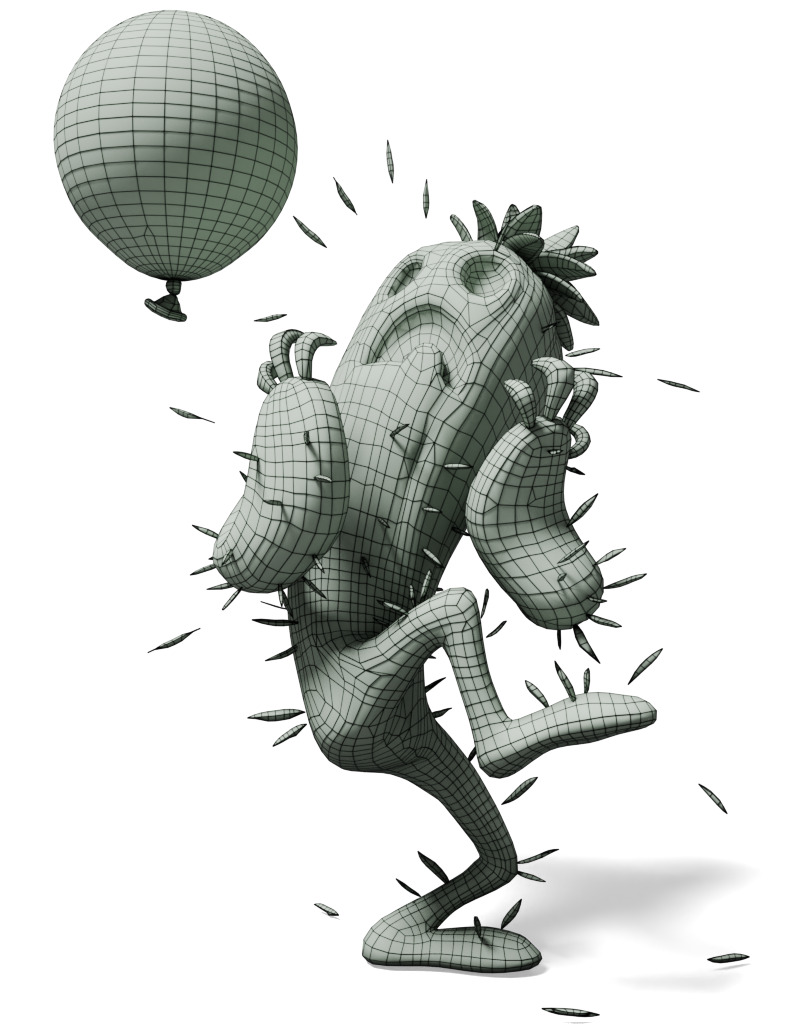}
        & \includegraphics[width=0.245\linewidth]{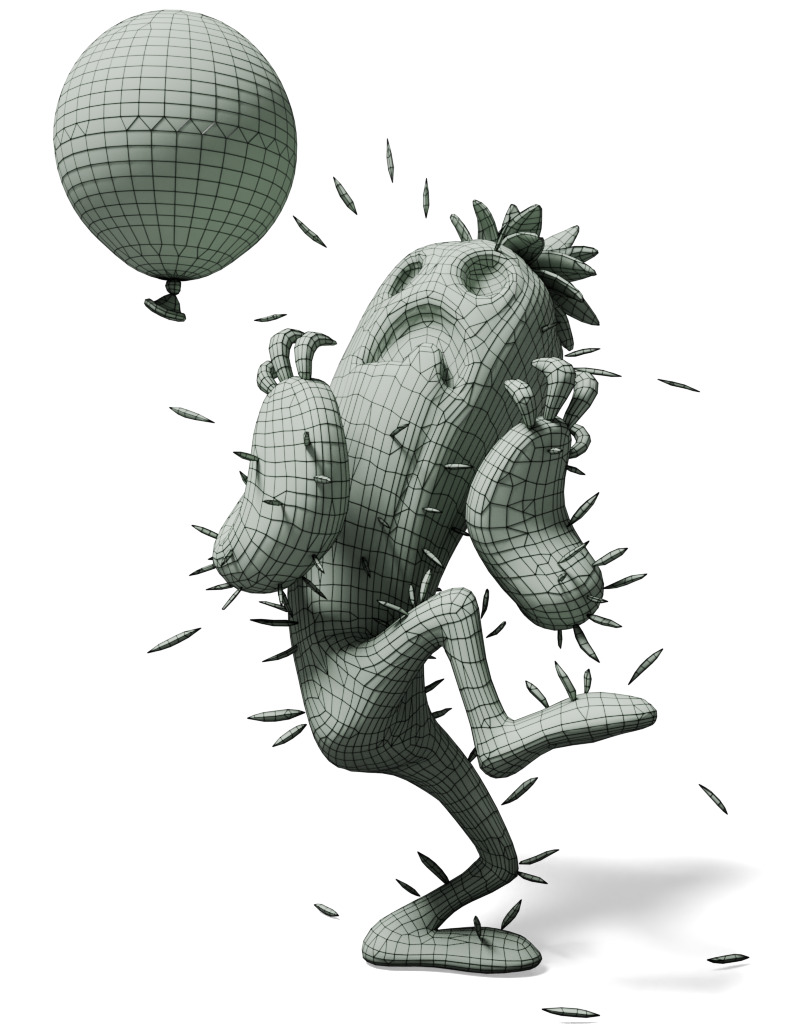}
        & \includegraphics[width=0.245\linewidth]{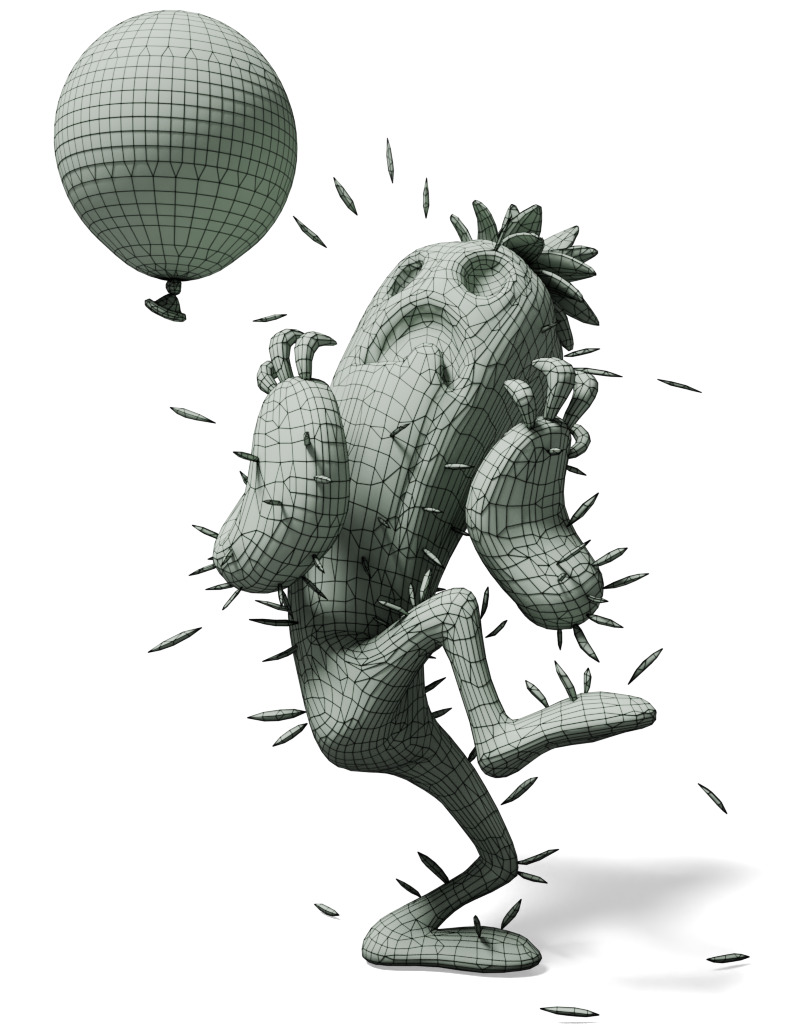} \\
        $\square/\triangle$: 38235/4 & 9368/389 & 8718/1705 & 7567/4000 \\
        Chamfer$^\downarrow$: & \num{3.512e-4} & \num{2.079e-4} & \num{1.635e-4} \\
        Hausdorff$^\downarrow$: & \num{4.817e-3} & \num{4.769e-3} & \num{2.815e-3} \\
    \end{tabular}
    \caption{We ablate the effect of varying $\epsilon_\text{abs}$ from Eq.~\ref{eq:abs-diff}. As $\epsilon_\text{abs}$ increases, there is stronger quad preservation, but a slight increase in the distance from the original mesh. \ccby felipegall.}
    \label{fig:abs-eps-ablation}
    \Description{Left: Input Cactus scared of a balloon, 2nd left, mostly quad mesh, 2nd right more tris are visible on but still mostly quads. Right: a mix of quads and tris.}
\end{figure}

We visualize the results of varying $\epsilon_\text{abs}$ in Fig.~\ref{fig:abs-eps-ablation}. Increasing $\epsilon_\text{abs}$ increases quad preservation, at the cost of lower geometric similarity, as can be seen when $\epsilon_\text{abs} = \num{1e-4}$, where the geometric distance increases by $\num{1.5e-4}$ from $\epsilon_\text{abs} = \num{1e-6}$ to $\num{1e-4}$. Such a small change is likely unnoticeable to a viewer, so depending on the downstream application it may be worthwhile to trade this geometric difference to preserve more quads.


\section{Discussion}

We develop modifications to the normal mesh reduction approach that are able to preserve the quad topology of input meshes at no cost, and demonstrate how these modifications can be extended to preserve symmetry and joint influences. We have also performed a thorough test demonstrating that \citep{qem_hoppe} is suitable for joint preservation, producing more geometrically similar articulated outputs than \citep{articulated_mesh_simplification} on most of our tested models.

For applications which require 100\% quad preservation, our approach cannot provide such a guarantee, as it will usually introduce a small number of triangles between regions where quads have different quadric error. Furthermore, it is also possible to devise an adversarial quad mesh where opposing quad edges vary significantly, which may defeat our algorithm. In practice though, we observe most meshes contain regular quad elements and are amenable to our approach.

Our symmetry preservation technique is suitable primarily for smaller meshes, due to quadratic scaling with time complexity O($|V||E|$). For many of our test meshes though, this is acceptable, since the constant factor is low, and each edge's symmetry weight can be computed in parallel. Another problem with our approach to symmetry preservation is that it is less common for static meshes to contain extrinsic symmetry, but may contain intrinsic symmetry. Our approach to symmetry preservation is unable to identify intrinsic symmetries and thus cannot preserve them. Animated meshes, on the other hand, often contain a T-pose with extrinsic symmetries, which our approach is well-suited for. We leave exploration of intrinsic topological symmetry preservation to future work.

\section{Conclusion}

In summary, we develop a robust and simple approach that converts existing triangle quadric mesh reduction algorithms into quad-dominant mesh reduction algorithms, while maintaining the same or better geometric quality. Our approach is easily extended to handle symmetric decimation, joint preserving decimation, or generic edge painting to preserve certain regions more. Our primary insight in this work is that local operations ordered entirely by geometric error define too strict of an ordering, and by relaxing the ordering we are able to easily impose an ordering which preserves quads, and by extension topology. We hope this work is adopted by tooling so that mesh reduction on all platforms can better preserve input topology.
\bibliographystyle{./ref/ACM-Reference-Format}
\bibliography{citations}
\appendix
\section{\label{sec:additional-results}Additional Results}

\begin{figure*}
    \renewcommand{\arraystretch}{0}
    \centering
    \begin{tabular}{c c c}
        Input & w/ Symmetry Preservation & w/o Symmetry Preservation  \\
        \includegraphics[width=0.33\linewidth]{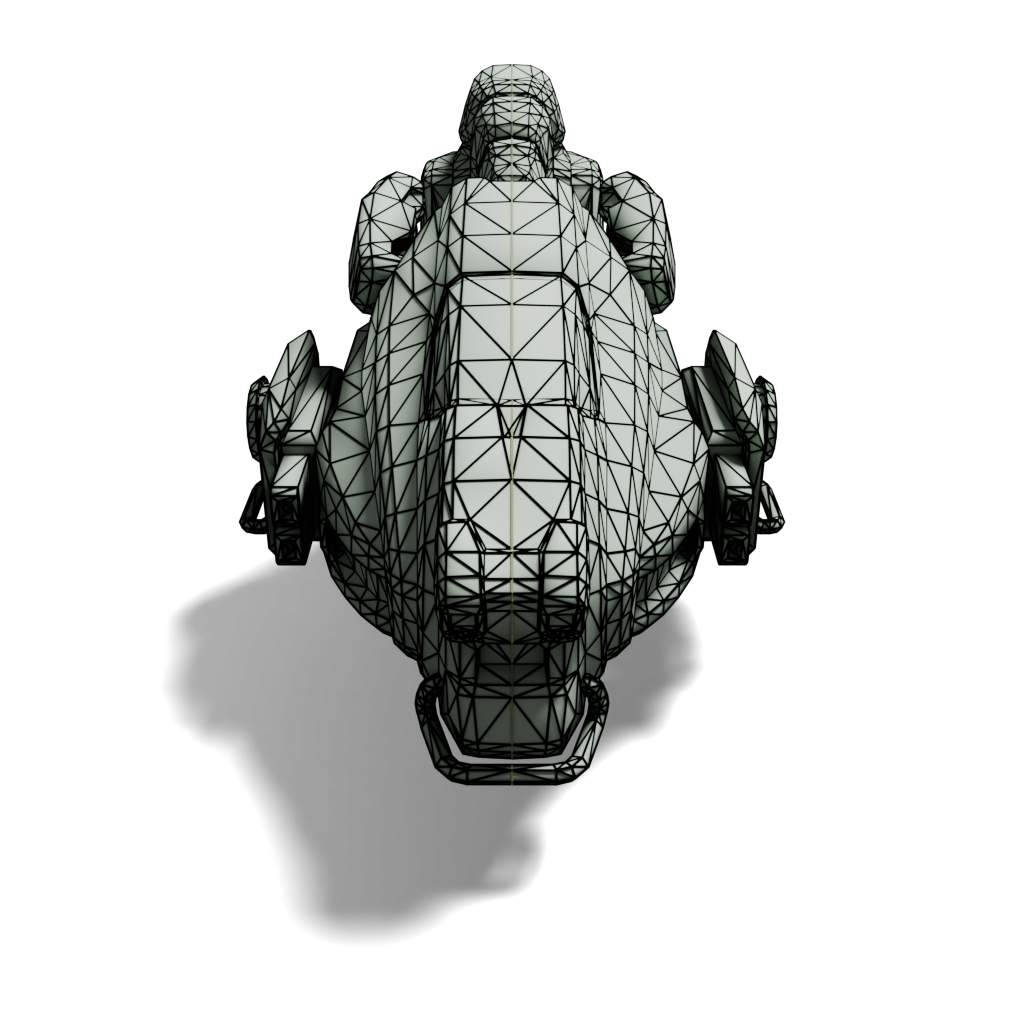} &
        \includegraphics[width=0.33\linewidth]{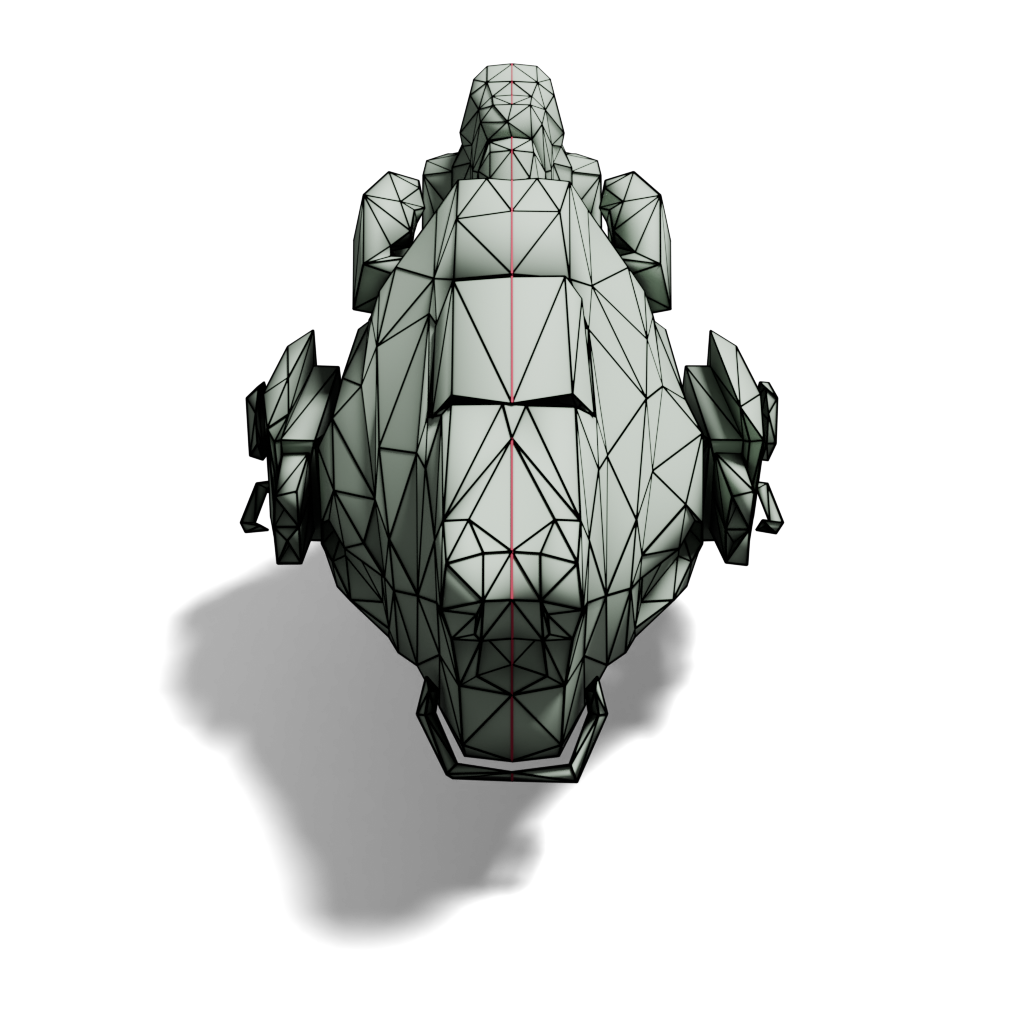} &
        \includegraphics[width=0.33\linewidth]{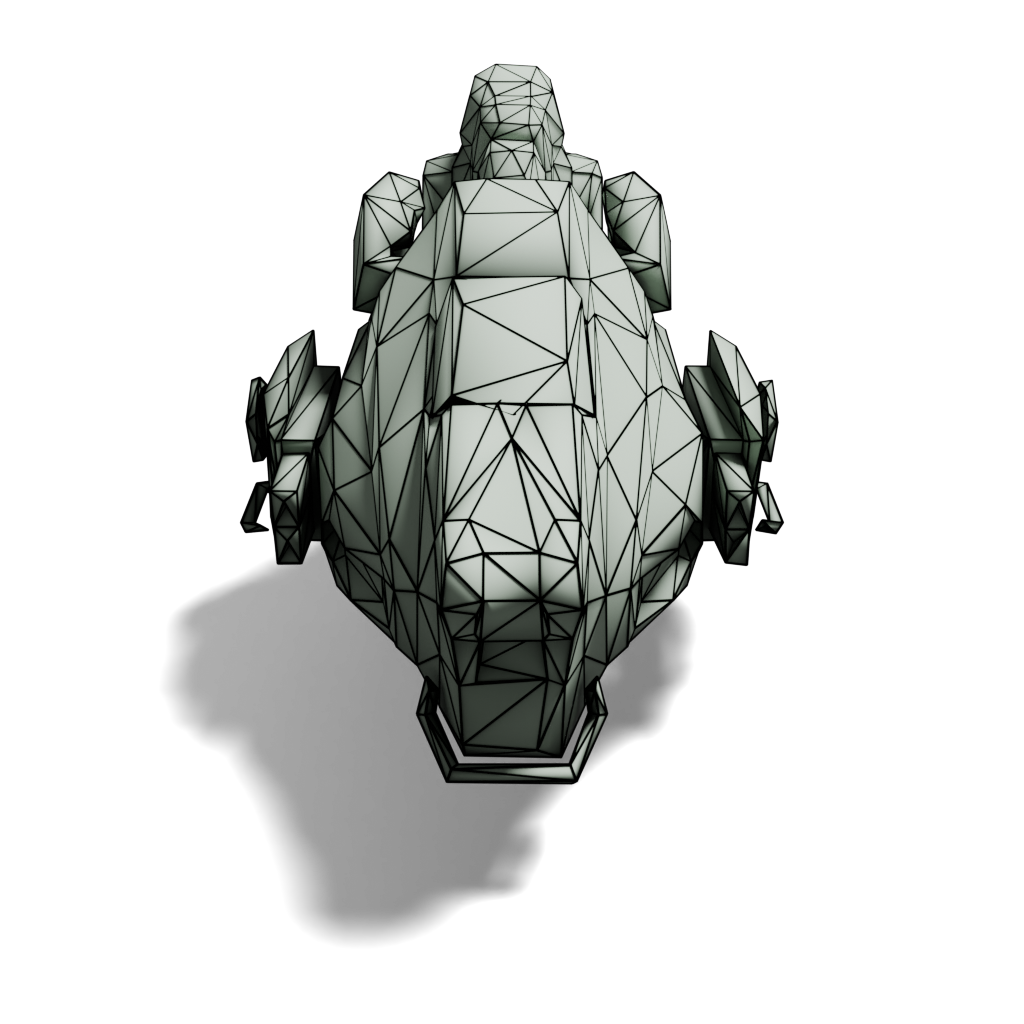}
        \\
        \includegraphics[width=0.33\linewidth]{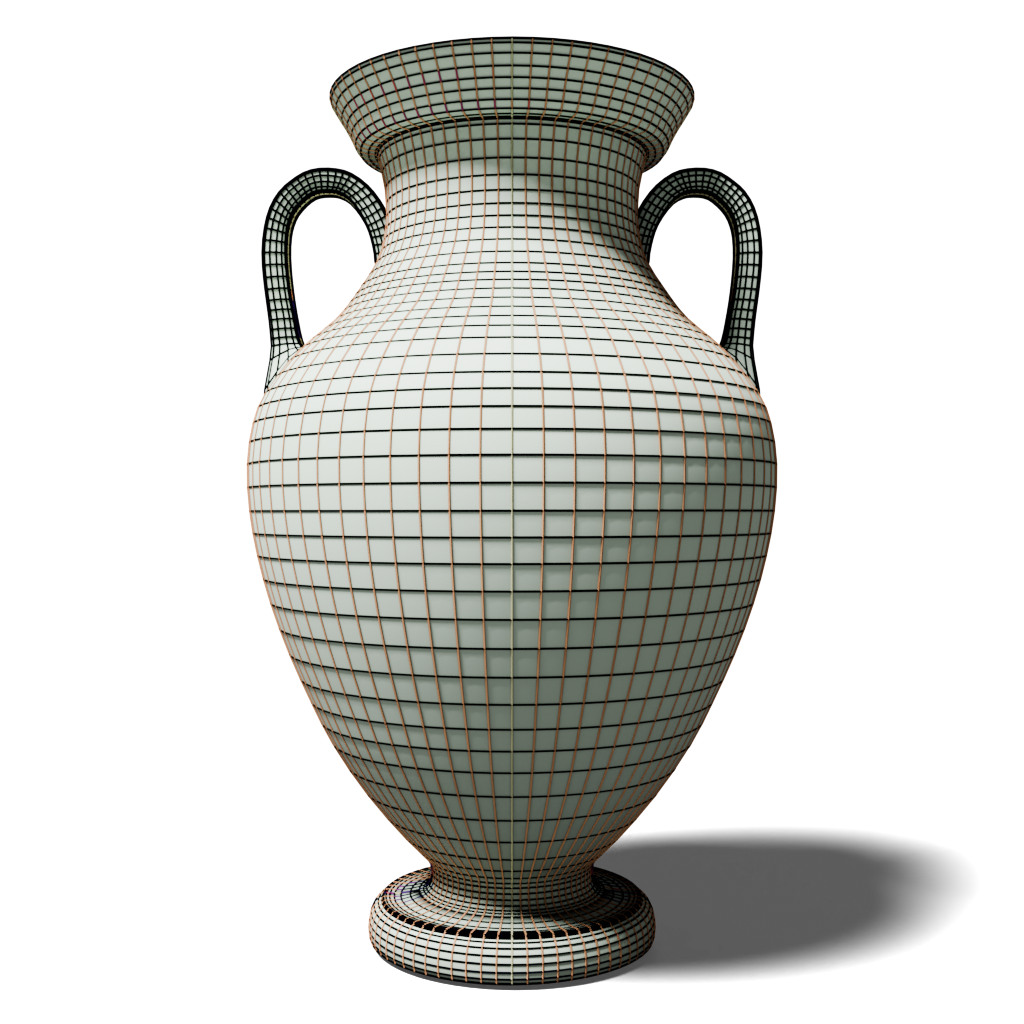} &
        \includegraphics[width=0.33\linewidth]{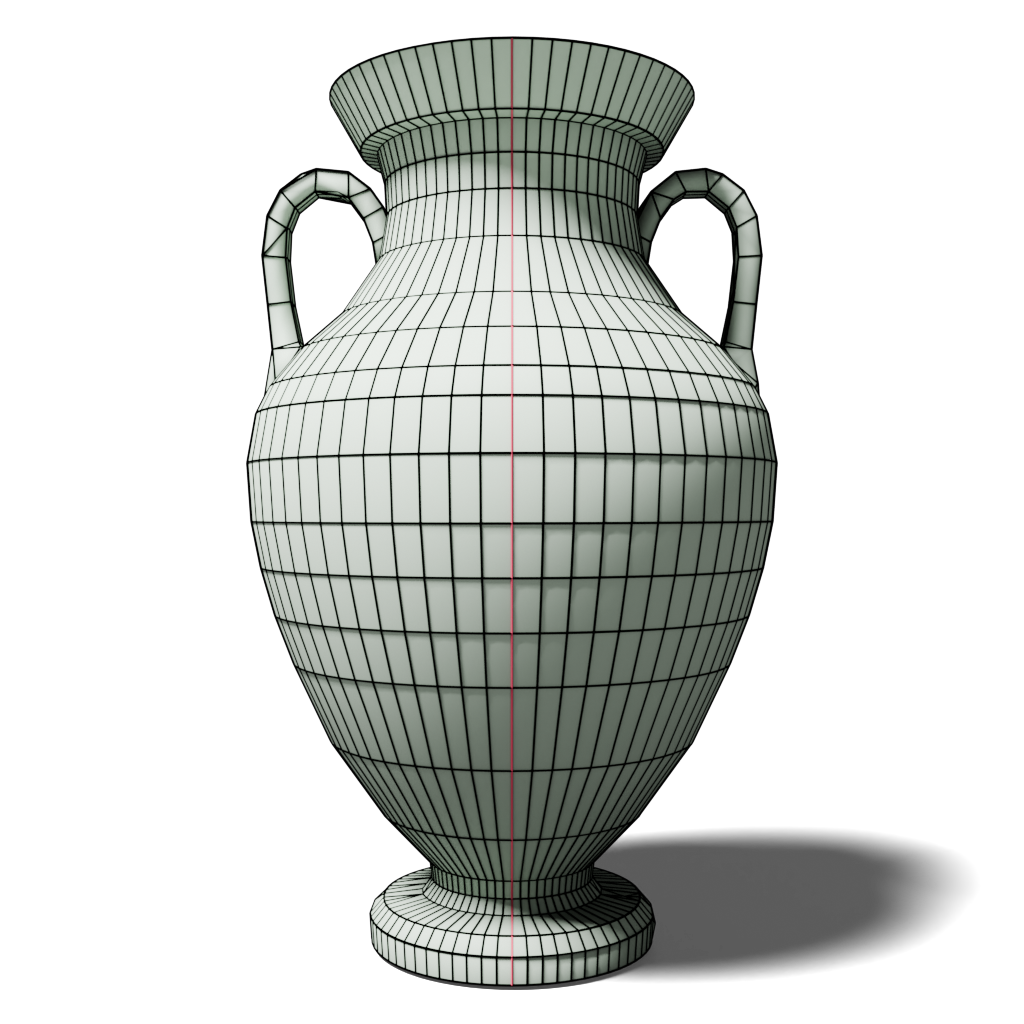} &
        \includegraphics[width=0.33\linewidth]{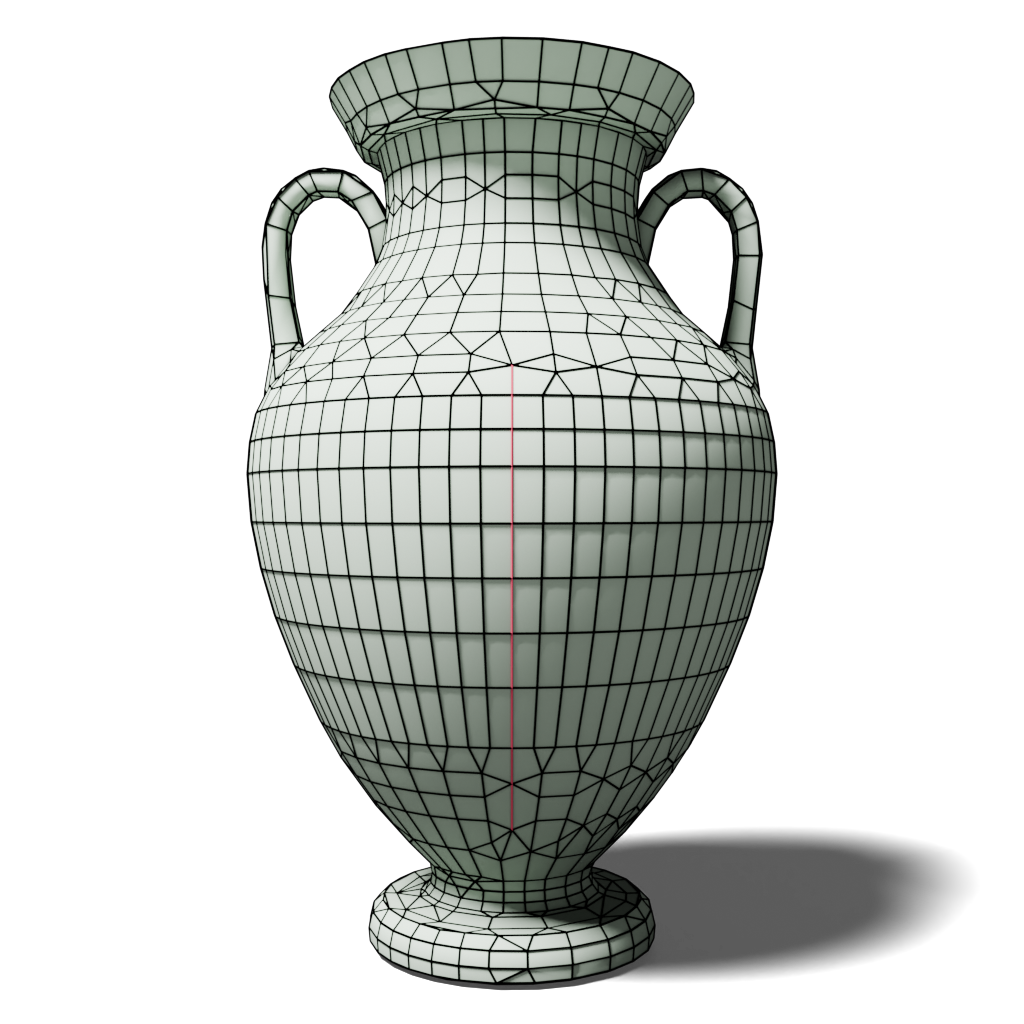}
        \\
        \includegraphics[width=0.33\linewidth]{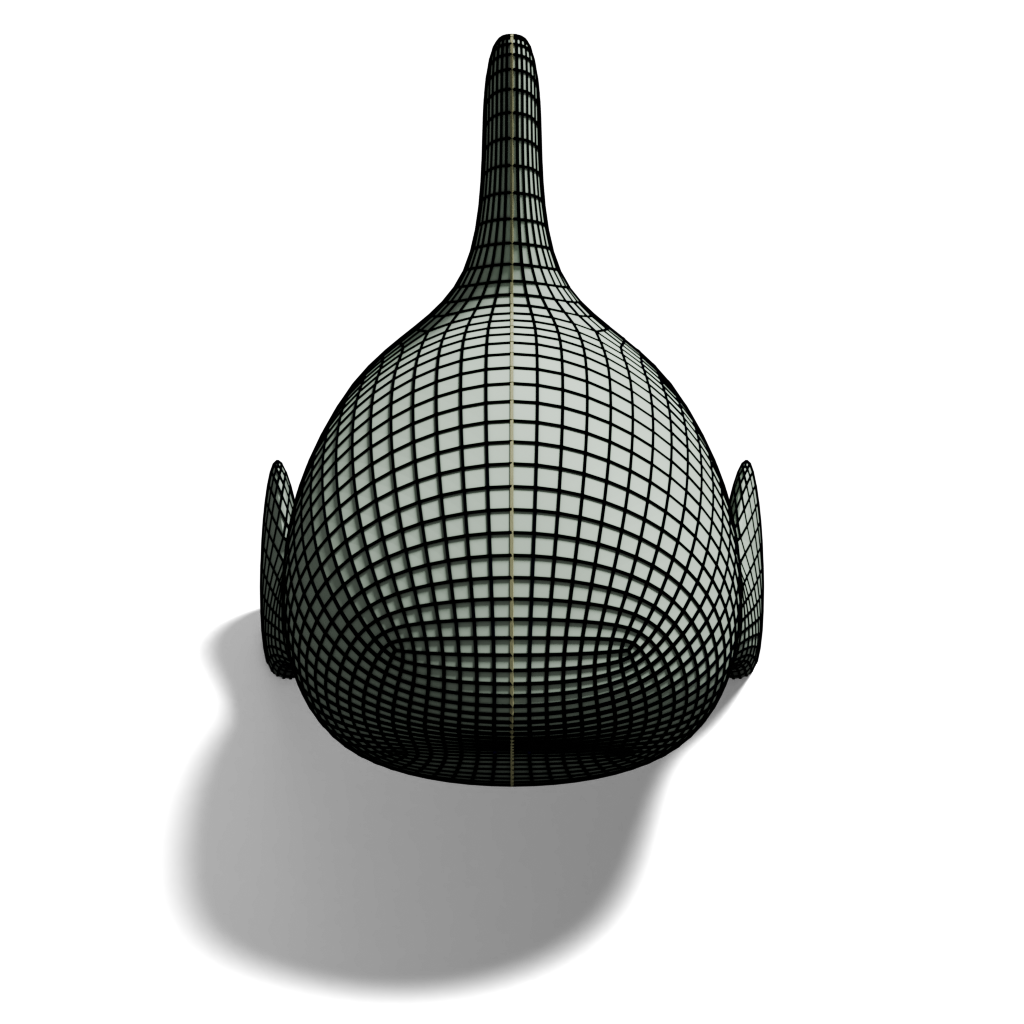} &
        \includegraphics[width=0.33\linewidth]{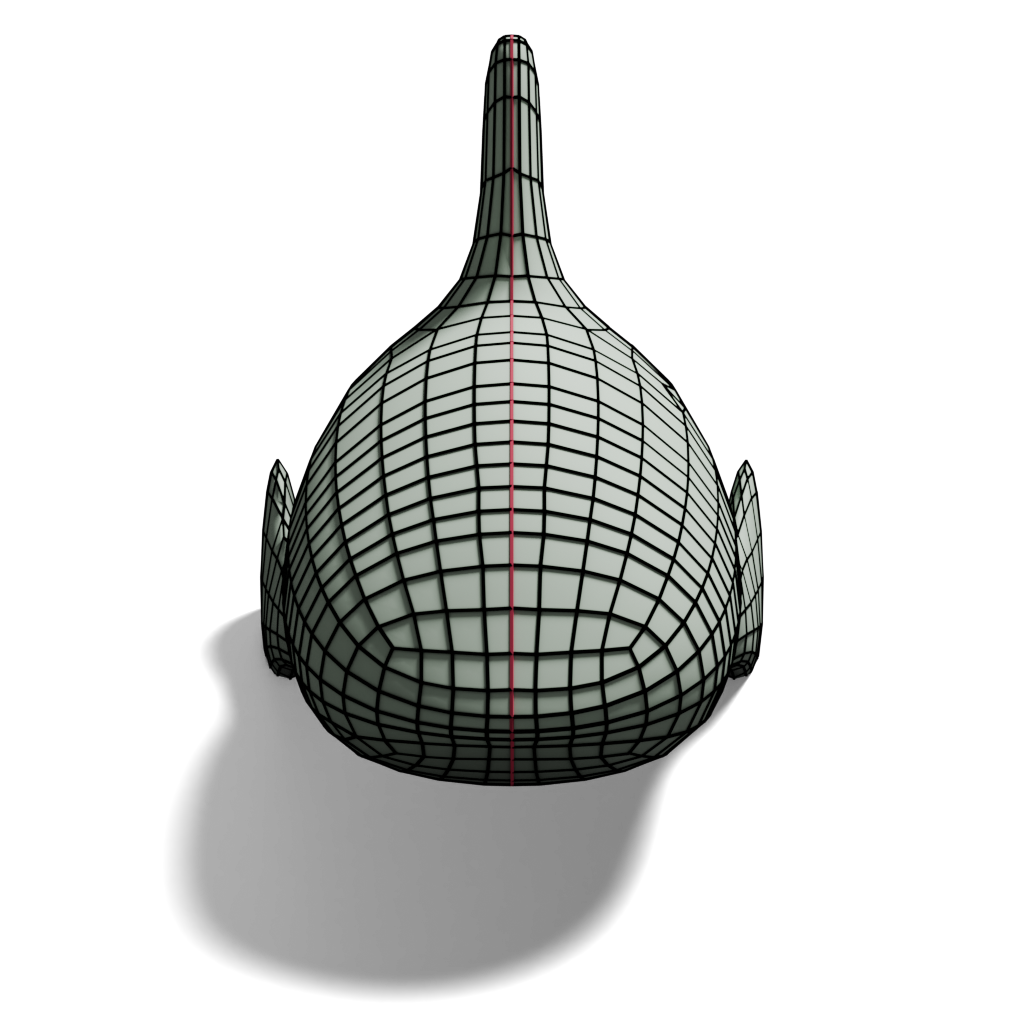} &
        \includegraphics[width=0.33\linewidth]{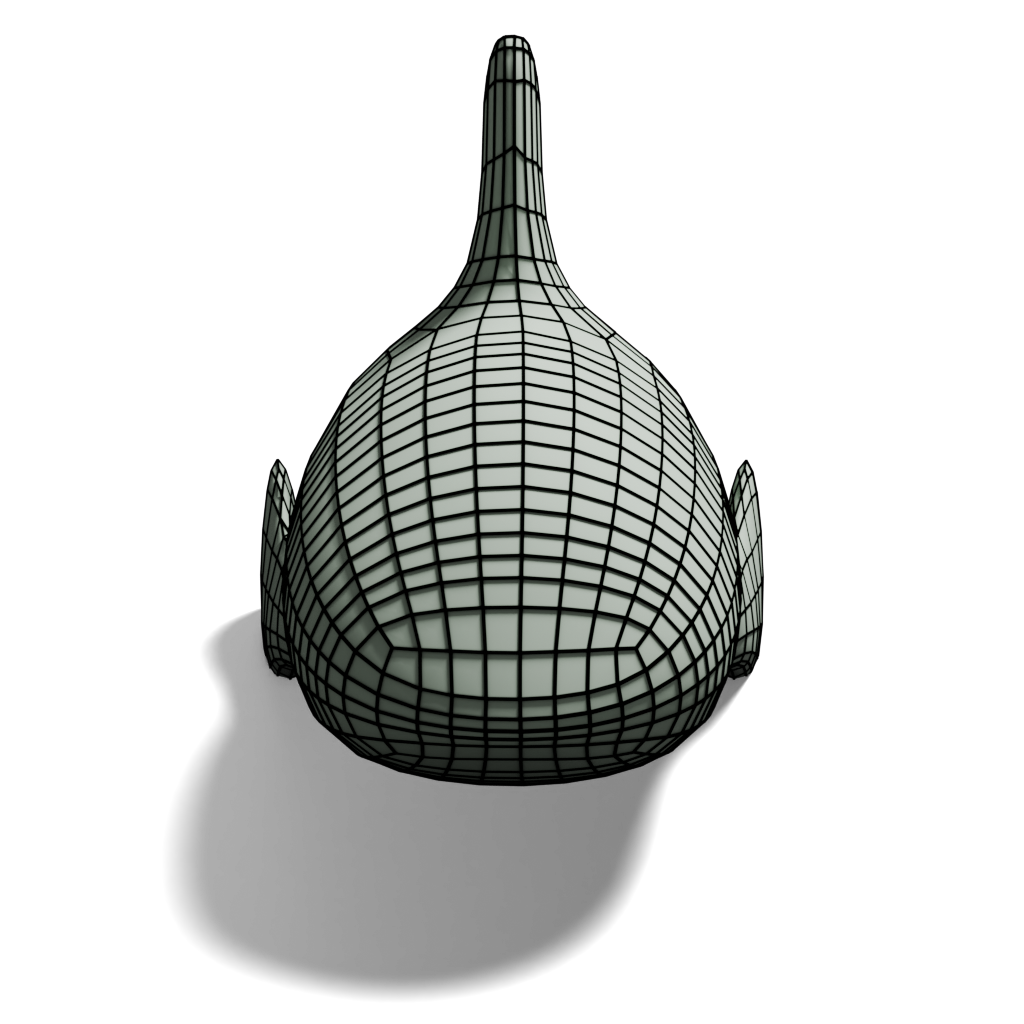}
        \\
        \includegraphics[width=0.33\linewidth]{diagrams/sym_colorbar.pdf} & \multicolumn{2}{c}{\textcolor{red}{Red} Along Axis of Symmetry} \\
    \end{tabular}
    \caption{We show additional examples of symmetry preservation on some example meshes. By increasing the weight of edges along planes of symmetry, our approach is able to preserve these edges on axes of symmetry in the output mesh topology. Note that for the bottom right mesh, even though it looks like the axis is preserved it is slightly offset from the center.}
    \label{fig:sym-additional-examples}
    \Description{
        First row: a hover bike triangle mesh.
        Second row: a greek vase quad mesh.
        Last row: a fish quad mesh.
        Left column: Input mesh with axes of symmetry highlighted, primarily through the yz plane.
        Center: symmetry is preserved with our approach which also generates cleaner topology.
        Right: Symmetry is not preserved as cleanly without symmetric weights, and the topology is much worse.
    }
\end{figure*}

\begin{figure*}
    \centering
    \begin{tabular}{c c}
        \textbf{Chamfer Distances$^\downarrow$} & \textbf{Hausdorff Distances$^\downarrow$} \\
        \put(-0.015\linewidth, 0.17\linewidth){\rotatebox{90}{\textbf{$50\% \triangle$}}}
        \includegraphics[width=0.45\linewidth]{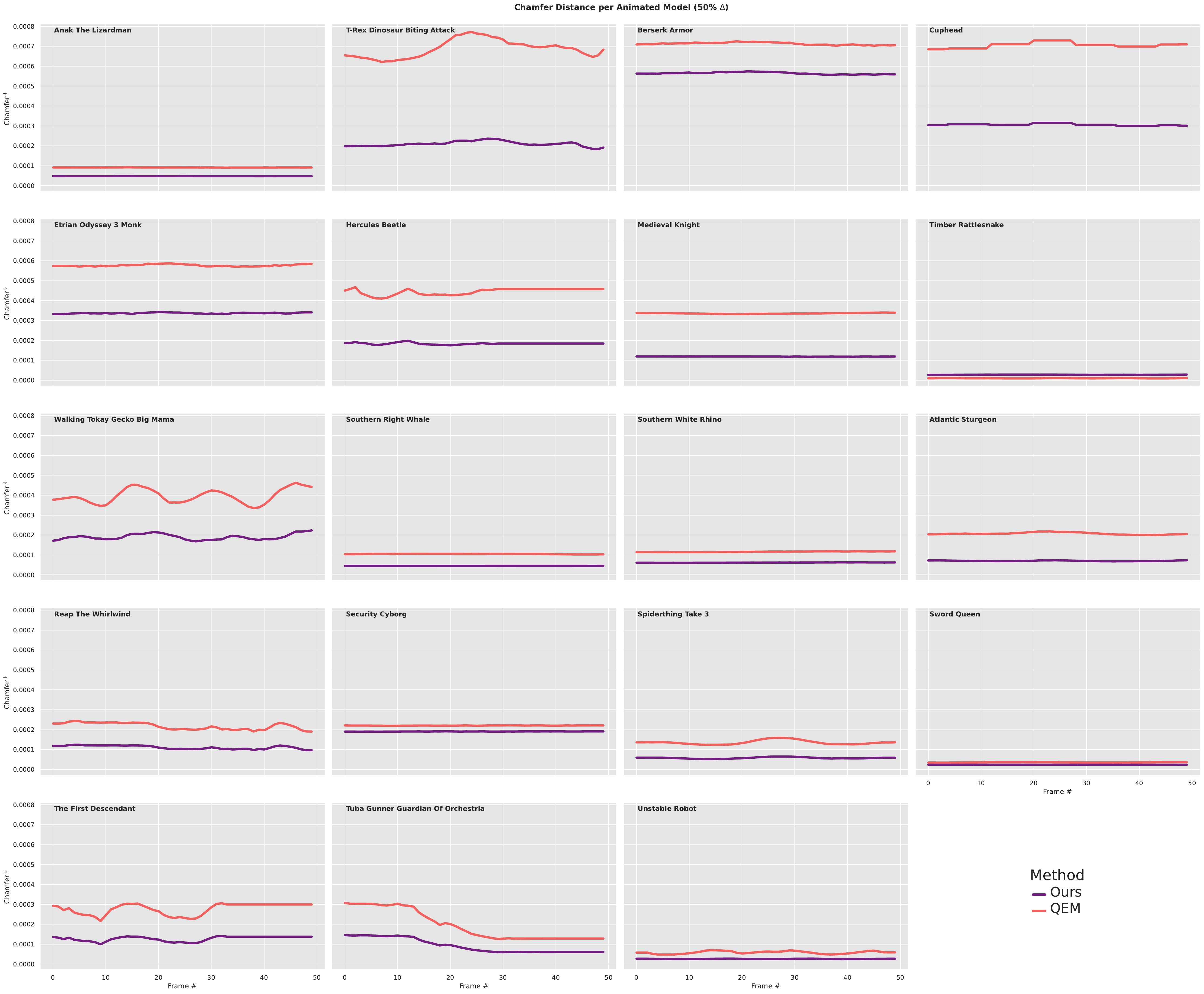} &
        \includegraphics[width=0.45\linewidth]{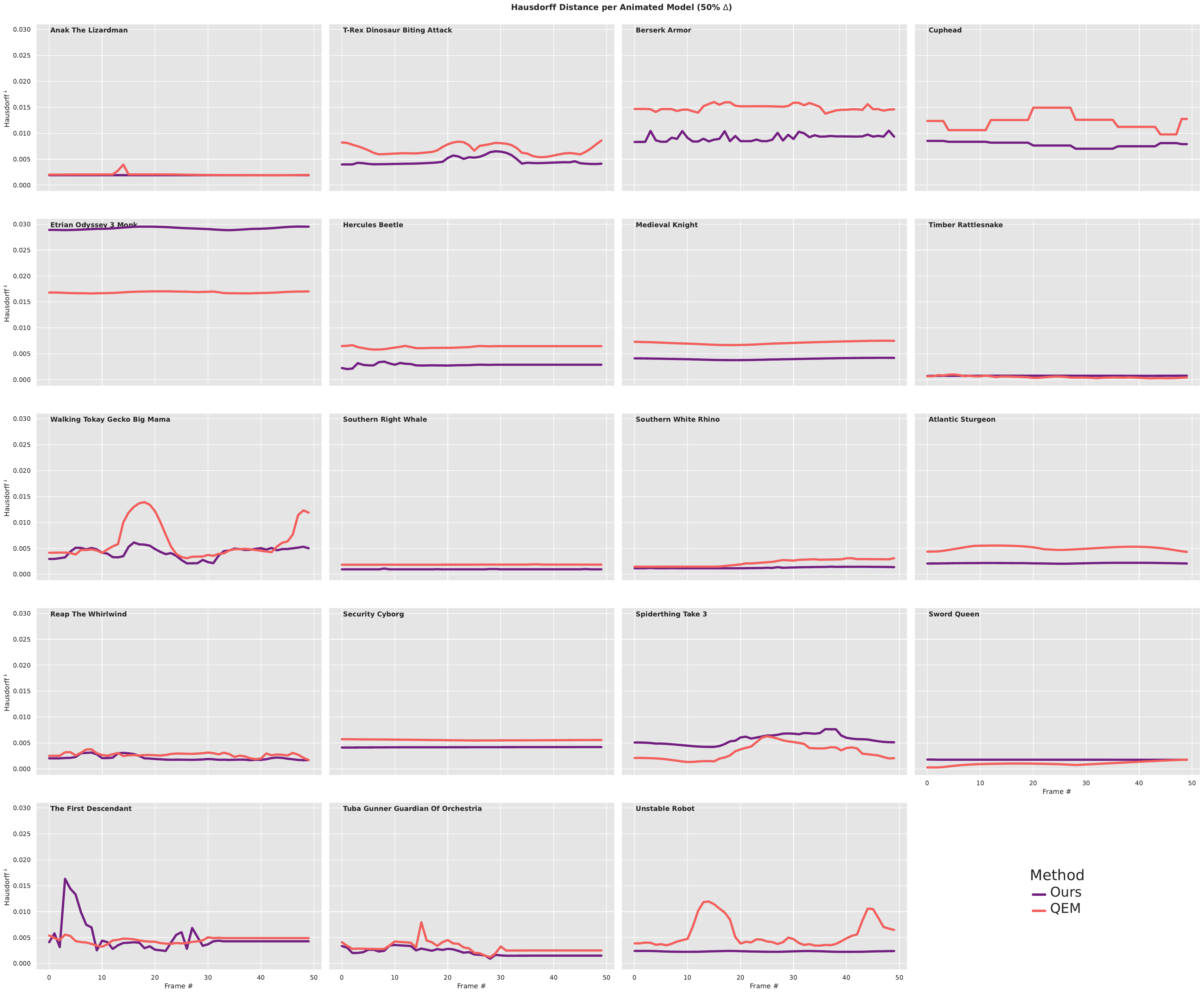} \\
        
        \put(-0.015\linewidth, 0.17\linewidth){\rotatebox{90}{\textbf{$25\% \triangle$}}}
        \includegraphics[width=0.45\linewidth]{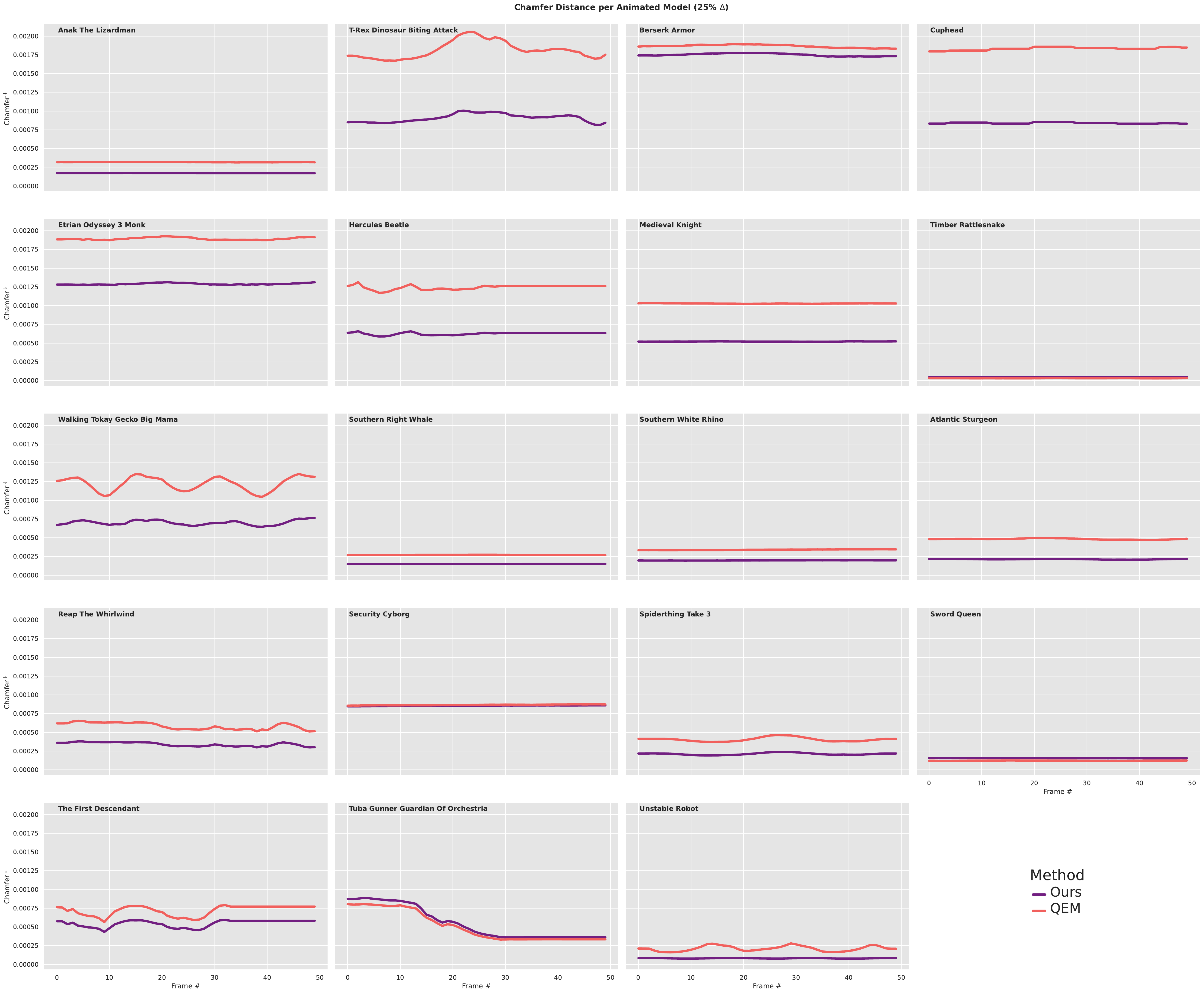} &
        \includegraphics[width=0.45\linewidth]{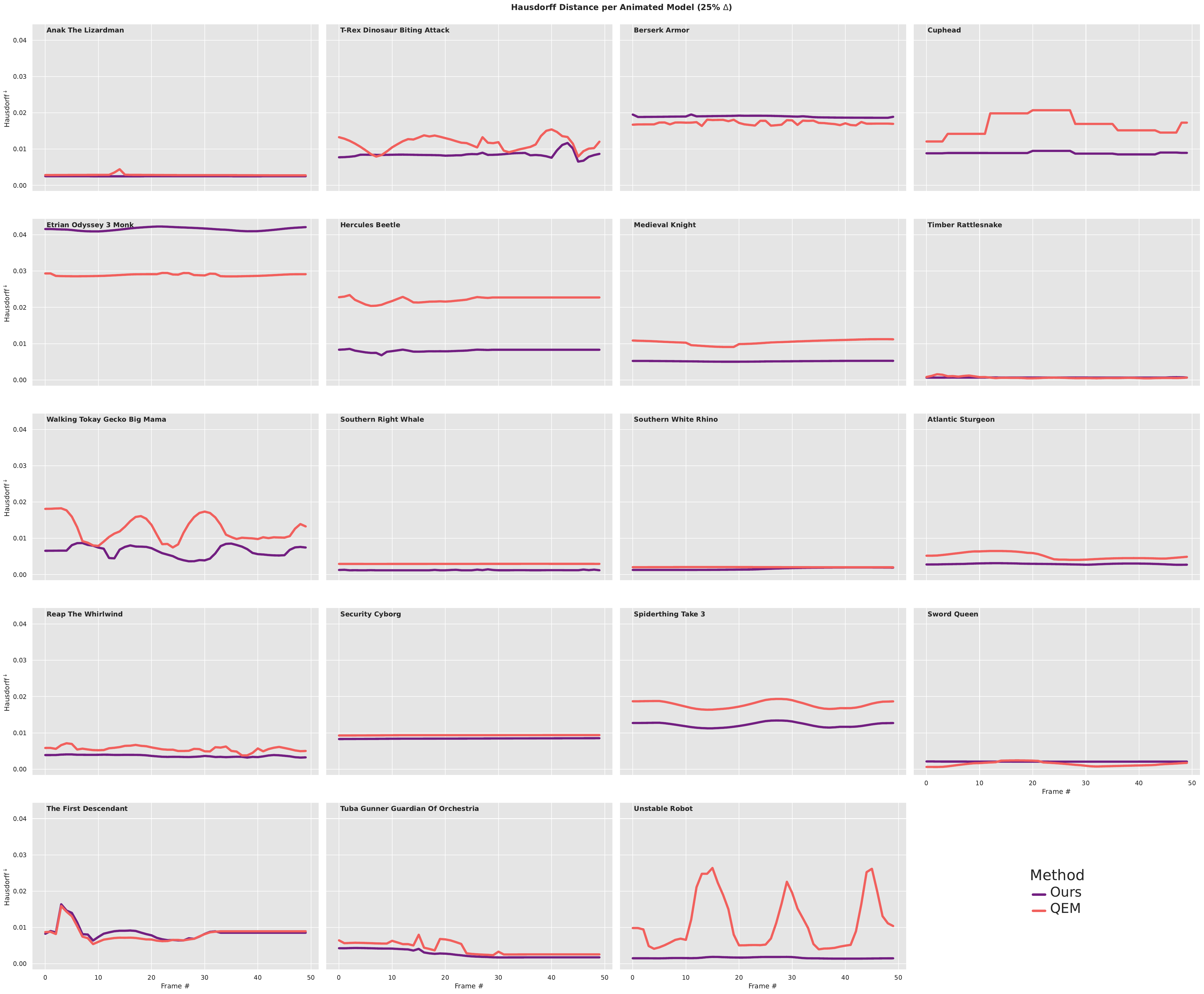} \\
        
        \put(-0.015\linewidth, 0.17\linewidth){\rotatebox{90}{\textbf{$10\% \triangle$}}}
        \includegraphics[width=0.45\linewidth]{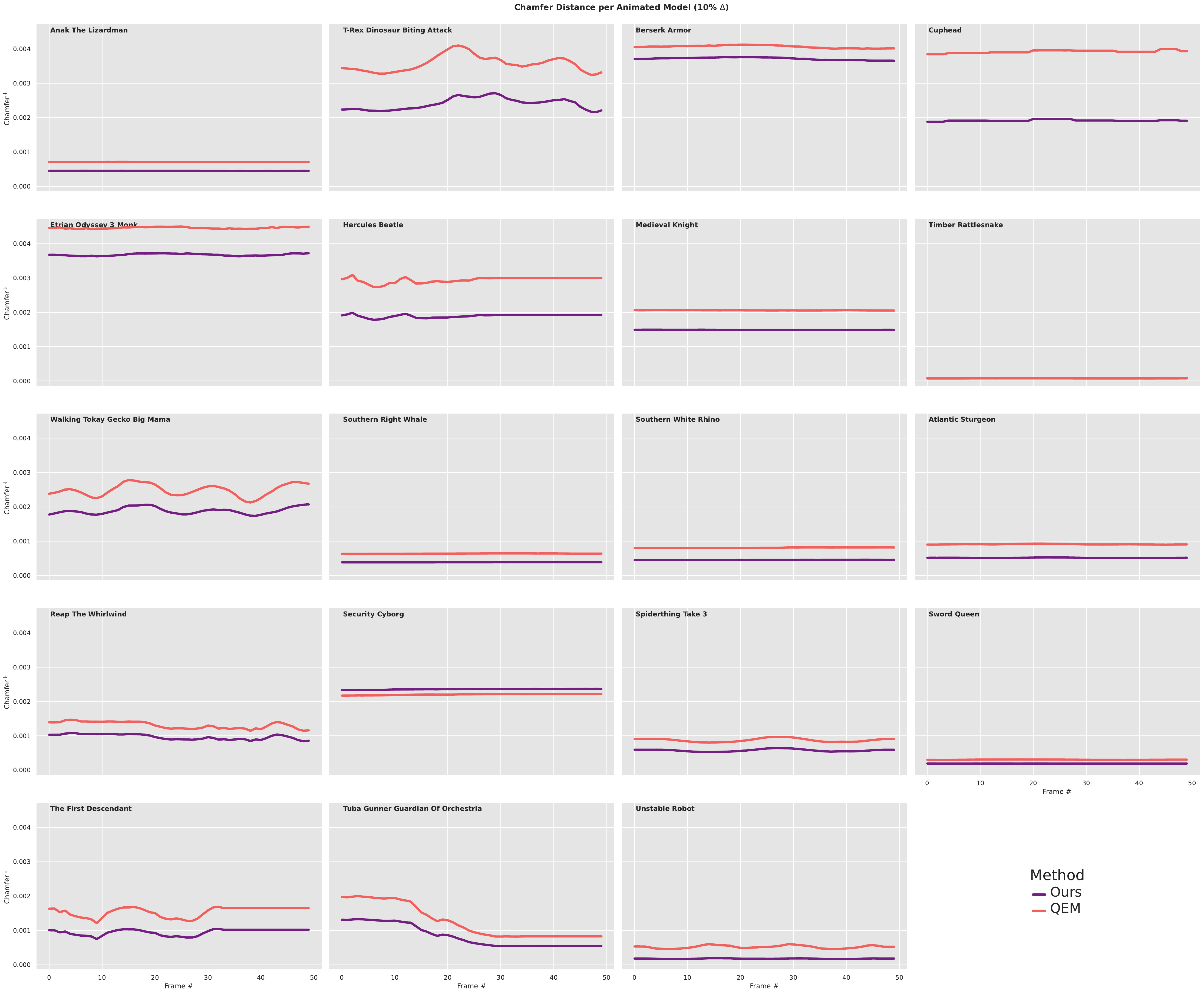} &
        \includegraphics[width=0.45\linewidth]{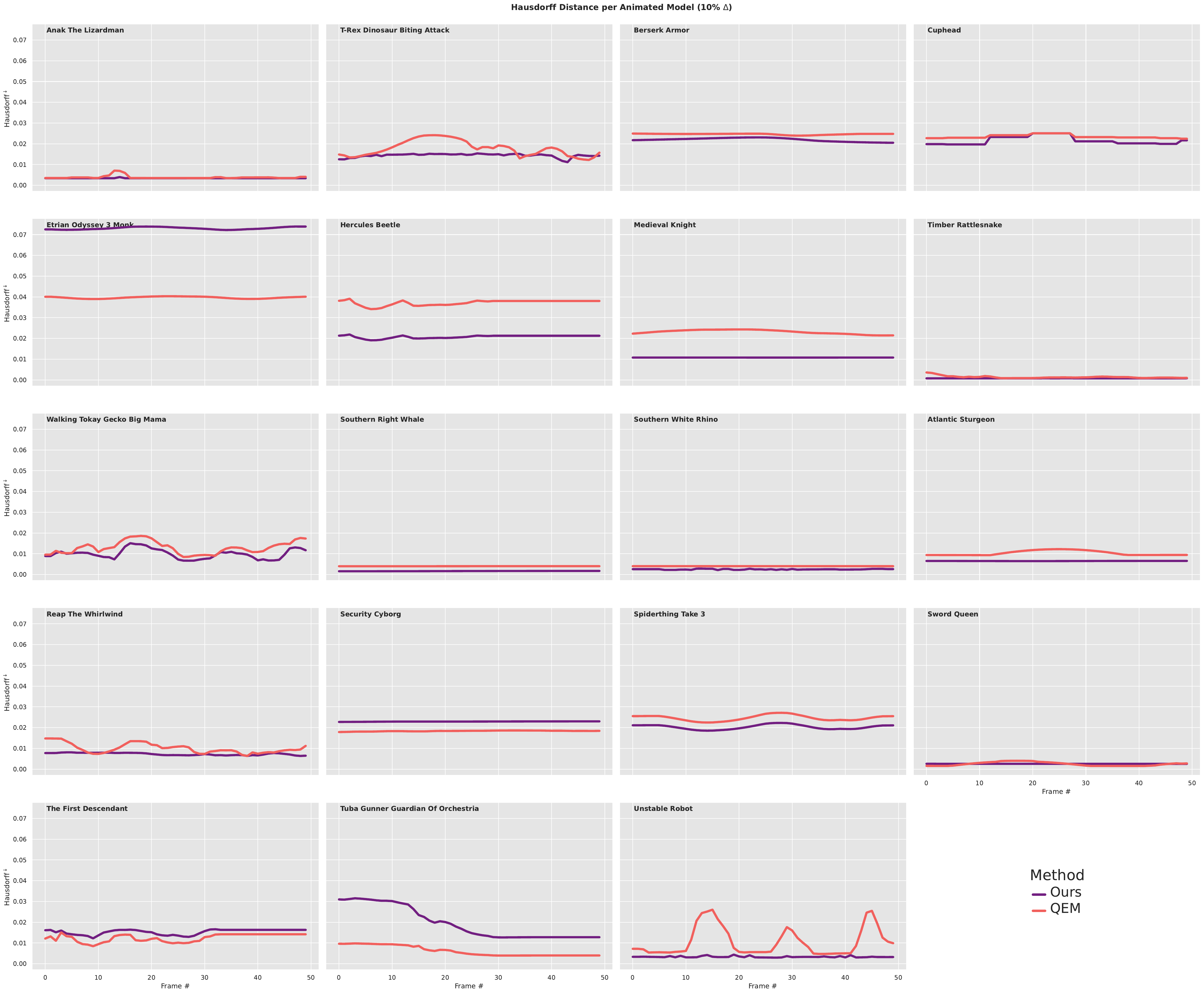} \\
    \end{tabular}
    \caption{Comparison of geometric distance for each of 19 animated models for our approach and ``QEM'' which implements \citep{articulated_mesh_simplification}. Our approach consistently outperforms prior work, especially when considering chamfer distance. On each chart, the x-axis indicates the frame of animation when the distance is measured. For both Chamfer and Hausdorff distances, lower on the y-axis indicates more similarity with the original model.}
    \label{fig:anim-mesh-comparison}
    
    \Description{A ton of time series plots. Left is Chamfer distances, right is Hausdorff distances. Each timeseries has one line for QEM and one line for our approach, indicating geometric distance at a specific frame. The line for our approach is generally lower than for QEM.}
\end{figure*}

In the appendix we include the algorithm for computing per edge symmetry in Alg~\ref{alg:symmetric-edge-weights}. Then, we include additional examples of symmetric preservation in Fig.~\ref{fig:sym-additional-examples}. We also include a table of summary statistics for our whole dataset in Tab.~\ref{tab:dataset_summary}, and complete results of reduction on each individual model in Tab.~\ref{tab:table50}, Tab.~\ref{tab:table25}, and Tab.~\ref{tab:table10}. Furthermore, we show Chamfer and Hausdorff distances on all animated models for the first 50 frames of animation in Fig.~\ref{fig:anim-mesh-comparison}. Finally, we show two visual comparisons of animated mesh reduction in Fig.~\ref{fig:anim-mesh-visual}.

\begin{algorithm}
\caption{Edge Symmetry Weights \label{alg:symmetric-edge-weights}}
\begin{algorithmic}[1]
\State\textbf{Input}: Vertices $V$, edges $E$, faces $F$, distance $\delta$
\State\textbf{Output}: Per edge weights w $\in [0,1]^{|E|}$
\For{e $ = (e_0, e_1)\in E$}
\State $d = \frac{e_0 - e_1}{\lVert e_0 - e_1 \rVert_2}$
\If{e adjacent to $f_0, f_1$} \Comment{Manifold Edges}
\State $n_\perp = \frac{n(f_0) + n(f_1)}{\lVert n(f_0) + n(f_1)\rVert_2}\times d$    
\ElsIf{e adjacent to $f_0$} \Comment{Boundary Edges}
\State $n_\perp = n(f_0)\times d$
\Else \Comment{Non-Manifold Edges}
\State w[e] = 0
\State \textbf{continue}
\EndIf
\State $\text{dists}_\text{edge}: \text{list of } \mathbb{R}^3$
\For{v $\in V$} \Comment{Local pos. of $v$ in plane defined by edge}
    \State $v_\text{local} = \text{local}(v, \text{origin} = e_0, x = n_\perp,  y = d, z = n_\perp\times d)$
    \If{$v_\text{local}[0] < \delta$} \Comment{dist to plane $< \delta$, match itself}
        \State $\text{matched}(e) \mathrel{+}= 1$
        \State\textbf{continue}
    \EndIf
    \State $\text{push } v_\text{local} \text{ into } \text{dists}_\text{edge}$
\EndFor
\State Sort $\text{dists}_\text{edge}$ by $\text{abs}(v[0])$
\While{not empty($\text{dists}_\text{edge}$)}
    \State $[d, x, y] = \text{dists}_\text{edge}$.pop\_back()
    \State Select last $k_0$ elements where $|v[0] - d| < \delta$
    \State Sort last $k_0$ elements by $v[1]$ (local x)
    \State Select last $k_1$ where $|v[1] - x| < \delta$
    \For{$v' = [d_n, x_n, y_n]\in\text{dists}_\text{edge}$}\\
        \Comment{Check if vertex is on opp. side of plane within $\delta$}
        \If{$\lVert (d_n + d)^2 + (x_n - x)^2 + (y_n - y)^2\rVert_2 < \delta$}
        \State $\text{matched}(e) \mathrel{+}= 2$
        \State Remove $v'$ from $\text{dists}_\text{edge}$
        \State\textbf{break}
        \EndIf
    \EndFor
    \State Sort last $k_0$ by $|v[0]|$ \Comment{Reset last $k_0$ to original state}
\EndWhile
\State w[e] = $\frac{\text{matched}(e)}{|V|}$
\EndFor
\Return w
\end{algorithmic}
\end{algorithm}

\begin{table*}
\renewcommand{\arraystretch}{0.7}
\centering
\footnotesize
\begin{tabular}{|c|c|c|c|c|c|}
\hline
Name & Input $|F|$ & Input $|V|$ & Input $\square$ & Input $\Delta$ & Source \\ \hline\hline
Door Panel & 64000 & 33599 & 0 & 64000 & \cczero Minneapolis Institute of Art \\\hline
2B & 1139322 & 920574 & 621539 & 517452 & \ccby MatheusEvan \\\hline
\textcolor{orange}{Anak The Lizardman} & 165550 & 88385 & 0 & 165550 & \ccbync Hawk \\\hline
Angelica & 78608 & 93049 & 64994 & 13574 & \ccby DarkNik \\\hline
Angkor Wat & 484200 & 293126 & 0 & 484200 & \ccbync GSXNet \\\hline
\textcolor{orange}{T-Rex Dinosaur Biting Attack} & 11938 & 6819 & 0 & 11938 & \ccby LasquetiSpice \\\hline
Arches & 14032 & 28540 & 14032 & 0 & \ccby Ayman.Ibrahim \\\hline
Armadillo & 99976 & 49990 & 0 & 99976 & The Stanford 3D Scanning Repository \\\hline
Armored Charizard & 4649 & 3273 & 12 & 4637 & \ccby projectmgame\\\hline
\textcolor{orange}{Berserk Armor} & 19210 & 22470 & 0 & 19210 & \ccby Angy97 \\\hline
Bicycle & 54995 & 102354 & 42464 & 10396 & \ccby wanderer \\\hline
Blub Quadrangulated & 7104 & 7317 & 7104 & 0 & \cczero Keenan Crane \\\hline
Botchling & 157549 & 164350 & 157140 & 409 & \ccby HD \\\hline
Casio Keyboard & 2804 & 3730 & 2649 & 0 & \ccby AleixoAlonso \\\hline
Chimney Pipe & 3404 & 4864 & 3404 & 0 & \ccby RubaQewar \\\hline
Chitinous Knight & 87258 & 89239 & 87212 & 46 & \ccby Ploobert \\\hline
Cube & 5756 & 7401 & 4236 & 1520 & \ccby Deslancer \\\hline
\textcolor{orange}{Cuphead} & 35240 & 19820 & 0 & 35240 & \ccbync AlmondFeather \\\hline
Cyberpunk Bike & 31916 & 37040 & 30488 & 1256 & \ccbync daCruz \\\hline
Dense Cube & 1014 & 1176 & 1014 & 0 & Handmade \\\hline
Dragon With Pearl & 999634 & 499803 & 0 & 999634 & \ccby Artec 3D \\\hline
Dreadroamer & 121482 & 81462 & 0 & 121482 & \ccby TooManyDemons \\\hline
Dry Tree & 149384 & 124158 & 0 & 149384 & \ccby Epic\_Tree\_Store \\\hline
Egyptian Vase & 7116 & 7557 & 7116 & 0 & \ccby Rijo \\\hline
\textcolor{orange}{Etrian Odyssey 3 Monk} & 22821 & 14201 & 0 & 22821 & \ccbync Chavafei \\\hline
Fandisk & 12946 & 6475 & 0 & 12946 & Public Domain \\\hline
Floral Pattern & 104752 & 118792 & 104752 & 0 & \ccby RitorDP \\\hline
French Halfbasket & 35288 & 41255 & 35288 & 0 & \ccby leeeck \\\hline
Globophobia & 38239 & 41366 & 38235 & 4 & \ccby felipegall \\\hline
Goblin Portrait & 54519 & 57415 & 54388 & 131 & \ccby Vadik An \\\hline
Greek Vase & 12545 & 14464 & 12544 & 0 & \ccby Davide Specchi\\\hline
\textcolor{orange}{Hercules Beetle} & 46892 & 140676 & 0 & 46892 & \ccby ArachnoBoy \\\hline
His Name Is Violence & 14416 & 18865 & 11778 & 2613 & \ccby Gleb Parshikov \\\hline
Hover Bike & 9327 & 7025 & 0 & 9327 & \ccbync Ptis \\\hline
Japanese House 2 & 233965 & 235330 & 233904 & 0 & \ccby neos\_nelia\\\hline
Japanese House 3 & 41462 & 42444 & 41298 & 132 & \ccby Karol \\\hline
Japanese Shinto Shrine & 6280 & 6638 & 5790 & 490 & \ccby Deerex \\\hline
Kagiya Edo Tokyo & 22819 & 18820 & 9855 & 12714 & \ccbynd t\_c. \\\hline
Lamborghini Countach & 126970 & 152102 & 119160 & 5342 & \ccbync Lexyc16 \\\hline
Lantern & 1360 & 1862 & 1328 & 32 & \ccby Blenderolokos \\\hline
Lego Figure & 3680 & 1938 & 0 & 3680 & \ccby Dirk.z \\\hline
Lekythos & 1075 & 1148 & 1075 & 0 & \ccby Global Digital Heritage and GDH-Afrika \\\hline
Lethe & 4782 & 8219 & 3986 & 789 & \ccby Klinepeter \\\hline
Marble Track & 32558 & 47029 & 26461 & 4566 & \ccby lin15101 \\\hline
Maze 45 & 37191 & 40663 & 37191 & 0 & \ccby Talaei-dev \\\hline
Maze & 37191 & 40663 & 37191 & 0 & \ccby Talaei-dev \\\hline
Mech & 8732 & 15771 & 5850 & 2872 & \ccby KhoaMinh \\\hline
\textcolor{orange}{Medieval Knight} & 41654 & 31935 & 0 & 41654 & \ccby by\_Rx \\\hline
Melon Pallet & 45016 & 45342 & 45016 & 0 & \ccby Duznot  \\\hline
Militech Robot & 52349 & 29403 & 0 & 52349 & \ccby Ludovic\_Jouault \\\hline
\textcolor{orange}{Timber Rattlesnake} & 240115 & 124297 & 0 & 240115 & \ccbync DigitalLife3D \\\hline
\textcolor{orange}{Walking Tokay Gecko Big Mama} & 7736 & 4491 & 0 & 7736 & \ccbync DigitalLife3D \\\hline
\textcolor{orange}{Southern Right Whale} & 36484 & 21547 & 0 & 36484 & \ccbync DigitalLife3D \\\hline
\textcolor{orange}{Southern White Rhino} & 64542 & 33743 & 0 & 64542 & \ccbync DigitalLife3D \\\hline
\textcolor{orange}{Atlantic Sturgeon} & 21856 & 11805 & 0 & 21856 & \ccbync DigitalLife3D \\\hline
Oni No Face Mask & 219131 & 109564 & 0 & 219131 & \ccby warpnik \\\hline
One Angery Dragon Boi & 67696 & 74786 & 67632 & 64 & \ccby ricocilliers \\\hline
Organic Armor & 47270 & 48638 & 47270 & 0 & \ccby Ploobert \\\hline
Pipe Wall & 1545 & 1791 & 1543 & 2 & \ccby Aerial\_Knight \\\hline
Prometheus & 418183 & 400330 & 379559 & 38442 & \ccby Lapshin Sergey (in Cyrillic) \\\hline
Radial & 10112 & 7338 & 0 & 10112 & \ccby Renafox \\\hline
\textcolor{orange}{Reap The Whirlwind} & 173179 & 109483 & 0 & 173179 & \ccby TooManyDemons \\\hline
Roman Soldier & 156714 & 148281 & 123893 & 32819 & \ccby Andy Woodhead \\\hline
Rope Bridge & 7256 & 7624 & 4444 & 2688 & \ccby lucq \\\hline
\textcolor{orange}{Security Cyborg} & 29928 & 24106 & 0 & 29928 & \ccby fletcherkinnear \\\hline
Shinto Watchtower & 131367 & 134770 & 123508 & 6164 & \ccby sin\_nass \\\hline
Shirakawago House & 15965 & 21329 & 15867 & 96 & \ccby Lokomotto \\\hline
Skyward Ddf Poco Ship & 183177 & 143771 & 12207 & 170968 & \ccby Simon Jennings \\\hline
Snowflake Orb & 77418 & 38395 & 0 & 77418 & \ccby Craleigh \\\hline
Space Fighter & 34965 & 55060 & 34057 & 40 & Free Standard Kerem Kavalci \\\hline
Speeder Bike & 11926 & 15152 & 11568 & 188 & \ccbync Tasha.Lime \\\hline
\textcolor{orange}{Spiderthing Take 3} & 67972 & 37862 & 0 & 67972 & \ccby fuzzy4fox \\\hline
Spiral Staircase & 6358 & 7437 & 1400 & 3949 & \ccby Hoody468 \\\hline
Spot Quadrangulated & 2928 & 3225 & 2928 & 0 & Keenan Crane \\\hline
Steel Guard -7Mb & 14744 & 14559 & 0 & 14744 & \ccby Mehdi Shahsavan \\\hline
\textcolor{orange}{Sword Queen} & 179602 & 119123 & 0 & 179602 & \ccbync Mateusz Woliński \\\hline
Tank & 15306 & 21338 & 15063 & 203 & \ccby maxpsr \\\hline
Teacup & 5728 & 3093 & 5728 & 0 & \ccby miranda.j.rice  \\\hline
The Dark Wraith & 119034 & 123271 & 80572 & 38462 & \ccby davidkong \\\hline
\textcolor{orange}{The First Descendant} & 106120 & 59569 & 0 & 106120 & \ccbync Catholomew \\\hline
\textcolor{orange}{Tuba Gunner Guardian Of Orchestria} & 66488 & 46167 & 0 & 66488 & \ccby Harri3D \\\hline
\textcolor{orange}{Unstable Robot} & 208568 & 124703 & 0 & 208568 & \ccby kalsloos \\\hline
Venice Mask & 260391 & 173954 & 35201 & 225190 & \ccbync DailyArt \\\hline
Vera Posed & 47790 & 66243 & 40269 & 7517 & \ccby JohnMesplay \\\hline
War Tank & 35126 & 68541 & 27758 & 7368 & \ccby larrynguyen \\\hline
Yeahright & 377344 & 377084 & 377344 & 0 & \cczero Keenan Crane \\\hline
\end{tabular}
\caption{Summary of our input dataset taken from meshes online. It consists of some full quad meshes, some full triangle meshes, and mix of hybrid meshes. Animated meshes are highlighted in \textcolor{orange}{orange}.}
\label{tab:dataset_summary}
\end{table*}

\begin{table*}
\renewcommand{\arraystretch}{0.8}
\sisetup{tight-spacing=true}
\setlength{\tabcolsep}{0.1pt}
\centering
\footnotesize
\begin{tabular}{|c|c|c|c|c|c|c|c|c|c|c|c|}
\hline
\multicolumn{3}{|c|}{50\% Reduction} & \multicolumn{3}{c|}{Ours} & \multicolumn{3}{c|}{QEM} & \multicolumn{3}{c|}{MeshLab} \\ \hline
Name & Out $\triangle$ & In $\square$ & Out $\square$ & Chamfer$^\downarrow$ & Hausdorff$^\downarrow$ & Out $\square$ & Chamfer$^\downarrow$ & Hausdorff$^\downarrow$ & Out $\square$ & Chamfer$^\downarrow$ & Hausdorff$^\downarrow$\\ \hline\hline
Door Panel & 32007 & 0 & 0 & $\cellcolor{blue!10} \num{1.008e-04}$ & $\num{1.637e-03}$ & 0 & $\num{2.049e-04}$ & $\num{3.336e-03}$ & 0 & $\num{1.610e-04}$ & $\cellcolor{blue!10} \num{1.404e-03}$ \\ \hline
2B & 880826 & 621539 & $\cellcolor{blue!10} 279096$ & $\num{3.421e-05}$ & $\num{1.959e-03}$ & 173585 & $\num{3.476e-05}$ & $\num{2.119e-03}$ & 0 & $\num{4.078e-05}$ & $\cellcolor{blue!10} \num{8.862e-04}$ \\ \hline
Angelica & 71861 & 64994 & $\cellcolor{blue!10} 26721$ & $\num{4.047e-04}$ & $\num{1.783e-02}$ & 23367 & $\num{4.007e-04}$ & $\num{2.371e-02}$ & 0 & $\cellcolor{blue!10} \num{2.501e-04}$ & $\cellcolor{blue!10} \num{7.024e-03}$ \\ \hline
Angkor Wat & 242128 & 0 & 0 & $\num{3.499e-05}$ & $\cellcolor{blue!10} \num{4.247e-04}$ & 0 & $\num{4.713e-05}$ & $\num{2.190e-03}$ & 0 & $\num{4.873e-05}$ & $\num{8.047e-04}$ \\ \hline
Arches & 14032 & 14032 & $\cellcolor{blue!10} 6860$ & $\cellcolor{blue!10} \num{2.043e-05}$ & $\cellcolor{blue!10} \num{2.440e-04}$ & 3302 & $\num{1.302e-03}$ & $\num{4.698e-03}$ & 0 & $\num{2.507e-04}$ & $\num{1.074e-03}$ \\ \hline
Armadillo & 49994 & 0 & 0 & $\num{1.321e-04}$ & $\cellcolor{blue!10} \num{5.599e-04}$ & 0 & $\num{1.949e-04}$ & $\num{2.758e-03}$ & 0 & $\num{1.242e-04}$ & $\num{9.217e-04}$ \\ \hline
Armored Charizard & 2332 & 12 & 12 & $\cellcolor{blue!10} \num{5.494e-04}$ & $\num{1.959e-02}$ & 12 & $\num{9.968e-04}$ & $\cellcolor{blue!10} \num{8.160e-03}$ & 0 & $\num{8.056e-04}$ & $\num{1.222e-02}$ \\ \hline
Bicycle & 60824 & 41618 & 20601 & $\num{5.517e-05}$ & $\num{3.485e-03}$ & $\cellcolor{blue!10} 25297$ & $\num{5.306e-05}$ & $\cellcolor{blue!10} \num{2.016e-03}$ & 0 & $\num{2.334e-04}$ & $\num{5.542e-03}$ \\ \hline
Blub Quadrangulated & 7104 & 7104 & $\cellcolor{blue!10} 3505$ & $\num{2.149e-04}$ & $\num{2.264e-03}$ & 2866 & $\num{2.672e-04}$ & $\num{7.983e-03}$ & 0 & $\num{1.726e-04}$ & $\cellcolor{blue!10} \num{1.467e-03}$ \\ \hline
Botchling & 157349 & 157140 & $\cellcolor{blue!10} 78206$ & $\num{7.155e-05}$ & $\num{2.838e-03}$ & 55701 & $\num{6.314e-05}$ & $\num{2.838e-03}$ & 0 & $\num{6.250e-05}$ & $\cellcolor{blue!10} \num{8.536e-04}$ \\ \hline
Casio Keyboard & 3448 & 2649 & 1018 & $\num{1.232e-04}$ & $\num{1.397e-02}$ & $\cellcolor{blue!10} 1206$ & $\num{7.605e-05}$ & $\cellcolor{blue!10} \num{5.985e-03}$ & 0 & $\num{4.232e-04}$ & $\num{1.537e-02}$ \\ \hline
Chimney Pipe & 3429 & 3404 & $\cellcolor{blue!10} 1633$ & $\cellcolor{blue!10} \num{2.128e-05}$ & $\cellcolor{blue!10} \num{1.099e-03}$ & 1186 & $\num{1.252e-04}$ & $\num{5.281e-03}$ & 0 & $\num{6.837e-04}$ & $\num{1.201e-02}$ \\ \hline
Chitinous Knight & 87234 & 87212 & $\cellcolor{blue!10} 43427$ & $\num{1.248e-04}$ & $\num{1.031e-03}$ & 26437 & $\num{1.301e-04}$ & $\num{1.109e-03}$ & 0 & $\num{1.265e-04}$ & $\cellcolor{blue!10} \num{8.789e-04}$ \\ \hline
Cube & 4998 & 4236 & $\cellcolor{blue!10} 1885$ & $\num{2.877e-04}$ & $\cellcolor{blue!10} \num{8.978e-04}$ & 1205 & $\cellcolor{blue!10} \num{2.351e-04}$ & $\num{1.521e-03}$ & 0 & $\num{1.078e-03}$ & $\num{2.095e-02}$ \\ \hline
Cyberpunk Bike & 31513 & 30488 & $\cellcolor{blue!10} 14602$ & $\num{3.656e-04}$ & $\num{1.567e-02}$ & 11502 & $\cellcolor{blue!10} \num{1.993e-04}$ & $\num{8.200e-03}$ & 0 & $\num{2.865e-04}$ & $\cellcolor{blue!10} \num{3.407e-03}$ \\ \hline
Dense Cube & 1014 & 1014 & $\cellcolor{blue!10} 390$ & $\num{2.122e-10}$ & $\num{1.517e-09}$ & 386 & $\num{3.642e-17}$ & $\num{4.095e-14}$ & 0 & $\num{9.377e-05}$ & $\num{5.942e-04}$ \\ \hline
Dragon With Pearl & 500166 & 0 & 0 & $\num{5.063e-05}$ & $\cellcolor{blue!10} \num{2.129e-04}$ & 0 & $\num{8.399e-05}$ & $\num{9.741e-04}$ & 0 & $\num{5.367e-05}$ & $\num{5.401e-04}$ \\ \hline
Dreadroamer & 63676 & 0 & 0 & $\cellcolor{blue!10} \num{5.936e-05}$ & $\num{2.732e-03}$ & 0 & $\num{1.307e-04}$ & $\cellcolor{blue!10} \num{1.907e-03}$ & 0 & $\num{1.380e-04}$ & $\num{2.111e-03}$ \\ \hline
Dry Tree & 74701 & 0 & 0 & $\num{2.814e-04}$ & $\num{5.847e-02}$ & 0 & $\num{7.536e-04}$ & $\num{6.081e-02}$ & 0 & $\cellcolor{blue!10} \num{1.951e-04}$ & $\cellcolor{blue!10} \num{4.280e-03}$ \\ \hline
Egyptian Vase & 7116 & 7116 & $\cellcolor{blue!10} 3368$ & $\cellcolor{blue!10} \num{1.447e-04}$ & $\cellcolor{blue!10} \num{8.356e-04}$ & 2081 & $\num{3.179e-04}$ & $\num{3.389e-03}$ & 0 & $\num{2.787e-04}$ & $\num{2.162e-03}$ \\ \hline
Fandisk & 6472 & 0 & 0 & $\num{5.102e-05}$ & $\num{4.440e-04}$ & 0 & $\num{1.625e-03}$ & $\num{1.225e-02}$ & 0 & $\num{2.058e-06}$ & $\cellcolor{blue!10} \num{5.116e-05}$ \\ \hline
Floral Pattern & 104752 & 104752 & $\cellcolor{blue!10} 52027$ & $\num{1.183e-05}$ & $\cellcolor{blue!10} \num{2.344e-04}$ & 41679 & $\num{5.210e-08}$ & $\num{1.088e-03}$ & 0 & $\num{2.128e-05}$ & $\num{4.580e-04}$ \\ \hline
French Halfbasket & 35289 & 35288 & $\cellcolor{blue!10} 17588$ & $\num{2.596e-05}$ & $\num{3.358e-04}$ & 11631 & $\num{1.524e-05}$ & $\num{2.781e-04}$ & 0 & $\num{1.983e-05}$ & $\num{2.450e-04}$ \\ \hline
Globophobia & 38245 & 38235 & $\cellcolor{blue!10} 18797$ & $\num{1.259e-04}$ & $\num{3.233e-03}$ & 13057 & $\num{1.456e-04}$ & $\cellcolor{blue!10} \num{1.355e-03}$ & 0 & $\num{1.123e-04}$ & $\num{3.771e-03}$ \\ \hline
Goblin Portrait & 54458 & 54388 & $\cellcolor{blue!10} 26332$ & $\num{2.100e-04}$ & $\num{3.183e-03}$ & 18645 & $\num{2.477e-04}$ & $\cellcolor{blue!10} \num{2.856e-03}$ & 0 & $\num{2.589e-04}$ & $\num{3.083e-03}$ \\ \hline
Greek Vase & 12575 & 12544 & $\cellcolor{blue!10} 6155$ & $\cellcolor{blue!10} \num{1.356e-04}$ & $\cellcolor{blue!10} \num{1.617e-03}$ & 5182 & $\num{7.366e-04}$ & $\num{2.423e-02}$ & 0 & $\num{2.673e-04}$ & $\num{4.101e-03}$ \\ \hline
His Name Is Violence & 13137 & 11778 & $\cellcolor{blue!10} 5485$ & $\num{3.610e-04}$ & $\num{6.078e-03}$ & 4097 & $\num{3.753e-04}$ & $\cellcolor{blue!10} \num{5.856e-03}$ & 0 & $\cellcolor{blue!10} \num{2.906e-04}$ & $\num{6.026e-03}$ \\ \hline
Hover Bike & 4662 & 0 & 0 & $\cellcolor{blue!10} \num{8.741e-04}$ & $\cellcolor{blue!10} \num{6.274e-03}$ & 0 & $\num{1.128e-03}$ & $\num{7.728e-03}$ & 0 & $\num{1.371e-03}$ & $\num{8.046e-03}$ \\ \hline
Japanese House 2 & 234803 & 233904 & $\cellcolor{blue!10} 116504$ & $\num{1.558e-05}$ & $\num{2.067e-04}$ & 85340 & $\num{1.331e-06}$ & $\num{1.711e-03}$ & 0 & $\num{5.733e-07}$ & $\cellcolor{blue!10} \num{9.506e-06}$ \\ \hline
Japanese House 3 & 41412 & 41298 & $\cellcolor{blue!10} 20374$ & $\num{1.167e-04}$ & $\cellcolor{blue!10} \num{2.141e-03}$ & 13169 & $\num{1.736e-04}$ & $\num{4.738e-03}$ & 0 & $\num{1.547e-04}$ & $\num{3.923e-03}$ \\ \hline
Japanese Shinto Shrine & 6104 & 5790 & $\cellcolor{blue!10} 2706$ & $\num{4.005e-05}$ & $\cellcolor{blue!10} \num{2.312e-03}$ & 2193 & $\num{4.513e-05}$ & $\num{2.600e-03}$ & 0 & $\num{6.975e-05}$ & $\num{2.813e-03}$ \\ \hline
Kagiya Edo Tokyo & 17192 & 9855 & 3911 & $\cellcolor{blue!10} \num{1.596e-04}$ & $\cellcolor{blue!10} \num{6.086e-03}$ & $\cellcolor{blue!10} 3982$ & $\num{2.743e-04}$ & $\num{1.020e-02}$ & 0 & $\num{2.338e-04}$ & $\num{1.951e-02}$ \\ \hline
Lamborghini Countach & 130216 & 119160 & 45823 & $\num{5.152e-05}$ & $\num{1.136e-02}$ & $\cellcolor{blue!10} 47205$ & $\num{4.759e-05}$ & $\num{1.479e-02}$ & 0 & $\num{1.380e-04}$ & $\cellcolor{blue!10} \num{2.885e-03}$ \\ \hline
Lantern & 1344 & 1328 & $\cellcolor{blue!10} 558$ & $\num{4.366e-05}$ & $\cellcolor{blue!10} \num{1.099e-03}$ & 402 & $\num{4.931e-05}$ & $\num{4.443e-03}$ & 0 & $\num{6.126e-05}$ & $\num{1.923e-03}$ \\ \hline
Lego Figure & 1839 & 0 & 0 & $\num{2.085e-04}$ & $\num{3.038e-03}$ & 0 & $\num{2.132e-04}$ & $\cellcolor{blue!10} \num{2.013e-03}$ & 0 & $\cellcolor{blue!10} \num{1.021e-04}$ & $\num{2.683e-03}$ \\ \hline
Lekythos & 1075 & 1075 & $\cellcolor{blue!10} 455$ & $\cellcolor{blue!10} \num{6.272e-04}$ & $\num{7.990e-03}$ & 318 & $\num{1.283e-03}$ & $\num{8.395e-03}$ & 0 & $\num{7.744e-04}$ & $\cellcolor{blue!10} \num{5.444e-03}$ \\ \hline
Lethe & 4395 & 3986 & $\cellcolor{blue!10} 1696$ & $\cellcolor{blue!10} \num{1.617e-04}$ & $\num{5.983e-03}$ & 1389 & $\num{2.232e-04}$ & $\cellcolor{blue!10} \num{3.999e-03}$ & 0 & $\num{3.113e-04}$ & $\num{4.510e-03}$ \\ \hline
Marble Track & 32376 & 26461 & 11444 & $\num{1.478e-04}$ & $\num{1.487e-02}$ & $\cellcolor{blue!10} 11578$ & $\num{1.424e-04}$ & $\num{1.487e-02}$ & 0 & $\num{4.012e-04}$ & $\num{2.699e-02}$ \\ \hline
Maze 45 & 37207 & 37191 & $\cellcolor{blue!10} 18441$ & $\num{2.313e-05}$ & $\num{4.135e-03}$ & 13217 & $\num{1.104e-04}$ & $\num{1.016e-02}$ & 0 & $\num{5.387e-08}$ & $\cellcolor{blue!10} \num{7.673e-04}$ \\ \hline
Maze & 37200 & 37191 & $\cellcolor{blue!10} 18432$ & $\num{3.098e-05}$ & $\num{5.571e-03}$ & 13227 & $\num{1.129e-04}$ & $\num{6.960e-03}$ & 0 & $\num{5.614e-08}$ & $\cellcolor{blue!10} \num{1.347e-03}$ \\ \hline
Mech & 7312 & 5850 & $\cellcolor{blue!10} 2208$ & $\cellcolor{blue!10} \num{3.259e-04}$ & $\cellcolor{blue!10} \num{4.357e-03}$ & 1965 & $\num{4.821e-04}$ & $\num{4.754e-03}$ & 0 & $\num{5.377e-04}$ & $\num{7.779e-03}$ \\ \hline
Melon Pallet & 45047 & 45016 & $\cellcolor{blue!10} 20636$ & $\num{2.242e-04}$ & $\num{1.074e-03}$ & 15452 & $\num{1.970e-04}$ & $\num{1.082e-03}$ & 0 & $\cellcolor{blue!10} \num{1.230e-04}$ & $\num{3.246e-03}$ \\ \hline
Militech Robot & 26181 & 0 & 0 & $\num{2.175e-04}$ & $\num{4.337e-03}$ & 0 & $\num{2.608e-04}$ & $\cellcolor{blue!10} \num{3.573e-03}$ & 0 & $\num{2.445e-04}$ & $\num{6.062e-03}$ \\ \hline
One Angery Dragon Boi & 67673 & 67632 & $\cellcolor{blue!10} 32971$ & $\num{1.104e-04}$ & $\cellcolor{blue!10} \num{1.572e-03}$ & 21239 & $\num{1.946e-04}$ & $\num{2.421e-03}$ & 0 & $\num{1.430e-04}$ & $\num{1.686e-03}$ \\ \hline
Oni No Face Mask & 109580 & 0 & 0 & $\num{4.541e-05}$ & $\num{2.390e-03}$ & 0 & $\num{8.324e-05}$ & $\num{2.391e-03}$ & 0 & $\num{3.087e-05}$ & $\num{2.370e-03}$ \\ \hline
Organic Armor & 47270 & 47270 & $\cellcolor{blue!10} 23584$ & $\num{1.183e-04}$ & $\num{6.573e-04}$ & 13875 & $\num{7.988e-05}$ & $\num{7.338e-04}$ & 0 & $\num{8.295e-05}$ & $\cellcolor{blue!10} \num{5.220e-04}$ \\ \hline
Pipe Wall & 1544 & 1543 & $\cellcolor{blue!10} 682$ & $\num{7.953e-04}$ & $\num{1.650e-02}$ & 446 & $\num{6.107e-04}$ & $\num{9.632e-03}$ & 0 & $\cellcolor{blue!10} \num{2.914e-04}$ & $\cellcolor{blue!10} \num{6.592e-03}$ \\ \hline
Prometheus & 399528 & 379559 & $\cellcolor{blue!10} 179616$ & $\num{3.606e-05}$ & $\num{2.362e-03}$ & 125942 & $\num{1.927e-05}$ & $\num{1.805e-03}$ & 0 & $\num{3.153e-05}$ & $\cellcolor{blue!10} \num{6.506e-04}$ \\ \hline
Radial & 5056 & 0 & 0 & $\cellcolor{blue!10} \num{6.918e-04}$ & $\cellcolor{blue!10} \num{6.173e-03}$ & 0 & $\num{7.935e-04}$ & $\num{1.185e-02}$ & 0 & $\num{1.440e-03}$ & $\num{8.309e-03}$ \\ \hline
Roman Soldier & 140324 & 123893 & $\cellcolor{blue!10} 60017$ & $\num{7.265e-05}$ & $\cellcolor{blue!10} \num{1.083e-03}$ & 38795 & $\num{7.583e-05}$ & $\num{1.617e-03}$ & 0 & $\num{9.637e-05}$ & $\num{1.624e-03}$ \\ \hline
Rope Bridge & 6160 & 4444 & 2422 & $\num{2.333e-04}$ & $\cellcolor{blue!10} \num{1.093e-03}$ & $\cellcolor{blue!10} 2563$ & $\num{1.100e-04}$ & $\num{1.718e-03}$ & 0 & $\num{7.980e-05}$ & $\num{6.571e-03}$ \\ \hline
Shinto Watchtower & 132166 & 123508 & $\cellcolor{blue!10} 60136$ & $\num{6.878e-05}$ & $\cellcolor{blue!10} \num{1.941e-03}$ & 54203 & $\num{1.005e-04}$ & $\num{2.867e-03}$ & 0 & $\num{6.460e-05}$ & $\num{4.144e-03}$ \\ \hline
Shirakawago House & 15919 & 15867 & $\cellcolor{blue!10} 7327$ & $\num{5.123e-04}$ & $\num{1.447e-02}$ & 5627 & $\num{7.896e-04}$ & $\cellcolor{blue!10} \num{1.093e-02}$ & 0 & $\cellcolor{blue!10} \num{2.747e-04}$ & $\num{1.622e-02}$ \\ \hline
Skyward Ddf Poco Ship & 97819 & 12207 & $\cellcolor{blue!10} 4884$ & $\num{9.907e-06}$ & $\num{1.903e-03}$ & 4079 & $\num{7.822e-06}$ & $\num{1.869e-03}$ & 0 & $\num{2.287e-05}$ & $\cellcolor{blue!10} \num{9.664e-04}$ \\ \hline
Snowflake Orb & 38708 & 0 & 0 & $\num{1.379e-04}$ & $\cellcolor{blue!10} \num{8.018e-04}$ & 0 & $\num{1.628e-04}$ & $\num{1.070e-03}$ & 0 & $\cellcolor{blue!10} \num{8.414e-05}$ & $\num{1.248e-03}$ \\ \hline
Space Fighter & 36326 & 34057 & 16311 & $\num{3.317e-05}$ & $\num{1.170e-02}$ & $\cellcolor{blue!10} 16499$ & $\num{1.799e-05}$ & $\cellcolor{blue!10} \num{6.564e-03}$ & 0 & $\num{9.965e-05}$ & $\num{1.242e-02}$ \\ \hline
Speeder Bike & 12133 & 11568 & $\cellcolor{blue!10} 5629$ & $\num{1.782e-04}$ & $\num{8.615e-03}$ & 3999 & $\num{2.829e-04}$ & $\cellcolor{blue!10} \num{5.627e-03}$ & 0 & $\num{2.177e-04}$ & $\num{6.727e-03}$ \\ \hline
Spiral Staircase & 7270 & 1400 & 680 & $\num{1.978e-04}$ & $\num{1.838e-01}$ & $\cellcolor{blue!10} 3329$ & $\num{1.672e-04}$ & $\cellcolor{blue!10} \num{4.361e-03}$ & 0 & $\num{1.666e-04}$ & $\num{6.631e-03}$ \\ \hline
Spot Quadrangulated & 2930 & 2928 & $\cellcolor{blue!10} 1243$ & $\cellcolor{blue!10} \num{4.262e-04}$ & $\cellcolor{blue!10} \num{2.733e-03}$ & 1087 & $\num{8.052e-04}$ & $\num{1.119e-02}$ & 0 & $\num{6.648e-04}$ & $\num{4.017e-03}$ \\ \hline
Steel Guard -7Mb & 7372 & 0 & 0 & $\num{4.162e-05}$ & $\cellcolor{blue!10} \num{2.044e-04}$ & 0 & $\num{5.118e-05}$ & $\num{8.723e-04}$ & 0 & $\num{1.225e-04}$ & $\num{3.812e-03}$ \\ \hline
Tank & 15229 & 15063 & $\cellcolor{blue!10} 7108$ & $\cellcolor{blue!10} \num{1.754e-04}$ & $\cellcolor{blue!10} \num{1.313e-03}$ & 4938 & $\num{3.182e-04}$ & $\num{5.119e-03}$ & 0 & $\num{3.228e-04}$ & $\num{5.113e-03}$ \\ \hline
Teacup & 5728 & 5728 & $\cellcolor{blue!10} 2673$ & $\cellcolor{blue!10} \num{2.003e-04}$ & $\cellcolor{blue!10} \num{2.658e-03}$ & 1968 & $\num{3.863e-04}$ & $\num{3.057e-03}$ & 0 & $\num{3.619e-04}$ & $\num{7.812e-03}$ \\ \hline
The Dark Wraith & 99833 & 80572 & $\cellcolor{blue!10} 36976$ & $\num{7.980e-05}$ & $\cellcolor{blue!10} \num{3.393e-03}$ & 25796 & $\num{1.168e-04}$ & $\num{4.388e-03}$ & 0 & $\num{1.126e-04}$ & $\num{3.744e-03}$ \\ \hline
Venice Mask & 147837 & 35201 & $\cellcolor{blue!10} 16699$ & $\num{1.310e-04}$ & $\num{2.131e-03}$ & 6497 & $\num{2.284e-04}$ & $\num{2.910e-03}$ & 0 & $\num{1.670e-04}$ & $\cellcolor{blue!10} \num{1.308e-03}$ \\ \hline
Vera Posed & 44056 & 40269 & $\cellcolor{blue!10} 18506$ & $\cellcolor{blue!10} \num{2.058e-04}$ & $\num{3.133e-03}$ & 13161 & $\num{2.675e-04}$ & $\cellcolor{blue!10} \num{2.661e-03}$ & 0 & $\num{3.017e-04}$ & $\num{3.115e-03}$ \\ \hline
War Tank & 31460 & 27758 & $\cellcolor{blue!10} 13108$ & $\num{8.595e-05}$ & $\cellcolor{blue!10} \num{2.656e-03}$ & 9875 & $\num{1.134e-04}$ & $\num{1.233e-02}$ & 0 & $\num{1.642e-03}$ & $\num{1.739e-02}$ \\ \hline
Yeahright & 377346 & 377344 & $\cellcolor{blue!10} 188536$ & $\num{3.682e-05}$ & $\num{4.155e-04}$ & 128378 & $\num{1.385e-05}$ & $\num{5.059e-04}$ & 0 & $\num{7.691e-06}$ & $\cellcolor{blue!10} \num{2.761e-04}$ \\ \hline
\hline
Average &  &  &  & $\num{1.775e-04}$ & $\num{7.692e-03}$ &  & $\num{2.881e-04}$ & $\num{6.081e-03}$ &  & $\num{2.661e-04}$ & $\num{4.977e-03}$ \\ \hline
\end{tabular}
\caption{\label{tab:table50} Complete results of our approach for static meshes with $50\%$ of the input triangles removed.}
\end{table*}
\begin{table*}
\centering
\renewcommand{\arraystretch}{0.8}
\sisetup{tight-spacing=true}
\setlength{\tabcolsep}{0.1pt}
\centering
\footnotesize
\begin{tabular}{|c|c|c|c|c|c|c|c|c|c|c|c|}
\hline
\multicolumn{3}{|c|}{25\% Reduction} & \multicolumn{3}{c|}{Ours} & \multicolumn{3}{c|}{QEM} & \multicolumn{3}{c|}{MeshLab} \\ \hline
Name & Out $\triangle$ & In $\square$ & Out $\square$ & Chamfer$^\downarrow$ & Hausdorff$^\downarrow$ & Out $\square$ & Chamfer$^\downarrow$ & Hausdorff$^\downarrow$ & Out $\square$ & Chamfer$^\downarrow$ & Hausdorff$^\downarrow$\\ \hline\hline
Door Panel & 16039 & 0 & 0 & $\cellcolor{blue!10} \num{1.946e-04}$ & $\num{1.989e-03}$ & 0 & $\num{3.378e-04}$ & $\num{7.587e-03}$ & 0 & $\num{2.807e-04}$ & $\cellcolor{blue!10} \num{1.652e-03}$ \\ \hline
2B  & 440761 & 621539 & $\cellcolor{blue!10} 151889$ & $\num{6.818e-05}$ & $\num{2.224e-03}$ & 52166 & $\num{1.277e-04}$ & $\num{2.380e-03}$ & 0 & $\num{1.073e-04}$ & $\cellcolor{blue!10} \num{1.787e-03}$ \\ \hline
Angelica  & 36211 & 64994 & $\cellcolor{blue!10} 8112$ & $\num{9.674e-04}$ & $\num{2.639e-02}$ & 7827 & $\num{1.404e-03}$ & $\num{2.491e-02}$ & 0 & $\cellcolor{blue!10} \num{8.762e-04}$ & $\cellcolor{blue!10} \num{1.341e-02}$ \\ \hline
Angkor Wat  & 121246 & 0 & 0 & $\num{7.128e-05}$ & $\cellcolor{blue!10} \num{9.543e-04}$ & 0 & $\num{1.074e-04}$ & $\num{3.108e-03}$ & 0 & $\num{1.027e-04}$ & $\num{1.540e-03}$ \\ \hline
Arches  & 7016 & 14032 & $\cellcolor{blue!10} 3254$ & $\cellcolor{blue!10} \num{9.316e-05}$ & $\cellcolor{blue!10} \num{2.009e-03}$ & 862 & $\num{2.600e-03}$ & $\num{7.268e-03}$ & 0 & $\num{3.432e-04}$ & $\num{1.028e-02}$ \\ \hline
Armadillo  & 25038 & 0 & 0 & $\num{2.659e-04}$ & $\cellcolor{blue!10} \num{9.818e-04}$ & 0 & $\num{4.744e-04}$ & $\num{4.352e-03}$ & 0 & $\num{2.657e-04}$ & $\num{1.465e-03}$ \\ \hline
Armored Charizard  & 1169 & 12 & 12 & $\cellcolor{blue!10} \num{1.962e-03}$ & $\num{2.836e-02}$ & 12 & $\num{2.694e-03}$ & $\cellcolor{blue!10} \num{1.968e-02}$ & 0 & $\num{3.202e-03}$ & $\num{4.216e-02}$ \\ \hline
Bicycle  & 30465 & 41618 & 7937 & $\num{1.382e-04}$ & $\cellcolor{blue!10} \num{3.485e-03}$ & $\cellcolor{blue!10} 9704$ & $\num{1.548e-04}$ & $\num{3.682e-03}$ & 0 & $\num{8.595e-04}$ & $\num{7.065e-03}$ \\ \hline
Blub Quadrangulated  & 3558 & 7104 & $\cellcolor{blue!10} 1441$ & $\cellcolor{blue!10} \num{3.576e-04}$ & $\cellcolor{blue!10} \num{2.134e-03}$ & 869 & $\num{7.649e-04}$ & $\num{1.152e-02}$ & 0 & $\num{4.538e-04}$ & $\num{2.874e-03}$ \\ \hline
Botchling  & 78739 & 157140 & $\cellcolor{blue!10} 34635$ & $\num{1.170e-04}$ & $\num{2.838e-03}$ & 17895 & $\num{1.840e-04}$ & $\num{2.837e-03}$ & 0 & $\num{1.541e-04}$ & $\cellcolor{blue!10} \num{2.521e-03}$ \\ \hline
Casio Keyboard  & 1732 & 2649 & 439 & $\num{5.059e-04}$ & $\cellcolor{blue!10} \num{1.338e-02}$ & $\cellcolor{blue!10} 623$ & $\num{4.607e-04}$ & $\num{1.461e-02}$ & 0 & $\num{8.921e-04}$ & $\num{2.516e-02}$ \\ \hline
Chimney Pipe  & 1702 & 3404 & $\cellcolor{blue!10} 733$ & $\num{3.713e-04}$ & $\cellcolor{blue!10} \num{6.450e-03}$ & 455 & $\num{3.810e-04}$ & $\num{7.586e-03}$ & 0 & $\num{1.509e-03}$ & $\num{1.785e-02}$ \\ \hline
Chitinous Knight  & 43657 & 87212 & $\cellcolor{blue!10} 18870$ & $\cellcolor{blue!10} \num{1.983e-04}$ & $\cellcolor{blue!10} \num{1.159e-03}$ & 6740 & $\num{3.032e-04}$ & $\num{1.932e-03}$ & 0 & $\num{2.578e-04}$ & $\num{1.470e-03}$ \\ \hline
Cube  & 2512 & 4236 & $\cellcolor{blue!10} 815$ & $\cellcolor{blue!10} \num{7.099e-04}$ & $\cellcolor{blue!10} \num{1.306e-03}$ & 333 & $\num{3.310e-03}$ & $\num{6.898e-02}$ & 0 & $\num{8.076e-03}$ & $\num{8.500e-02}$ \\ \hline
Cyberpunk Bike  & 15841 & 30488 & $\cellcolor{blue!10} 5535$ & $\num{6.163e-04}$ & $\num{1.567e-02}$ & 4950 & $\cellcolor{blue!10} \num{5.319e-04}$ & $\num{1.047e-02}$ & 0 & $\num{5.843e-04}$ & $\cellcolor{blue!10} \num{5.152e-03}$ \\ \hline
Dense Cube  & 506 & 1014 & $\cellcolor{blue!10} 211$ & $\num{2.311e-10}$ & $\num{1.517e-09}$ & 120 & $\num{7.735e-17}$ & $\num{5.674e-14}$ & 0 & $\num{1.664e-04}$ & $\num{1.018e-02}$ \\ \hline
Dragon With Pearl  & 251698 & 0 & 0 & $\num{1.075e-04}$ & $\cellcolor{blue!10} \num{3.907e-04}$ & 0 & $\num{1.921e-04}$ & $\num{1.784e-03}$ & 0 & $\num{1.162e-04}$ & $\num{9.499e-04}$ \\ \hline
Dreadroamer  & 33380 & 0 & 0 & $\cellcolor{blue!10} \num{2.487e-04}$ & $\cellcolor{blue!10} \num{3.165e-03}$ & 0 & $\num{4.761e-04}$ & $\num{4.110e-03}$ & 0 & $\num{4.775e-04}$ & $\num{5.095e-03}$ \\ \hline
Dry Tree  & 37434 & 0 & 0 & $\num{6.452e-04}$ & $\num{5.841e-02}$ & 0 & $\num{1.976e-03}$ & $\num{9.498e-02}$ & 0 & $\cellcolor{blue!10} \num{4.283e-04}$ & $\cellcolor{blue!10} \num{7.455e-03}$ \\ \hline
Egyptian Vase  & 3566 & 7116 & $\cellcolor{blue!10} 1201$ & $\cellcolor{blue!10} \num{4.449e-04}$ & $\cellcolor{blue!10} \num{1.729e-03}$ & 541 & $\num{9.039e-04}$ & $\num{6.196e-03}$ & 0 & $\num{9.441e-04}$ & $\num{4.269e-03}$ \\ \hline
Fandisk  & 3236 & 0 & 0 & $\num{9.921e-05}$ & $\num{9.996e-04}$ & 0 & $\num{2.523e-03}$ & $\num{1.527e-02}$ & 0 & $\cellcolor{blue!10} \num{9.716e-06}$ & $\cellcolor{blue!10} \num{2.283e-04}$ \\ \hline
Floral Pattern  & 52380 & 104752 & $\cellcolor{blue!10} 24658$ & $\num{3.309e-05}$ & $\cellcolor{blue!10} \num{2.764e-04}$ & 10830 & $\num{1.360e-05}$ & $\num{2.429e-03}$ & 0 & $\num{5.533e-05}$ & $\num{3.001e-03}$ \\ \hline
French Halfbasket  & 17757 & 35288 & $\cellcolor{blue!10} 7518$ & $\num{3.792e-05}$ & $\cellcolor{blue!10} \num{2.863e-04}$ & 3151 & $\num{4.361e-05}$ & $\num{5.695e-04}$ & 0 & $\num{6.204e-05}$ & $\num{4.846e-04}$ \\ \hline
Globophobia  & 19155 & 38235 & $\cellcolor{blue!10} 7163$ & $\cellcolor{blue!10} \num{1.756e-04}$ & $\cellcolor{blue!10} \num{1.726e-03}$ & 3804 & $\num{3.640e-04}$ & $\num{3.330e-03}$ & 0 & $\num{2.990e-04}$ & $\num{7.217e-03}$ \\ \hline
Goblin Portrait  & 27331 & 54388 & $\cellcolor{blue!10} 6910$ & $\cellcolor{blue!10} \num{3.722e-04}$ & $\num{5.069e-03}$ & 5633 & $\num{6.509e-04}$ & $\num{3.900e-03}$ & 0 & $\num{6.035e-04}$ & $\cellcolor{blue!10} \num{3.614e-03}$ \\ \hline
Greek Vase  & 6309 & 12544 & $\cellcolor{blue!10} 2369$ & $\cellcolor{blue!10} \num{4.882e-04}$ & $\cellcolor{blue!10} \num{1.629e-03}$ & 1847 & $\num{1.664e-03}$ & $\num{2.724e-02}$ & 0 & $\num{7.140e-04}$ & $\num{8.103e-03}$ \\ \hline
His Name Is Violence  & 6637 & 11778 & $\cellcolor{blue!10} 2092$ & $\num{9.448e-04}$ & $\cellcolor{blue!10} \num{6.665e-03}$ & 1416 & $\num{9.599e-04}$ & $\num{9.267e-03}$ & 0 & $\num{1.538e-03}$ & $\num{2.599e-02}$ \\ \hline
Hover Bike  & 2331 & 0 & 0 & $\num{2.584e-03}$ & $\num{1.572e-02}$ & 0 & $\num{2.633e-03}$ & $\cellcolor{blue!10} \num{1.112e-02}$ & 0 & $\num{3.395e-03}$ & $\num{1.488e-02}$ \\ \hline
Japanese House 2  & 117401 & 233904 & $\cellcolor{blue!10} 56987$ & $\num{5.703e-05}$ & $\num{1.179e-03}$ & 28476 & $\num{1.834e-05}$ & $\num{2.066e-03}$ & 0 & $\num{8.944e-06}$ & $\cellcolor{blue!10} \num{1.748e-04}$ \\ \hline
Japanese House 3  & 20708 & 41298 & $\cellcolor{blue!10} 5773$ & $\num{3.187e-04}$ & $\cellcolor{blue!10} \num{2.111e-03}$ & 4021 & $\num{3.283e-04}$ & $\num{5.221e-03}$ & 0 & $\num{3.717e-04}$ & $\num{4.071e-03}$ \\ \hline
Japanese Shinto Shrine  & 3025 & 5790 & $\cellcolor{blue!10} 1200$ & $\num{2.715e-04}$ & $\num{4.168e-02}$ & 957 & $\num{2.606e-04}$ & $\cellcolor{blue!10} \num{5.111e-03}$ & 0 & $\num{3.924e-04}$ & $\num{1.574e-02}$ \\ \hline
Kagiya Edo Tokyo  & 8611 & 9855 & $\cellcolor{blue!10} 1764$ & $\cellcolor{blue!10} \num{5.702e-04}$ & $\cellcolor{blue!10} \num{8.303e-03}$ & 1235 & $\num{7.073e-04}$ & $\num{1.027e-02}$ & 0 & $\num{1.069e-03}$ & $\num{2.821e-02}$ \\ \hline
Lamborghini Countach  & 65761 & 119160 & $\cellcolor{blue!10} 15461$ & $\num{1.345e-04}$ & $\num{1.409e-02}$ & 14162 & $\num{1.393e-04}$ & $\num{1.429e-02}$ & 0 & $\num{3.950e-04}$ & $\cellcolor{blue!10} \num{1.266e-02}$ \\ \hline
Lantern  & 676 & 1328 & $\cellcolor{blue!10} 246$ & $\num{1.237e-03}$ & $\cellcolor{blue!10} \num{2.920e-03}$ & 179 & $\cellcolor{blue!10} \num{1.825e-04}$ & $\num{7.984e-03}$ & 0 & $\num{4.266e-04}$ & $\num{5.975e-03}$ \\ \hline
Lego Figure  & 930 & 0 & 0 & $\num{8.633e-04}$ & $\num{8.914e-03}$ & 0 & $\num{7.354e-04}$ & $\cellcolor{blue!10} \num{6.224e-03}$ & 0 & $\cellcolor{blue!10} \num{4.914e-04}$ & $\num{1.477e-02}$ \\ \hline
Lekythos  & 541 & 1075 & $\cellcolor{blue!10} 172$ & $\cellcolor{blue!10} \num{1.871e-03}$ & $\num{1.208e-02}$ & 80 & $\num{3.780e-03}$ & $\num{1.792e-02}$ & 0 & $\num{3.734e-03}$ & $\num{1.204e-02}$ \\ \hline
Lethe  & 2205 & 3986 & $\cellcolor{blue!10} 597$ & $\num{1.669e-03}$ & $\num{1.351e-01}$ & 541 & $\cellcolor{blue!10} \num{7.650e-04}$ & $\cellcolor{blue!10} \num{8.717e-03}$ & 0 & $\num{1.302e-03}$ & $\num{1.467e-02}$ \\ \hline
Marble Track  & 16465 & 26461 & $\cellcolor{blue!10} 4146$ & $\num{3.885e-04}$ & $\num{1.487e-02}$ & 2408 & $\cellcolor{blue!10} \num{3.099e-04}$ & $\num{1.487e-02}$ & 0 & $\num{6.366e-04}$ & $\num{2.690e-02}$ \\ \hline
Maze 45  & 18606 & 37191 & $\cellcolor{blue!10} 8088$ & $\num{2.121e-04}$ & $\num{3.500e-03}$ & 3418 & $\num{2.222e-04}$ & $\num{1.200e-02}$ & 0 & $\cellcolor{blue!10} \num{2.143e-06}$ & $\cellcolor{blue!10} \num{3.366e-03}$ \\ \hline
Maze  & 18606 & 37191 & $\cellcolor{blue!10} 8039$ & $\num{2.947e-04}$ & $\num{5.200e-03}$ & 3523 & $\num{2.447e-04}$ & $\num{1.701e-02}$ & 0 & $\cellcolor{blue!10} \num{2.605e-06}$ & $\cellcolor{blue!10} \num{4.629e-03}$ \\ \hline
Mech  & 3706 & 5850 & $\cellcolor{blue!10} 767$ & $\num{1.338e-03}$ & $\cellcolor{blue!10} \num{1.101e-02}$ & 745 & $\num{1.361e-03}$ & $\num{1.539e-02}$ & 0 & $\num{2.150e-03}$ & $\num{1.197e-02}$ \\ \hline
Melon Pallet  & 22838 & 45016 & $\cellcolor{blue!10} 7331$ & $\num{4.165e-04}$ & $\cellcolor{blue!10} \num{2.776e-03}$ & 4071 & $\num{5.394e-04}$ & $\num{5.196e-03}$ & 0 & $\cellcolor{blue!10} \num{2.956e-04}$ & $\num{4.788e-03}$ \\ \hline
Militech Robot  & 13145 & 0 & 0 & $\num{7.468e-04}$ & $\cellcolor{blue!10} \num{6.058e-03}$ & 0 & $\num{8.301e-04}$ & $\num{7.448e-03}$ & 0 & $\cellcolor{blue!10} \num{6.931e-04}$ & $\num{6.965e-03}$ \\ \hline
One Angery Dragon Boi  & 33859 & 67632 & $\cellcolor{blue!10} 14669$ & $\cellcolor{blue!10} \num{2.825e-04}$ & $\num{2.752e-03}$ & 6068 & $\num{5.199e-04}$ & $\num{3.868e-03}$ & 0 & $\num{4.019e-04}$ & $\cellcolor{blue!10} \num{2.651e-03}$ \\ \hline
Oni No Face Mask  & 54844 & 0 & 0 & $\num{1.014e-04}$ & $\num{2.419e-03}$ & 0 & $\num{2.212e-04}$ & $\num{2.495e-03}$ & 0 & $\num{8.108e-05}$ & $\num{2.421e-03}$ \\ \hline
Organic Armor  & 23646 & 47270 & $\cellcolor{blue!10} 11265$ & $\num{1.510e-04}$ & $\cellcolor{blue!10} \num{7.198e-04}$ & 3469 & $\num{1.847e-04}$ & $\num{1.272e-03}$ & 0 & $\num{1.776e-04}$ & $\num{1.013e-03}$ \\ \hline
Pipe Wall  & 774 & 1543 & $\cellcolor{blue!10} 245$ & $\num{3.548e-03}$ & $\num{1.278e-01}$ & 135 & $\num{1.951e-03}$ & $\cellcolor{blue!10} \num{1.797e-02}$ & 0 & $\cellcolor{blue!10} \num{1.812e-03}$ & $\num{1.278e-01}$ \\ \hline
Prometheus  & 200452 & 379559 & $\cellcolor{blue!10} 80874$ & $\num{6.088e-05}$ & $\num{2.362e-03}$ & 41858 & $\num{5.792e-05}$ & $\num{1.899e-03}$ & 0 & $\num{8.706e-05}$ & $\cellcolor{blue!10} \num{1.339e-03}$ \\ \hline
Radial  & 2528 & 0 & 0 & $\num{2.914e-03}$ & $\cellcolor{blue!10} \num{1.421e-02}$ & 0 & $\cellcolor{blue!10} \num{2.443e-03}$ & $\num{2.055e-02}$ & 0 & $\num{7.336e-03}$ & $\num{2.997e-02}$ \\ \hline
Roman Soldier  & 70260 & 123893 & $\cellcolor{blue!10} 26162$ & $\cellcolor{blue!10} \num{1.454e-04}$ & $\cellcolor{blue!10} \num{1.083e-03}$ & 13148 & $\num{2.407e-04}$ & $\num{3.640e-03}$ & 0 & $\num{2.507e-04}$ & $\num{1.940e-03}$ \\ \hline
Rope Bridge  & 3080 & 4444 & 912 & $\num{1.045e-03}$ & $\num{6.704e-03}$ & $\cellcolor{blue!10} 1061$ & $\num{3.612e-04}$ & $\cellcolor{blue!10} \num{6.306e-03}$ & 0 & $\cellcolor{blue!10} \num{1.654e-04}$ & $\num{6.567e-03}$ \\ \hline
Shinto Watchtower  & 66193 & 123508 & $\cellcolor{blue!10} 22678$ & $\num{4.236e-04}$ & $\cellcolor{blue!10} \num{2.376e-03}$ & 17016 & $\num{4.396e-04}$ & $\num{3.724e-03}$ & 0 & $\num{1.042e-03}$ & $\num{7.490e-03}$ \\ \hline
Shirakawago House  & 7997 & 15867 & $\cellcolor{blue!10} 3288$ & $\num{1.228e-03}$ & $\cellcolor{blue!10} \num{2.083e-02}$ & 2068 & $\num{1.806e-03}$ & $\num{2.694e-02}$ & 0 & $\cellcolor{blue!10} \num{1.015e-03}$ & $\num{2.089e-02}$ \\ \hline
Skyward Ddf Poco Ship  & 53668 & 12207 & $\cellcolor{blue!10} 3573$ & $\num{3.090e-05}$ & $\num{1.902e-03}$ & 2252 & $\num{4.774e-05}$ & $\num{1.893e-03}$ & 0 & $\num{8.694e-05}$ & $\num{3.181e-03}$ \\ \hline
Snowflake Orb  & 19360 & 0 & 0 & $\num{4.783e-04}$ & $\cellcolor{blue!10} \num{3.925e-03}$ & 0 & $\num{6.225e-04}$ & $\num{4.489e-03}$ & 0 & $\cellcolor{blue!10} \num{3.863e-04}$ & $\num{5.454e-03}$ \\ \hline
Space Fighter  & 18179 & 34057 & 7087 & $\num{1.417e-04}$ & $\cellcolor{blue!10} \num{1.170e-02}$ & $\cellcolor{blue!10} 7273$ & $\num{1.261e-04}$ & $\num{2.351e-02}$ & 0 & $\num{3.248e-04}$ & $\num{1.242e-02}$ \\ \hline
Speeder Bike  & 6117 & 11568 & $\cellcolor{blue!10} 2177$ & $\cellcolor{blue!10} \num{5.052e-04}$ & $\num{1.269e-02}$ & 1290 & $\num{7.277e-04}$ & $\cellcolor{blue!10} \num{7.305e-03}$ & 0 & $\num{7.627e-04}$ & $\num{1.069e-02}$ \\ \hline
Spiral Staircase  & 3666 & 1400 & 431 & $\num{3.344e-04}$ & $\cellcolor{blue!10} \num{3.181e-03}$ & $\cellcolor{blue!10} 1175$ & $\num{3.311e-04}$ & $\num{5.227e-03}$ & 0 & $\num{2.837e-04}$ & $\num{1.017e-02}$ \\ \hline
Spot Quadrangulated  & 1474 & 2928 & 319 & $\cellcolor{blue!10} \num{1.174e-03}$ & $\cellcolor{blue!10} \num{3.752e-03}$ & $\cellcolor{blue!10} 321$ & $\num{2.221e-03}$ & $\num{1.805e-02}$ & 0 & $\num{1.882e-03}$ & $\num{8.820e-03}$ \\ \hline
Steel Guard -7Mb  & 3686 & 0 & 0 & $\num{3.056e-04}$ & $\cellcolor{blue!10} \num{8.985e-04}$ & 0 & $\cellcolor{blue!10} \num{1.779e-04}$ & $\num{2.313e-03}$ & 0 & $\num{6.544e-04}$ & $\num{1.470e-02}$ \\ \hline
Tank  & 7753 & 15063 & $\cellcolor{blue!10} 2668$ & $\cellcolor{blue!10} \num{5.878e-04}$ & $\cellcolor{blue!10} \num{3.168e-03}$ & 1819 & $\num{8.630e-04}$ & $\num{6.261e-03}$ & 0 & $\num{1.068e-03}$ & $\num{8.947e-03}$ \\ \hline
Teacup  & 2872 & 5728 & $\cellcolor{blue!10} 1078$ & $\cellcolor{blue!10} \num{5.922e-04}$ & $\cellcolor{blue!10} \num{3.523e-03}$ & 559 & $\num{9.836e-04}$ & $\num{9.353e-03}$ & 0 & $\num{1.069e-03}$ & $\num{1.324e-02}$ \\ \hline
The Dark Wraith  & 49951 & 80572 & $\cellcolor{blue!10} 16367$ & $\cellcolor{blue!10} \num{1.871e-04}$ & $\cellcolor{blue!10} \num{4.231e-03}$ & 7945 & $\num{3.172e-04}$ & $\num{4.387e-03}$ & 0 & $\num{3.110e-04}$ & $\num{5.200e-03}$ \\ \hline
Venice Mask  & 74056 & 35201 & $\cellcolor{blue!10} 5805$ & $\cellcolor{blue!10} \num{2.870e-04}$ & $\cellcolor{blue!10} \num{1.651e-03}$ & 1901 & $\num{4.972e-04}$ & $\num{4.702e-03}$ & 0 & $\num{3.605e-04}$ & $\num{2.025e-03}$ \\ \hline
Vera Posed  & 22178 & 40269 & $\cellcolor{blue!10} 5877$ & $\cellcolor{blue!10} \num{5.464e-04}$ & $\cellcolor{blue!10} \num{4.052e-03}$ & 4038 & $\num{6.937e-04}$ & $\num{5.271e-03}$ & 0 & $\num{8.483e-04}$ & $\num{1.379e-02}$ \\ \hline
War Tank  & 15764 & 27758 & $\cellcolor{blue!10} 4984$ & $\num{7.279e-04}$ & $\cellcolor{blue!10} \num{3.481e-03}$ & 3388 & $\num{7.236e-04}$ & $\num{1.329e-02}$ & 0 & $\num{4.366e-03}$ & $\num{2.899e-02}$ \\ \hline
Yeahright  & 188676 & 377344 & $\cellcolor{blue!10} 93770$ & $\num{5.681e-05}$ & $\cellcolor{blue!10} \num{3.109e-04}$ & 37774 & $\num{4.919e-05}$ & $\num{9.043e-04}$ & 0 & $\num{2.998e-05}$ & $\num{7.253e-04}$ \\ \hline
\hline
Average &  &  &  & $\num{5.830e-04}$ & $\num{1.067e-02}$ &  & $\num{7.950e-04}$ & $\num{1.069e-02}$ &  & $\num{9.485e-04}$ & $\num{1.221e-02}$ \\ \hline
\end{tabular}
\caption{\label{tab:table25} Complete results of our approach for static meshes with $25\%$ of the input triangles removed.}
\end{table*}
\begin{table*}
\centering
\renewcommand{\arraystretch}{0.8}
\sisetup{tight-spacing=true}
\setlength{\tabcolsep}{0.1pt}
\centering
\footnotesize
\begin{tabular}{|c|c|c|c|c|c|c|c|c|c|c|c|}
\hline
\multicolumn{3}{|c|}{10\% Reduction} & \multicolumn{3}{c|}{Ours} & \multicolumn{3}{c|}{QEM} & \multicolumn{3}{c|}{MeshLab} \\ \hline
Name & Out $\triangle$ & In $\square$ & Out $\square$ & Chamfer$^\downarrow$ & Hausdorff$^\downarrow$ & Out $\square$ & Chamfer$^\downarrow$ & Hausdorff$^\downarrow$ & Out $\square$ & Chamfer$^\downarrow$ & Hausdorff$^\downarrow$\\ \hline\hline
Door Panel & 6471 & 0 & 0 & $\cellcolor{blue!10} \num{3.508e-04}$ & $\num{3.604e-03}$ & 0 & $\num{6.001e-04}$ & $\num{2.088e-02}$ & 0 & $\num{4.200e-04}$ & $\cellcolor{blue!10} \num{2.743e-03}$ \\ \hline
2B & 176780 & 621539 & $\cellcolor{blue!10} 59923$ & $\cellcolor{blue!10} \num{1.658e-04}$ & $\num{1.529e-02}$ & 9406 & $\num{5.500e-03}$ & $\num{8.592e-02}$ & 0 & $\num{2.169e-04}$ & $\cellcolor{blue!10} \num{2.911e-03}$ \\ \hline
Angelica & 14858 & 64994 & 1834 & $\num{2.318e-03}$ & $\num{3.814e-02}$ & $\cellcolor{blue!10} 2226$ & $\num{3.749e-03}$ & $\num{3.325e-02}$ & 0 & $\cellcolor{blue!10} \num{1.914e-03}$ & $\cellcolor{blue!10} \num{2.522e-02}$ \\ \hline
Angkor Wat & 48952 & 0 & 0 & $\cellcolor{blue!10} \num{1.481e-04}$ & $\cellcolor{blue!10} \num{2.332e-03}$ & 0 & $\num{2.690e-04}$ & $\num{1.025e-02}$ & 0 & $\num{2.152e-04}$ & $\num{3.530e-03}$ \\ \hline
Arches & 2806 & 14032 & $\cellcolor{blue!10} 1182$ & $\cellcolor{blue!10} \num{5.677e-04}$ & $\num{1.464e-02}$ & 247 & $\num{4.073e-03}$ & $\cellcolor{blue!10} \num{1.076e-02}$ & 0 & $\num{5.310e-03}$ & $\num{1.651e-02}$ \\ \hline
Armadillo & 10090 & 0 & 0 & $\num{5.953e-04}$ & $\cellcolor{blue!10} \num{1.900e-03}$ & 0 & $\num{1.136e-03}$ & $\num{7.254e-03}$ & 0 & $\num{6.007e-04}$ & $\num{3.457e-03}$ \\ \hline
Armored Charizard & 471 & 12 & 12 & $\cellcolor{blue!10} \num{4.895e-03}$ & $\cellcolor{blue!10} \num{3.601e-02}$ & 12 & $\num{5.485e-03}$ & $\num{4.389e-02}$ & 0 & $\num{1.127e-02}$ & $\num{6.829e-02}$ \\ \hline
Bicycle & 12167 & 41618 & 2196 & $\num{5.912e-04}$ & $\num{4.784e-02}$ & $\cellcolor{blue!10} 2330$ & $\cellcolor{blue!10} \num{4.959e-04}$ & $\cellcolor{blue!10} \num{7.003e-03}$ & 0 & $\num{2.208e-03}$ & $\num{1.375e-02}$ \\ \hline
Blub Quadrangulated & 1426 & 7104 & $\cellcolor{blue!10} 389$ & $\cellcolor{blue!10} \num{8.329e-04}$ & $\cellcolor{blue!10} \num{2.621e-03}$ & 137 & $\num{1.931e-03}$ & $\num{2.477e-02}$ & 0 & $\num{1.211e-03}$ & $\num{7.290e-03}$ \\ \hline
Botchling & 31534 & 157140 & $\cellcolor{blue!10} 11039$ & $\cellcolor{blue!10} \num{2.842e-04}$ & $\cellcolor{blue!10} \num{2.942e-03}$ & 3411 & $\num{4.792e-04}$ & $\num{3.428e-03}$ & 0 & $\num{4.000e-04}$ & $\num{3.021e-03}$ \\ \hline
Casio Keyboard & 682 & 2649 & 163 & $\num{2.146e-03}$ & $\cellcolor{blue!10} \num{1.387e-02}$ & 163 & $\num{1.950e-03}$ & $\num{4.485e-02}$ & 0 & $\cellcolor{blue!10} \num{1.661e-03}$ & $\num{4.509e-02}$ \\ \hline
Chimney Pipe & 714 & 3404 & $\cellcolor{blue!10} 235$ & $\cellcolor{blue!10} \num{3.147e-03}$ & $\cellcolor{blue!10} \num{8.976e-03}$ & 123 & $\num{3.492e-03}$ & $\num{2.293e-02}$ & 0 & $\num{5.191e-03}$ & $\num{4.793e-02}$ \\ \hline
Chitinous Knight & 17476 & 87212 & $\cellcolor{blue!10} 5810$ & $\cellcolor{blue!10} \num{3.976e-04}$ & $\num{2.995e-03}$ & 945 & $\num{6.607e-04}$ & $\num{3.989e-03}$ & 0 & $\num{5.423e-04}$ & $\cellcolor{blue!10} \num{2.517e-03}$ \\ \hline
Cube & 1018 & 4236 & $\cellcolor{blue!10} 336$ & $\cellcolor{blue!10} \num{2.307e-03}$ & $\cellcolor{blue!10} \num{3.164e-03}$ & 52 & $\num{8.905e-03}$ & $\num{1.103e-01}$ & 0 & $\num{2.912e-02}$ & $\num{1.273e-01}$ \\ \hline
Cyberpunk Bike & 6375 & 30488 & $\cellcolor{blue!10} 1891$ & $\num{1.172e-03}$ & $\num{1.536e-02}$ & 1114 & $\num{1.209e-03}$ & $\cellcolor{blue!10} \num{1.166e-02}$ & 0 & $\num{1.515e-03}$ & $\num{1.195e-02}$ \\ \hline
Dense Cube & 202 & 1014 & $\cellcolor{blue!10} 93$ & $\num{3.429e-10}$ & $\num{1.517e-09}$ & 12 & $\num{5.677e-07}$ & $\num{1.562e-05}$ & 0 & $\num{2.105e-03}$ & $\num{3.316e-02}$ \\ \hline
Dragon With Pearl & 101220 & 0 & 0 & $\num{2.140e-04}$ & $\cellcolor{blue!10} \num{8.429e-04}$ & 0 & $\num{4.738e-04}$ & $\num{3.016e-03}$ & 0 & $\num{2.300e-04}$ & $\num{1.500e-03}$ \\ \hline
Dreadroamer & 15182 & 0 & 0 & $\cellcolor{blue!10} \num{8.440e-04}$ & $\num{6.997e-03}$ & 0 & $\num{1.034e-03}$ & $\cellcolor{blue!10} \num{6.257e-03}$ & 0 & $\num{1.185e-03}$ & $\num{1.040e-02}$ \\ \hline
Dry Tree & 15242 & 0 & 0 & $\num{1.953e-03}$ & $\num{9.429e-02}$ & 0 & $\num{4.735e-03}$ & $\num{1.202e-01}$ & 0 & $\cellcolor{blue!10} \num{9.845e-04}$ & $\cellcolor{blue!10} \num{1.051e-02}$ \\ \hline
Egyptian Vase & 1430 & 7116 & $\cellcolor{blue!10} 316$ & $\cellcolor{blue!10} \num{1.363e-03}$ & $\cellcolor{blue!10} \num{3.962e-03}$ & 57 & $\num{1.838e-03}$ & $\num{8.640e-03}$ & 0 & $\num{2.815e-03}$ & $\num{1.070e-02}$ \\ \hline
Fandisk & 1298 & 0 & 0 & $\num{2.581e-04}$ & $\num{2.468e-03}$ & 0 & $\num{5.515e-03}$ & $\num{2.452e-02}$ & 0 & $\cellcolor{blue!10} \num{5.092e-05}$ & $\cellcolor{blue!10} \num{6.444e-04}$ \\ \hline
Floral Pattern & 20976 & 104752 & $\cellcolor{blue!10} 9180$ & $\num{6.129e-05}$ & $\cellcolor{blue!10} \num{1.047e-03}$ & 1301 & $\num{6.657e-05}$ & $\num{4.118e-03}$ & 0 & $\num{1.470e-04}$ & $\num{3.539e-03}$ \\ \hline
French Halfbasket & 7066 & 35288 & $\cellcolor{blue!10} 3392$ & $\num{1.025e-04}$ & $\cellcolor{blue!10} \num{5.269e-04}$ & 458 & $\num{9.936e-05}$ & $\num{1.381e-03}$ & 0 & $\num{2.241e-04}$ & $\num{1.391e-03}$ \\ \hline
Globophobia & 7671 & 38235 & $\cellcolor{blue!10} 2447$ & $\cellcolor{blue!10} \num{4.932e-04}$ & $\num{1.320e-02}$ & 661 & $\num{8.700e-04}$ & $\cellcolor{blue!10} \num{5.204e-03}$ & 0 & $\num{7.158e-04}$ & $\num{1.060e-02}$ \\ \hline
Goblin Portrait & 11008 & 54388 & $\cellcolor{blue!10} 1762$ & $\cellcolor{blue!10} \num{7.723e-04}$ & $\cellcolor{blue!10} \num{8.097e-03}$ & 880 & $\num{1.382e-03}$ & $\num{8.175e-03}$ & 0 & $\num{1.250e-03}$ & $\num{9.116e-03}$ \\ \hline
Greek Vase & 2520 & 12544 & $\cellcolor{blue!10} 732$ & $\cellcolor{blue!10} \num{1.150e-03}$ & $\cellcolor{blue!10} \num{3.896e-03}$ & 337 & $\num{3.101e-03}$ & $\num{3.301e-02}$ & 0 & $\num{2.166e-03}$ & $\num{1.079e-02}$ \\ \hline
His Name Is Violence & 2700 & 11778 & $\cellcolor{blue!10} 614$ & $\num{2.333e-03}$ & $\cellcolor{blue!10} \num{1.489e-02}$ & 364 & $\cellcolor{blue!10} \num{2.105e-03}$ & $\num{1.774e-02}$ & 0 & $\num{4.400e-03}$ & $\num{3.653e-02}$ \\ \hline
Hover Bike & 939 & 0 & 0 & $\num{6.194e-03}$ & $\num{3.323e-02}$ & 0 & $\cellcolor{blue!10} \num{5.400e-03}$ & $\num{2.478e-02}$ & 0 & $\num{6.840e-03}$ & $\cellcolor{blue!10} \num{2.381e-02}$ \\ \hline
Japanese House 2 & 46964 & 233904 & $\cellcolor{blue!10} 21572$ & $\num{1.374e-04}$ & $\num{1.412e-03}$ & 9400 & $\num{1.859e-04}$ & $\num{3.645e-03}$ & 0 & $\cellcolor{blue!10} \num{6.991e-05}$ & $\cellcolor{blue!10} \num{1.076e-03}$ \\ \hline
Japanese House 3 & 8294 & 41298 & $\cellcolor{blue!10} 2282$ & $\num{9.295e-04}$ & $\cellcolor{blue!10} \num{4.238e-03}$ & 1363 & $\num{7.680e-04}$ & $\num{1.167e-02}$ & 0 & $\cellcolor{blue!10} \num{6.551e-04}$ & $\num{7.385e-03}$ \\ \hline
Japanese Shinto Shrine & 1310 & 5790 & $\cellcolor{blue!10} 472$ & $\num{2.789e-03}$ & $\num{4.795e-02}$ & 388 & $\cellcolor{blue!10} \num{1.376e-03}$ & $\cellcolor{blue!10} \num{4.030e-02}$ & 0 & $\num{3.911e-03}$ & $\num{7.180e-02}$ \\ \hline
Kagiya Edo Tokyo & 3450 & 9855 & $\cellcolor{blue!10} 889$ & $\num{1.531e-03}$ & $\cellcolor{blue!10} \num{1.503e-02}$ & 320 & $\num{1.534e-03}$ & $\num{2.876e-02}$ & 0 & $\num{1.927e-03}$ & $\num{5.719e-02}$ \\ \hline
Lamborghini Countach & 26826 & 119160 & 3444 & $\num{4.780e-04}$ & $\num{3.353e-02}$ & $\cellcolor{blue!10} 3868$ & $\cellcolor{blue!10} \num{3.224e-04}$ & $\cellcolor{blue!10} \num{1.440e-02}$ & 0 & $\num{2.308e-03}$ & $\num{4.675e-02}$ \\ \hline
Lantern & 272 & 1328 & $\cellcolor{blue!10} 84$ & $\num{5.683e-03}$ & $\cellcolor{blue!10} \num{1.480e-02}$ & 68 & $\cellcolor{blue!10} \num{3.496e-03}$ & $\num{4.947e-02}$ & 0 & $\num{1.388e-02}$ & $\num{5.030e-02}$ \\ \hline
Lego Figure & 371 & 0 & 0 & $\num{3.840e-03}$ & $\cellcolor{blue!10} \num{1.952e-02}$ & 0 & $\num{2.986e-03}$ & $\num{2.064e-02}$ & 0 & $\cellcolor{blue!10} \num{2.264e-03}$ & $\num{2.999e-02}$ \\ \hline
Lekythos & 221 & 1075 & $\cellcolor{blue!10} 42$ & $\cellcolor{blue!10} \num{6.265e-03}$ & $\cellcolor{blue!10} \num{1.550e-02}$ & 8 & $\num{7.899e-03}$ & $\num{2.752e-02}$ & 0 & $\num{9.839e-03}$ & $\num{2.193e-02}$ \\ \hline
Lethe & 897 & 3986 & $\cellcolor{blue!10} 196$ & $\num{4.691e-03}$ & $\num{1.781e-01}$ & 173 & $\cellcolor{blue!10} \num{2.339e-03}$ & $\cellcolor{blue!10} \num{2.008e-02}$ & 0 & $\num{5.248e-03}$ & $\num{3.005e-02}$ \\ \hline
Marble Track & 7044 & 26461 & $\cellcolor{blue!10} 1002$ & $\num{1.824e-03}$ & $\num{2.042e-02}$ & 971 & $\cellcolor{blue!10} \num{6.080e-04}$ & $\cellcolor{blue!10} \num{1.487e-02}$ & 0 & $\num{1.110e-03}$ & $\num{2.737e-02}$ \\ \hline
Maze 45 & 7461 & 37191 & $\cellcolor{blue!10} 3320$ & $\num{2.295e-04}$ & $\num{3.931e-03}$ & 681 & $\num{2.794e-04}$ & $\num{1.071e-02}$ & 0 & $\cellcolor{blue!10} \num{1.056e-04}$ & $\cellcolor{blue!10} \num{3.551e-03}$ \\ \hline
Maze & 7465 & 37191 & $\cellcolor{blue!10} 3222$ & $\num{3.180e-04}$ & $\num{5.151e-03}$ & 801 & $\num{3.894e-04}$ & $\num{6.815e-03}$ & 0 & $\cellcolor{blue!10} \num{1.451e-04}$ & $\cellcolor{blue!10} \num{4.883e-03}$ \\ \hline
Mech & 1550 & 5850 & $\cellcolor{blue!10} 212$ & $\num{3.803e-03}$ & $\num{1.739e-02}$ & 210 & $\cellcolor{blue!10} \num{3.197e-03}$ & $\cellcolor{blue!10} \num{1.644e-02}$ & 0 & $\num{5.292e-03}$ & $\num{2.443e-02}$ \\ \hline
Melon Pallet & 9178 & 45016 & $\cellcolor{blue!10} 1609$ & $\num{9.265e-04}$ & $\cellcolor{blue!10} \num{2.800e-03}$ & 701 & $\num{8.442e-04}$ & $\num{9.450e-03}$ & 0 & $\cellcolor{blue!10} \num{6.854e-04}$ & $\num{5.820e-03}$ \\ \hline
Militech Robot & 5309 & 0 & 0 & $\num{2.221e-03}$ & $\num{9.526e-03}$ & 0 & $\cellcolor{blue!10} \num{1.880e-03}$ & $\num{9.546e-03}$ & 0 & $\num{2.223e-03}$ & $\num{4.477e-02}$ \\ \hline
One Angery Dragon Boi & 13606 & 67632 & $\cellcolor{blue!10} 3581$ & $\cellcolor{blue!10} \num{7.260e-04}$ & $\cellcolor{blue!10} \num{2.652e-03}$ & 1042 & $\num{1.159e-03}$ & $\num{9.430e-03}$ & 0 & $\num{8.773e-04}$ & $\num{6.936e-03}$ \\ \hline
Oni No Face Mask & 22072 & 0 & 0 & $\num{2.044e-04}$ & $\num{2.480e-03}$ & 0 & $\num{5.213e-04}$ & $\num{4.911e-03}$ & 0 & $\num{1.750e-04}$ & $\cellcolor{blue!10} \num{2.288e-03}$ \\ \hline
Organic Armor & 9460 & 47270 & $\cellcolor{blue!10} 3939$ & $\cellcolor{blue!10} \num{3.127e-04}$ & $\cellcolor{blue!10} \num{1.252e-03}$ & 459 & $\num{4.088e-04}$ & $\num{2.845e-03}$ & 0 & $\num{4.127e-04}$ & $\num{2.140e-03}$ \\ \hline
Pipe Wall & 312 & 1543 & $\cellcolor{blue!10} 59$ & $\num{9.273e-03}$ & $\num{1.297e-01}$ & 40 & $\cellcolor{blue!10} \num{4.413e-03}$ & $\cellcolor{blue!10} \num{7.112e-02}$ & 0 & $\num{5.385e-03}$ & $\num{1.278e-01}$ \\ \hline
Prometheus & 80992 & 379559 & $\cellcolor{blue!10} 27777$ & $\num{1.449e-04}$ & $\num{1.210e-02}$ & 8636 & $\num{1.733e-04}$ & $\cellcolor{blue!10} \num{1.999e-03}$ & 0 & $\num{2.211e-04}$ & $\num{2.275e-03}$ \\ \hline
Radial & 1012 & 0 & 0 & $\num{9.064e-03}$ & $\cellcolor{blue!10} \num{4.366e-02}$ & 0 & $\cellcolor{blue!10} \num{8.057e-03}$ & $\num{5.451e-02}$ & 0 & $\num{2.482e-02}$ & $\num{8.721e-02}$ \\ \hline
Roman Soldier & 28201 & 123893 & $\cellcolor{blue!10} 8416$ & $\cellcolor{blue!10} \num{3.586e-04}$ & $\cellcolor{blue!10} \num{2.455e-03}$ & 2372 & $\num{6.014e-04}$ & $\num{3.994e-03}$ & 0 & $\num{5.988e-04}$ & $\num{3.418e-03}$ \\ \hline
Rope Bridge & 1234 & 4444 & $\cellcolor{blue!10} 474$ & $\num{5.797e-03}$ & $\num{4.262e-02}$ & 293 & $\cellcolor{blue!10} \num{8.610e-04}$ & $\cellcolor{blue!10} \num{8.431e-03}$ & 0 & $\num{1.103e-03}$ & $\num{2.202e-02}$ \\ \hline
Shinto Watchtower & 26964 & 123508 & $\cellcolor{blue!10} 10531$ & $\num{1.235e-03}$ & $\cellcolor{blue!10} \num{3.226e-03}$ & 4154 & $\num{1.275e-03}$ & $\num{8.236e-03}$ & 0 & $\num{1.902e-03}$ & $\num{1.466e-02}$ \\ \hline
Shirakawago House & 3238 & 15867 & $\cellcolor{blue!10} 982$ & $\cellcolor{blue!10} \num{3.033e-03}$ & $\cellcolor{blue!10} \num{2.430e-02}$ & 636 & $\num{4.001e-03}$ & $\num{6.284e-02}$ & 0 & $\num{4.743e-03}$ & $\num{6.843e-02}$ \\ \hline
Skyward Ddf Poco Ship & 19831 & 12207 & $\cellcolor{blue!10} 1519$ & $\num{1.222e-04}$ & $\cellcolor{blue!10} \num{2.718e-03}$ & 653 & $\num{1.642e-04}$ & $\num{2.935e-03}$ & 0 & $\num{4.220e-04}$ & $\num{8.469e-03}$ \\ \hline
Snowflake Orb & 7802 & 0 & 0 & $\cellcolor{blue!10} \num{3.618e-03}$ & $\cellcolor{blue!10} \num{7.027e-03}$ & 0 & $\num{4.045e-03}$ & $\num{2.033e-02}$ & 0 & $\num{4.113e-03}$ & $\num{1.861e-02}$ \\ \hline
Space Fighter & 7437 & 34057 & 2472 & $\num{6.579e-04}$ & $\num{1.365e-02}$ & $\cellcolor{blue!10} 2487$ & $\cellcolor{blue!10} \num{4.859e-04}$ & $\num{2.472e-02}$ & 0 & $\num{1.161e-03}$ & $\cellcolor{blue!10} \num{1.242e-02}$ \\ \hline
Speeder Bike & 2486 & 11568 & $\cellcolor{blue!10} 721$ & $\num{1.744e-03}$ & $\num{1.695e-01}$ & 360 & $\num{1.699e-03}$ & $\num{2.706e-02}$ & 0 & $\num{2.383e-03}$ & $\cellcolor{blue!10} \num{1.816e-02}$ \\ \hline
Spiral Staircase & 1496 & 1400 & 215 & $\num{1.682e-03}$ & $\cellcolor{blue!10} \num{8.283e-03}$ & $\cellcolor{blue!10} 327$ & $\cellcolor{blue!10} \num{7.642e-04}$ & $\num{1.520e-02}$ & 0 & $\num{8.742e-04}$ & $\num{3.462e-02}$ \\ \hline
Spot Quadrangulated & 602 & 2928 & $\cellcolor{blue!10} 73$ & $\cellcolor{blue!10} \num{2.744e-03}$ & $\cellcolor{blue!10} \num{8.350e-03}$ & 40 & $\num{5.393e-03}$ & $\num{3.987e-02}$ & 0 & $\num{4.828e-03}$ & $\num{1.762e-02}$ \\ \hline
Steel Guard -7Mb & 1490 & 0 & 0 & $\num{1.238e-03}$ & $\cellcolor{blue!10} \num{3.627e-03}$ & 0 & $\cellcolor{blue!10} \num{5.901e-04}$ & $\num{5.150e-03}$ & 0 & $\num{7.474e-03}$ & $\num{6.117e-02}$ \\ \hline
Tank & 3148 & 15063 & $\cellcolor{blue!10} 916$ & $\cellcolor{blue!10} \num{1.715e-03}$ & $\cellcolor{blue!10} \num{9.359e-03}$ & 534 & $\num{1.951e-03}$ & $\num{1.232e-02}$ & 0 & $\num{2.382e-03}$ & $\num{1.841e-02}$ \\ \hline
Teacup & 1162 & 5728 & $\cellcolor{blue!10} 234$ & $\cellcolor{blue!10} \num{1.342e-03}$ & $\cellcolor{blue!10} \num{4.688e-03}$ & 91 & $\num{2.030e-03}$ & $\num{1.248e-02}$ & 0 & $\num{2.683e-03}$ & $\num{1.831e-02}$ \\ \hline
The Dark Wraith & 20134 & 80572 & $\cellcolor{blue!10} 4155$ & $\cellcolor{blue!10} \num{4.387e-04}$ & $\cellcolor{blue!10} \num{4.865e-03}$ & 1676 & $\num{6.707e-04}$ & $\num{5.159e-03}$ & 0 & $\num{7.361e-04}$ & $\num{8.545e-03}$ \\ \hline
Venice Mask & 29883 & 35201 & $\cellcolor{blue!10} 1608$ & $\cellcolor{blue!10} \num{5.726e-04}$ & $\num{4.691e-03}$ & 389 & $\num{9.735e-04}$ & $\num{7.530e-03}$ & 0 & $\num{7.487e-04}$ & $\cellcolor{blue!10} \num{4.059e-03}$ \\ \hline
Vera Posed & 8995 & 40269 & $\cellcolor{blue!10} 1858$ & $\num{1.421e-03}$ & $\cellcolor{blue!10} \num{8.336e-03}$ & 1005 & $\num{1.464e-03}$ & $\num{1.027e-02}$ & 0 & $\num{1.943e-03}$ & $\num{1.738e-02}$ \\ \hline
War Tank & 6578 & 27758 & $\cellcolor{blue!10} 1647$ & $\num{2.413e-03}$ & $\cellcolor{blue!10} \num{2.748e-02}$ & 915 & $\cellcolor{blue!10} \num{1.953e-03}$ & $\num{3.493e-02}$ & 0 & $\num{7.018e-03}$ & $\num{3.555e-02}$ \\ \hline
Yeahright & 75484 & 377344 & $\cellcolor{blue!10} 36959$ & $\num{1.327e-04}$ & $\cellcolor{blue!10} \num{6.005e-04}$ & 6600 & $\num{1.322e-04}$ & $\num{1.741e-03}$ & 0 & $\num{1.033e-04}$ & $\num{3.103e-03}$ \\ \hline
\hline
Average &  &  &  & $\num{1.811e-03}$ & $\num{1.970e-02}$ &  & $\num{2.066e-03}$ & $\num{2.135e-02}$ &  & $\num{3.160e-03}$ & $\num{2.366e-02}$ \\ \hline
\end{tabular}
\caption{\label{tab:table10} Complete results of our approach for static meshes with $10\%$ of the input triangles removed.}
\end{table*}

\begin{figure*}
    \setlength{\tabcolsep}{0pt}
    \centering
    \begin{tabular}{c c c}
        Input & QEM & Ours \\
        $\triangle$: 11938 & 5976 & 5976 \\
        \put(-0.02\linewidth,0.08\linewidth){\rotatebox{90}{Frame 5}}
        \includegraphics[width=0.33\linewidth]{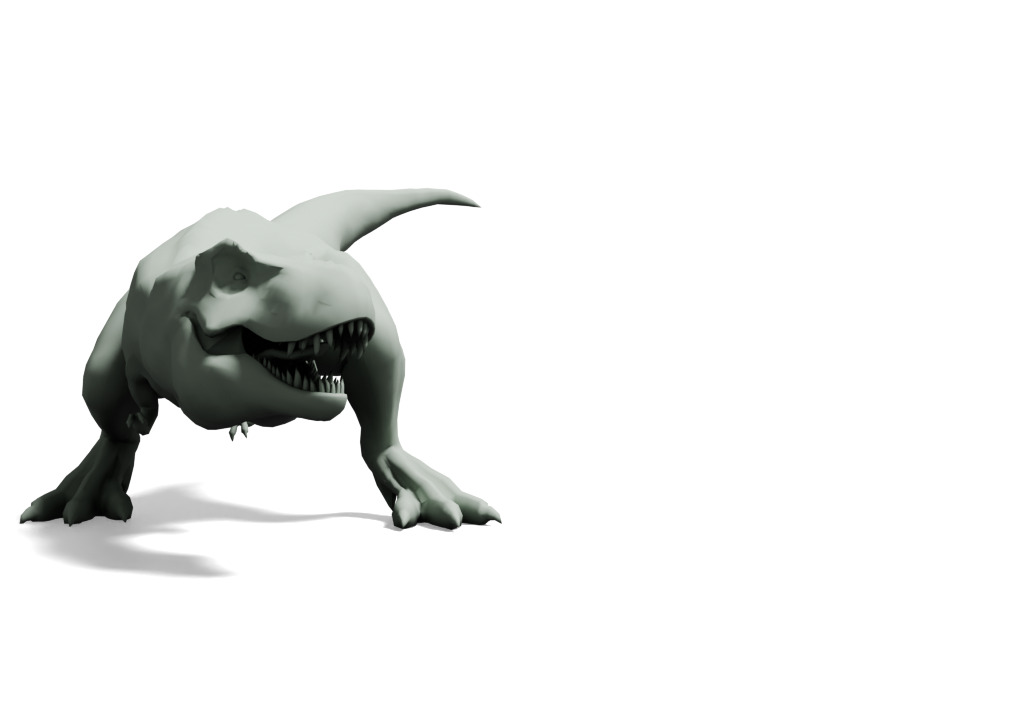} &
        \includegraphics[width=0.33\linewidth]{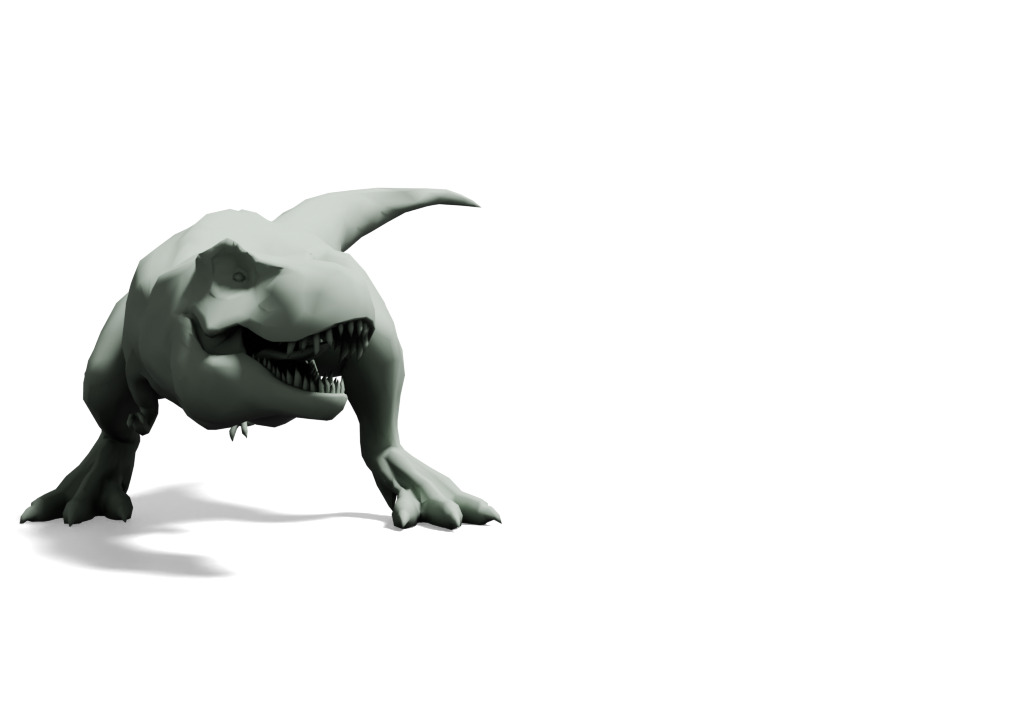} &
        \includegraphics[width=0.33\linewidth]{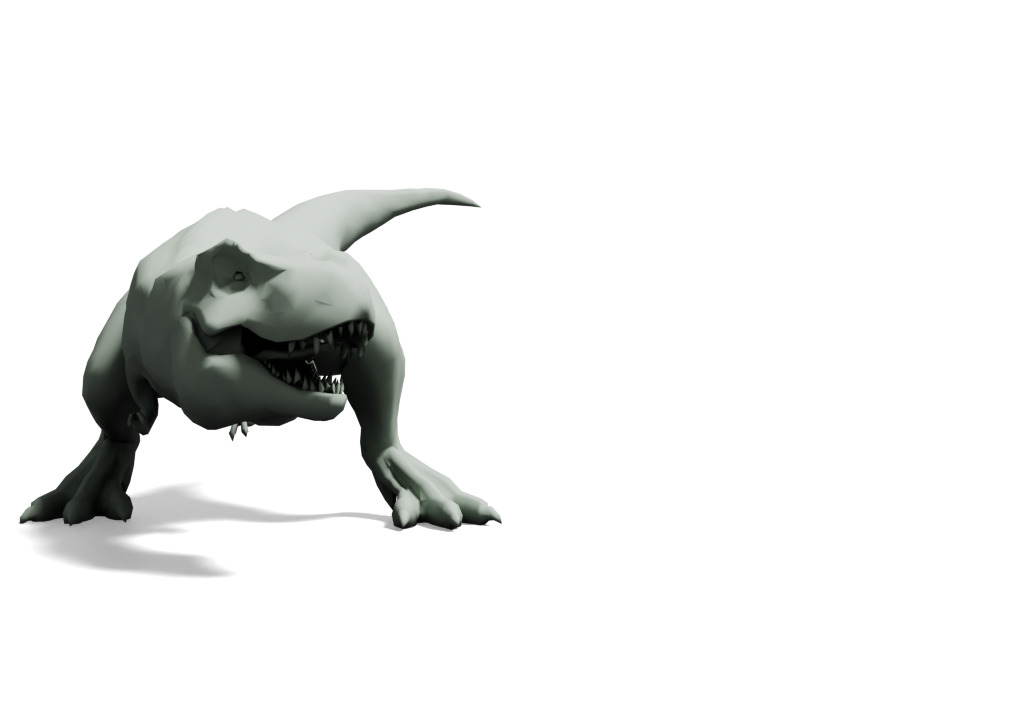} \\
        Chamfer Distance (At frame)$^\downarrow$: & \num{6.410e-4} & \num{1.99e-4} \\
        
        \put(-0.02\linewidth,0.08\linewidth){\rotatebox{90}{Frame 40}}
        \includegraphics[width=0.33\linewidth]{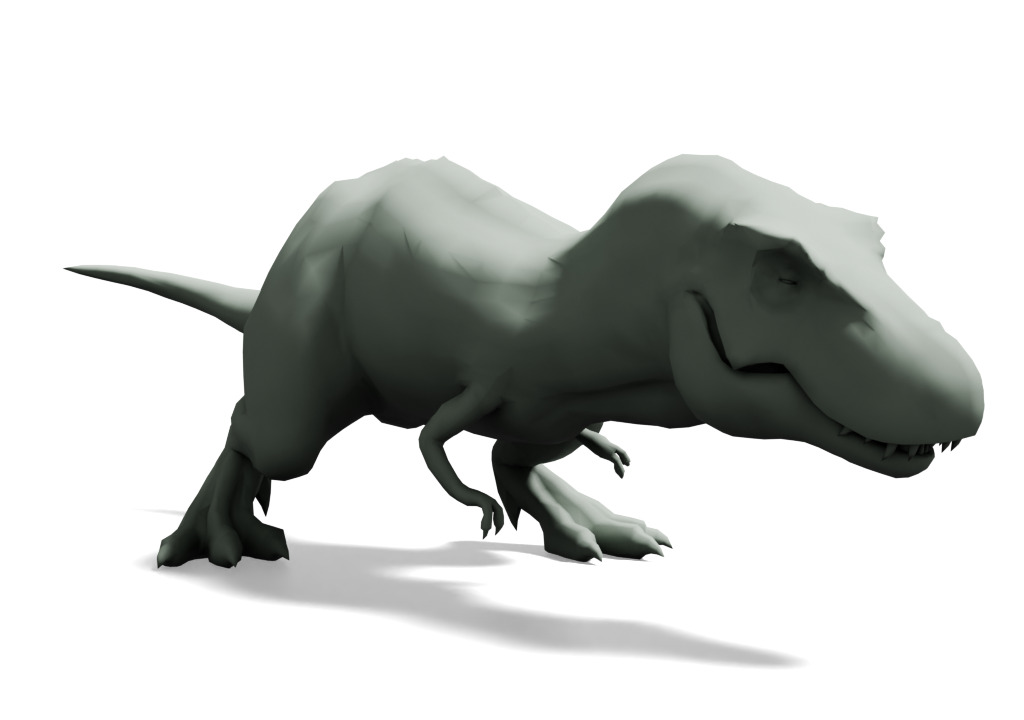} &
        \includegraphics[width=0.33\linewidth]{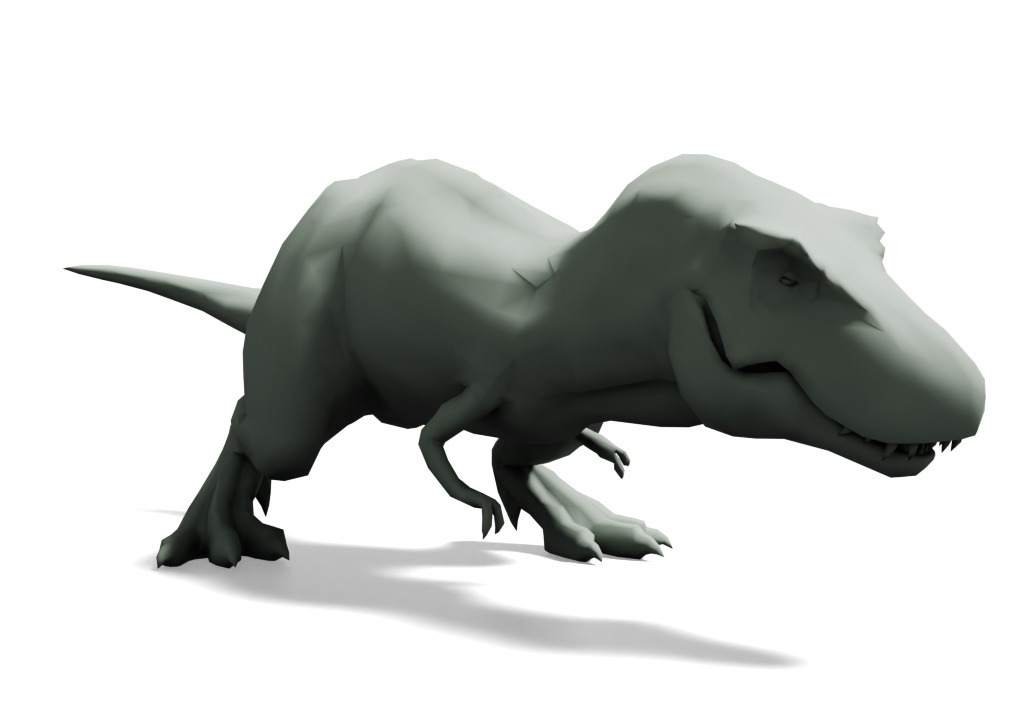} &
        \includegraphics[width=0.33\linewidth]{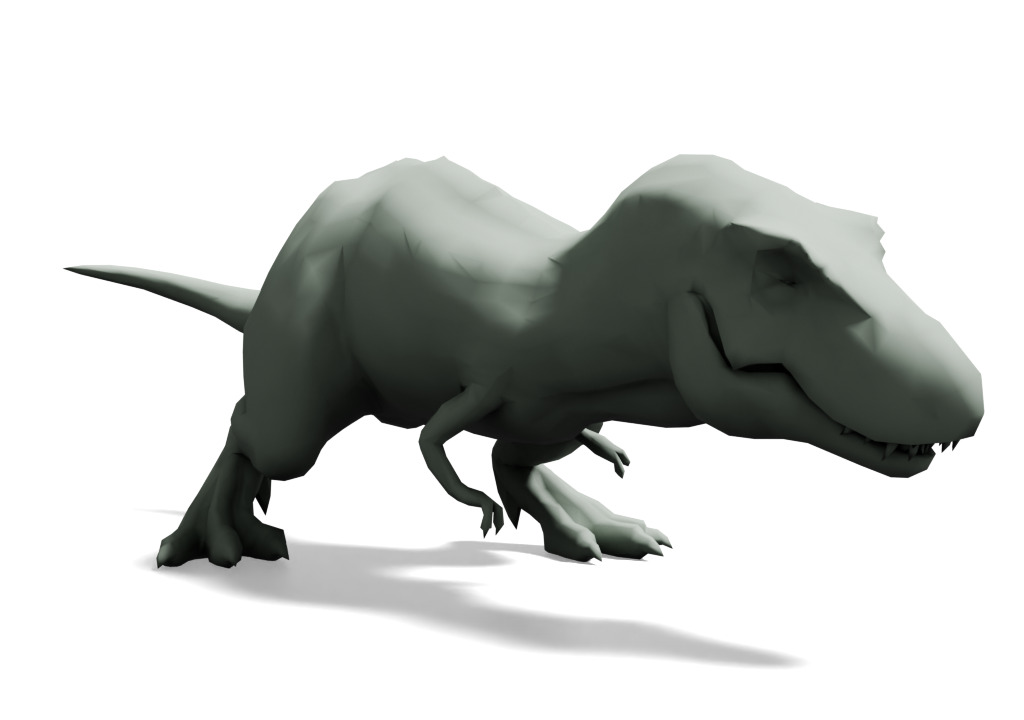} \\
        Chamfer Distance (At frame)$^\downarrow$: & \num{7.018e-4} & \num{2.099e-4} \\

        \hline
        $\triangle$: 19210 & 4811 & 4812 \\
        \put(-0.02\linewidth,0.12\linewidth){\rotatebox{90}{Frame 20}}
        \includegraphics[width=0.3\linewidth]{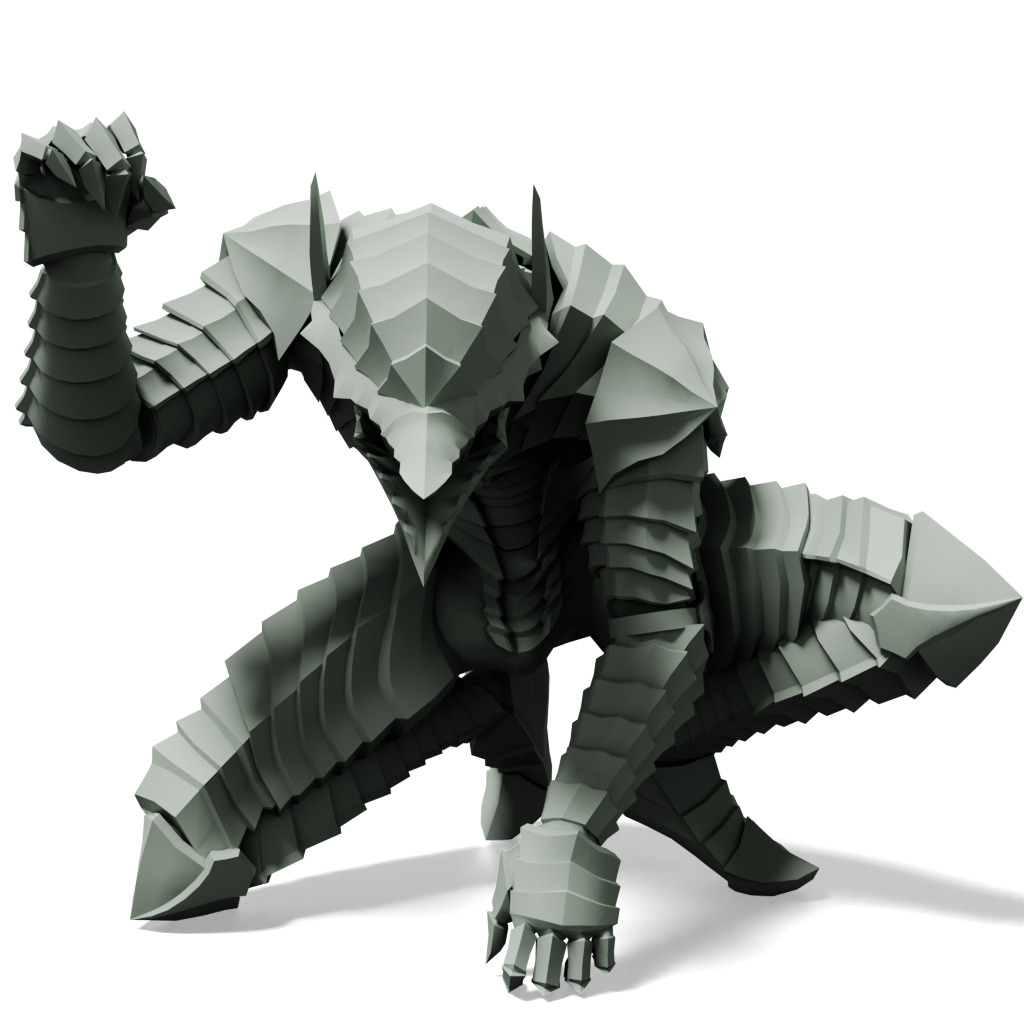} &
        \includegraphics[width=0.3\linewidth]{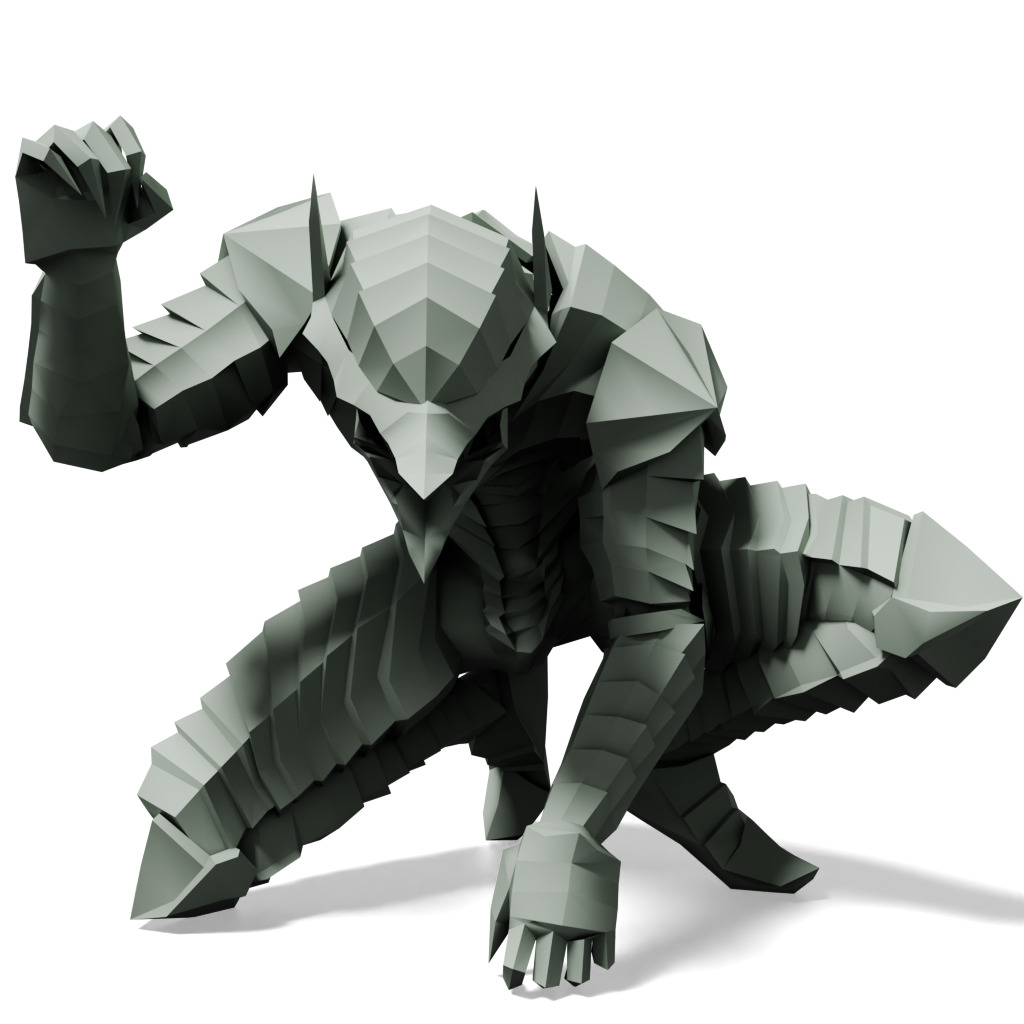} &
        \includegraphics[width=0.3\linewidth]{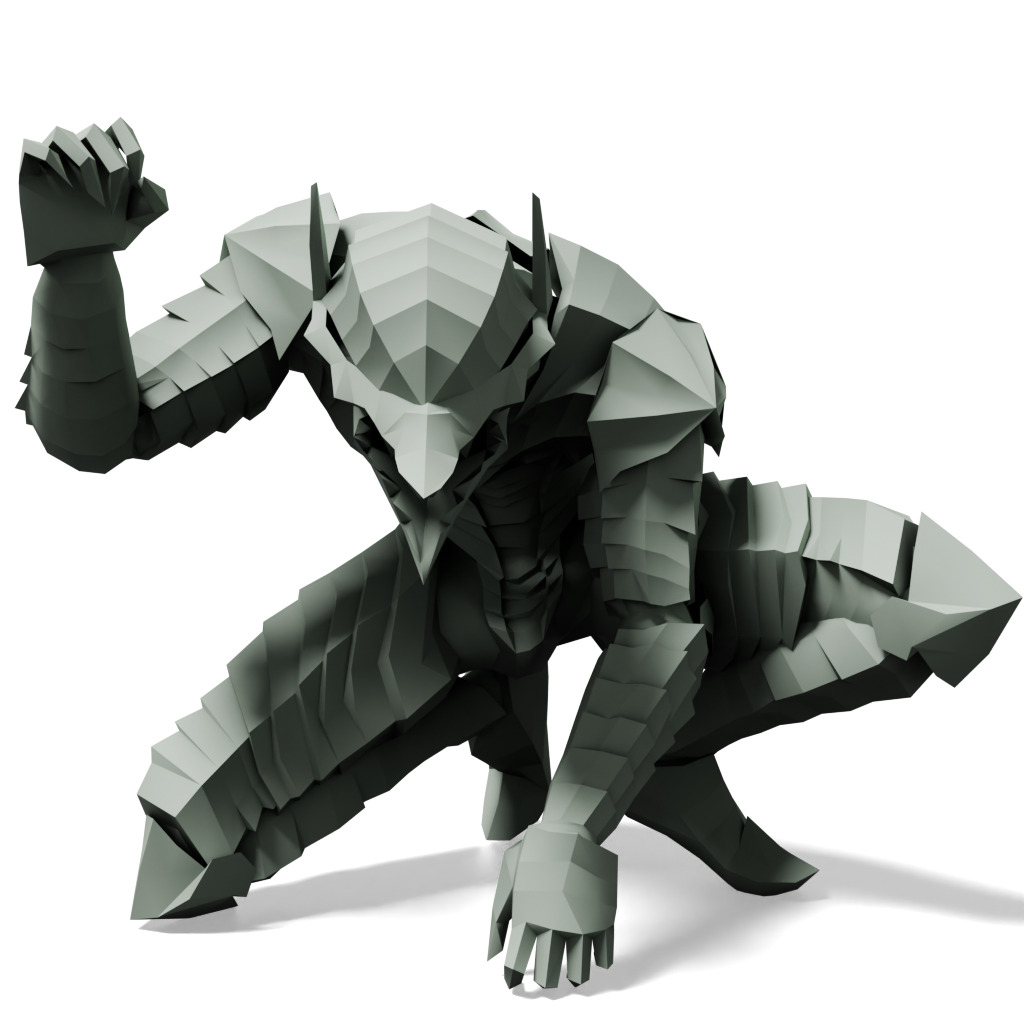} \\
        Chamfer Distance (At frame)$^\downarrow$: & \num{1.889e-3} & \num{1.776e-3} \\
        
        \put(-0.02\linewidth,0.12\linewidth){\rotatebox{90}{Frame 40}}
        \includegraphics[width=0.3\linewidth]{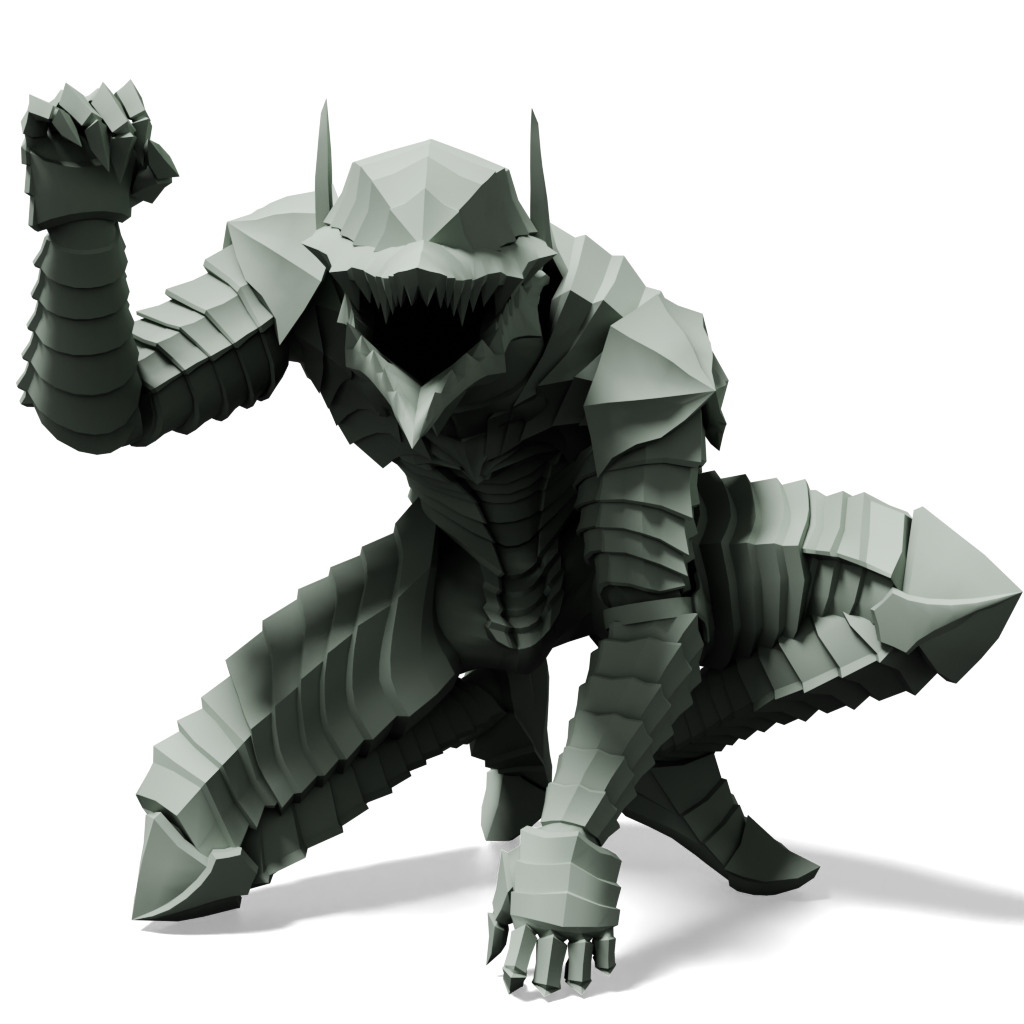} &
        \includegraphics[width=0.3\linewidth]{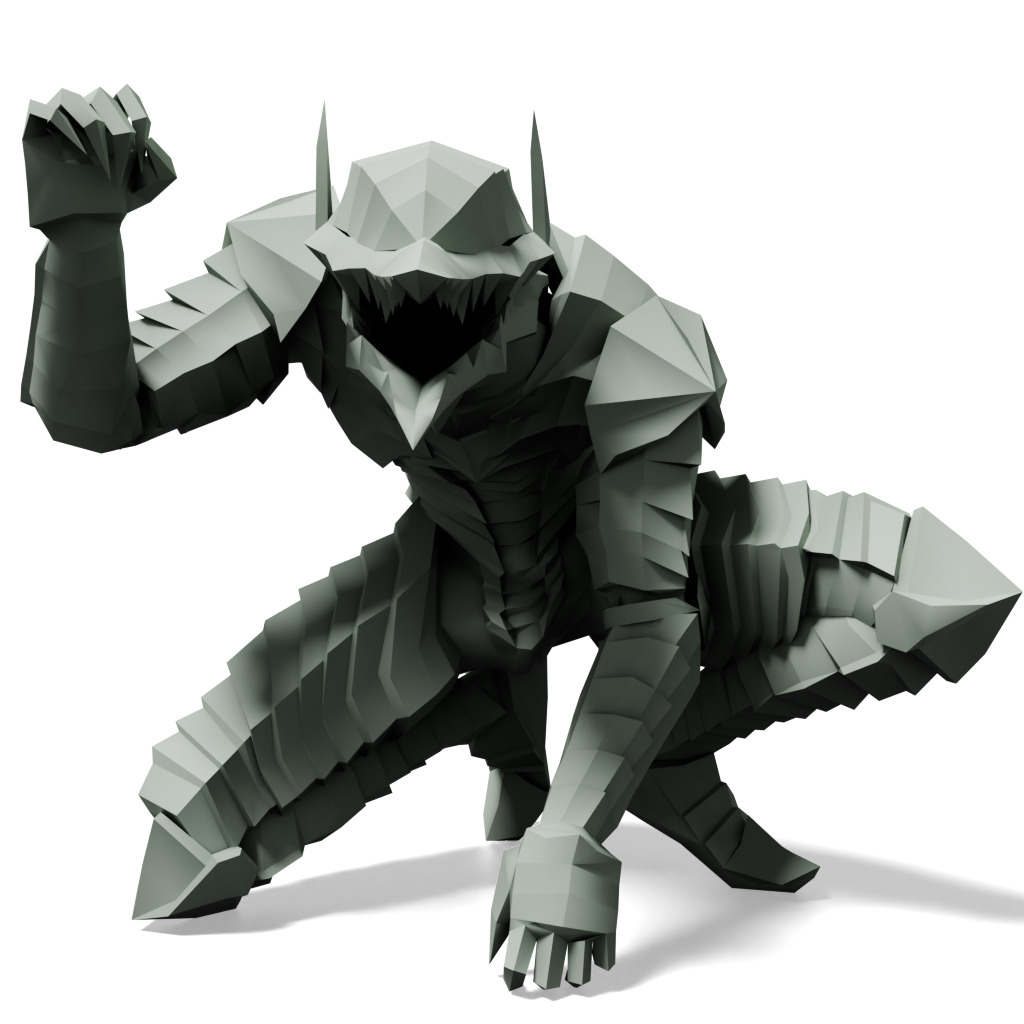} &
        \includegraphics[width=0.3\linewidth]{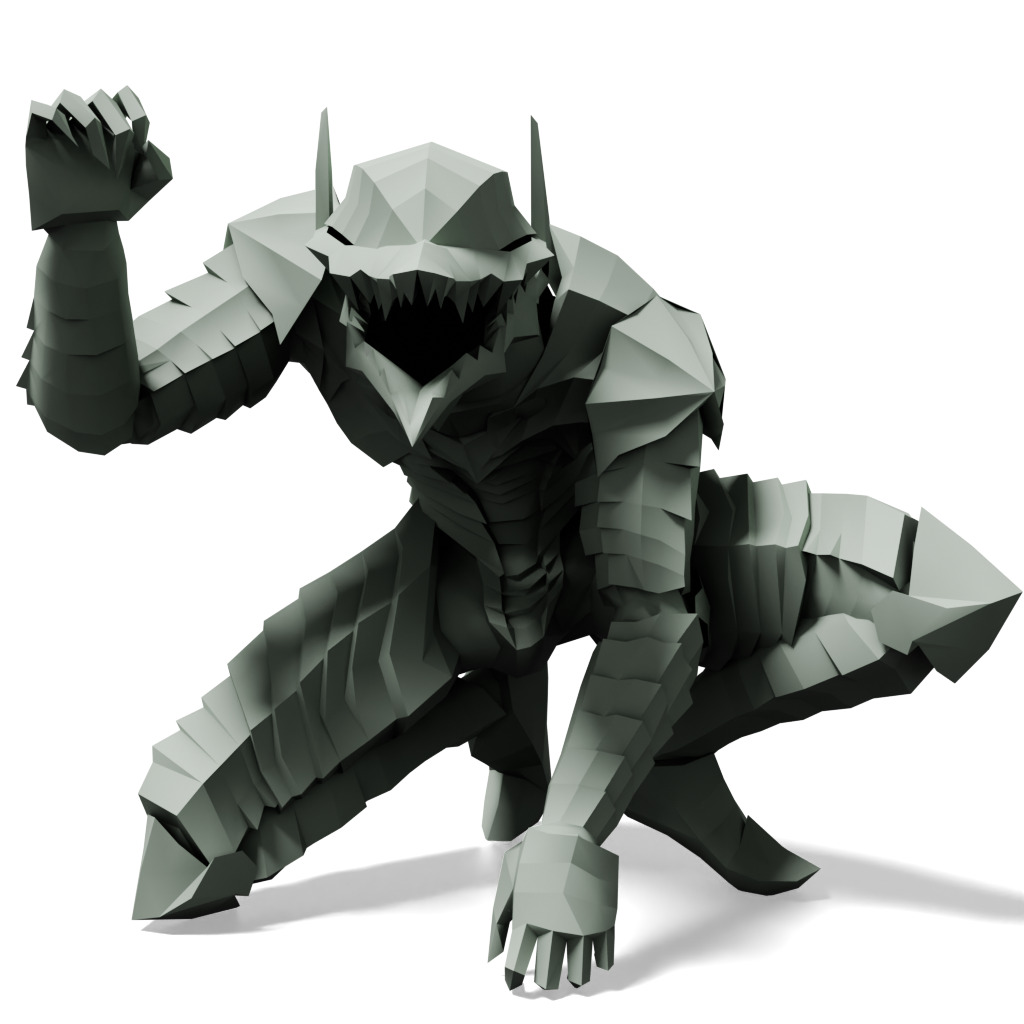} \\
        Chamfer Distance (At frame)$^\downarrow$: & \num{1.844e-3} & \num{1.730e-3} \\
    \end{tabular}
    \caption{Visual comparison of our approach against QEM, which implements ~\citep{articulated_mesh_simplification}, on two animated meshes. For each mesh, the camera is at a single position, but it appears different as the T-rex travels a large distance during animation.\label{fig:anim-mesh-visual}}

    \Description{Animated mesh of a t-rex. First row is in a crouching pose, second row is a biting pose. Third row is man in armor looking downwards, fourth row is armored man looking outwards.}
\end{figure*}

\end{document}